\documentclass[12pt]{article}  
\usepackage{cite}
\usepackage{epsfig}
\usepackage{graphicx}
\usepackage{amsmath}
\usepackage{amssymb}
\usepackage{ulem}
\usepackage{mdwlist} 
\usepackage{color}  
\usepackage[OT2, T1]{fontenc}
\usepackage[russian,english]{babel}

\usepackage{a41}
\usepackage{color}
\usepackage[rflt]{floatflt}
\usepackage{float}
\usepackage{slashed}

\newcommand{\HarmonicSumsP}{\texttt{HarmonicSums}}

\DeclareMathOperator{\arcsinh}{arcsinh}
\DeclareMathOperator{\arctanh}{arctanh}

\usepackage{array}

\usepackage[english]{babel}

\usepackage{url}

\usepackage{amsmath, amsthm, amssymb}
\newtheorem{thm}{Theorem}[section]

\newtheorem{definition}[thm]{Definition}

 \newcommand{\GeV}{\mathrm{GeV}}

 \newcommand{\MS}{\overline{\sf MS}}

 \newcommand{\Ahathat}{\hat{\hspace*{0mm}\hat{A}}}
 \newcommand{\Athathat}{\hat{\hspace*{0mm}\hat{\tilde{A}}}}
  \newcommand{\Atildehathat}{\hat{\hspace*{0mm}\hat{\tilde{A}}}}

\newcommand{\lnMa}[1]{L_1^{#1}}
\newcommand{\lnMb}[1]{L_2^{#1}}

\newcommand{\NN}{\nonumber}

\newcommand{\N}{\mathbb N}

\newcommand{\Li}{{\rm Li}}
\newcommand{\HA}{{\rm H}}

\newcommand{\Mvec}{{\rm\bf M}}

\newcommand{\ie}{i.e.,\ }

\newcommand{\ep}{\varepsilon}
\newcommand{\sig}{\sigma}
\newcommand{\dd}{\mathrm{d}}

\let\set\mathbb

\usepackage{rotating}

\usepackage{graphicx}

\newcounter{mmacnt}
\def\restartmma{\setcounter{mmacnt}{0}}
\restartmma \catcode`|=\active
\def|#1|{\mathrm{#1}}
\catcode`|=12
\newenvironment{mma}{
 \par\smallskip
 \catcode`|=\active
 \parskip=0pt\parindent=0pt 
 \small
 \def\In##1\\{%
\def\linebreak{\hfill\break\null\qquad}%
\refstepcounter{mmacnt}
\hangindent=2.5em\hangafter=0
\leavevmode
\llap{\tiny\sffamily n[\arabic{mmacnt}]:=\kern.5em}%
\mathversion{bold}\footnotesize$\displaystyle##1$\normalsize
\mathversion{normal}\par
 }%
 \def\Print##1\\{%
\def\linebreak{\hfill\break}%
\hangindent=2.5em\hangafter=0
\leavevmode ##1\par}%
 \def\Out##1\\{%
\def\linebreak{$\hfill\break\null\hfill$}%
\kern\abovedisplayskip\par
\hangindent=2.5em\hangafter=0
\leavevmode
\llap{\tiny\sffamily Out[\arabic{mmacnt}]=\kern.5em}
\footnotesize$\displaystyle##1$\normalsize\hfill\null\par
\kern\belowdisplayskip
 }%
 \def\Warning##1##2\\{%
\def\linebreak{\hfill\break}%
\hangindent=2.5em\hangafter=0
\leavevmode
{\scriptsize##1 : ##2}\par}%
}{%
 \par\smallskip
}

\usepackage{color}

\newenvironment{fshaded}{%
\MakeFramed {\FrameRestore}
}%
{\endMakeFramed}

\allowdisplaybreaks[4]

\begin{document}
\setlength{\baselineskip}{0.515cm}
\sloppy
\thispagestyle{empty}
\begin{flushleft}
DESY 18--016
\\
DO--TH 18/02\\
March 2018\\
\end{flushleft}

\mbox{}
\vspace*{\fill}
\begin{center}

{\LARGE\bf The Two-mass Contribution to the Three-Loop}

\vspace*{2mm}
{\LARGE\bf \boldmath Gluonic Operator Matrix Element $A_{gg,Q}^{(3)}$}

\vspace{3cm}
\large
J.~Ablinger$^a$, 
J.~Bl\"umlein$^b$, 
A.~De Freitas$^b$, 
A.~Goedicke$^c$, 

\vspace*{2mm}
C.~Schneider$^a$,   and  K.~Sch\"onwald$^b$ 

\vspace{1.cm}
\normalsize
{\it $^a$~Research Institute for Symbolic Computation (RISC),\\
  Johannes Kepler University, Altenbergerstra{\ss}e 69,
  A--4040, Linz, Austria}\\

\vspace*{3mm}
{\it  $^b$ Deutsches Elektronen--Synchrotron, DESY,}\\
{\it  Platanenallee 6, D-15738 Zeuthen, Germany}

\vspace*{3mm}
{\it  $^c$ Institut f\"ur Theoretische Teilchenphysik
Campus S\"ud,} \\ 
{\it Karlsruher Institut f\"ur Technologie (KIT) D-76128 Karlsruhe, Germany}
\\

\end{center}
\normalsize
\vspace{\fill}
\begin{abstract}
\noindent
We calculate the two-mass QCD contributions to the massive operator matrix element  $A_{gg,Q}$ 
at $\mathcal{O} (\alpha_s^3)$ in analytic form in Mellin $N$- and $z$-space, maintaining the 
complete dependence on the heavy quark mass ratio. These terms are important ingredients for the 
matching relations of the variable flavor number scheme in the presence of two heavy quark flavors, 
such as charm and bottom. In Mellin $N$-space the result is given in the form of nested harmonic, 
generalized harmonic, cyclotomic and binomial sums, with arguments depending on the mass ratio. 
The Mellin inversion of these quantities to $z$-space gives rise to generalized iterated integrals 
with square root valued letters in the alphabet, depending on the mass ratio as well. Numerical 
results are presented.
\end{abstract}

\vspace*{\fill}
\noindent
\numberwithin{equation}{section}
\newpage 
\section{Introduction}
\label{sec:1}

\vspace*{1mm}
\noindent
The massive operator matrix elements~(OMEs) $A_{ij}$ for partonic transitions rule the matching conditions
in the variable  flavor number scheme. They start to receive contributions from two different massive quarks
starting at two-loop order through one-particle reducible graphs and at three-loop order due to irreducible
contributions. Since the mass ratio between charm and bottom is not sufficiently small ($m_c^2/m_b^2 \sim 
0.1$\footnote{Here the on-shell values for the charm quark mass $m_c = 1.59$~GeV and bottom quark mass $m_b = 
4.78$~GeV are used  
\cite{Alekhin:2012vu,Agashe:2014kda}. Throughout this paper we will use the on-shell masses, as the 
calculation has been performed in this scheme. The transformation to the $\overline{\sf MS}$ scheme for 
the quark masses is given in Section~\ref{sec:63}.}), there is no single heavy quark dominance in the mass 
region of charm and bottom and one has to account for both mass effects at the same time, cf. also 
\cite{Ablinger:2017err}.

In the present paper we calculate these two-mass contributions to the OME $A_{gg,Q}$ up to 
$\mathcal{O}(\alpha_s^3)$. 
The $\mathcal{O}(T_F^2)$ single mass contributions have been computed in \cite{Ablinger:2014tla} and the 
$\mathcal{O}(T_F^2 N_F)$ terms in \cite{Blumlein:2012vq}. The complete single mass OME is given in 
\cite{AGG}. 
Previously, fixed moments of the OME for the Mellin variable $N=2,4,6$ as a series expansion up to 
$\mathcal{O} ( \eta^3 \ln^2(\eta) )$ with $\eta = m_c^2/m_b^2  < 1$, the ratio of the heavy quark 
masses 
squared, were calculated in Ref.~\cite{Ablinger:2017err,MOM} using the package 
\texttt{Q2E/EXP}~\cite{Harlander:1997zb,Seidensticker:1999bb}. There, also all contributing scalar 
prototype graphs were calculated in $z$-space, and the results were converted to $N$-space by 
an automated  Mellin transform. Furthermore, the analytic results for $A_{qq,Q}^{\text{NS},(3)}$, 
$A_{qq,Q}^{\text{NS-TR},(3)}$ and $A_{gq,Q}^{(3)}$ both in $N$- and $z$-space have been obtained. Recently, 
also the $z$-space result in the pure-singlet case $A_{Qq}^{\text{PS},3}$ has been calculated in 
Ref.~\cite{Ablinger:2017xml}. Additionally, various three-loop single mass OMEs have been completed, 
cf.~\cite{Ablinger:2014lka,Ablinger:2014uka,Ablinger:2014vwa,Blumlein:2016xcy,Behring:2016hpa,Ablinger:2010ty,
Blumlein:2012vq,Behring:2015zaa,Behring:2015roa}. The logarithmic contributions to all OMEs have been 
computed to 3-loop
 order in \cite{Behring:2014eya}. For $A_{Qg}^{(3)}$, all contributions which can be 
expressed via first order factorizable differential or difference equations in $N$- or $z$-space have been 
obtained in 
Refs.~\cite{Ablinger:2016swq,Ablinger:2015tua,PROC1}.

We perform the calculation of the two-mass contribution to $A_{gg,Q}^{(3)}$ first in $N$-space and then use an 
automated
inverse Mellin-transform to arrive at the $z$-space result. This is a change of paradigm from earlier work on 
scalar 
prototype diagrams \cite{Ablinger:2017err}, where first a representation in $z$-space was derived and the 
$N$-space solution was found by a Mellin-transform. We have checked the present method for the scalar diagrams 
and found agreement with the previous results. Since both the corresponding difference equations in Mellin 
$N$-space and the differential equations in $z$-space for the contributing diagrams are first order 
factorizable, one can choose to perform the calculation either in $N$- or $z$-space without further 
difficulties, cf.~\cite{Ablinger:2015tua}. There is even no need to refer to special bases.

The paper is organized as follows. In Section~\ref{sec:2} we recall essentials for the representation of the 
renormalized OME $A_{gg,Q}$ in the two-mass case, cf.~\cite{Ablinger:2017err}. In Section~\ref{sec:3} the 
fixed moments for the constant part of the two-mass contributions are given in complete form for 
later comparison\footnote{In Ref.~\cite{Ablinger:2017err}, moments were presented for the irreducible 
contributions only.}. We outline the general steps of the computation in Section~\ref{sec:4} and illustrate
them in detail for a particular diagram in Section~\ref{sec:5}. In Section~\ref{sec:6} the result of 
the two-mass contributions to $A_{gg,Q}$ are given both in $N$- and $z$-space. In the latter case, we use
in part single-valued integral representations in order to still allow for root-valued iterated integrals
(appearing as integrands) that have representations in terms of harmonic polylogarithms (HPLs) of more involved
arguments. These integrals can be performed in principle, but lead to voluminous functional 
expressions for parts of which new numerical implementations would have to be developed as two-variable
functions, which we tried to avoid. Numerical results are presented comparing the two-mass contribution to the
whole $T_F^2$ term for the heavy quark contributions. The conclusions are given in Section~\ref{sec:7}. 
Special integrals and functions
of the momentum fraction $z$ and the mass ratio squared $\eta$ appearing in intermediate and the final result 
of the calculation are listed in the Appendix, also for further use in other two-mass projects.
Here we have expressed the appearing iterated integrals in terms of harmonic polylogarithms at more 
complicated arguments, which is thoroughly possible in these cases. This allows particular fast 
numerical implementations. We also present the renormalized OME $\tilde{A}_{gg,Q}$ in $N$ and $z$-space.
\section{The renormalized two-mass OME $\boldsymbol{\tilde{A}_{gg,Q}}$}
\label{sec:2}

\vspace*{1mm}
\noindent
The complete renormalization and factorization procedure for all OMEs up to $\mathcal{O}(\alpha_s^3)$ in the 
case of a single massive quark has been presented in Ref.~\cite{MOM}, while that for the two-mass case has been 
derived 
in Ref.~\cite{Ablinger:2017err}. In order to perform a separated treatment for the two-mass effects, one splits 
the OMEs into parts stemming from graphs with only one massive flavor and one part containing the 
contributions from both via
\begin{eqnarray}
\hat{\hat{A}}_{ij} \left( \frac{m_1^2}{\mu^2}, \frac{m_2^2}{\mu^2} \right) 
&=& \hat{\hat{A}}_{ij} \left( \frac{m_1^2}{\mu^2} \right) 
+ \hat{\hat{A}}_{ij} \left( \frac{m_2^2}{\mu^2} \right) 
+ \hat{\hat{\tilde{A}}}_{ij} \left( \frac{m_1^2}{\mu^2}, \frac{m_2^2}{\mu^2} \right)~.
\end{eqnarray}
Here we briefly summarize the main results for $A_{gg,Q}$.
In the following we use the mass ratio $\eta$ given by
\begin{eqnarray}
    \eta = \frac{m_2^2}{m_1^2} < 1. 
\end{eqnarray}
We abbreviate a series of logarithms by
\begin{eqnarray}
    L_1 = \ln \left( \frac{m_1^2}{\mu^2} \right) , && 
    L_2 = \ln \left( \frac{m_2^2}{\mu^2} \right),~~~~
    L_\eta = \ln \left( \eta \right).
\end{eqnarray}
Here $\mu^2$ denotes the renormalization and factorization scales, which we set equal. The physical masses 
are denoted by $m_i$, while the unrenormalized quantities are denoted by $\hat{m}_i$.\footnote{Note that we use 
the symbols $\eta$, $L_1$ and $L_2$ synonymously for renormalized and unrenormalized masses, since no 
confusion is expected.}

The operator matrix element $A_{gg,Q}$ receives two-mass contributions beginning at two-loop order.
At $\mathcal{O}(\alpha_s^2)$ these contributions stem from one-particle reducible contributions only,
while from 3-loop order onwards, genuine two-mass effects appear. The unrenormalized OME is given by
\begin{eqnarray}
\Atildehathat_{gg,Q} = \sum_{k = 2}^\infty a_s^k \Atildehathat_{gg,Q}^{(k)},
\end{eqnarray}
with $a_s = \alpha_s/(4\pi) = g_s^2/(4\pi)^2$ and $g_s$ the strong coupling constant.
The generic pole structure of $A_{gg,Q}$ up to three-loop order is given by \cite{Ablinger:2017err}
\begin{eqnarray}
\Atildehathat_{gg,Q}^{(2)} \left( \frac{\hat{m}_1^2}{\mu^2}, \frac{\hat{m}_2^2}{\mu^2} \right) &=& \left( \frac{\hat{m}_1 \hat{m}_2}{\mu} \right) ^\ep \frac{8 \beta_{0,Q}^2}{\ep^2} \text{exp} \left( 2 \sum\limits_{i=2}^{\infty} \frac{\zeta_i}{i} \left( \frac{\ep}{2} \right)^i \right) , \\
\Atildehathat_{gg,Q}^{(3)} \left( \frac{\hat{m}_1^2}{\mu^2}, \frac{\hat{m}_2^2}{\mu^2} \right) &=&
\frac{1}{\ep^3} \Biggl[
-\frac{5}{3} \hat{\gamma}_{qg}^{(0)} \beta_{0,Q} \gamma_{gq}^{(0)}
-\frac{56}{3} \beta_{0} \beta_{0,Q}^2
-\frac{28}{3} \beta_{0,Q}^2 \gamma_{gg}^{(0)}
-48 \beta_{0,Q}^3
\Biggr] 
\NN\\&&
+\frac{1}{\ep^2} \Biggl[
\Biggl\{
-7 \beta_{0,Q}^2 \gamma_{gg}^{(0)}
-14 \beta_{0} \beta_{0,Q}^2
-\frac{5}{4} \hat{\gamma}_{qg}^{(0)} \beta_{0,Q} \gamma_{gq}^{(0)}
-36 \beta_{0,Q}^3
\Biggr\} \left(L_1+L_2\right)
\NN\\&&
+\frac{1}{12} \hat{\gamma}_{qg}^{(0)} \hat{\gamma}_{gq}^{(1)}
-\frac{7}{3} \beta_{0,Q} \hat{\gamma}_{gg}^{(1)}
+\frac{4}{3} \beta_{1,Q} \beta_{0,Q}
-20 \delta m_1^{(-1)} \beta_{0,Q}^2
\Biggr]
\NN\\&&
+\frac{1}{\ep} \Biggl[
\Biggl\{
\frac{1}{16} \hat{\gamma}_{qg}^{(0)} \hat{\gamma}_{gq}^{(1)}
-15 \delta m_1^{(-1)} \beta_{0,Q}^2
-\frac{7}{4} \beta_{0,Q} \hat{\gamma}_{gg}^{(1)}
+\beta_{1,Q} \beta_{0,Q}
\Biggr\} \left(L_1+L_2\right)
\NN\\&&
+\Biggl\{
-15 \beta_{0,Q}^3
-\frac{11}{16} \hat{\gamma}_{qg}^{(0)} \beta_{0,Q} \gamma_{gq}^{(0)}
-\frac{13}{2} \beta_{0} \beta_{0,Q}^2
-\frac{13}{4} \beta_{0,Q}^2 \gamma_{gg}^{(0)}
\Biggr\}
\left(L_1^2+L_2^2\right)
\NN\\&&
+\Biggl\{
-4 \beta_{0,Q}^2 \gamma_{gg}^{(0)}
-24 \beta_{0,Q}^3
-8 \beta_{0} \beta_{0,Q}^2
-\frac{1}{2}\hat{\gamma}_{qg}^{(0)} \beta_{0,Q} \gamma_{gq}^{(0)}
\Biggr\} L_1 L_2
\NN\\&&
-\frac{1}{2} \beta_{0,Q}^2 \zeta_2 \gamma_{gg}^{(0)}
+\frac{2}{3} \gamma_{gg}^{(2),N_F^2}
-12 \beta_{0,Q} a_{gg,Q}^{(2)}
-18 \beta_{0,Q}^3 \zeta_2
+\frac{1}{8} \beta_{0,Q} \zeta_2 \gamma_{gq}^{(0)} \hat{\gamma}_{qg}^{(0)}
\NN\\&&
-\beta_{0} \beta_{0,Q}^2 \zeta_2
-16 \delta m_1^{(0)} \beta_{0,Q}^2
+4 \beta_{0,Q} \tilde{\delta} m_2^{(-1)}
\Biggr]
+\tilde{a}_{gg,Q}^{(3)}\left(m_1^2,m_2^2,\mu^2\right)~.
\end{eqnarray}
Here we used the notation\footnote{In Eqs.~(3.137, 3.138) of Ref.~\cite{Ablinger:2017err} unfortunately 
only the shift $N_F + 1 \rightarrow N_F$ has been used, which we correct here.} 
\begin{eqnarray}
    \hat{\gamma}_{ij} &=& \gamma_{ij} (N_F+2) - \gamma_{ij} (N_F),
\end{eqnarray}
$\gamma_{ij}^{(l),N_F^2}$ denotes the coefficient of the term proportional to $N_F^2$.\footnote{In Eqs.~(3.137) of Ref.~\cite{Ablinger:2017err} the notation 
$\hat{\tilde{\gamma}}_{ij}$ was used, which does not reproduce the $N_F^2$ term if the anomalous dimension starts 
at $\mathcal{O} \left( N_F^0 \right) $.} 
$\gamma_{ij}^{(l)}$ are anomalous 
dimensions at $(l+1)$-loops.
The quantities $a_{gg,Q}^{(2)}$ and $\bar{a}_{gg,Q}^{(2)}$ denote the $\mathcal{O}(\ep^0)$ and $\mathcal{O}(\ep)$ 
terms of the two-loop OME $\Ahathat_{gg,Q}^{(2)}$, cf. Refs.~\cite{Buza:1995ie,Buza:1996wv,Bierenbaum:2007qe,Bierenbaum:2008yu,Bierenbaum:2009zt}, and
\begin{eqnarray}
\beta_{0,Q} &=& - \frac{4}{3} T_F, \\
\delta m_1 &=&  C_F \left[ \frac{6}{\ep} - 4 + \left( 4 + \frac{3}{4} \zeta_2 \right) + \mathcal{O}(\ep^2) \right] \\
    &\equiv& \frac{\delta m_1^{(-1)}}{\ep} + \delta m_1^{(0)} + \ep \delta m_1^{(1)} + \mathcal{O}(\ep^2),\\
\tilde{\delta}{m_2}^{i} &=& C_F T_F \Biggl\{
  \frac{8}{\ep^2}
 -\frac{14}{\ep}
 +8 r_i^4 \HA_{0}^2(r_i)
 -8 (r_i+1)^2 \left(r_i^2-r_i+1\right) \HA_{-1,0}(r_i)
 \NN\\&&
 +8 (r_i-1)^2 \left(r_i^2+r_i+1\right) \HA_{1,0}(r_i)
 +8 r_i^2 \HA_0(r_i)
 +\frac{3}{2} \left(8 r_i^2+15\right)
 \NN\\&&
+2 \Bigl[4 r_i^4-12 r_i^3-12 r_i+5\Bigr] \zeta_2 \Biggr\} + \mathcal{O}(\ep)
 \\
               &\equiv&  \frac{\tilde{\delta}{m_2}^{(-2)}}{\ep^2}
                        +\frac{\tilde{\delta}{m_2}^{(-1)}}{\ep}
                        +\tilde{\delta}{m_2}^{i,(0)} + \mathcal{O}(\ep)~, \label{delm2mixexp}
\end{eqnarray}
Ref.~\cite{Gray:1990yh}, with
\begin{eqnarray}
\label{eq:r12}
r_1 = \sqrt{\eta} &\text{and}& r_2 = \frac{1}{\sqrt{\eta}}~.
\end{eqnarray}
We work in $D = 4 + \varepsilon$ space-time dimensions.
Furthermore, $\zeta_i = \sum\limits_{j=1}^{\infty} \frac{1}{j^i}, i \in \mathbb{N}, i \geq 2$ denotes the Riemann 
$\zeta$-function, 
$N_F$ denotes the number of massless flavors and $C_A$, $C_F$ and $T_F$ are the color factors which for a general $SU(N)$ take the values $T_F=\frac{1}{2}$, $C_A=N$ and $C_F = \frac{N^2-1}{2N}$.
The $\HA_i(x)$ are simple harmonic polylogarithms \cite{Remiddi:1999ew}.
\begin{eqnarray}
\HA_{b,\vec{a}}(x) &=& \int_0^x dy f_b(y) \HA_{\vec{a}}(y),~~\HA_\emptyset = 1,~~~ f_{b} \in 
\Bigl\{\frac{1}{x},\frac{1}{1-x},\frac{1}{1+x}\Bigr\}.
\end{eqnarray}
In the following the relation
\begin{eqnarray}
\tilde{\delta} m_2^{(-2)} &=& - \delta m_1^{(-1)} \beta_{0,Q} .
\end{eqnarray}
is used to shorten the expressions.

In the $\MS$-scheme, renormalizing the heavy masses on-shell, the renormalized expressions are given by
\begin{eqnarray}
\label{eq:renOME2}
\tilde{A}_{gg,Q}^{(2), \MS}
&=& 2 \beta_{0,Q}^2 L_1 L_2, \\
\label{eq:renOME3}
\tilde{A}_{gg,Q}^{(3), \MS}
&=&
\Bigg\{
\frac{25}{24} \beta_{0,Q}^2 \gamma_{gg}^{(0)}
+\frac{25}{12} \beta_{0} \beta_{0,Q}^2
+\frac{9}{2} \beta_{0,Q}^3
+\frac{23}{96} \hat{\gamma}_{qg}^{(0)} \beta_{0,Q} \gamma_{gq}^{(0)}
\Biggr\} \left(L_1^3+L_2^3\right)
+\Biggl\{
\frac{1}{8} \hat{\gamma}_{qg}^{(0)} \beta_{0,Q} \gamma_{gq}^{(0)}
\NN\\&&
+\beta_{0,Q}^2 \gamma_{gg}^{(0)}
+2 \beta_{0} \beta_{0,Q}^2
+6 \beta_{0,Q}^3
\Biggr\} \left(L_1^2 L_2+L_2^2 L_1\right)
+\Biggl\{
-\frac{1}{4} \beta_{1,Q} \beta_{0,Q}
+\frac{13}{16} \beta_{0,Q} \hat{\gamma}_{gg}^{(1)}
\NN\\&&
+\frac{29}{4} \delta m_1^{(-1)} \beta_{0,Q}^2
-\frac{1}{64} \hat{\gamma}_{qg}^{(0)} \hat{\gamma}_{gq}^{(1)}
\Biggr\} \left(L_1^2+L_2^2\right)
+8 L_2 L_1 \delta m_1^{(-1)} \beta_{0,Q}^2
+\Biggl\{
 \frac{9}{4} \beta_{0} \beta_{0,Q}^2 \zeta_2
\NN\\&&
+\frac{27}{2} \beta_{0,Q}^3 \zeta_2
-3 \beta_{0,Q} \tilde{\delta}m_2^{(-1)}
+\frac{9}{8} \zeta_2 \beta_{0,Q}^2 \gamma_{gg}^{(0)}
+12 \delta m_1^{(0)} \beta_{0,Q}^2
+\frac{3}{32} \beta_{0,Q} \zeta_2 \gamma_{gq}^{(0)} \hat{\gamma}_{qg}^{(0)}
\NN\\&&
+6 \beta_{0,Q} a_{gg,Q}^{(2)}
\Biggr\} \left(L_2+L_1\right)
-\frac{1}{32} \hat{\gamma}_{qg}^{(0)} \zeta_2 \hat{\gamma}_{gq}^{(1)}
+\frac{1}{8} \beta_{0,Q} \zeta_2 \hat{\gamma}_{gg}^{(1)}
+\frac{1}{3} \beta_{0} \beta_{0,Q}^2 \zeta_3
+12 \beta_{0,Q} \overline{a}_{gg,Q}^{(2)}
\NN\\&&
+6 \beta_{0,Q}^3 \zeta_3+16 \delta m_1^{(1)} \beta_{0,Q}^2
+\frac{1}{6} \beta_{0,Q}^2 \zeta_3 \gamma_{gg}^{(0)}
-2 \beta_{0,Q} \left(
\tilde{\delta} {m_2}^{1,(0)}
+\tilde{\delta} {m_2}^{2,(0)}
\right)
\NN\\&&
+\frac{9}{2} \delta m_1^{(-1)} \beta_{0,Q}^2 \zeta_2
-\frac{1}{24} \zeta_3 \beta_{0,Q} \gamma_{gq}^{(0)} \hat{\gamma}_{qg}^{(0)}
-\frac{1}{2} \zeta_2 \beta_{0,Q} \beta_{1,Q}
+\tilde{a}_{gg,Q}^{(3)}\left(m_1^2,m_2^2,\mu^2\right)~. \label{Agg3QMSren}
\end{eqnarray}

These expressions already include contributions from one-particle reducible graphs.
They can be written by lower order terms of $A_{gg,Q}$ and the gluon vacuum polarization, defined via
\begin{eqnarray}
\hat{\Pi}_{\mu\nu}^{ab}(p^2,\hat{m}_1^2,\hat{m}_2^2,\mu^2,\hat{a}_s) &=& i\delta^{ab}
                            \left[-g_{\mu\nu}p^2 +p_\mu p_\nu\right] 
                            \hat{\Pi}(p^2,\hat{m}_1^2,\hat{m}_2^2,\mu^2,\hat{a}_s)~,  \\
\hat{\Pi}(p^2,\hat{m}_1^2,\hat{m}_2^2,\mu^2,\hat{a}_s) &=&
                            \sum_{k=1}^{\infty}\hat{a}_s^k
                            \hat{\Pi}^{(k)}(p^2,\hat{m}_1^2,\hat{m}_2^2,\mu^2)
                            ~.
\end{eqnarray}
We split the gluon self-energy into parts depending only on one mass and one part stemming from graphs  
containing both masses, the same way as we did in the case of the massive OMEs
\begin{eqnarray}
\hat{\Pi}^{(k)}\left(p^2,\hat{m}_1^2,\hat{m}_2^2,\mu^2\right)&=&
\hat{\Pi}^{(k)}\Bigl(p^2,\frac{\hat{m}_1^2}{\mu^2}\Bigr)+
\hat{\Pi}^{(k)}\Bigl(p^2,\frac{\hat{m}_2^2}{\mu^2}\Bigr)
+\hat{\tilde{\Pi}}^{(k)}\left(p^2,\hat{m}_1^2,\hat{m}_2^2,\mu^2\right)~.
\end{eqnarray}
Up to two-loop order the gluon self-energy does not obtain contributions from graphs with two masses
\begin{eqnarray}
\hat{\tilde{\Pi}}^{(k)}(p^2,\hat{m}_1^2,\hat{m}_2^2,\mu^2)&=&0~~\text{for}~~k\in\{1,2\}~.
\end{eqnarray}

The single-mass on-shell vacuum polarization of the gluon is given by\footnote{$\Pi_2$ and $\Pi_3$ need a minus 
sign, w.r.t. 
terms given in Ref.~\cite{Ablinger:2017err}, to fit our definition of the self-energy.} 
\begin{eqnarray}
  \label{eqPI1}
   \hat{\Pi}^{(1)}\Bigl(0,\frac{\hat{m}^2}{\mu^2}\Bigr)&=&
            T_F\Bigl(\frac{\hat{m}^2}{\mu^2}\Bigr)^{\ep/2}
                        \left[
             -\frac{8}{3\ep}
              \exp \Bigl(\sum_{i=2}^{\infty}\frac{\zeta_i}{i}
                       \Bigl(\frac{\ep}{2}\Bigr)^{i}\Bigr)
             \right]~,
               ~\label{GluSelf1} 
               \\
   \hat{\Pi}^{(2)}\Bigl(0,\frac{\hat{m}^2}{\mu^2}\Bigr)&=&
      T_F\Bigl(\frac{\hat{m}^2}{\mu^2}\Bigr)^{\ep}\Biggl\{
      -\frac{4}{\ep^2} C_A + \frac{1}{\ep} \left(5 C_A-12 C_F\right) 
      + C_A \Bigl(\frac{13}{12} -\zeta_2\Bigr)
      - \frac{13}{3} C_F   
\NN\\ &&
      + \ep \left[C_A \Bigl(\frac{169}{144} + \frac{5}{4} \zeta_2 - 
      \frac{\zeta_3}{3} \Bigr) 
     - C_F \Bigl(\frac{35}{12}+3 \zeta_2\Bigr) 
     \right]\Biggr\}
       + 
            O(\ep^2)~,  \label{GluSelf2}
\\
             \hat{\Pi}^{(3)}\Bigl(0,\frac{\hat{m}^2}{\mu^2}\Bigr)&=&
       T_F\Bigl(\frac{\hat{m}^2}{\mu^2}\Bigr)^{3\ep/2}\Biggl\{
                        \frac{1}{\ep^3}\left[
                                 -\frac{32}{9}T_F C_A \left(2 N_F+1\right)
                                 + \frac{164}{9} C_A^2      
                                       \right]
\NN\\ &&
                       +\frac{1}{\ep^2}\left[
                               \frac{80}{27} (C_A-6 C_F) N_FT_F
                              +\frac{8}{27} (35 C_A-48 C_F) T_F
                              -\frac{781}{27} C_A^2 \right.                                                 
\NN\\ &&
                       \left.       +\frac{712}{9}C_AC_F
                                       \right]
                       +\frac{1}{\ep}\biggl[ 
                                \frac{4}{27}\big(
                                                       C_A(-101-18\zeta_2)
                                                      -62C_F
                                                \big)N_FT_F 
\NN\\ &&
                              -\frac{2}{27}   \big(
                                                       C_A(37+18 \zeta_2)
                                                       +80 C_F
                                                \big) T_F
                              +C_A^2            \Bigl(
                                                  -12\zeta_3
                                                  +\frac{41}{6}\zeta_2
                                                  +\frac{3181}{108}
                                                \Bigr)
\NN\\ &&
                              +C_A C_F           \Bigl(
                                                   16\zeta_3
                                                  -\frac{1570}{27}
                                                \Bigr)
                              +\frac{272}{3}C_F^2
                                       \biggr]
\NN\\ &&
                       +N_FT_F    \biggl[
                                       C_A\Bigl(
                                             \frac{56}{9}\zeta_3
                                            +\frac{10}{9}\zeta_2
                                            -\frac{3203}{243}
                                          \Bigr)
                                      -C_F\Bigl(
                                            \frac{20}{3}\zeta_2
                                            +\frac{1942}{81}
                                          \Bigr)
                                       \biggr]
\NN\\ &&
                       +T_F      \biggl[
                                       C_A\Bigl(
                                            -\frac{295}{18}\zeta_3
                                            +\frac{35}{9}\zeta_2
                                            +\frac{6361}{486}
                                          \Bigr)
                                      -C_F\Bigl(
                                            7\zeta_3
                                            +\frac{16}{3}\zeta_2
                                            +\frac{218}{81}
                                          \Bigr)
                                   \biggr]
\NN\\ &&
                       +C_A^2      \biggl(
                                       4{\sf B_4}
                                      -27\zeta_4
                                      +\frac{1969}{72}\zeta_3
                                      -\frac{781}{72}\zeta_2
                                      +\frac{42799}{3888}
                                   \biggr)
\NN\\ &&
                       +C_A C_F      \biggl(
                                      -8{\sf B_4}
                                      +36\zeta_4
                                      -\frac{1957}{12}\zeta_3
                                                                            +\frac{89}{3}\zeta_2
                                      +\frac{10633}{81}
                                   \biggr)
\NN\\ &&
                       +C_F^2      \biggl(
                                      \frac{95}{3}\zeta_3
                                      +\frac{274}{9}
                                   \biggr)
                                                   \Biggr\} + O(\ep)~.
                                          \label{GluSelf3}
\end{eqnarray}
Here 
\begin{eqnarray}
    B_4 = -4\zeta_2 \ln^2(2) + \frac{2}{3} \ln^4(2) - \frac{13}{2}\zeta_4 +16 \text{Li}_4 \left( \frac{1}{2} \right) \approx -1.762800093\dots
\end{eqnarray}
is a constant frequently encountered in massive calculations \cite{MOM,B4cite} and $\Li_n(z)$ denotes the 
polylogarithm \cite{LEWIN}.

The two-mass contributions to $\hat{\Pi}^{(3)}$ have been given before as power series up to $\mathcal{O}(\eta^3 \ln^2(\eta))$ in \cite{Ablinger:2017err}.
The full dependence on $\eta$ can be implicitly found in \cite{Pi32Ma} and has been independently calculated in Ref.~\cite{Pi32Mb}.
The quantity is given by
\begin{eqnarray}
\hat{\tilde{\Pi}}^{(3)}(0,m_1^2,m_2^2,\mu^2) &=& 
C_F T_F^2 \Biggl\{
\frac{256}{9 \varepsilon^2}
+\frac{64}{3 \varepsilon} \left[\lnMa{}+\lnMb{}+\frac{5}{9}\right]
-5 \eta -\frac{5}{\eta}
\nonumber \\ &&
+\left(-\frac{5 \eta }{8}-\frac{5}{8 \eta }+\frac{51}{4}\right) \ln ^2(\eta )+\left(\frac{5}{2 \eta }-\frac{5 \eta }{2}\right) \ln (\eta)
+\frac{32 \zeta_2}{3}
\nonumber \\ &&
+32 \lnMa{} \lnMb{}
+\frac{80}{9} \lnMa{}+\frac{80}{9} \lnMb{}
+\frac{1246}{81}
\nonumber \\ &&
+\left(\frac{5 \eta^{3/2}}{2}
+\frac{5}{2 \eta^{3/2}}
+\frac{3 \sqrt{\eta}}{2}
+\frac{3}{2 \sqrt{\eta}}\right) \Biggl[\frac{1}{8} \ln \left(\frac{1+\sqrt{\eta}}{1-\sqrt{\eta}}\right) \ln^2(\eta)
\nonumber \\ &&
-{\rm Li}_3\left(-\sqrt{\eta}\right)
+{\rm Li}_3\left(\sqrt{\eta}\right)
-\frac{1}{2} \ln (\eta) \left({\rm Li}_2\left(\sqrt{\eta}\right)
-{\rm Li}_2\left(-\sqrt{\eta}\right)\right)\Biggr]
\Biggr\}
\nonumber \\ &&
-C_A T_F^2 \Biggl\{
\frac{64}{9 \varepsilon^3} 
+\frac{16}{3 \varepsilon^2} \Biggl[ \left(\lnMa{}+\lnMb{}\right)
-\frac{35}{9}\Biggr]
+\frac{4}{\varepsilon} \Biggl[
\lnMa{2}+\lnMb{2}
-\frac{35}{9} \lnMa{}
\nonumber \\ &&
-\frac{35}{9} \lnMb{}
+\frac{2}{3} \zeta_2
+\frac{37}{27}
\Biggr]
+2 \left(\lnMa{3}+\lnMb{3}\right)
-\frac{70}{3} \lnMa{} \lnMb{}
-\frac{4}{9} \ln^3(\eta)
\nonumber \\ &&
+\left(2  \zeta_2+\frac{37}{9}\right) \left(\lnMa{}+\lnMb{}\right)
+\left[\frac{8}{3} \ln(1-\eta)- \frac{2}{3} \left(\eta+\frac{1}{\eta}\right) -\frac{179}{18}\right] 
\nonumber \\ && \times \ln^2(\eta)
-\frac{16}{3} \left(\eta+\frac{1}{\eta}\right)
-\frac{70}{9} \zeta_2
-\frac{56}{9} \zeta_3 
-\frac{3769}{243}
\nonumber \\ &&
+ \frac{8}{3} \left(\frac{1}{\eta}-\eta\right) \ln(\eta)
+\frac{16}{3} \big({\rm Li}_2(\eta) \ln(\eta)-{\rm Li}_3(\eta)\big)
\nonumber \\ &&
+\left[ 8 \frac{1+\eta^3}{3 \eta^{3/2}}+ 10  \frac{1+\eta}{\sqrt{\eta}}\right]
\Biggl[\frac{1}{8} \ln \left(\frac{1+\sqrt{\eta}}{1-\sqrt{\eta }}\right) \ln^2(\eta)
-{\rm Li}_3\left(-\sqrt{\eta}\right)
\nonumber \\ &&
+{\rm Li}_3\left(\sqrt{\eta}\right)
-\frac{1}{2} \ln(\eta) \left({\rm Li}_2\left(\sqrt{\eta}\right)-{\rm Li}_2\left(-\sqrt{\eta}\right)\right)\Biggr] + \mathcal{O}(\ep)
\Biggr\}.
\label{eq:pi2m}
\end{eqnarray}

With these ingredients we can split the reducible contributions as follows 
\begin{eqnarray}
\Atildehathat_{gg}^{(2)}\Bigl(\frac{\hat{m}_1^2}{\mu^2},\frac{\hat{m}_2^2}{\mu^2}\Bigl)&=&
 -\Ahathat_{gg}^{(1)} \Bigl(\frac{\hat{m}_1^2}{\mu^2}\Bigr)  \hat{\Pi}^{(1)}\Bigl(0,\frac{\hat{m}_2^2}{\mu^2}\Bigr) -\Ahathat_{gg}^{(1)} \Bigl(\frac{\hat{m}_2^2}{\mu^2}\Bigr)  \hat{\Pi}^{(1)}\Bigl(0,\frac{\hat{m}_1^2}{\mu^2}\Bigr) , \\
  \Athathat_{gg}^{(3)}\Bigl(\frac{\hat{m}_1^2}{\mu^2},\frac{\hat{m}_2^2}{\mu^2}\Bigl)&=&
\Athathat_{gg}^{(3),\rm{irr}}\Bigl(\frac{\hat{m}_1^2}{\mu^2},\frac{\hat{m}_2^2}{\mu^2}\Bigl)
-\hat{\tilde{\Pi}}^{(3)}\left(0,\hat{m}_1^2,\hat{m}_2^2,\mu^2\right)
 \NN\\&&
 -\Ahathat_{gg}^{\prime \ (2),\rm{irr}}\left(\frac{\hat{m}_1^2}{\mu^2}\right) 
  \hat{\Pi}^{(1)}\Bigl(0,\frac{\hat{m}_2^2}{\mu^2}\Bigr)
 -\Ahathat_{gg}^{\prime \ (2),\rm{irr}}\left(\frac{\hat{m}_2^2}{\mu^2}\right) 
  \hat{\Pi}^{(1)}\Bigl(0,\frac{\hat{m}_1^2}{\mu^2}\Bigr)
 \NN\\&&
 -2 \Ahathat_{gg}^{(1)}\left(\frac{\hat{m}_1^2}{\mu^2}\right) 
  \hat{\Pi}^{(2)}\Bigl(0,\frac{\hat{m}_2^2}{\mu^2}\Bigr)
 -2 \Ahathat_{gg}^{(1)}\left(\frac{\hat{m}_2^2}{\mu^2}\right) 
  \hat{\Pi}^{(2)}\Bigl(0,\frac{\hat{m}_1^2}{\mu^2}\Bigr)
 \NN\\&&
 +\Ahathat_{gg}^{(1)}\left(\frac{\hat{m}_1^2}{\mu^2}\right) 
  \left[2 \hat{\Pi}^{(1)}\Bigl(0,\frac{\hat{m}_1^2}{\mu^2}\Bigr)+ \hat{\Pi}^{(1)}\Bigl(0,\frac{\hat{m}_2^2}{\mu^2}\Bigr)\right]
  \hat{\Pi}^{(1)}\Bigl(0,\frac{\hat{m}_2^2}{\mu^2}\Bigr)
 \NN\\&&
 +\Ahathat_{gg}^{(1)}\left(\frac{\hat{m}_2^2}{\mu^2}\right) 
  \left[2 \hat{\Pi}^{(1)}\Bigl(0,\frac{\hat{m}_2^2}{\mu^2}\Bigr)+\hat{\Pi}^{(1)}\Bigl(0,\frac{\hat{m}_1^2}{\mu^2}\Bigr)\right]
  \hat{\Pi}^{(1)}\Bigl(0,\frac{\hat{m}_1^2}{\mu^2}\Bigr)~.
  \label{eq:ReducibleContr}
\end{eqnarray}
Here $\Ahathat_{gg}^{\prime \ (2),\rm{irr}}$ denotes the irreducible part of the unrenormalized two-loop 
OME $\Ahathat_{gg}^{(2)}$ with gluons in the initial and final state only.
When using the projector, which will be introduced in Eq.~\eqref{eq:Projectors}, also diagrams 
with a ghost in the initial and final state contribute to $\Ahathat_{gg}^{(2)}$.
These, however, must not be included in the reducible contributions for the three-loop OME.
This statement does also directly apply to the one loop OME $\Ahathat_{gg}^{(1)}$, but since no 
ghost contributions are present here, we can identify $\Ahathat_{gg}^{\prime \ (1)}=\Ahathat_{gg}^{(1)}$.
\section{Fixed moments of $\boldsymbol{\hat{\hat{\tilde{A}}}_{gg,Q}^{(3)}}$}
\label{sec:3}

\vspace*{1mm}
\noindent
In Ref.~\cite{Ablinger:2017err} the fixed moments $N=2,4,6$ of all two-mass OMEs at 3-loop order were 
presented as series expansions up to $\mathcal{O} (\eta^3 L_\eta^2 )$. For the constant part of 
$\hat{\hat{\tilde{A}}}_{gg,Q}^{(3)}$, the irreducible contributions were given. To allow for a direct 
comparison with the general $N$ results presented later, we list in the following these terms, including 
the reducible parts.
They are given by
\begin{eqnarray}
\lefteqn{\tilde{a}_{gg,Q}^{(3)} \bigl( N=2 \bigr) = } \NN \\ && 
C_F T_F^2 
\biggl\{
        -\frac{25556}{729}
        +\biggl(
                -\frac{512}{9}
                +\frac{160}{9} L_1
                +\frac{160}{9} L_2
        \biggr) \zeta_2
        -\frac{1408}{81} \zeta_3
        -\frac{3484}{81} L_1
        -\frac{1336}{27} L_1^2
        +\frac{992}{81} L_1^3
\nonumber \\ &&
        -\frac{16820}{243} L_2
        -\frac{1936}{27} L_1 L_2
        +\frac{64}{27} L_1^2 L_2
        -\frac{1336}{27} L_2^2
        +\frac{320}{27} L_1 L_2^2
        +\frac{736}{81} L_2^3
        +\eta  
	\biggl(
                \frac{758944}{30375}+\frac{22976}{2025}L_{\eta }
\nonumber \\ &&
		-\frac{448}{135} L_{\eta }^2
	\biggr)
        +\eta ^2 
	\biggl(
                -\frac{169892864}{10418625}+\frac{1028192}{99225} L_{\eta } -\frac{4768}{945}  L_{\eta }^2 
	\biggr)
        +\eta ^3 
	\biggl(
                -\frac{826805984}{843908625}+\frac{5893184}{2679075}L_{\eta }
\nonumber \\ &&
		-\frac{23872}{8505} L_{\eta }^2
	\biggr)
\biggr\}
+C_A T_F^2 
\biggl\{
        -\frac{59314}{2187}
        +\biggl(
                \frac{1340}{81}
                -\frac{308}{9} L_1
                -\frac{308}{9} L_2
        \biggr) \zeta_2
        +\frac{176}{81} \zeta_3
        -\frac{6844}{243} L_1
\nonumber \\ &&
        +\frac{1090}{81} L_1^2
        -\frac{1276}{81} L_1^3
        +12 L_2
        +\frac{1840}{81} L_1 L_2
        -\frac{440}{27} L_1^2 L_2
        +\frac{1090}{81} L_2^2
        -\frac{616}{27} L_1 L_2^2
        -\frac{1100}{81} L_2^3
\nonumber \\ &&
        +\eta  
	\biggl(
                -\frac{256304}{10125}+\frac{7184}{675}L_{\eta }+\frac{8}{45}L_{\eta }^2
	\biggr)
        +\eta ^2 
	\biggl(
                -\frac{1565036}{496125}+\frac{6008}{4725}L_{\eta }+\frac{8}{45}L_{\eta }^2
	\biggr)
\nonumber \\ &&
        +\eta ^3 
	\biggl(
                -\frac{56086736}{843908625}-\frac{164464}{2679075}L_{\eta }+\frac{2552}{8505}L_{\eta }^2
	\biggr)
\biggr\}
+ T_F^3
\biggl\{
        \biggl(
                32 L_1
                +32 L_2
        \biggr) \zeta_2
\nonumber \\ &&
        +\frac{128}{9} \zeta_3
        +\frac{32}{3} L_1^3
        +\frac{64}{3} L_1^2 L_2
        +\frac{64}{3} L_1 L_2^2
        +\frac{32}{3} L_2^3
\biggr\}
+ \mathcal{O} \Bigl( \eta^4 L_\eta^3 \Bigr),
\\
\lefteqn{\tilde{a}_{gg,Q}^{(3)} \bigl( N=4 \bigr) =} \NN \\ && 
C_F T_F^2 
\biggl\{
        -\frac{934723727}{21870000}
        +\biggl(
                -\frac{226583}{4050}
                +\frac{121}{45} L_1
                +\frac{121}{45} L_2
        \biggr) \zeta_2
        -\frac{5324}{2025} \zeta_3
        -\frac{9432079}{243000} L_1
\nonumber \\ &&
        -\frac{2051797 }{40500}L_1^2
        +\frac{3751 }{2025}L_1^3
        -\frac{3415493}{81000} L_2
        -\frac{673474 }{10125}L_1 L_2
        +\frac{242}{675} L_1^2 L_2
        -\frac{2051797 }{40500}L_2^2
\nonumber \\ &&
        +\frac{242}{135} L_1 L_2^2
        +\frac{2783 }{2025}L_2^3
        +\eta  
	\biggl(
                \frac{1556008}{253125}+\frac{18544 }{5625}L_{\eta }-\frac{416 }{1125}L_{\eta }^2
	\biggr)
	+\eta ^2 
	\biggl(
                -\frac{92973466}{17364375}+\frac{160036 }{55125}L_{\eta }
\nonumber \\ &&
		-\frac{428 }{315}L_{\eta }^2
	\biggr)
        +\eta ^3 
	\biggl(
                -\frac{1109454088}{4219543125}+\frac{4900048 }{13395375}L_{\eta }-\frac{35648 }{42525}L_{\eta }^2
	\biggr)
\biggr\}
+C_A T_F^2 
\biggl\{
        -\frac{518340979}{1822500}
\nonumber \\ &&
        +\biggl(
                -\frac{32182}{675}
                -\frac{3304}{45} L_1
                -\frac{3304}{45} L_2
        \biggr) \zeta_2
        +\frac{1888}{405} \zeta_3
        -\frac{13735499}{60750} L_1
        -\frac{31169}{675} L_1^2
        -\frac{13688}{405} L_1^3
\nonumber \\ &&
        -\frac{811661 }{6750}L_2
        -\frac{34208}{675} L_1 L_2
        -\frac{944}{27} L_1^2 L_2
        -\frac{31169 }{675} L_2^2
        -\frac{6608}{135} L_1 L_2^2
        -\frac{2360 }{81}L_2^3
\nonumber \\ &&
        +\eta  
	\biggl(
                -\frac{22204183}{303750}+\frac{697393 }{20250}L_{\eta }-\frac{7303 }{2700}L_{\eta }^2
	\biggr)
        +\eta ^2 
	\biggl(
                -\frac{94581301}{10418625}+\frac{492763 }{99225}L_{\eta }-\frac{205 }{189}L_{\eta }^2
	\biggr)
\nonumber \\ &&
        +\eta ^3 
	\biggl(
                -\frac{692255687}{1687817250}+\frac{3118727 }{5358150}L_{\eta }-\frac{7319}{34020} L_{\eta }^2
	\biggr)
\biggr\}
+ T_F^3
\biggl\{
        \biggl(
                32 L_1
                +32 L_2
        \biggr) \zeta_2
        +\frac{128}{9} \zeta_3
\nonumber \\ &&
        +\frac{32 }{3}L_1^3
        +\frac{64}{3} L_1^2 L_2
        +\frac{64}{3} L_1 L_2^2
        +\frac{32 }{3}L_2^3
\biggr\} 
+ \mathcal{O} \Bigl( \eta^4 L_\eta^3 \Bigr) ,
\\
\lefteqn{\tilde{a}_{gg,Q}^{(3)} \bigl( N=6 \bigr) =} \NN \\ && 
C_A T_F^2 
\biggl\{
        -\frac{68860626799}{187535250}
        +\biggl(
                -\frac{193394}{2835}
                -\frac{806}{9} L_1
                -\frac{806}{9} L_2
        \biggr) \zeta_2
        +\frac{3224}{567} \zeta_3
        -\frac{9618442}{33075} L_1
\nonumber \\ &&
        -\frac{1294861}{19845} L_1^2
        -\frac{23374}{567} L_1^3
        -\frac{1919194}{11907} L_2
        -\frac{1471552}{19845} L_1 L_2
        -\frac{8060}{189} L_1^2 L_2
        -\frac{1294861}{19845} L_2^2
\nonumber \\ &&
        -\frac{1612}{27} L_1 L_2^2
        -\frac{20150}{567} L_2^3
        +\eta  
	\biggl(
                -\frac{488831873}{5315625}+\frac{14655008}{354375} L_{\eta }-\frac{12167}{3375} L_{\eta }^2
	\biggr)
\nonumber \\ &&
        +\eta ^2 
	\biggl(
                -\frac{469449112}{52093125}+\frac{2525176}{496125} L_{\eta }-\frac{232}{225} L_{\eta }^2
	\biggr)
        +\eta ^3 
	\biggl(
                -\frac{1795386647}{4219543125}+\frac{8701352}{13395375} L_{\eta }
\nonumber \\ &&
		-\frac{4819}{42525} L_{\eta }^2
	\biggr)
\biggr\}
+C_F T_F^2 
\biggl\{
        -\frac{705306787007}{15315378750}
        +\biggl(
                -\frac{4410376}{77175}
                +\frac{484}{441} L_1
                +\frac{484}{441} L_2
        \biggr) \zeta_2
        -\frac{21296}{19845}\zeta_3
\nonumber \\ &&
        -\frac{2991682411}{72930375} L_1
        -\frac{12017984}{231525} L_1^2
        +\frac{15004}{19845} L_1^3
        -\frac{334770739}{8103375} L_2
        -\frac{15657416}{231525} L_1 L_2
        +\frac{968}{6615} L_1^2 L_2
\nonumber \\ &&
        -\frac{12017984}{231525} L_2^2
        +\frac{968}{1323} L_1 L_2^2
        +\frac{11132}{19845} L_2^3
	+\eta  
	\biggl(
                \frac{3661888}{826875}+\frac{10784}{3375} L_{\eta }-\frac{1216}{11025} L_{\eta }^2
	\biggr)
\nonumber \\ &&
        +\eta ^2 
	\biggl(
                -\frac{930064}{180075}+\frac{589024}{231525} L_{\eta }-\frac{1504}{1225} L_{\eta }^2
	\biggr)
        +\eta ^3 
	\biggl(
                -\frac{283956224}{1181472075}+\frac{2587744}{18753525} L_{\eta }-\frac{251008}{297675} L_{\eta }^2
	\biggr)
\biggr\}
\nonumber \\ &&
+  T_F^3
\biggl\{
        \biggl(
                32 L_1
                +32 L_2
        \biggr) \zeta_2
        +\frac{128}{9} \zeta_3
        +\frac{32}{3} L_1^3
        +\frac{64}{3} L_1^2 L_2
        +\frac{64}{3} L_1 L_2^2
        +\frac{32}{3} L_2^3
\biggr\}
+ \mathcal{O} \Bigl( \eta^4 L_\eta^3 \Bigr) .
\end{eqnarray}
\section{Details of the calculation}
\label{sec:4}

\begin{figure}[ht]
\begin{center}
\begin{minipage}[c]{0.20\linewidth}
  \includegraphics[width=1\textwidth]{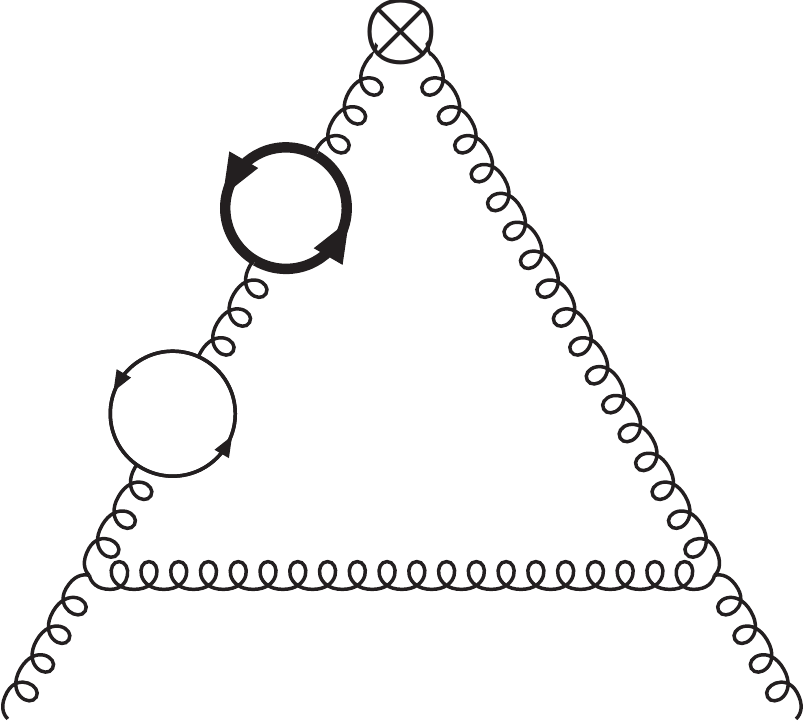}
\vspace*{-11mm}
\begin{center}
{\footnotesize (1)}
\end{center}
\end{minipage}
\hspace*{1mm}
\begin{minipage}[c]{0.20\linewidth}
  \includegraphics[width=1\textwidth]{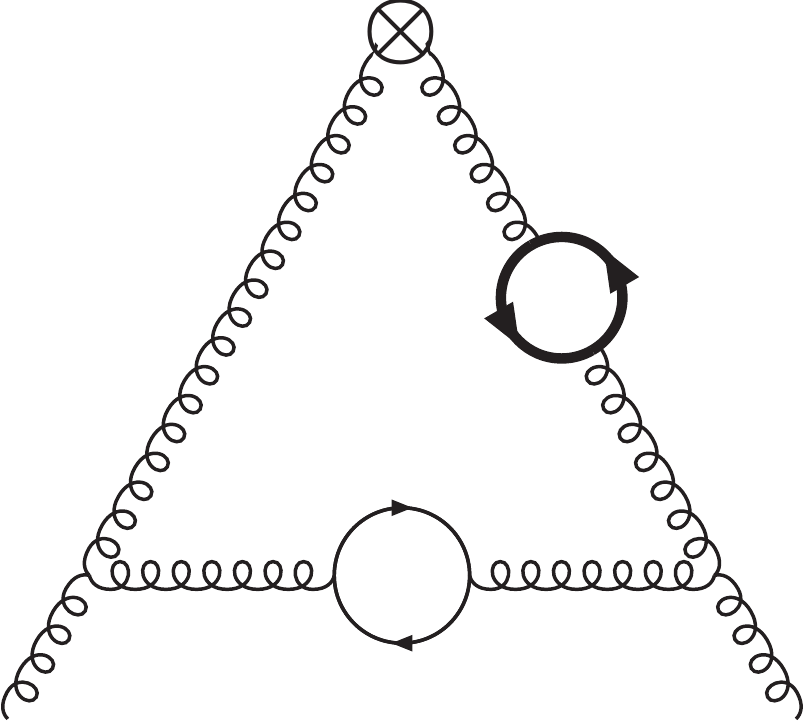}
\vspace*{-11mm}
\begin{center}
{\footnotesize (2)}
\end{center}
\end{minipage}
\hspace*{1mm}
\begin{minipage}[c]{0.20\linewidth}
  \includegraphics[width=1\textwidth]{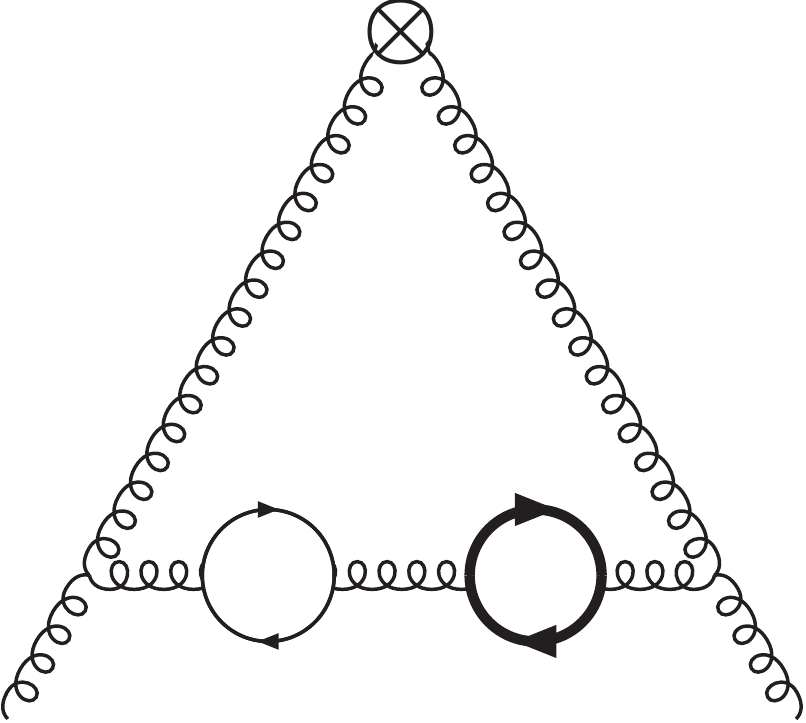}
\vspace*{-11mm}
\begin{center}
{\footnotesize (3)}
\end{center}
\end{minipage}
\hspace*{1mm}
\begin{minipage}[c]{0.20\linewidth}
  \includegraphics[width=1\textwidth]{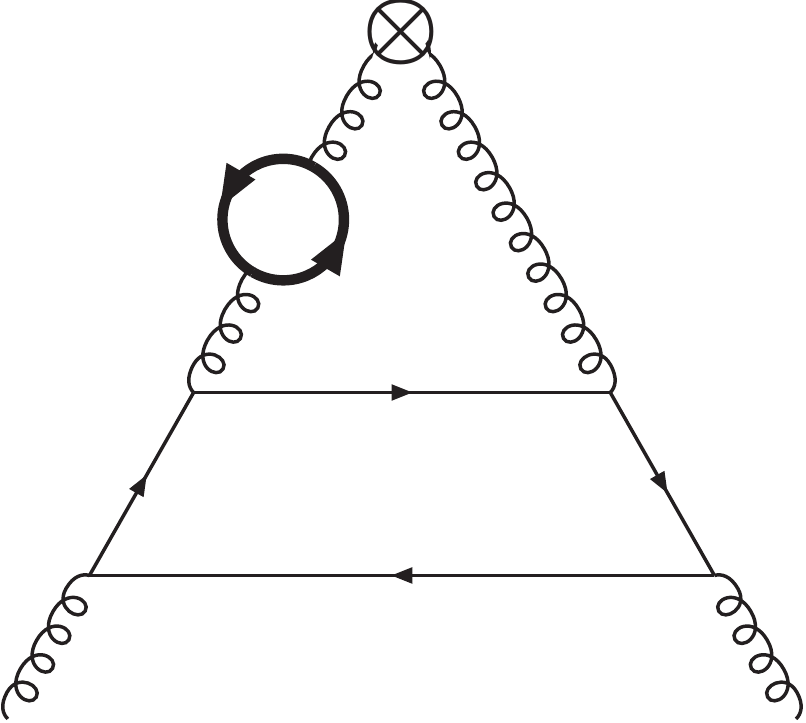}
\vspace*{-11mm}
\begin{center}
{\footnotesize (4)}
\end{center}
\end{minipage}

\vspace*{5mm}

\begin{minipage}[c]{0.20\linewidth}
  \includegraphics[width=1\textwidth]{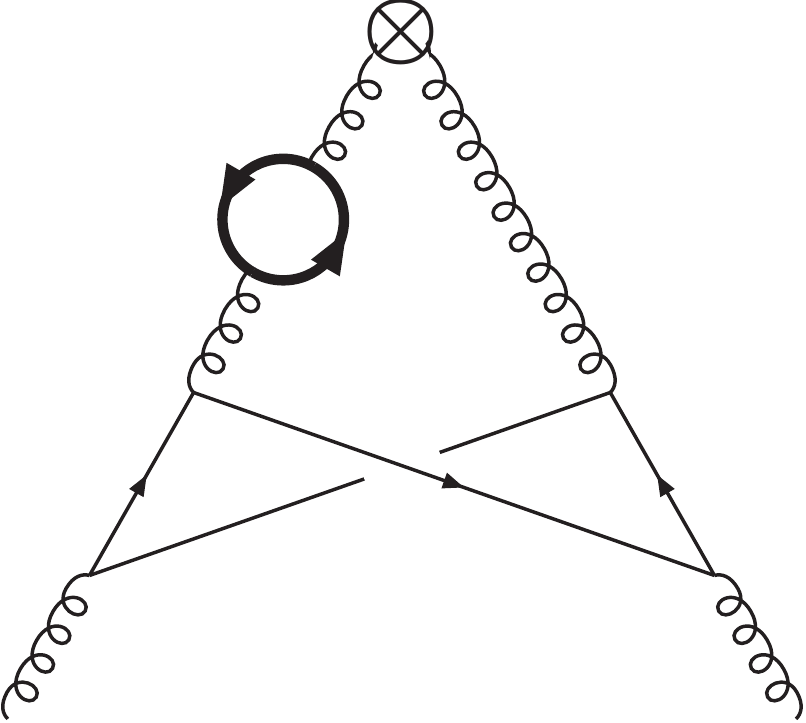}
\vspace*{-11mm}
\begin{center}
{\footnotesize (5)}
\end{center}
\end{minipage}
\hspace*{1mm}
\begin{minipage}[c]{0.20\linewidth}
  \includegraphics[width=1\textwidth]{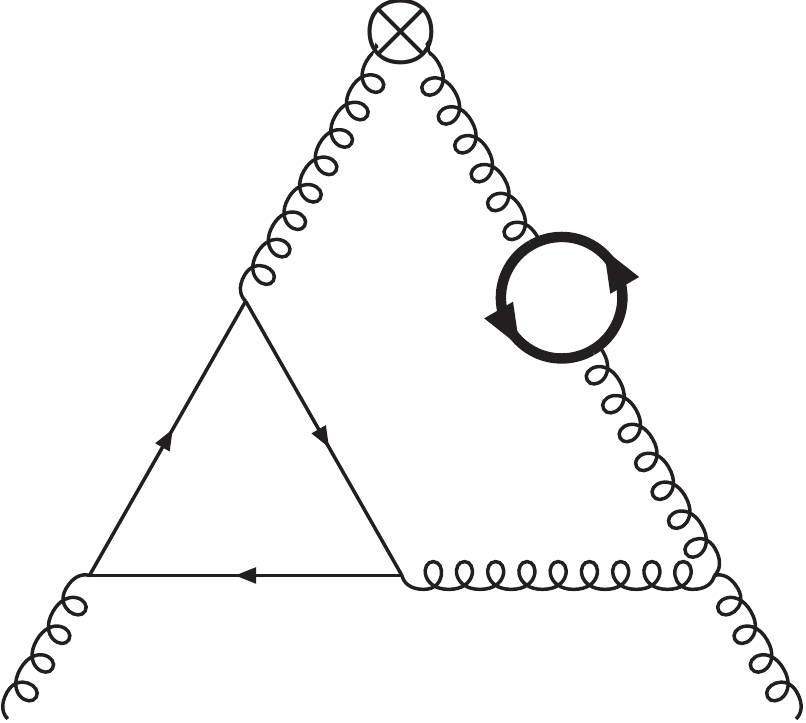}
\vspace*{-11mm}
\begin{center}
{\footnotesize (6)}
\end{center}
\end{minipage}
\hspace*{1mm}
\begin{minipage}[c]{0.20\linewidth}
  \includegraphics[width=1\textwidth]{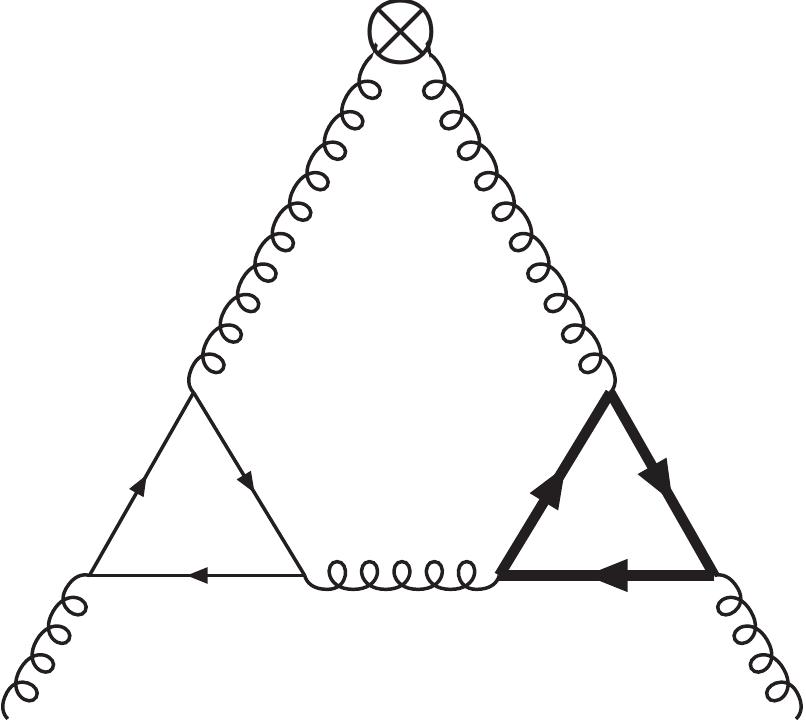}
\vspace*{-11mm}
\begin{center}
{\footnotesize (7)}
\end{center}
\end{minipage}
\hspace*{1mm}
\begin{minipage}[c]{0.20\linewidth}
  \includegraphics[width=1\textwidth]{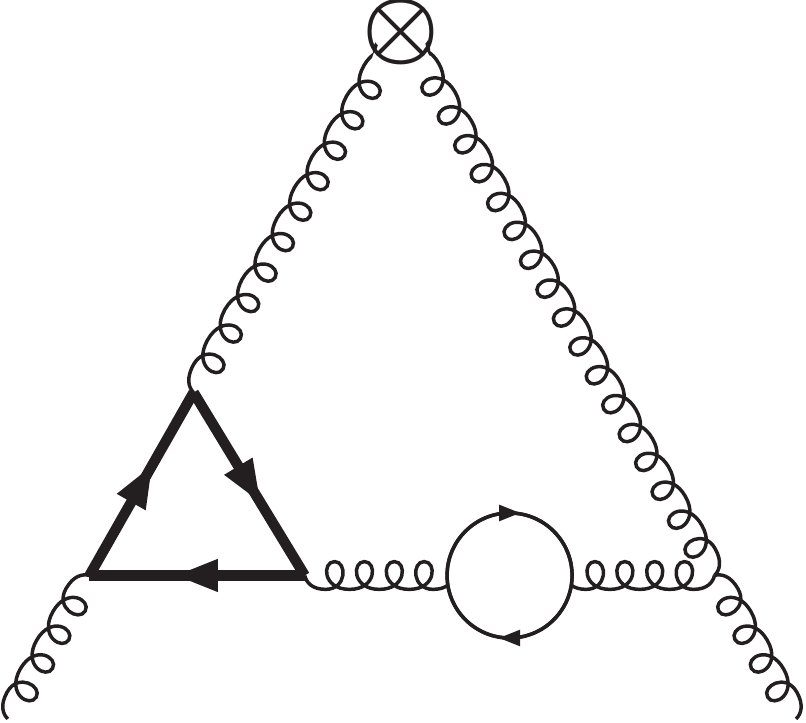}
\vspace*{-11mm}
\begin{center}
{\footnotesize (8)}
\end{center}
\end{minipage}

\vspace*{5mm}

\begin{minipage}[c]{0.20\linewidth}
  \includegraphics[width=1\textwidth]{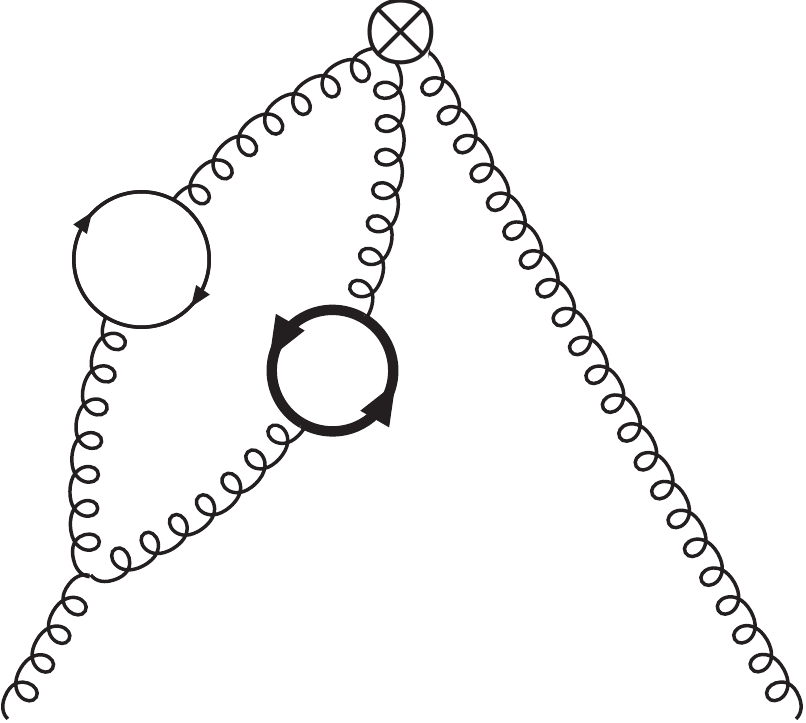}
\vspace*{-11mm}
\begin{center}
{\footnotesize (9)}
\end{center}
\end{minipage}
\hspace*{1mm}
\begin{minipage}[c]{0.20\linewidth}
  \includegraphics[width=1\textwidth]{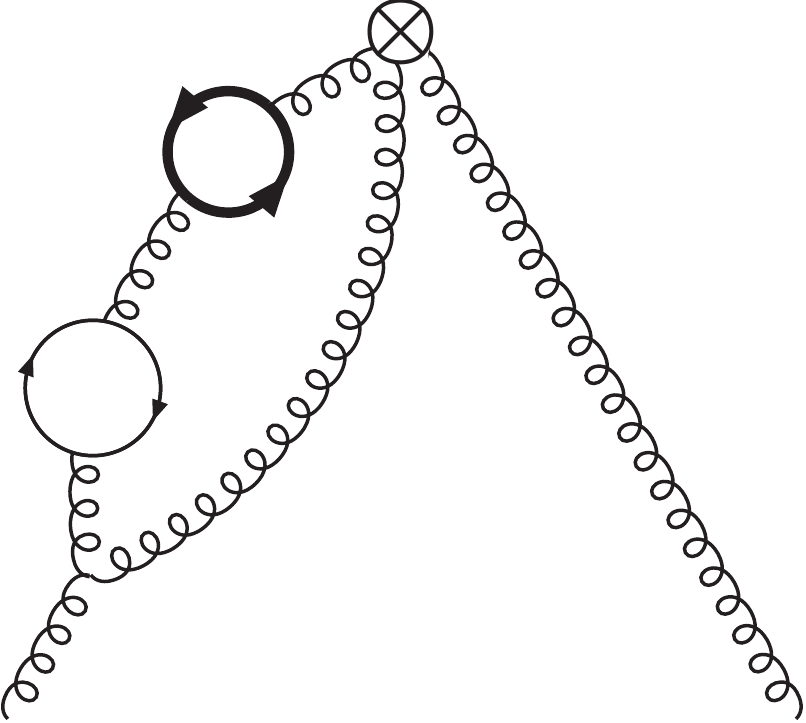}
\vspace*{-11mm}
\begin{center}
{\footnotesize (10)}
\end{center}
\end{minipage}
\hspace*{1mm}
\begin{minipage}[c]{0.20\linewidth}
  \includegraphics[width=1\textwidth]{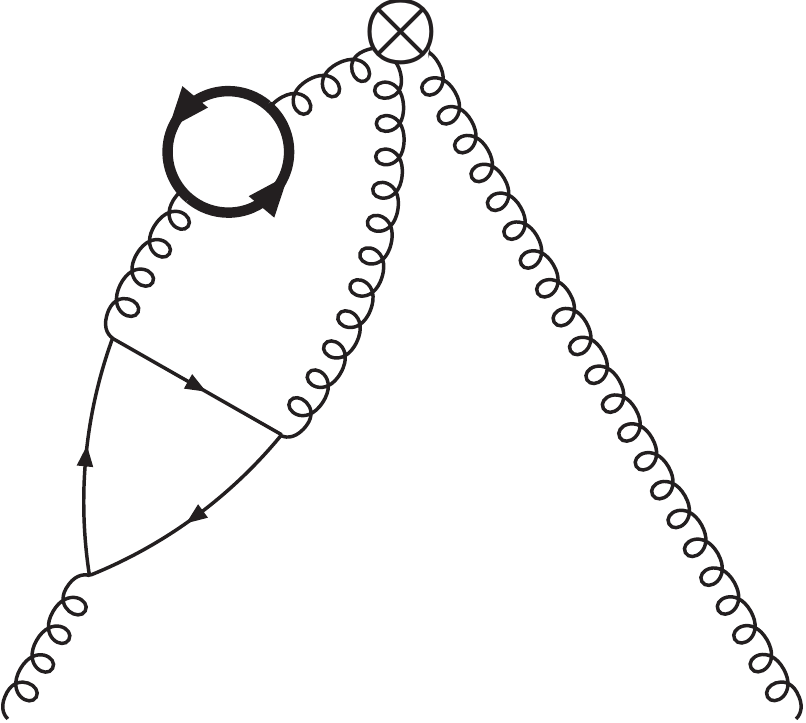}
\vspace*{-11mm}
\begin{center}
{\footnotesize (11)}
\end{center}
\end{minipage}
\hspace*{1mm}
\begin{minipage}[c]{0.20\linewidth}
  \includegraphics[width=1\textwidth]{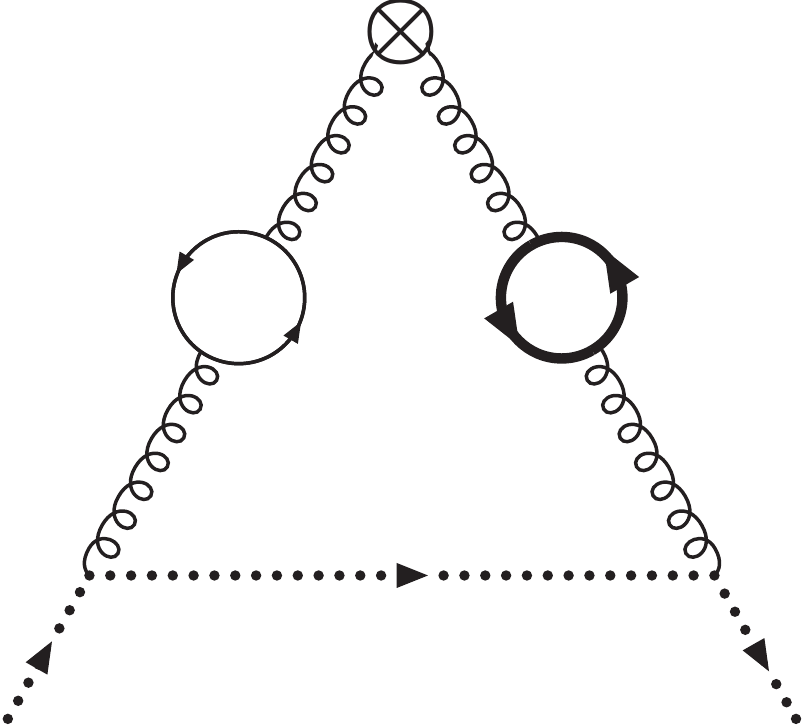}
\vspace*{-11mm}
\begin{center}
{\footnotesize (12)}
\end{center}
\end{minipage}
\caption{\small The twelve different topologies for $\tilde{A}_{gg,Q}^{(3)}$. Curly lines: gluons; dotted 
lines: 
ghosts; thin arrow lines: lighter massive quark; thick arrow lines: heavier massive quark; the symbol 
$\otimes$ represents the corresponding local operator insertion, cf.~\cite{Bierenbaum:2009mv}.} 
\label{fig:GGdiagrams}
\end{center}
\end{figure}

\vspace*{1mm}
\noindent
There are 76 irreducible diagrams contributing to the OME $\tilde{A}_{gg,Q}^{(3)}$, out of which 6 contain 
external ghost lines. Since the value of a diagram is not changed by moving the operator insertion to a 
different gluon line with the same momentum, we are left with 12 topologically different diagrams.
We checked these identities for fixed moments $N=2,4,6$ with the help 
of \texttt{Q2E/EXP}~\cite{Harlander:1997zb,Seidensticker:1999bb}.  Half of these diagrams are symmetric 
under the exchange $m_1 \leftrightarrow m_2$, while the other half has to be evaluated for both possible mass 
assignments. One representative of each of the twelve topologies is shown in Figure~\ref{fig:GGdiagrams}.

The unrenormalized OME is obtained by applying the gluonic or ghost projector
\begin{eqnarray}
     P_g^{\mu\nu} = - \frac{\delta_{ab}}{N_c^2 - 1} \frac{ g^{\mu\nu} }{ D-2 } \left( \Delta.p \right)^{-N} &,&
     P_{gh}^{\mu\nu} = - \frac{\delta_{ab}}{N_c^2 - 1} \frac{ 1 }{ D-2 } \left( \Delta.p \right)^{-N}
     \label{eq:Projectors}
\end{eqnarray}
to the Green's functions with external gluon or ghost lines, respectively, adding all contributions up, 
including  the one-particle reducible contributions from Eq. \eqref{eq:ReducibleContr} as well.
For the Feynman rules we follow Ref.~\cite{BeiglboCk:2006lfa}.
Here $p$ denotes the momentum of the external on-shell gluon, $\Delta$ is a light-like $D$-vector (\ie $p^2=\Delta^2=0$) and $a,b$ are the color indices of the external gluons (ghosts). 
For the ghost diagrams an additional factor of 2 has to be included.
Furthermore, special care has to be taken when including the reducible contributions. Here the irreducible 
two-loop contribution $\Ahathat_{gg,Q}^{(2),irr}$ in Eq.~\eqref{eq:ReducibleContr} has also to be calculated using 
the projector of Eq.~\eqref{eq:Projectors}, excluding the ghost contributions which enter the complete two-loop result.

\vspace*{1mm}
\noindent
\subsection{Computation Strategy}
\label{sec:41}

\vspace*{1mm}
\noindent
Since the number of diagrams we have to calculate is small and the reduction to master integrals can introduce 
spurious terms which only cancel in the final result, we aim at computing the diagrams without reduction to 
master integrals. Moreover, the reduction to master integrals in the two-mass case with local operator
insertions requires a substantial computational time. The direct calculation of the Feynman diagrams in 
$N$-space will require the treatment of a large amount of terms as well, due to large numerator 
structures in the gluonic case. The result is obtained directly, without having to calculate two-parameter
master integrals, e.g. by solving differential equations. As we will see later, in this way we would
obtain very involved expressions, which can be avoided by introducing efficient one-dimensional integral
representations, see also Ref.~\cite{Ablinger:2017xml}. They can be found most easily working in Mellin $N$ 
space.
 
Furthermore, our first goal, in contrast to the treatment in Ref.~\cite{Ablinger:2017xml}, is to derive 
first the $N$-space solution and to obtain the $z$-space result via an analytic  Mellin-inversion thereafter.
This is possible since all occurring difference equations turn out to be first order factorizable, so 
closed form solutions of these sums can be found using established difference field techniques using the 
package {\tt Sigma} \cite{SIG1,SIG2}.
In the following paragraph, we outline the basic computational strategy to calculate the diagrams.     
After that, we give a more detailed description of the calculation of a particular diagram as an example.

The 76 contributing irreducible diagrams have been generated using \texttt{QGRAF} \cite{QGRAF}, in the 
version given in Ref.~\cite{Bierenbaum:2009mv} which includes local operator insertions. After 
identifying the 
12 different topologies, we set up dedicated {\tt FORM} \cite{FORM} routines to perform the Dirac algebra 
and traces. The color algebra is done using the {\tt FORM} program \texttt{COLOR} \cite{vanRitbergen:1998pn}.
For fermionic bubble insertions we use the identity
\begin{eqnarray}
    \Pi^{\mu\nu}_{ab} ( k ) = - \frac{ 8 T_F g^2}{(4\pi)^{D/2}} \delta_{ab} ( k^2 g^{\mu\nu} - k^\mu k^\nu ) \int\limits_{0}^{1} \text{d}x \frac{ \Gamma(2-D/2) \left( x(1-x) \right)^{D/2-1} }{ \left( -k^2 + \frac{m^2}{x(1-x)} \right)^{2-D/2} } .
\end{eqnarray}
Next, the Feynman parameterization was performed on the full numerator and denominator structure,
\ie we do not cancel structures appearing in the numerator against the denominator.
This provides us with a uniform Feynman parameterization for the whole diagram.
At last the momentum integrations were performed one after another, starting from loops without the operator 
insertion. The resulting tensor integrals were reduced to scalar ones according to the rules stated in 
Appendix~\ref{app:A} and thus mapped to the basic one-loop integral~\cite{BeiglboCk:2006lfa}
\begin{eqnarray}
\int \frac{\dd^d k}{(2\pi)^d} \frac{ (k^2)^m }{ ( k^2 + R^2 )^n } 
&=& \frac{1}{(4\pi)^{D/2}} \frac{\Gamma(n-m-D/2)}{\Gamma(n)} \frac{\Gamma(m+D/2)}{\Gamma(D/2)} 
\left( R^2 \right)^{m-n+D/2} .
\end{eqnarray}
It is important to perform the integration of the momentum with the operator insertion as the last one. 
In this way only the additional scalar product $p.k$ can appear, which simplifies the reduction to scalar 
integrals drastically, since only a single term of the binomial decomposition of $(k.\Delta + R_0 p.\Delta )^N$ 
can contribute to the integral. 

After these steps we are left with a linear combination of up to 7-fold Feynman parameter integrals, with 
the general structure
\begin{eqnarray}
    \prod\limits_{j=1}^{i} \int\limits_0^1 \mathrm{d} x_i \ x_i^{a_i} (1-x_i)^{b_i} \ R_0^N \ \left[ R_1 \ m_a^2 + R_2 \ m_b^2 \right]^{-s}.
\end{eqnarray} 
Here $R_1$ and $R_2$ are simple rational functions of $x_i$ and $1-x_i$ and $R_0$ is a polynomial in $x_i$ 
stemming from the local operator insertion. In the next step we split the rightmost factor by means of 
a Mellin-Barnes integral \cite{MB1a,MB1b,MB2,MB3,MB4}
\begin{eqnarray}
\frac{1}{(A+B)^{s}} &=& \frac{1}{2\pi i} \frac{1}{\Gamma (s)} B^{-s} \int\limits_{-i\infty}^{+i\infty} 
\mathrm{d} \sigma \left( \frac{A}{B} \right)^{\sigma} \Gamma (-\sigma) \Gamma( \sigma + s ) ,
\end{eqnarray}
where the real part of the integration contour has to be chosen such that the ascending poles are separated
from the descending ones.
Our next aim is to compute the Feynman parameter integrals.
To do this, the operator polynomial $R_0$ can be decomposed with the help of the binomial theorem
\begin{eqnarray}
    (A + B)^N &=& \sum\limits_{i=0}^{N} \binom{N}{i} A^i B^{N-i}. 
\label{eq:I2}
\end{eqnarray}
This splitting has to be performed as often as necessary to obtain hyperexponential terms in $x_i$ and 
$1-x_i$ only. In the present case, we had to split the polynomial up to three times.
Attempts to combine the expression into a linear combination of higher transcendental functions in order 
to keep the additional summations as few as possible have failed, because overlapping divergencies of the 
$\Gamma$-functions appeared, preventing to choose a proper path for the Mellin-Barnes integral. 
This indicates that these transformations cannot be performed naively after the Mellin-Barnes representation 
has been applied. Applying these transformations, all Feynman parameter integrals can be expressed by Euler's 
Beta-functions
\begin{eqnarray}
    \text{B} (a,b) = \int\limits_0^1 \text{d} z z^{a-1} (1-z)^{b-1} = \frac{\Gamma(a)\Gamma(b)}{\Gamma(a+b)}.
\end{eqnarray}

For example, we encountered the integral
\begin{eqnarray}
    I &=& \Gamma\left( -\tfrac{3\ep}{2} \right) \int\limits_0^1 \left( \prod\limits_{i=1}^7 \dd z_i \right) z_1^2 \left( z_2 (1-z_2) \right)^{\tfrac{\ep}{2}} z_3^2 \left( z_4 (1-z_4) \right)^{\tfrac{\ep}{2}} (1-z_5) \left( z_6 (1-z_6) \right)^{\tfrac{\ep}{2}} z_7^{1+\tfrac{\ep}{2}} \NN \\ && 
\times (1-z_7)^2 \bigl( z_7 ( z_1 z_6 + z_3 ( 1 - z_6 ) ) + z_5 ( 1 - z_7 ) \bigr)^{N-4} \left( \frac{z_6 \ m_a^2}{z_2(1-z_2)} + \frac{(1-z_6) \ m_b^2}{z_4(1-z_4)}  \right)^{\tfrac{3\ep}{2}}  
\end{eqnarray}
for the computation of diagram~7 in Figure~\ref{fig:GGdiagrams}.
Here we can decompose the operator polynomial as 
\begin{eqnarray}
    \lefteqn{ \hspace*{-2cm}
\bigl( z_7 ( z_1 z_6 + z_3 ( 1 - z_6 ) ) + z_5 ( 1 - z_7 ) \bigl)^{N-4} =} \NN \\
    &&  \sum\limits_{j=0}^{N-4} \sum\limits_{i=0}^{j} \binom{N-4}{j} \binom{j}{i} z_7^j \, 
z_1^i \, z_6^i \, z_3^{j-i} \, (1-z_6)^{j-i} \, z_5^{N-4-j} \, (1-z_7)^{N-4-j}~.  
\end{eqnarray}
After applying the Mellin-Barnes integral and integrating the Feynman parameters we find
\begin{eqnarray}
    I &=& \frac{(m_b^2)^{\tfrac{3\ep}{2}}}{2\pi i} \sum\limits_{j=0}^{N-4} \sum\limits_{i=0}^{j} 
\binom{N-4}{j} \binom{j}{i} \frac{\Gamma(3+i) \Gamma(3-i+j) \Gamma(N-j-3)}{\Gamma(4+i) 
\Gamma(4-i+j) \Gamma(N+1+\tfrac{\ep}{2}) } \int\limits_{-i\infty}^{+i\infty} \dd \sigma 
\left( \frac{m_a^2}{m_b^2} \right)^\sig 
\NN \\
     && \times \Gamma(-\sig) \Gamma(-\tfrac{3\ep}{2} + \sig) \Gamma(1-\tfrac{\ep}{2}+i+\sig) 
\Gamma(1+\ep-i+j-\sig)  \NN \\
     && \times \frac{ \Gamma(1+\tfrac{\ep}{2}-\sig) \Gamma(3+\tfrac{\ep}{2}-\sig)  \Gamma(1-\ep-\sig) 
\Gamma(3-\ep+\sig) }{\Gamma(4+\ep-2\sig) \Gamma(4-2\ep+2\sig) }~.
\end{eqnarray}
Note that the summands arising from the binomial decomposition in Eq. \eqref{eq:I2} appear naturally in 
nested form. We have not yet specified $m_a$ or $m_b$ to the physical masses, since there are diagrams 
with both possibilities. In the following we choose to exploit the symmetry of the Mellin-Barnes integral to 
arrive at two different representations either proportional to $(m_a^2/m_b^2)^\sig$ or to $(m_b^2/m_a^2)^\sig$.
In this way we can choose $m_a^2/m_b^2=\eta$ or $m_b^2/m_a^2=\eta$ and close the contour to the right in 
both cases.
At this point we could have followed earlier approaches by applying the packages \texttt{MB} \cite{MB} 
and \texttt{MBresolve} \cite{MBr} to resolve the singularity structure of the integrals and expand the 
final integral in $\ep$.
However, 
the additional dependence on $N$ and up to four summation quantifiers renders the automated finding 
of a suitable integration contour non-trivial.
Therefore, we calculated these integrals by summing up the residues of the ascending poles of the integrand 
keeping the $\ep$-dependence and are expanding afterwards. In general, residues had to be taken at $\sig = k$, 
$\sig = k + \ep/2$ and $\sig = k + \ep$, where $k$ is an integer larger than an integral specific minimum.
In the end, each integral is represented by a linear combination of three infinite sums, over which additional 
binomial sums have to be performed. Nevertheless, we used the packages \texttt{MB} and \texttt{MBresolve} to 
check our sum representations for fixed values of the Mellin variable $N$.

The final multi-sum can now be handled by the packages \texttt{Sigma} \cite{SIG1,SIG2}, 
\texttt{EvaluateMultiSums} and \texttt{SumProduction} \cite{EMSSP}. Here additionally \texttt{HarmonicSums} 
\cite{HARMONICSUMS,Ablinger:2011te,Ablinger:2013cf} was used for limiting procedures and operations on special 
functions and numbers. The sum representation of each integral, which can take up to 
$\mathcal{O}(4\rm{MB})$, was crushed to a optimal representation using \texttt{SumProduction}.  
This representation contains constants from taking out points from summation boundaries and multi-sums with 
large summand structures. These multi-sums were then handled by \texttt{EvaluateMultiSums}, which uses 
\texttt{Sigma} and \texttt{HarmonicSums}. The results were expressed in terms of nested harmonic-, generalized 
harmonic-, cyclotomic- and binomial-sums. Furthermore, generalized harmonic- and cyclotomic-sums at infinity 
contribute. These can be expressed in terms of HPLs depending on $\eta$ in the argument with 
the help of \texttt{HarmonicSums}.

Prior to the solution for general values of $N$, our sum representations also allow to calculate fixed
even moments, without expanding in the parameter $\eta$, cf.~Section~\ref{sec:3}. They also serve as input 
values for the general $N$-solution.

\subsection{Summation techniques}\label{Sec:SummationTools}

\vspace*{1mm}
\noindent
Using all the transformations of Subsection~\ref{sec:41}, the integrals under consideration can be given in terms 
of thousands of multi-sums with size up to $\mathcal{O}(4\rm{MB})$. 
For instance, for diagram~7 one gets an expression of 65 MB size that is given in terms of 10262 multi-sums. 9122 of these sums are triple sums of the form
\begin{align}\label{Equ:Diag7TripleSumCases}
\sum_{j=0}^{N-l}\sum_{i=0}^j\sum_{k=0}^{\infty}f(\ep,\eta,N,j,i,k),&&
\sum_{j=0}^{N-l}\sum_{i=0}^j\sum_{k=0}^{i}g(\ep,\eta,N,j,i,k),
\end{align}
for some nonnegative integers $l$, and the remaining 1140 sums consist of double sums of similar type. One of these typical triple sums is 
\begin{multline}\label{Equ:SimpleTripleSum}
T(\ep,\eta,N)=\sum_{j=0}^{N}\sum_{i=0}^j\sum_{k=0}^{i}f(\ep,\eta,N,j,i,k)
=\sum_{j=0}^{N}\sum_{i=0}^j\sum_{k=0}^{i}\tfrac{(4+\ep) (-2+N) (-1+N) N \pi (-1)^{2-k}}{2+\ep}\times\\
\times 2^{-2+\ep} e^{-\frac{3 \ep \gamma }{2}} \eta^{k}\tfrac{        \Gamma (
                1
                -\frac{\ep}{2}
                -i
                +j
                +k
        )\Gamma (
        -1-\frac{\ep}{2}) \Gamma (
        2+\frac{\ep}{2}) \Gamma (1+N) \Gamma (1
+\ep
+i
-k
) \Gamma (
        -\frac{3 \ep}{2}
        +k
) \Gamma (1
-\ep
+k
) \Gamma (3
-\ep
+k
) \Gamma (
        -\frac{1}{2}
        -\frac{\ep}{2}
        +k
)}{\Gamma (
        -\frac{3}{2}-\frac{\ep}{2}) \Gamma (
        \frac{5}{2}+\frac{\ep}{2}) \Gamma (2+i) \Gamma (1+k) \Gamma (2
-i
+j
) \Gamma (2
-\ep
+k
) \Gamma (
        \frac{5}{2}
        -\ep
        +k
) \Gamma (
        -\frac{\ep}{2}
        +k
) \Gamma (
        5
        +\frac{\ep}{2}
        +N
)}.
\end{multline}

In the following we will present our summation toolbox that enables one to compute the $\ep$-expansions of the arising sums and thus of the desired integrals by summing up all these $\ep$-expansions. More precisely, we will utilize summation algorithms that succeed in representing the coefficients of the $\ep$-expansion of sums like~\eqref{Equ:SimpleTripleSum} in terms of 
hypergeometric products and indefinite nested sums defined over such products that can be defined as follows.

\medskip

\noindent\textit{Definition.}
Let $f(N)$ be an expression that evaluates at non-negative 
integers (from a certain point on) to elements of a field\footnote{In our setting, we are given a rational function field $\mathbb{K}=\mathbb{K}'(\eta)$ in terms of the variable $\eta$ where $\mathbb{K}'$ is a subfield containing the rational numbers and various constants such as $\zeta_2$.} $\mathbb{K}$ of characteristic $0$. Then $f(N)$ 
is called an indefinite nested sum (over hypergeometric products) w.r.t.\ $N$ if it is composed by elements from the rational function field $\set K(N)$, the three operations 
($+,-,\cdot$), hypergeometric products of the form $\prod_{k=l}^Nh(k)$ with $l\in\set N$ and $h(k)$ being a rational function in $k$ and being free of $N$, and sums of the form $\sum_{k=l}^Nh(k)$ with $l\in\set N$ and with 
$h(k)$ being an indefinite nested (over hypergeometric products) w.r.t.\ $k$ and being free of $N$. 

\medskip

Note that this class covers as special cases the harmonic sums~\cite{HSUM},
\begin{eqnarray}\label{Equ:HarmonicSumsIntro}
S_{b,\vec{a}}(N) &=& \sum_{k=1}^N \frac{({\rm sign}(b))^k}{k^{|b|}} S_{\vec{a}}(k),
a_i,b \in \mathbb{Z} \backslash \{0\}, N \in \mathbb{N}, S_\emptyset 
=1,
\end{eqnarray}
the generalized harmonic sums~\cite{Moch:2001zr,Ablinger:2013cf}
\begin{eqnarray}
S_{b,\vec{a}}(d,\vec{c})(N) &=& \sum_{k=1}^N \frac{d^k}{k^{|b|}} S_{\vec{a}}(\vec{c})(k),
a_i,b \in \mathbb{N} \backslash \{0\}, c_i,d \in \mathbb{C}(\sqrt{\eta}) \backslash \{0\}, N \in \mathbb{N}, 
S_\emptyset 
=1,
\end{eqnarray}
cyclotomic harmonic sums~\cite{Ablinger:2011te} or finite nested binomial 
sums~\cite{Ablinger:2014bra}.\footnote{For surveys on these quantities see e.g. \cite{REF1}.
Infinite nested binomial sums have also been considered in \cite{BinSums}.} Here the variable $d$ may depend on 
$\sqrt{\eta}$.
Furthermore, generalized harmonic- and cyclotomic-sums at infinity contribute. These 
can be expressed in terms of HPLs depending on $\eta$  in the argument with the help of \texttt{HarmonicSums}.

In Subsection~\ref{Sec:DefiniteSummation} the basic summation mechanism of simplification for such definite sums to indefinite nested sums is presented using the packages \texttt{Sigma} \cite{SIG1,SIG2} and \texttt{EvaluateMultiSums}. 
As it turns out, this general tactic is not sufficient for the explicitly given expressions: the expressions are 
scattered into too many pieces of sums and in the intermediate calculation steps for the individual sums, clumsy 
sums arise that cannot be handled properly with our summation tools. Therefore, 
we will utilize in Subsection~\ref{Sec:MergingSums} in addition the package \texttt{SumProduction} 
\cite{EMSSP}, which merges the input sums to appropriate forms that can be handled with our summation techniques.

\subsubsection{Definite summation tools}\label{Sec:DefiniteSummation}

\vspace*{1mm}
\noindent

In the following we present a survey of the crucial summation tools that assist in the calculation of an $\ep$-expansion for the triple sum~\eqref{Equ:SimpleTripleSum}. First, we compute the first coefficients of the $\ep$-expansion of the summand
$$f(\ep,\eta,N,j,i,k)=f_{-1}(\eta,N,j,i,k)\,\ep^{-1}+f_{0}(\eta,N,j,i,k)\,\ep^{0}+O(\ep),$$
with
$$f_{-1}=\frac{8 (2+k) (-2+N) (-1+N) N}{(-1+2 k) (1+2 k) (3+2 k) (1+N) (2+N) (3+N) (4+N)} \frac{(-1)^k {\eta}^{k} (i
-k
)! (-i
+j
+k
)!}{(1+i)! (1
-i
+j
)!}$$
and $f_0$ in terms of such factorials, the harmonic sums $S_1({i-k}),S_1({k}),S_1({-i+j+k}),S_1({N})$ and the 
cyclotomic harmonic sum
$$S_{\{2,1,1\}}({k})=\sum_{i=1}^k\frac{1}{2k+1}.$$
Next, we apply the summations over each coefficient, and get the $\ep$-expansion of the triple sum:
\begin{equation}\label{Equ:SumExpansionD7Instance}
T(\ep,\eta,N)=T_{-1}(\eta,N)\ep^{-1}+T_0(\eta,N)\ep^{0}+O(\ep^1),
\end{equation}
with
\begin{equation}\label{Equ:CoefficientSums}
T_r(\eta,N)=\sum_{j=0}^{N}\sum_{i=0}^j\sum_{k=0}^{i}f_{r}(\eta,N,j,i,k).
\end{equation}
One is now faced with the task of simplifying $T_{-1}$ and $T_0$, both being free of $\ep$. The simpler 
coefficient $T_{-1}(N)$ 
can now be simplified by the summation machinery of \texttt{Sigma}~\cite{SIG1,SIG2} based on 
difference ring theory~\cite{DRTheory}. Namely, one transforms from inside to outside the arising objects in~\eqref{Equ:CoefficientSums} to the desired indefinite nested sum form. E.g., for the innermost sum $h(\eta,N,j,i)=\sum_{k=0}^{i}f_{-1}(\eta,N,j,i,k)$ of $T_{-1}(\eta,N)$ we proceed as follows:
\begin{enumerate}
 \item[(1)] We compute a linear recurrence of $h(\eta,N,j,i)$ in $i$ of order $2$:
 \begin{multline}\label{Equ:RecForInnerSum}
 a_0(\eta,N,j,i) h(\eta,N,j,i)+a_1(\eta,N,j,i) h(\eta,N,j,i+1)\\
 +a_2(\eta,N,j,i) h(\eta,N,j,i+2)=r(\eta,N,j,i),
 \end{multline}
 with polynomial coefficients $a_0,a_2,a_2$ in $\eta,N,j,i$. The right hand side $r$ is given in terms of a  
linear combination of hypergeometric products depending on $\eta,N,j,i$. This machinery is based on the 
creative telescoping paradigm~\cite{AequalB} in the setting of difference rings~\cite{SIG1,SIG2,DRTheory}.
 \item[(2)] Next, we solve the found recurrence~\eqref{Equ:RecForInnerSum} in terms of indefinite nested 
sums~\cite{SIG1,SIG2,dAlembert}: we find $2$ linearly independent solutions of the homogeneous version of the recurrence and one particular solution of the recurrence itself. 
One of the most complicated indefinite nested sums w.r.t. $i$ is:
\begin{equation}\label{Equ:InnerSumSumExp}
\sum_{h=1}^i \frac{(-1)^h {\eta}^{h}}{\binom{2 h}{h} (-1
-2 h
+2 j
) h! (-3
-2 h
+2 j
)! (-h
+j
)_h}\sum_{k=1}^h \frac{(-1)^k {\eta}^{-k} k!}{(j
-k
)_k},
\end{equation}
where $(x)_k=x(x+1)\dots(x+k-1)$ denotes the Pochhammer symbol.

 \item[(3)] Finally, we compute the $2$ initial values $h(\eta,N,j,l)$ with $l=0,1$ and combine the solutions of the recurrence such that they match with the given initial values. This yields an alternative expression of the sum $h(\eta,N,j,i)$ where by construction the occurring objects are indefinite nested w.r.t.\ $i$. 
\end{enumerate}
For all three steps, 195 seconds are used and we obtain an expression where $5$ indefinite nested sums w.r.t.\ $i$ 
appear; one of them is~\eqref{Equ:InnerSumSumExp}. 
Now we apply the next summation quantifier $\sum_{i=0}^j$ to this expression and repeat the same machinery: 
compute a linear recurrence of this new sum w.r.t.\ the $j$ (which is the summation variable of the final sum), 
solve the recurrence in terms of indefinite nested sums w.r.t.\ $j$ and combine the solutions to find an alternative representation of the double sum which now is indefinite nested w.r.t.\ $j$. For this calculation step, 
1210 seconds are needed and  9 indefinite nested sums w.r.t.\ $j$ arose in the found representation. Finally, we repeat this 
once more in 295 seconds and obtain an expression of the single pole term
$T_{-1}(\eta,N)$
in terms of 19 indefinite nested sums w.r.t.\ $N$. Summarizing, we needed about 1700 seconds to transform the 
triple sum~\eqref{Equ:SimpleTripleSum} to an expression in terms of indefinite nested sums which turn out to be harmonic sums, generalized harmonic sums and generalized binomial sums, like, e.g., 
$$\sum_{h=1}^N 2^{-2 h} \big(
        1-{\eta}\big)^h \binom{2 h}{h} 
\sum_{k=1}^h \frac{2^{2 k}}{k^2 \binom{2 k}{k}}.$$
In general, this summation mechanism of recurrence finding and recurrence solving (see the summation steps 
(1)--(3) from above) has to be applied slightly more carefully:
\begin{itemize}
 \item If there are poles at the summation bounds (coming from the internal summation representations), the summation range has to be be refined and extra terms have to be treated by another round of our definite summation tools.
 \item If the summand is too large (e.g., more than 100 MB of size or composed by more than 300 indefinite nested sums), it is split into appropriate smaller parts.  Then the summation mechanism is applied separately to them, and the results are combined properly before the next summation quantifier is applied.
 \item In the case of infinite summation quantifiers (see, e.g., the second sum 
in~\eqref{Equ:Diag7TripleSumCases}), also 
limits have to be handled. Here asymptotic expansions are computed using the summation package \texttt{HarmonicSums}~\cite{HARMONICSUMS}.
\end{itemize}

\noindent
All these steps are skillfully combined within the package \texttt{EvaluateMultiSums}~\cite{EMSSP} using the 
difference ring machinery of \texttt{Sigma} and the special function algorithms of \texttt{HarmonicSums}. E.g., executing the command 

\footnotesize
$$\tt\bf EvaluateMultiSum[f_{-1},\{\{k,0,i\},\{i,0,j\},\{j,0,N\},\{N\},\{1\},SplitSums\to False]$$
\normalsize
the single pole term $T_{-1}(\eta,N)$ of~\eqref{Equ:SumExpansionD7Instance} 
is simplified to an expression in terms of indefinite nested sums as sketched above.

Performing all 9122 triple sums in this way (and ignoring double sums) indicates that already the single pole 
terms for all these sums require more than 180 days of calculation time. Calculating the constant term in this way 
seems hopeless. Besides, the following intrinsic 
problem arose: when we tried to calculate the constant term for our triple sum~\eqref{Equ:SimpleTripleSum}, we 
encountered internally definite sums that are not expressible in terms of indefinite nested sums\footnote{More precisely, we obtained recurrences of definite sums where not all solutions are expressible in terms of indefinite nested sums and where the found solutions cannot be combined accordingly.}.
However, such alien sums can be avoided by merging all these scattered sums as much as possible and 
treating them in one stroke. This observation will be utilized in the next subsection.

\subsubsection{The full tactic}\label{Sec:MergingSums}

\vspace*{1mm}
\noindent
In order to cure the problems mentioned at the end of the last subsection, we proceed by performing the following 
three steps.

\begin{description}
\item[Step I:] The arising sums are crushed to an optimal representation using \texttt{SumProduction}. In this way, one only obtains very few master sums that have to be treated.

\item[Step II:] These remaining multi-sums are then handled by \texttt{EvaluateMultiSums}, which uses \texttt{Sigma} and \texttt{HarmonicSums}.

\item[Step III:] The results of the master sums are combined to get the final result. Since the calculations of 
the master sums are carried out independently, the found indefinite nested sums between different master sums are not synchronized, i.e., many relations among them exist. Thus all relations among the arising sums are computed with \texttt{Sigma} and the final result is given in terms of indefinite nested sums that are all algebraically independent among each other. As a consequence, most of the arising sums vanish and a rather compact expression remains.
\end{description}

\noindent In the following we work out further details for the two most challenging diagrams: 
diagram 7 and diagram 11b, i.e. with the bubble-mass for the heaviest quark $m = m_2$. 

\medskip

\noindent\textbf{Details for Diagram 7}

\medskip

\noindent We proceed with the input expression of diagram~7 introduced already in 
Subsection~\ref{Sec:SummationTools}.

\medskip

\noindent Step~I: Using the package \texttt{SumProduction}~\cite{EMSSP} the 10262 sums are merged to the following six sums
\begin{align*}
F_1(\ep,\eta,N)&=\sum_{j=0}^{N-4}\sum_{i=0}^j\sum_{k=0}^{i}h_1(\ep,\eta,N,j,i,k),&\\
F_2(\ep,\eta,N)&=\sum_{j=0}^{N-4}\sum_{i=0}^jh_2(\ep,\eta,N,j,i),&
F_3(\ep,\eta,N)&=\sum_{j=0}^{N-3}h_3(\ep,\eta,N,j),\\
F_4(\ep,\eta,N)&=\sum_{j=0}^{N-4}\sum_{i=0}^j\sum_{k=0}^{\infty}h_4(\ep,\eta,N,j,i,k),\\
F_5(\ep,\eta,N)&=\sum_{j=0}^{N-4}\sum_{k=0}^{\infty}h_5(\ep,\eta,N,j,k),&
F_6(\ep,\eta,N)&=\sum_{k=0}^{\infty}h_5(\ep,\eta,N,k),
\end{align*}

\noindent
plus extra terms that only depend on hypergeometric products. More precisely, if the {\tt Mathematica} variable 
$\text{D7}$ contains the input expression of D7, the \texttt{SumProduction}-command

\footnotesize
\begin{multline*}
\tt\bf ReduceMultiSums[D7, \{N\}, \{1\}, \{\infty\},\\ 
\tt\bf MergeSummand \to True, 
 AlwaysMerge \to True, SynchronizeBounds \to True]
\end{multline*}
\normalsize
synchronizes the summation ranges, and maps the arising hypergeometric products ($\Gamma$-functions, factorials, 
Pochhammer symbols, powers) to products that are algebraically independent among each other. This computation required 37892 seconds (10.5 hours) and produced an alternative expression of diagram 7 with 156 GB size.

\noindent Next, we apply in Step~II our summation technologies to these 6 sums and compute the $\ep$-expansions up to 
the constant term in terms of indefinite nested sums, and combine in Step~III the found coefficients to obtain the complete $\ep$-expansion of diagram 7. Here the arising indefinite nested sums are algebraically independent among each other. To perform these last two steps, we needed 1 hour for the triple pole term, 2 hours for the double pole term,
2 days for the single pole term, and 20 days for the constant term.
Together with step (I), this amounts to 23 days of computation time to obtain the desired sum representation of diagram 7.

In the following we give some further details of Steps~II and~III for the computation of the constant term of the $\ep$-expansion. 

\medskip

\noindent Step~II: For instance, consider the sum $F_1$ whose sum  
requires $35.6$ GB memory; after expanding the summand in $\ep$, the constant term uses 47 MB of memory. 
Then activating the summation machinery from above to the given triple sum, one needs 605563 seconds (7 days) to 
obtain the constant term of the $\ep$-expansion of $F_1$. The result can be given in terms of $280$ indefinite 
nested sums. This information of $F_1$ and of the other sums $F_2,\dots,F_6$ can be also found in 
Table~\ref{fig:ReducedSumsDiag7}.

\begin{table}[ht]
\begin{center}
\begin{tabular}{|r|r|r|rr|r|}
\hline
sum & size of sum & summand size of & time of & & number of\\
&(with $\ep$)& constant term &calculation&&indefinite sums\\
\hline
\hline
$F_1$ & 35.6 MB & 47 MB &605563 s & (7 days) & 280\\
$F_2$ & 18.8 MB & 24.1 MB & 128207 s & (1.5 days) & 139\\
$F_3$ & 2.1 MB & 1.9 MB & 11190 s & (3.1 hours) & 51\\
\hline
$F_4$ & 62.0 MB & 767 MB &560604 s & (6.5 days) & 360\\
$F_5$ & 31.0 MB & 349 MB & 313111 s & (3.6 days) & 173\\
$F_6$ & 6.1 MB & 22.1 MB & 31825 s & (8.8 hours) & 40\\
\hline
\end{tabular}
\caption{\sf \small Summary of the calculation of the master sums for the constant term of diagram~7.}
\label{fig:ReducedSumsDiag7}
\end{center}
\end{table}
\noindent In summary, the total computation time for the simplification of all 6 sums required 19 days. 

\noindent Step~III: Combining the constant coefficients of all 6 master sums yields an expression using 23 MB memory consisting of 788 sums and products. Finally, we eliminate all algebraic relations 
among the arising sums. Namely, if the derived expression is given in the {\tt Mathematica} variable 
\texttt{unreducedExpr}, this job is carried out with the \texttt{Sigma} command

\footnotesize
$$\tt\bf SigmaReduceList[unreducedExpr,\{N\}].$$
\normalsize
In total, we needed 26 hours to rewrite the found expression in terms of only 46~basis sums that are all 
algebraically independent from each other. The algebraic independence follows by difference ring 
theory~\cite{DRTheory}; for a connection to the underlying quasi-shuffle algebras of the arising sums we refer also to~\cite{AS:18}.
As a consequence, 
the expression of 788 sums collapsed in the last step to an expression in terms of 46 sums that requires in total only 0.365 MB. Here one of the most complicated sums is
$$
\sum_{h=1}^N 2^{-2 h}(
        1-{\eta})^h \binom{2 h}{h} \Big(
        \sum_{i=1}^h \tfrac{2^{2 i}(
                1-{\eta})^{-i}}{i \binom{2 i}{i}}
\Big)
\Big(
        \sum_{i=1}^h \tfrac{(
                1-{\eta})^i \binom{2 i}{i}}{2^{2 i}}\Big) 
\Bigg(\sum_{i=1}^h \tfrac{2^{2 i} (
        1-{\eta})^{-i}}{i \binom{2 i}{i}}\sum_{j=1}^i \tfrac{1}{j}\sum_{k=1}^j\tfrac{(
        1-{\eta})^k}{k}\Bigg).$$

\medskip

\noindent\textbf{Details for Diagram 11b}

\medskip
        
\noindent After carrying out the transformations of Subsection~\ref{sec:41}, diagram 11b can be represented by an 
expression in terms of 14865 sums that requires in total 95 MB of memory. More precisely, the expression consists 
of 150 single sums, 1000 double sums, 12160 triple sums and 1555 quadruple sums.

\noindent Step~I: We utilize first the package \texttt{SumProduction}  and crunch in 8640 seconds the expression to an expression of 377 MB size consisting only of 8 sums where the summation ranges are given in the first column 
in Table~\ref{fig:ReducedSumsDiag11b}.
        
\begin{table}[ht]
\begin{center}
\begin{tabular}{|r|r|r|rr|r|}
\hline
sum & size of sum & summand size of & time of & & number of\\
&(with $\ep$)& constant term &calculation&&indefinite sums\\
\hline
\hline
{\tiny$\displaystyle\sum_{i_4=2}^{N-3}\sum_{i_3=0}^{i_4-2}\sum_{i_2=0}^{i_3}\sum_{i_1=0}^{\infty}$} &  17.7 MB & 266.3 MB MB & 177529 s & (2.1 days) & 1188\\
{\tiny$\displaystyle\sum_{i_3=3}^{N-4}\sum_{i_2=0}^{i_3-1}\sum_{i_1=0}^{\infty}$} & 232 MB &  1646.4 MB &980756 s & (11.4 days) & 747\\
{\tiny$\displaystyle\sum_{i_2=3}^{N-4}\sum_{i_1=0}^{\infty}$}   & 67.7 MB & 458 MB & 524485 s & (6.1 days) & 557\\
{\tiny$\displaystyle\sum_{i_1=0}^{\infty}$} & 38.2 MB & 90.5 MB &689100 s & (8.0 days) & 44\\
\hline
{\tiny$\displaystyle\sum_{i_4=2}^{N-3}\sum_{i_3=0}^{i_4-2}\sum_{i_2=0}^{i_3}\sum_{i_1=0}^{i_2}$} & 1.3 MB & 6.5 MB & 305718 s & (3.5 days) & 1933\\
{\tiny$\displaystyle\sum_{i_3=3}^{N-4}\sum_{i_2=0}^{i_3-1}\sum_{i_1=0}^{i_2}$} & 11.6 MB & 32.4 MB & 710576 s & (8.2 days) & 621\\
{\tiny$\displaystyle\sum_{i_2=3}^{N-4}\sum_{i_1=0}^{i_2}$}   & 4.5 MB & 5.5 MB & 435640 s  & (5.0 days) & 536\\
{\tiny$\displaystyle\sum_{i_1=3}^{N-4}$} & 0.7 MB & 1.3 MB & 9017s & (2.5 hours) & 68\\
\hline
\end{tabular}
\caption{\sf \small Summary of the calculation of the master sums for the constant term diagram~11b.}
\label{fig:ReducedSumsDiag11b}
\end{center}
\end{table}

Next, we calculate the $\ep$-expansions for each of the 8 sums and combine the results to get the $\ep$-expansion 
of diagram 11b itself. For the triple pole term, this amounts to 89 minutes, for the double pole term to 19 hours, for the single pole term to 6.9 days, and for the constant term  to 77.7 days. In the following some extra information is given for the calculation of the constant term. 

\medskip

\noindent Step~II: In Table~\ref{fig:ReducedSumsDiag11b} further details are given for the treatment of the 8 
multi-sums. E.g., for the quadruple sum involving one infinite sum (first row) the input summand uses 17.7 MB of memory. After its expansion the constant term requires 266 MB of memory and the total computation time to transform this sum in terms of indefinite nested sums requires 2.1 days (using the package \texttt{EvaluateMultiSums} utilizing \texttt{Sigma} and \texttt{HarmonicSums}). Let us be even a bit more specific: carrying out the infinite sum requires 32.1 hours and leads to an expression of 8.6 MB size in terms of 41 indefinite nested sums. Carrying out the next sum quantifier 
$\sum_{i_2=0}^{i_3}$ needs 37 minutes and leads to an expression of size 3.4 MB in terms of 64 indefinite sums. 
Dealing with the summation sign $\sum_{i_3=0}^{i_4-2}$ produces in 18.9 hours an expression of size 4.1 MB in terms of 222 indefinite nested sums, and finally, processing the last summation quantifier 
$\sum_{i_4=2}^{N-3}$ produces in 18.6 hours the final result of 
15.2 MB size in terms of 1188 indefinite nested sums\footnote{While processing the last
summation quantifier, we skipped the task to eliminate algebraic relations among the derived 1188 sums. This
challenge will be shifted to Step~3 of our general procedure.}.

\medskip

\noindent Step~III: Combining the constant coefficients of all 8 master sums yields an expression using 154 MB of 
memory consisting of 4110 sums and products. Finally, the elimination of all algebraic relations among the arising 
sums needed 32.5 days and yield a compact expression in terms of 74~sums/products that requires in total only 8.3~MB of memory. 

\medskip

We remark that during the calculations of diagram 7 and in particular of diagram 11b the hardest calculations arose that the summation packages \texttt{Sigma} and \texttt{EvaluateMultiSum} have faced so far. Various sub-algorithms and sub-routines had to be improved and optimized in order to compute recurrences for such gigantic summands and to solve the found recurrences efficiently in terms of indefinite nested sums and products. In particular, the elimination of algebraic relations among the arising sums where pushed to the limit: the underling tower of difference rings contained up to 1500 extension variables (so-called $R\Pi\Sigma$-extensions~\cite{DRTheory}) and the underlying algorithms were heavily optimized to work with them efficiently. 
\subsection{Deriving the \boldmath $z$-space solution}
\label{sec:52}

\vspace*{1mm}
\noindent
The general method to go from the $N$-space to the $z$-space is elaborated in \cite{InvMellin1,InvMellin2}. The 
main idea is to find a recurrence in $N$ for the quantity under consideration and from that to derive a differential equation for 
the solution in $z$-space which can be afterwards solved. In the frame of the current project, we used an improved 
version of the method presented in \cite{InvMellin1,InvMellin2}, which we will sketch in the following.
A detailed description of this enhanced method will be given in \cite{InvMellin3}.

For a given nested sum of the form
\begin{eqnarray}\label{invmelsum}
 F(N):=F_0(N)\sum_{i_1=1}^N F_1(i_1)\sum_{i_2=1}^{i_1} F_2(i_2)\cdots \sum_{i_k=1}^{i_{k-1}} F_k(i_k)
\end{eqnarray}
we look for a representation in the form
\begin{eqnarray}\label{melsum}
  G(N)=\sum_{j=0}^k v_j^N \int_0^1 dx (x^N-a_j^N) \sum_{i=1}^{b_j}d_{i,j}f_{i,j}(x)
\end{eqnarray}
such that $F(N)=G(N)$ for all $N\in \set \N$ with $N>N_0$ for some $N_0\in \set N$,
where in our cases $v_j,\ a_j,\ d_{i,j}\in \set K(\eta)$ and $f_{i,j}(z)$ are expressions of the form 
$$
p(\eta,z)\;g(z)
$$ with $p(\eta,z)\in \set K(\eta)(z)$ and $g(z)$ is an iterated integral. 
For $0\leq j \leq k$ we define
\begin{eqnarray}
\bar{F}_j(N):=F_0(N)\sum_{i_1=1}^N F_1(i_1)\sum_{i_2=1}^{i_1} F_2(i_2)\cdots \sum_{i_j=1}^{i_{j-1}} F_j(i_j).
\end{eqnarray}
Hence for example $\bar{F}_k(N)=F(N),\ \bar{F}_{k-1}(N)$ is the original sum with the innermost summation quantifier dropped, $\bar{F}_{1}(N)=F_0(N)\sum_{i_1=1}^N F_1(i_1)$ and $\bar{F}_{0}(N)=F_0(N).$

Note that for the sums under consideration the $v_j$ from (\ref{melsum}) can be read off from $F(N)$, and that it 
is straightforward to find recurrences $R_j$ (with shifts in $N$) such that $v_j^{-N}\bar{F}_j(N)$ is a solution of $R_j.$

After deriving such recurrences $R_j,$ we can use the algorithm from \cite{InvMellin1} to derive a 
differential 
equations $D_j$ (with differentiation in $z$) for the inverse Mellin transforms of $v_j^{-N}\bar{F}_j(N).$ 
The $f_{i,j}(z)$ are precisely the solutions of the differential equations $D_j.$
Hence, after solving the differential equations it remains to fix the $d_{i,j}$ by checking a sufficient amount of 
initial values.

This method is implemented in \HarmonicSumsP\ and with the help of the \HarmonicSumsP-command

\footnotesize
$$\tt\bf GeneralInvMellin[Expr,N,x,Method\to2,Assumptions\to0<\eta<1],$$
\normalsize

\noindent we find for instance:
\begin{eqnarray}
4^{-N} \binom{2 N}{N} 
\sum_{\tau=1}^N \frac{4^{\tau} \big(
        \frac{1}{1-\eta }\big)^{\tau} \big(
        \tau!\big)^2}{\big(
        2 \tau \big)! \tau}
&=&
-\frac{2 \left(\sqrt{\eta }+\eta ^{3/2}+4 {G}\left(\sqrt{1-\eta -\tau } \sqrt{-\tau };1\right)
\right)}{(1-\eta )^2 \sqrt{\eta } \pi } \int_0^1 \frac{x^N}{\sqrt{1-x} \sqrt{x}}  dx
\nonumber\\
&& -\frac{(1-\eta )^{-N}}{\sqrt{\eta }}\int_0^1 \frac{x^N}{\sqrt{1-\eta -x} 
\sqrt{-x}} \, dx,
\end{eqnarray}

\begin{eqnarray}
\sum_{i_1=1}^N \frac{2^{-2 i_1} \binom{2 i_1}{i_1} 
\sum_{i_2=1}^{i_1} \frac{2^{2 i_2} 
\sum_{i_3=1}^{i_2} \frac{1}{i_3}}{\binom{2 i_2}{i_2} i_2^2}}{i_1}&=&\nonumber\\
&&\hspace{-6cm}\int_0^1dx\frac{(x^n-1)\sqrt{x} (4 x-2)}{\sqrt{1-x}}
\biggl(
\pi  \log (2)-8\; {G}\left({\scriptstyle \sqrt{1-\tau } \sqrt{\tau },\frac{1}{1-\tau }};x\right)-4\; 
{G}\left({\scriptstyle \sqrt{1-\tau } \sqrt{\tau }};x\right)-\frac{7 \zeta (3)}{2 \pi }
\biggr)+\nonumber\\
&&\hspace{-6cm}\int_0^1dx\frac{x^n-1}{1-x}\biggl(
\frac{-21 x^2+32 x^3-18 x^4}{12}+8\; {G}\left({\scriptstyle\sqrt{1-\tau } 
\sqrt{\tau },\sqrt{1-\tau } \sqrt{\tau }};x\right)+16\; {G}\left({\scriptstyle\sqrt{1-\tau } 
\sqrt{\tau },\sqrt{1-\tau } \sqrt{\tau },\frac{1}{1-\tau }};x\right)\nonumber\\
&&\hspace{-5cm}+\frac{(x-5 x^2+8 x^3-4 x^4){G}\left({\scriptstyle\frac{1}{1-\tau }};x\right)}{2}-\frac{\; 
{G}\left({\scriptstyle\sqrt{1-\tau } \sqrt{\tau }};x\right)(2 \pi ^2  \log (2)-7 \zeta (3))}{\pi }
\biggr).
\end{eqnarray}
Here $G$ denotes an iterated integral defined in (\ref{eq:GG}).
In total we computed the inverse Mellin transforms for about 50 sums using this method, which took around 2 hours on a standard desktop PC. The following sum is one of the most complicated sums we had to consider:
\begin{eqnarray}
\sum_{\tau_1=1}^N \frac{4^{-\tau_1} \binom{2 \tau_1}{\tau_1} 
\displaystyle{\sum_{\tau_2=1}^{\tau_1} \frac{4^{\tau_2} \big(
        \frac{\eta }{\eta -1}\big)^{\tau_2} 
\displaystyle{\sum_{\tau_3=1}^{\tau_2} \frac{\big(
        \frac{\eta -1}{\eta }\big)^{\tau_3} 
\displaystyle{\sum_{\tau_4=1}^{\tau_3} \frac{1}{\tau_4}}}{\tau_3}}}{\tau_2}}}{\tau_1}.
\end{eqnarray}
Note that similar to the final representation of Ref.~\cite{Ablinger:2017xml}, we do not include all polynomial 
pre-factors in $N$ 
but leave these to be included by a Mellin convolution.
In this way the inner generalized iterated integrals can be evaluated as HPLs with involved arguments.
\section{An Explanatory Example}
\label{sec:5}

\vspace*{1mm}
\noindent
In this section we want to illustrate the computational steps in more detail considering diagram~2 
of 
Figure~\ref{fig:GGdiagrams}.
The small numerator structure of this diagram allows to present the calculation in detail.
Since here the $\eta$ and $N$ structures do not factorize, they give rise to more involved structures 
compared to the single mass case.

After inserting the Feynman rules, applying the gluonic projector, performing the Dirac-algebra and combining 
the denominators via Feynman parameters, one obtains
\begin{eqnarray}
    D_2^{A(B)} &=& - C_A T_F^2 \frac{1+(-1)^N}{2} \frac{a_s^3}{(4\pi)^{3\ep/2}} \frac{64}{2+\ep} \frac{1}{2\pi i}\Bigl[ (10+4\ep) J_1^{A(B)} (N-1) + (2+\ep) J_1(N)^{A(B)} 
\NN \\ &&
- 4(3+\ep) J_2^{A(B)} (N-1) + 4(2+\ep) J_2^{A(B)} (N) + 2(5+2\ep) J_2^{A(B)}(N-2) 
\NN \\ &&
+ 2 J_3^{A(B)} (N-1) - (2+\ep) J_3^{A(B)} (N-2) \Bigr]~,
\end{eqnarray}
where $A$($B$) represent different mass assignments.
We normalize the functions $J_i$ according to
\begin{eqnarray}
J_i^{A(B)} (n) &=& \left( \frac{m_1^2}{\mu^2} \right)^{\tfrac{3\ep}{2}} j_i^{A(B)}(n)~.
\end{eqnarray}
In the following we use the short hand notation
\begin{eqnarray}
\Gamma \begin{bmatrix} a_1^{b_1}, a_2^{b_2}, \dots \\ c_1^{d_1}, c_2^{d_2}, \dots \end{bmatrix} \equiv 
\frac{\Gamma^{b_1}(a_1) \Gamma^{b_2}(a_2) \dots}{\Gamma^{d_1}(c_1)\Gamma^{d_2}(c_2) \dots}~.
\end{eqnarray}

\subsection{The \boldmath $N$-space solution}
\label{sec:51}

\vspace*{1mm}
\noindent
The functions $J_1$ to $J_3$ are given by the following functions 
\begin{eqnarray}
j_1^A (n) &=&  \int\limits_{-i\infty}^{+i\infty} \mathrm{d} \sigma \, \eta^\sigma \, \Gamma \begin{bmatrix} -\sig, \sig-\tfrac{3\ep}{2},(2+\tfrac{\ep}{2}-\sig)^2, (2-\ep+\sig)^2,\ep-\sig,n-\tfrac{\ep}{2}+\sig \\ 4+\ep-2\sig,4-2\ep+2\sig,2+\ep+n \end{bmatrix} , 
\NN \\
j_1^B (n) &=&  \int\limits_{-i\infty}^{+i\infty} \mathrm{d} \sigma \, \eta^\sigma \, \Gamma \begin{bmatrix} -\sig, \sig -\tfrac{3\ep}{2} , \sig -\tfrac{\ep}{2} , (2-\ep+\sig)^2,(2+\tfrac{\ep}{2}-\sig)^2,n+\ep-\sig \\ 2+n+\tfrac{\ep}{2},4+\ep-2\sig,4-2\ep+2\sig \end{bmatrix}, \NN \\
j_2^A (n) &=& \int\limits_{-i\infty}^{+i\infty} \mathrm{d} \sigma \, \eta^\sigma \, \Gamma \begin{bmatrix} -\sig , \sig -\tfrac{3\ep}{2} , \ep - \sig , 1 -\tfrac{\ep}{2} + n + \sig , \sig -\tfrac{\ep}{2} , (2-\ep+\sig)^2 , (2 + \tfrac{\ep}{2} - \sig)^2 \\ 4 + \ep - 2 \sig , 4 - 2\ep + 2 \sig, 1-\tfrac{\ep}{2}+\sig, \tfrac{\ep}{2}+2+n \end{bmatrix}, 
\NN \\
j_2^B (n) &=&  \int\limits_{-i\infty}^{+i\infty} \mathrm{d} \sigma \, \eta^\sigma \, \Gamma \begin{bmatrix} -\sig, \sig -\tfrac{3\ep}{2} , \sig -\tfrac{\ep}{2}, (2-\ep-\sig)^2,(2+\tfrac{\ep}{2}-\sig)^2,\ep-\sig,1+n+\ep-\sig \\ 1+\ep-\sig,2+n+\tfrac{\ep}{2},4+\ep-2\sig,4-2\ep+2\sig \end{bmatrix}, 
\NN \\
j_3^A (n) &=&  \int\limits_{-i\infty}^{+i\infty} \mathrm{d} \sigma \, \eta^\sigma \, 
\Gamma \begin{bmatrix} -\sig, \sig -\tfrac{3\ep}{2},\ep-\sig,(\tfrac{\ep}{2}+2-\sig)^2,(2-\ep+\sig)^2,
\sig -\tfrac{\ep}{2},2-\tfrac{\ep}{2}+n+\sig \\ \ep+4-2\sig, -2\ep+4+2\sig,\tfrac{\ep}{2}+2+n, 2 - \tfrac{\ep}{2} + 
\sigma  \end{bmatrix}, 
\NN \\
j_3^B (n) &=& \int\limits_{-i\infty}^{+i\infty} \mathrm{d} \sigma \, \eta^\sigma \, \Gamma \begin{bmatrix} -\sig,\sig-\tfrac{3\ep}{2},\sig-\tfrac{\ep}{2},(2-\ep+\sig)^2,(2+\tfrac{\ep}{2}-\sig)^2,\ep-\sig,2+n+\ep-\sig \\ 2+\ep-\sig,2+n+\tfrac{\ep}{2},4+\ep-2\sig,4-2\ep+2\sig \end{bmatrix}.
\NN \\
\\ \NN
\end{eqnarray}

The contour integrals are evaluated by taking residues at the ascending poles, which are added up. One obtains
\begin{eqnarray}
J_1^A (n) = \left( \frac{m_1^2}{\mu^2} \right)^{3\ep/2} \sum\limits_{k=0}^\infty \, \eta^k \, \left( T_{1,1}(n) 
+ T_{1,2}(n) + T_{1,3}(n) \right), \\
J_2^A (n) = \left( \frac{m_1^2}{\mu^2} \right)^{3\ep/2} \sum\limits_{k=0}^\infty \, \eta^k \, \left( T_{2,1}(n) 
+ T_{2,2}(n) + T_{2,3}(n) \right), \\
J_3^A (n) = \left( \frac{m_1^2}{\mu^2} \right)^{3\ep/2} \sum\limits_{k=0}^\infty \, \eta^k \, \left( T_{3,1}(n) 
+ T_{3,2}(n) + T_{3,3}(n) \right),
\end{eqnarray}
where $T_{i,1}$ follows from the residue at $\sig = k $, $T_{i,2}$ from the residue at 
$\sig = \ep + k$ and $T_{i,3}$ from the residue at $\sig = \ep/2 + 2 + k $.  The explicit expressions read
\begin{eqnarray}
T_{1,1}(n) &=& \frac{2^\ep \pi}{64} \, \Gamma \begin{bmatrix} -\tfrac{\ep}{2}-2,-\ep, 
\tfrac{\ep}{2}+3,\ep+1,k-\tfrac{3\ep}{2},2-\ep+k, k-\tfrac{\ep}{2}-\tfrac{3}{2}, n-\tfrac{\ep}{2}+k \\ 
-\tfrac{\ep}{2}-\tfrac{5}{2}, \tfrac{\ep}{2}+\tfrac{7}{2},\tfrac{\ep}{2}+n+2,1+k,1-\ep+k,\tfrac{5}{2}-\ep+k,
k-\tfrac{\ep}{2}-1 \end{bmatrix},
\\ 
T_{1,2}(n) &=& \frac{2^{\ep} \pi \eta^\ep}{64} \, \Gamma \begin{bmatrix} 3-\tfrac{\ep}{2},1-\ep, 
\tfrac{\ep}{2}-2,\ep,2+k,k-\tfrac{\ep}{2}, k+\tfrac{\ep}{2}-\tfrac{3}{2}, \tfrac{\ep}{2}+n+k \\ 
\tfrac{7}{2}-\tfrac{\ep}{2}, \tfrac{\ep}{2}-\tfrac{5}{2},\tfrac{\ep}{2}+n+2,1+k,\tfrac{5}{2}+k,\ep+1+k,\tfrac{\ep}{2}-1+k \end{bmatrix},
\\
T_{1,3}(n) &=& -\frac{2^{\ep} \eta^{\tfrac{\ep}{2}+2} }{64} \, \Gamma \begin{bmatrix} 
-\tfrac{\ep}{2}-1,2-\tfrac{\ep}{2},\tfrac{\ep}{2}-1,\tfrac{\ep}{2}+2,\tfrac{1}{2}+k,2+n+k,2-\ep+k,4-\tfrac{\ep}{2}+k \\ \tfrac{\ep}{2}+2+n , 1+k, 3-\tfrac{\ep}{2}+k,\tfrac{9}{2}-\tfrac{\ep}{2}+k,\tfrac{\ep}{2}+3+k\end{bmatrix},
\nonumber\\
\\
T_{2,1}(n) &=& \frac{2^\ep \pi}{64} \, \Gamma \begin{bmatrix} -2-\tfrac{\ep}{2},3+\tfrac{\ep}{2},-\ep,1+\ep \\ -\tfrac{5}{2}-\tfrac{\ep}{2},\tfrac{7}{2}+\tfrac{\ep}{2},2+n+\tfrac{\ep}{2} \end{bmatrix} \NN \\ && \times
\Gamma \begin{bmatrix}k-\tfrac{3\ep}{2},2-\ep+k,k-\tfrac{3}{2}-\tfrac{\ep}{2},k-\tfrac{\ep}{2},1+n-\tfrac{\ep}{2}+k \\ 1+k,1-\ep+k,\tfrac{5}{2}-\ep+k,k-1-\tfrac{\ep}{2},1-\tfrac{\ep}{2}+k \end{bmatrix},
\\ 
T_{2,2}(n) &=& \frac{2^{\ep} \pi \eta^\ep}{64} \, \Gamma \begin{bmatrix} 1-\ep,3-\tfrac{\ep}{2},\tfrac{\ep}{2}-2,\ep \\ \tfrac{7}{2}-\tfrac{\ep}{2},\tfrac{\ep}{2}-\tfrac{5}{2},2+n+\tfrac{\ep}{2} \end{bmatrix} \NN \\ && \times
\Gamma \begin{bmatrix} 2+k,k-\tfrac{\ep}{2},\tfrac{\ep}{2}-\tfrac{3}{2}+k,\tfrac{\ep}{2}+k,1+n+\tfrac{\ep}{2} + k 
\\ 1+k,\tfrac{5}{2}+k,\tfrac{\ep}{2}-1+k,1+\tfrac{\ep}{2}+k,1+\ep+k \end{bmatrix},
\\
T_{2,3}(n) &=& -\frac{2^{\ep} \eta^{\tfrac{\ep}{2}+2} }{64} \, \Gamma \begin{bmatrix} -1-\tfrac{\ep}{2},
2-\tfrac{\ep}{2},2+\tfrac{\ep}{2}, -1 + \tfrac{\ep}{2} \\ 2+n+\tfrac{\ep}{2} \end{bmatrix} \NN \\ &&
\times \Gamma \begin{bmatrix} \tfrac{1}{2}+k,2+k,3+n+k,2-\ep+k,4-\tfrac{\ep}{2}+k \\ 
1+k,3+k,3-\tfrac{\ep}{2}+k,\tfrac{9}{2}-\tfrac{\ep}{2}+k,3+\tfrac{\ep}{2}+k \end{bmatrix},
\\
T_{3,1}(n) &=& \frac{2^\ep \pi}{64} \, \Gamma \begin{bmatrix} -2-\tfrac{\ep}{2},3+\tfrac{\ep}{2},-\ep,1+\ep \\  -\tfrac{5}{2}-\tfrac{\ep}{2},\tfrac{7}{2}+\tfrac{\ep}{2},2+n+\tfrac{\ep}{2}\end{bmatrix} \NN \\ && \times
\Gamma \begin{bmatrix} k-\tfrac{3\ep}{2},2-\ep+k,k-\tfrac{3}{2}-\tfrac{\ep}{2},k-\tfrac{\ep}{2},2+n-\tfrac{\ep}{2}+k \\1+k,1-\ep+k,\tfrac{5}{2}-\ep+k,k-1-\tfrac{\ep}{2},2-\tfrac{\ep}{2}+k \end{bmatrix},
\\ 
T_{3,2}(n) &=& \frac{2^{\ep} \pi \eta^\ep}{64} \, \Gamma \begin{bmatrix} 1-\ep,3-\tfrac{\ep}{2},\tfrac{\ep}{2}-2,\ep \\ \tfrac{7}{2}-\tfrac{\ep}{2},\tfrac{\ep}{2}-\tfrac{5}{2},2+n+\tfrac{\ep}{2} \end{bmatrix} \NN \\ && \times 
\Gamma \begin{bmatrix} 2+k,k-\tfrac{\ep}{2},\tfrac{\ep}{2}-\tfrac{3}{2}+k,\tfrac{\ep}{2}+k,2+n+\tfrac{\ep}{2}+k \\ 1+k,\tfrac{5}{2}+k,\tfrac{\ep}{2}-1+k,2+\tfrac{\ep}{2}+k,1+\ep+k \end{bmatrix},
\\
T_{3,3}(n) &=& -\frac{2^{\ep} \eta^{\tfrac{\ep}{2}+2} }{64} \, \Gamma \begin{bmatrix} -1-\tfrac{\ep}{2},2-\tfrac{\ep}{2},\tfrac{\ep}{2}-1,2+\tfrac{\ep}{2} \\ 2+n+\tfrac{\ep}{2} \end{bmatrix} \NN \\ && \times
\Gamma \begin{bmatrix} \tfrac{1}{2}+k,2+k,4+n+k,2-\ep+k,4-\tfrac{\ep}{2}+k \\ 1+k,4+k,3-\tfrac{\ep}{2}+k,\tfrac{9}{2}-\tfrac{\ep}{2}+k,3+\tfrac{\ep}{2}+k \end{bmatrix}.
\end{eqnarray}
Here we applied Legendre's duplication
\begin{eqnarray}
\Gamma ( s + 2 k ) &=& \frac{ (-1)^{1-s-2k} }{ \pi } \Gamma \left( k + \frac{s}{2} \right) \Gamma \left( k + 
\frac{s}{2} + \frac{1}{2} \right),
\end{eqnarray}
and Euler's reflection formula
\begin{eqnarray}
\Gamma ( s - k ) &=& (-1)^{k-1} \frac{ \Gamma ( -s ) \Gamma ( s + 1 ) }{ \Gamma ( k + 1 - s) } , \ \text{for} \ k\in\mathbb{N} \ \text{and} \ s \notin \mathbb{Z}
\end{eqnarray}
to the $\Gamma$-ratios. 

The expressions for $J_i^B$ look similar.
In the following we concentrate on the calculation of $D_2^A$.
It is worth mentioning, however, that care is needed at taking the residues for the other mass assignment.
Here structures like
\begin{eqnarray}
\frac{\Gamma(\ep-\sig)\Gamma(2+n+\ep-\sig)}{\Gamma(2+\ep-\sig)}
\end{eqnarray}
develop residues at isolated boundary points, \ie in this example the residues at $\sig=\ep,1+\ep$ have to be 
treated differently than the ones at $\sig=2+n+\ep+k$ with $k \in \mathbb{N}$.
Therefore, the final representation for $D_2^B$ does not only contain sums but also terms from separately 
taken residues.  

In Mellin $N$-space we use harmonic sums \cite{HSUM} and generalized harmonic sums 
\cite{Moch:2001zr,Ablinger:2013cf} to represent the result. 
In $z$-space the corresponding functions are harmonic polylogarithms $\HA_{\vec{a}}(z)$ 
\cite{Remiddi:1999ew} and 
generalized iterated integrals, $G[\{\vec{b}\},z]$ over alphabets of the kind discussed 
in \cite{Ablinger:2014bra}, 
which we find algorithmically \cite{Ablinger:2013cf,Ablinger:2014bra}, 
and special values thereof. The sum representation, moreover, also contains harmonic polylogarithms of the 
mass ratio $\sqrt{\eta}$.
\begin{eqnarray}
\label{eq:GG}
G\left[\left\{g(x),\vec{h}(x)\right\},z\right] &=& \int_0^z dy g(y) 
G\left[\left\{\vec{h}(x)\right\},y\right].
\end{eqnarray}
Here the functions $g_i,h$ are arbitrary functions for which the respective integral (\ref{eq:GG}) exists.

The full expression for $D_2^A$ can now be handled with \texttt{SumProduction}, \texttt{EvaluateMultiSums}, \texttt{Sigma} and \texttt{HarmonicSums}.
For the complete diagram we obtain
\begin{eqnarray}
D_2^A &=& 
C_A T_F^2 \frac{1+(-1)^N}{2} a_s^3 S_\ep^3 \left( \frac{m_1^2}{\mu^2} \right)^{\tfrac{3\ep}{2}}\Biggl\{ 
\frac{256 P_8}{27 \ep^3 (N-1) N (N+1)}
+\frac{1}{\ep^2}
\biggl[
\frac{64 P_ 4}{81 (N-1)^2 N^2 (N+1)^2}
\nonumber \\ &&
+\frac{64 P_8}{9 (N-1) N (N+1)} \HA_0 (\eta )
-\frac{64 P_8}{27 (N-1) N (N+1)} S_1
\biggr]
+\frac{1}{\ep}
\biggl[
\frac{32 P_ 6}{81 (N-1)^3 N^3 (N+1)^3}
\nonumber \\ &&
+\frac{32 P_ 3}{27 (N-1)^2 N^2 (N+1)^2} \HA_0(\eta )
+\frac{32 P_8}{9 (N-1) N (N+1)} \HA_0^2(\eta )
\nonumber \\ &&
-\frac{32 P_ 5}{81 (N-1)^2 N^2 (N+1)^2} S_1
+\frac{32 P_8}{27 (N-1) N (N+1)} S_1^2
+\frac{32 P_8}{9 (N-1) N (N+1)} \zeta_2
\biggr]
\nonumber \\ &&
-\frac{8 P_{11}}{729 (N-1)^4 N^4 (N+1)^4 (2 N - 5) (2 N - 3) (2 N - 1) \eta }
\nonumber \\ &&
+ \frac{2 P_ 7 (1-\eta)^{-N} }{27 (N-1)^2 N^2 (N+1) (2 N - 5) (2 N - 3) (2 N - 1) \eta }
\biggl(
        \frac{1}{2} \HA_0(\eta )^2
\nonumber \\ &&
        + H_ 0(\eta ) S_1(1-\eta,N)
        - S_ 2(1-\eta,N)
        + S_{1,1}(1-\eta ,1,N)
\biggr)
\nonumber \\ &&
-\frac{4 P_ 9}{27 (N-1)^3 N^3 (N+1)^3 (2 N - 5) (2 N - 3) (2 N - 1) \eta } \HA_0(\eta )
\nonumber \\ &&
+\frac{8 P_ 2}{27 (N-1)^2 N^2 (N+1)^2} \HA_0^2(\eta )
+\frac{32 P_8}{27 (N-1) N (N+1)} \HA_0^3(\eta )
\nonumber \\ &&
-\frac{8 P_8}{9 (N-1) N (N+1)}  \HA_0^2(\eta ) \HA_1(\eta )
+\frac{16 P_8}{9 (N-1) N (N+1)} \HA_0(\eta ) \HA_{0,1}(\eta )
\nonumber \\ &&
-\frac{16 P_8}{9 (N-1) N (N+1)} \HA_{0,0,1}(\eta )
+\biggl(
	\frac{8 P_8}{9 (N-1) N (N+1)} \HA_0^2(\eta )
\nonumber \\ &&
	-\frac{16 P_ 1}{9 (N-1)^2 N^2 (N+1)} \HA_0(\eta )
        -\frac{4 P_{10}}{81 (N-1)^3 N^3 (N+1)^3 (2N-5) (2N-3) (2N-1) \eta }
\nonumber \\ &&
        -\frac{8 P_8}{9 (N-1) N (N+1)} S_2
\biggr) S_1
+\biggl(
        \frac{8 P_ 4}{81 (N-1)^2 N^2 (N+1)^2}
\nonumber \\ &&
        +\frac{8 P_8}{9 (N-1) N (N+1)} \HA_0(\eta )
\biggr) S_1^2
-\frac{8 P_8}{81 (N-1) N (N+1)} S_1^3
+\biggl(
        \frac{8 P_ 1}{9 (N-1)^2 N^2 (N+1)}
\nonumber \\ &&
        +\frac{8 P_8}{9 (N-1) N (N+1)} \HA_0(\eta )
\biggr) S_2
-\frac{112 P_8}{81 (N-1) N (N+1)} S_3
\nonumber \\ &&
-\frac{16 P_8}{9 (N-1) N (N+1)} 
\biggl(
        \frac{1}{2} \HA_0^2(\eta )
        + S_{1,1}(1-\eta ,1,N)
\biggr) S_ 1\big(\frac{1}{1-\eta },N\big)
\nonumber \\ &&
-\frac{16 P_8}{9 (N-1) N (N+1)}
\biggl(
	\HA_0(\eta ) S_{1,1}\big(\frac{1}{1-\eta },1-\eta ,N\big)
	- S_{1,2}\big(\frac{1}{1-\eta },1-\eta ,N\big)
\nonumber \\ &&
	+ S_{1,2}\big(1-\eta ,\frac{1}{1-\eta },N\big)
	- S_{1,1,1}\big(1-\eta ,1,\frac{1}{1-\eta },N\big)
	- S_{1,1,1}\big(1-\eta ,\frac{1}{1-\eta },1,N\big)
\biggr)
\nonumber \\ &&
-\frac{4^{-N} P_{12}}{54 \eta^{3/2} (N+1) (2N-5) (2N-3) (2N-1)} \binom{2 N}{N}
\biggl(
        \HA_0^2(\eta ) \bigl[ \HA_{-1}\big(\sqrt{\eta }\big) + \HA_1\big(\sqrt{\eta }\big) \bigr]
\nonumber \\ &&
        -  4 \HA_0(\eta ) \bigl[ \HA_{0,1}\big(\sqrt{\eta }\big) + \HA_{0,-1}\big(\sqrt{\eta }\big) \bigr]
        +  8 \bigl[ \HA_ {0,0,1}\big(\sqrt{\eta }\big) + \HA_{0,0,-1}\big(\sqrt{\eta }\big) \bigr]
\biggr)
\nonumber \\ &&
-\frac{4^{-N} P_{12}}{27 (N+1) (2N-5) (2N-3) (2N-1) \eta } \binom{2 N}{N} 
	\sum_{i=1}^N  \frac{4^i}{ \binom{2 i}{i} } \biggl( \frac{1}{i^3} - \frac{1}{i^2} \HA_0(\eta) - \frac{1}{i^2} S_1(i) 
\nonumber \\ &&
		+ \frac{(1-\eta)^{-i}}{i} \biggl[ \frac{1}{2} \HA_0^2(\eta) + S_ 1(1-\eta ,i) H_ 0(\eta ) - S_ 2(1-\eta ,i) + S_{1,1}(1-\eta ,1,i) \biggr]
	\biggr) 
\nonumber \\ && 
+\biggl(
        \frac{8 P_ 4}{27 (N-1)^2 N^2 (N+1)^2}
        +\frac{8 P_8}{3 (N-1) N (N+1)} \HA_0(\eta )
        -\frac{8 P_8}{9 (N-1) N (N+1)} S_1
\biggr) \zeta_2
\nonumber \\ &&
-\frac{32 P_8}{27 (N-1) N (N+1)} \zeta_3
\biggr\},
\end{eqnarray}
with the polynomials
\begin{eqnarray}
P_1 &=& 
N^5-N^4+2 N^3-14 N^2-4 N+6
~,
\\
P_2 &=& 
N^6-36 N^5-33 N^4+12 N^3+224 N^2+66 N-54
~,
\\
P_3 &=& 
2 N^6-18 N^5-15 N^4-12 N^3+85 N^2+36 N-18
~,
\\
P_4 &=& 
7 N^6-36 N^5-27 N^4-60 N^3+116 N^2+78 N-18
~,
\\
P_5 &=& 
8 N^6-18 N^5-9 N^4-84 N^3-23 N^2+48 N+18
~,
\\
P_6 &=& 
30 N^9-94 N^8-112 N^7-43 N^6+300 N^5+56 N^4-272 N^3-99 N^2+30 N-36
~,
\\
P_7 &=& 
-8 N^9 \eta ^2
-4 N^8 \eta  (28-23 \eta )
-2 N^7 \big(15-566 \eta +87 \eta ^2\big)
+3 N^6 \big(35-1162 \eta -185 \eta ^2\big)
\nonumber \\ &&
-2 N^5 \big(30-1605 \eta -1099 \eta ^2\big)
-4 N^4 \big(75-367 \eta +608 \eta ^2\big)
+2 N^3 \big(255+127 \eta +12 \eta ^2\big)
\nonumber \\ &&
-45 N^2 \big(5+202 \eta -35 \eta ^2\big)
+8064 N \eta 
-2160 \eta 
~,
\\
P_8 &=& 
N^3 -3 N^2 -2 N -6
~,
\\
P_9 &=& 
8 N^{13} \eta ^2
-12 N^{12} \eta  (46+7 \eta )
-2 N^{11} \big(15-1970 \eta -37 \eta ^2\big)
+N^{10} \big(75-7772 \eta +813 \eta ^2\big)
\nonumber \\ &&
+3 N^9 \big(25+298 \eta -575 \eta ^2\big)
-N^8 \big(435-2834 \eta +495 \eta ^2\big)
+N^7 \big(165+19500 \eta +3511 \eta ^2\big)
\nonumber \\ &&
+N^6 \big(645-26320 \eta -1833 \eta ^2\big)
-N^5 \big(435+9526 \eta +1823 \eta ^2\big)
\nonumber \\ &&
-N^4 \big(285-23566 \eta -2679 \eta ^2\big)
+15 N^3 \big(15+40 \eta -3 \eta ^2\big)
-36 N^2 \eta  (281+30 \eta )
\nonumber \\ &&
+5472 N \eta 
-1080 \eta 
~,
\\
P_{10} &=&  
24 N^{13} \eta ^2
+12 N^{12} (22-21 \eta ) \eta 
-2 N^{11} \big(45+1418 \eta -111 \eta ^2\big)
\nonumber \\ &&
+N^{10} \big(225+7628 \eta +2439 \eta ^2\big)
+3 N^9 \big(75-2002 \eta -1725 \eta ^2\big)
\nonumber \\ &&
-5 N^8 \big(261-2030 \eta +297 \eta ^2\big)
+3 N^7 \big(165-8900 \eta +3511 \eta ^2\big)
\nonumber \\ &&
+N^6 \big(1935+3064 \eta -5499 \eta ^2\big)
-N^5 \big(1305-28030 \eta +5469 \eta ^2\big)
\nonumber \\ &&
-N^4 \big(855+5686 \eta -8037 \eta ^2\big)
+3 N^3 \big(225-4312 \eta -45 \eta ^2\big)
\nonumber \\ &&
+60 N^2 \eta  (49-54 \eta )
-432 N \eta 
+1080 \eta
~,
\\
P_{11} &=&  
216 N^{16} \eta ^2
-4 N^{15} \eta  (836+567 \eta )
+6 N^{14} \big(135+6466 \eta +297 \eta ^2\big)
\nonumber \\ &&
-3 N^{13} \big(675+34454 \eta -8073 \eta ^2\big)
-3 N^{12} \big(945-11644 \eta +16191 \eta ^2\big)
\nonumber \\ &&
+2 N^{11} \big(6885-8819 \eta -17658 \eta ^2\big)
-6 N^{10} \big(405-72572 \eta -23562 \eta ^2\big)
\nonumber \\ &&
-2 N^9 \big(14580+147371 \eta +14418 \eta ^2\big)
+6 N^8 \big(2700-111523 \eta -24003 \eta ^2\big)
\nonumber \\ &&
+162 N^7 \big(155+3061 \eta +527 \eta ^2\big)
-6 N^6 \big(2970-92344 \eta -5571 \eta ^2\big)
\nonumber \\ &&
-N^5 \big(7695+547820 \eta +50463 \eta ^2\big)
+3 N^4 \big(2025+7994 \eta +10125 \eta ^2\big)
\nonumber \\ &&
+90 N^3 \eta  (730+81 \eta )
-108 N^2 \eta  (526+135 \eta )
+35964 N \eta 
-4860 \eta
~,
\\
P_{12} &=& 
-16 N^6 \eta ^3
-72 N^5 \eta ^2 (3-2 \eta )
-12 N^4 \eta  \big(27-135 \eta -4 \eta ^2\big)
\nonumber \\ &&
-6 N^3 \big(5-270 \eta +351 \eta ^2+222 \eta ^3\big)
+N^2 \big(45-2349 \eta -2673 \eta ^2+1129 \eta ^3\big)
\nonumber \\ &&
+12 N \big(5+216 \eta +72 \eta ^2+77 \eta ^3\big)
-45 (1-\eta ) \big(5+104 \eta -13 \eta ^2\big)
~.
\end{eqnarray}
The diagram explicitly fulfills the symmetry 
\begin{eqnarray}
D_2^A \left(m_1,m_2,\eta\right) = D_2^B\left(m_2,m_1,\frac{1}{\eta}\right).
\end{eqnarray}
We calculated all diagrams which differ for the different mass assignments separately and checked that the 
symmetry relation holds.
For mass symmetric diagrams, we checked the independence of the mass assignment explicitly.
\subsection{The \boldmath $z$-space solution}
\label{sec:52a}

\vspace*{1mm}
\noindent
The result in $z$-space for diagram~2, split into the usual contributions, reads:
\begin{eqnarray}
        D_2^A (z) &=& C_A T_F^2 \frac{1+(-1)^N}{2} \biggl[D_2^{A,\delta} 
\delta (1-z) + D_2^{A,+} (z) + D_2^{A,\text{reg}} (z) + \Mvec^{-1} \left[ n g_1 (n) \right] (z) 
\NN \\ &&
+ \Mvec^{-1} \left[ n^2 g_2 (n) \right] (z) \biggr] .
\end{eqnarray}
Here we use the notations $n=N-1$, 
\begin{eqnarray}
\Mvec\left[ g(z) \right] &=& \int_{0}^1 \mathrm{d}z z^n g(z) = g(n) \\
\Mvec^{-1} \left[ g(n) \right] &=& g(z),
\end{eqnarray}
and $\Mvec$ denotes the Mellin transform.
Terms of the type 
\begin{eqnarray}
\Mvec^{-1} \left[ n^l g_l (n) \right] (z),~~l = 1,2,
\end{eqnarray}
which will not contribute in the final result of all diagrams are dropped in the following expressions.
The rational pre-factors can be absorbed by applying the relations
\begin{eqnarray}
\label{eq:abs1}
        \frac{1}{(n+a)^i} \int_{0}^{1} \text{d}z \ z^n \ f(z) &=& \int_{0}^{1} \text{d}z \ z^n \biggl\{ \int_z^1 
\text{d} y \ (-1)^{i-1} \left( \frac{y}{z} \right)^{a} \left[ \HA_0 \left( \frac{y}{z} \right) \right]^{i-1}  f(y)    
\biggr\} \\ \label{eq:abs2}
        n \int_{0}^{1} \text{d}z \ z^n \ f(z) &=& \left. ( z^n - 1 ) z f(z) \right|_{0}^{1} - \int_{0}^{1} (z^n - 
1) \frac{d}{dz} \left( z f(z) \right). 
\end{eqnarray}
Furthermore, we will set $\mu=m_1$ for brevity to shorten the expression. 
The logarithmic dependence on the mass can be easily restored by using the full $N$-space result and will be 
entirely given in terms of HPLs. One obtains

\begin{eqnarray}
D_2^{A,\delta,\ep^0} &\propto& 
\frac{4}{729} (836+243 \eta )
+\frac{2}{9} (46+3 \eta ) \HA_0(\eta )
+\frac{8}{27} \HA_0^2(\eta )
+\frac{32}{27} \HA_0^3(\eta )
-\frac{8}{9} \HA_0^2(\eta ) \HA_1(\eta )
\nonumber \\ &&
+\frac{16}{9} \HA_0(\eta ) \HA_{0,1}(\eta )
-\frac{16}{9} \HA_{0,0,1}(\eta )
+\frac{8}{27} \left( 7+9 \HA_0(\eta ) \right) \zeta_2           
-\frac{32}{27} \zeta_3
~,\\
D_2^{A,+,\ep^0} (z) &\propto& 
\frac{1}{1-z}
\biggl[
\frac{2}{27} (22-9 \eta )
+\frac{16}{9} \HA_0(\eta )
-\frac{8}{9} \HA_0^2(\eta )
+\frac{16}{81} \HA_0
+\frac{8}{27} \HA_0^2
\nonumber \\ &&
-\biggl(
        \frac{16}{27} \HA_0
        +\frac{16}{81} \big(7+9 \HA_0(\eta )\big)
\biggr) \HA_1
+\frac{8}{27} \HA_1^2
+\frac{16}{9} \HA_{0,1}
\biggr]
\nonumber \\ &&
- \frac{(27-8 \eta ) \sqrt{\eta }}{108 \pi(1-z)^{3/2} \sqrt{z}}
\biggl[
          \HA_0^2(\eta ) \biggl( \HA_{-1}\big(\sqrt{\eta }\big) +  \HA_1\big(\sqrt{\eta }\big) \biggr)
\nonumber \\ &&
        -4 \HA_0(\eta ) \biggl( \HA_{0,1}\big(\sqrt{\eta }\big) + \HA_{0,-1}\big(\sqrt{\eta }\big) \biggr)
        +8 \biggl( \HA_{0,0,1}\big(\sqrt{\eta }\big) + \HA_{0,0,-1}\big(\sqrt{\eta }\big) \biggr)
\biggr]
\nonumber \\ &&
-\frac{8}{27 (1-z)} \zeta_2
- F_1^{D_2} (z)
+ F_{+}^{D_2} (z)
~,\\
D_2^{A,\text{reg},\ep^0} (z) &\propto& 
\frac{2 \HA_0(\eta ) Q_1}{81 \eta  z}
+\frac{2 Q_4}{729 \eta  z^2}
+\frac{10 (3-2 \eta )}{81 \eta  z^{5/2}}
+ \frac{\big(45-10 \eta -54 z-810 \eta  z\big)}{81 \eta z^{5/2}} \biggl( \HA_0(\eta ) 
\nonumber \\ &&
+ \HA_1 + 2 \HA_{-1}\big(\sqrt{z}\big) - 2 \ln(2) \biggr)
- \frac{Q_5}{108 \eta ^{3/2} \pi z^{5/2} \sqrt{1-z}}
\biggl\{
	\HA_0^2(\eta) \bigl[ \HA_{-1}\big(\sqrt{\eta }\big) 
\nonumber \\ &&
	+ \HA_1\big(\sqrt{\eta }\big)  \bigr]
	- 4 \HA_0(\eta) \bigl[ \HA_{0,1}\big(\sqrt{\eta }\big) + \HA_{0,-1}\big(\sqrt{\eta }\big)  \bigr]
	+ 8 \bigl[ \HA_{0,0,1}\big(\sqrt{\eta }\big) 
\nonumber \\ &&
+ \HA_{0,0,-1}\big(\sqrt{\eta }\big) \bigr]
\biggr\} 
-\frac{8 \big(89-84 z+28 z^2\big)}{27 z} \HA_0^2(\eta )
-\frac{32 Q_6}{27 z} \HA_0^3(\eta )
+\biggl[
        \frac{2 Q_2}{81 \eta  z}
\nonumber \\ &&
        -\frac{16 \big(37-36 z+11 z^2\big)}{27 z} \HA_0(\eta )
        -\frac{8 Q_6}{3 z} \HA_0^2(\eta )
\biggr] \HA_0
\nonumber \\ &&
-\biggl[
        \frac{8 \big(59-60 z+22 z^2\big)}{81 z}
        +\frac{8 Q_6}{9 z} \HA_0(\eta )
\biggr] \HA_0^2
-\frac{8 Q_6}{81 z} \HA_0^3
+\frac{8 Q_6}{9 z} \HA_0^2(\eta ) \HA_1(\eta )
\nonumber \\ &&
+\biggl[
        \frac{2 Q_3}{81 \eta  z}
        +\frac{16 \big(-5+4 z+2 z^2\big)}{9 z} \HA_0(\eta )
        -\frac{8 Q_6}{9 z} \HA_0^2(\eta )
\nonumber \\ &&
        -\frac{16 (-2+z) (-26+11 z)}{81 z} \HA_0
        +\frac{8 Q_6}{27 z} \HA_0^2
\biggr] \HA_1
-\frac{16 Q_6}{9 z} \HA_0(\eta ) \HA_{0,1}(\eta )
\nonumber \\ &&
-\biggl[
        \frac{8 \big(59-60 z+40 z^2\big)}{81 z}
        +\frac{8 Q_6}{9 z} \HA_0(\eta )
        +\frac{8 Q_6}{27 z} \HA_0
\biggr] \HA_1^2
\nonumber \\ &&
+\biggl[
        \frac{16 (5-4 z)}{9 z}
        -\frac{16 Q_6}{9 z} \HA_0(\eta )
        -\frac{16 Q_6}{9 z} \HA_0
        +\frac{16 Q_6}{9 z} \HA_1
\biggr] \HA_{0,1}
\nonumber \\ &&
+\frac{8 Q_6}{81 z} \HA_1^3
+\frac{16 Q_6}{9 z} \HA_{0,0,1}(\eta )
+\frac{32 Q_6}{9 z} \HA_{0,0,1}
-\frac{32 Q_6}{9 z} \HA_{0,1,1}
\nonumber \\ &&
-\biggl[
        \frac{8 \big(163-156 z+98 z^2\big)}{81 z}
        +\frac{8 Q_6}{9 z} \HA_0(\eta )
        +\frac{40 Q_6}{27 z} \HA_0
        +\frac{8 Q_6}{27 z} \HA_1
\biggr] \zeta_2
\nonumber \\ &&
-\frac{32 Q_6}{27 z} \zeta_3
+ F_7^{D_2} (z)
+ \int\limits_{z}^{1} \mathrm{d}y \biggl[ \frac{\sqrt{y}}{2 z^{3/2}} F_2^{D_2} (y)
                                        + \frac{y^{3/2} }{2 z^{5/2}} F_3^{D_2} (y)
                                        + \frac{1}{y} F_4^{D_2} (y)
\nonumber \\ &&
                                        + \frac{z}{y^2} F_5^{D_2} (y)
                                        - \frac{1}{z} \HA_0\left(\frac{z}{y}\right) F_6^{D_2} (y)
                                        - \frac{1}{z} F_6^{D_2} (y)
                                        - \biggl( 6-\frac{6}{y}-4 z+\frac{4 z}{y^2} \biggr) F_{+}^{D_2} (y)
\biggr]
\nonumber \\ &&
- \biggl( 6-\frac{5}{z}-4 z \biggr) \int\limits_{0}^{z} \mathrm{d}y F_{+}^{D_2} (y)
~,
\end{eqnarray}
with the polynomials
\begin{eqnarray}
Q_1 &=&
-1600 \eta +3 \left(39 \eta ^2+710 \eta +15\right) z +6 \left(4 \eta ^2-221 \eta -3\right) z^2
~,
\\
Q_2 &=&
-176 \eta +9 \left(13 \eta ^2+66 \eta +5\right) z +2 \left(12 \eta ^2-199 \eta -9\right) z^2
~,
\\
Q_3 &=&
1248 \eta +3 \left(39 \eta ^2-314 \eta +15\right) z + 2 \left(12 \eta ^2+265 \eta -9\right) z^2
~,
\\
Q_4 &=&
45 (2 \eta -9)+(351-17000 \eta ) z+6 \left(315 \eta ^2+2761 \eta -108\right) z^2
\nonumber \\ &&
+2 \left(324 \eta ^2-6017 \eta +81\right) z^3
~,
\\
Q_5 &=&
-10 +(270 \eta +23) z +\left(9 \eta ^3+783 \eta ^2-729 \eta -39\right) z^2 
\nonumber \\ &&
+\left(62 \eta ^3-810 \eta ^2+810 \eta +34\right) z^3 + 8 \left(4 \eta ^3+27 \eta ^2-54 \eta -1\right) z^4
~,
\\ 
Q_6 &=& 5-6 z+4 z^2
~.
\end{eqnarray}
The functions $F_k$ are given by
\begin{eqnarray}
F_1^{D_2}(z) &=&
-\frac{2 z R_1}{27 (1-z)}
-\frac{2 R_2}{27}
- \frac{2 (27-8 \eta ) }{27 \sqrt{z} (1-z)^{3/2} } G_1(z)
\biggl\{
	2 (1-\eta ) 
	+ (1+\eta ) H_ 0(\eta )
\biggr\}
\nonumber \\ &&
-\frac{5 (1+\eta ) (27-8 \eta ) }{81 \pi \sqrt{z} (1-z)^{3/2} }
-\frac{2 (27-8 \eta ) \big(1+\eta +\eta ^2\big)  }{81 (1-\eta ) \pi \sqrt{z} (1-z)^{3/2} } \HA_0(\eta )
\nonumber \\ &&
-\frac{\eta  (1+\eta ) (27-8 \eta ) }{54 (1-\eta )^2 \pi \sqrt{z} (1-z)^{3/2} } \HA_0^2(\eta)
- \frac{(27-8 \eta ) }{54 \sqrt{z} (1-z)^{3/2}}
\biggl\{ 
	4 (1+\eta) \biggl[ G_6(z) + G_7(z) 
\nonumber \\ &&
	- \frac{8}{\pi} \biggl( K_{19} + K_{20} \biggr) \biggr]
        - (1-\eta)^2 \biggl[ G_{12}(z) + G_{13}(z) - K_{13} - K_{14} + \HA_0(\eta) 
\nonumber \\ &&
\times \biggl( G_4(z) - K_6 \biggr) 
	+ \frac{8}{\pi} \biggl( K_{21} + K_{22} + K_{23} + K_{24} + \HA_0(\eta) K_{15} \biggr)
	\biggr]     
\biggr\} 
\nonumber \\ &&
+\frac{R_3}{27 (1-z+\eta  z)} 
\bigl[ \HA_0(\eta ) + \HA_0 + \HA_1 \bigr]
+ \frac{(27-8 \eta )}{36 \pi \sqrt{z}(1-z)^{3/2}} \zeta_2
\nonumber\\ && \times
\biggl\{
	 2 (1-\eta )
	+ (1+\eta ) \HA_0(\eta )
\biggr\} 
~,
\\
F_2^{D_2}(y) &=&
\frac{4 R_4}{3 \eta ^2}
+ \frac{4(1+15 \eta )}{3 \eta ^2 \sqrt{1-y} \sqrt{y}} G_1(y)
\biggl\{
	2 (1-\eta ) 
	+ (1+\eta ) \HA_0(\eta )
\biggr\} 
+ \frac{10 (1+\eta ) (1+15 \eta )}{9 \eta ^2 \pi \sqrt{1-y} \sqrt{y}}
\nonumber \\ &&
+\frac{4 (1+15 \eta ) \big(1+\eta +\eta ^2\big)}{9 (1-\eta ) \eta ^2 \pi \sqrt{1-y} \sqrt{y} } \HA_0(\eta )
+\frac{(1+\eta ) (1+15 \eta )}{3 (1-\eta )^2 \eta  \pi \sqrt{1-y} \sqrt{y}} \HA_0^2(\eta )
\nonumber \\ &&
+ \frac{1+15 \eta }{3 \eta ^2 \sqrt{1-y} \sqrt{y}}
\biggl\{
	4 (1+\eta) \biggl[ G_6(y) + G_7(y) - \frac{8}{\pi} \biggl( K_{19} + K_{20} \biggr) \biggr] 
	- (1-\eta)^2 
\nonumber \\ &&
\times \biggl[ G_{12}(y) + G_{13}(y)
	- K_{13} - K_{14} 
	+ \HA_0(\eta) \biggl( G_4(y) - K_6 \biggr) 
	+ \frac{8}{\pi} \biggl( K_{21} + K_{22} + K_{23} 
\nonumber \\ &&
+ K_{24} + \HA_0(\eta) K_{15} \biggr) \biggr]       
\biggr\} 
-\frac{2 R_5}{3 \eta ^2 (1-y+\eta  y)} \bigl[ \HA_0(\eta ) + \HA_0(y) + \HA_1(y) \bigr]
\nonumber \\ &&
- \frac{1+15 \eta }{2 \eta ^2 \pi \sqrt{1-y} \sqrt{y}} \zeta_2
\biggl\{
        2 (1-\eta ) 
        + (1+\eta ) \HA_0(\eta )
\biggr\} 
~,
\\
F_3^{D_2}(y) &=&
\frac{10 (\eta +y-\eta  y)}{9 \eta ^2}
- \frac{10}{9 \eta ^2 \sqrt{1-y} \sqrt{y}} G_1(y)
\biggl\{
	2 (1-\eta )
	+ (1+\eta ) \HA_0(\eta )
\biggr\}
\nonumber \\ &&
-\frac{25 (1+\eta )}{27 \eta ^2 \pi \sqrt{1-y} \sqrt{y}}
-\frac{10 \big(1+\eta +\eta ^2\big)}{27 (1-\eta ) \eta ^2 \pi \sqrt{1-y} \sqrt{y}} \HA_0(\eta )
-\frac{5 (1+\eta )}{18 (1-\eta )^2 \eta  \pi \sqrt{1-y} \sqrt{y} } \HA_0^2(\eta )
\nonumber \\ &&
+ \frac{1}{18 \eta^2 \sqrt{1-y} \sqrt{y}}
\biggl\{      
        - 20 (1+\eta) \biggl[ G_6(y) + G_7(y) - \frac{8}{\pi} \biggl( K_{19} + K_{20} \biggr) \biggr]
	+ 5 (1-\eta)^2 
\nonumber \\ &&
\times \biggl[ G_{12}(y) + G_{13}(y) 
	- K_{13} - K_{14} 
	+ \HA_0(\eta) \biggl( G_4(y) - K_6 \biggr) 
	+ \frac{8}{\pi} \biggl( K_{21} + K_{22} + K_{23} 
\nonumber \\ &&
+ K_{24} + \HA_0(\eta) K_{15} \biggr) \biggr]        
\biggr\} 
+\frac{5 R_6}{27 \eta ^2 (1-y+\eta  y)} \bigl[ \HA_0(\eta ) + \HA_0(y) + \HA_1(y) \bigr]
\nonumber \\ &&
+ \frac{5}{12 \eta ^2 \pi \sqrt{1-y} \sqrt{y}} \zeta_2
\biggl\{
        2 (1-\eta )
        + (1+\eta ) \HA_0(\eta )
\biggr\} 
~,
\\
F_4^{D_2}(y) &=&
-\frac{2 R_7}{9 \eta ^2}
- \frac{2 (1-\eta ) \big(5+104 \eta -13 \eta ^2\big) }{9 \eta ^2 \sqrt{1-y} \sqrt{y}} G_1(y)
\biggl\{
	2 (1-\eta )
	+  (1+\eta ) \HA_0(\eta )
\biggr\} 
\nonumber \\ &&
-\frac{5 (1-\eta^2 ) \big(5+104 \eta -13 \eta ^2\big)}{27 \eta ^2 \pi \sqrt{1-y} \sqrt{y} }
-\frac{2 \big(1+\eta +\eta ^2\big)\big(5+104 \eta -13 \eta ^2\big)}{27 \eta ^2 \pi \sqrt{1-y} \sqrt{y} } \HA_0(\eta )
\nonumber \\ &&
-\frac{(1+\eta ) \big(-5-104 \eta +13 \eta ^2\big)}{18 (-1+\eta ) \eta  \pi \sqrt{1-y} \sqrt{y} } \HA_0^2(\eta )
- \frac{(1-\eta ) \big(5+104 \eta -13 \eta ^2\big) }{18 \eta ^2 \sqrt{1-y} \sqrt{y}}
\biggl\{
         4 (1+\eta) 
\nonumber \\ &&
\times \biggl[ G_6(y) + G_7(y) 
	- \frac{8}{\pi} \biggl( K_{19} + K_{20} \biggr) \biggr]
	- (1-\eta)^2 \biggl[ G_{12}(y) + G_{13}(y) - K_{13} 
\nonumber \\ &&
- K_{14} 
	+ \HA_0(\eta) \biggl( G_4(y) - K_6 \biggr) 
	+ \frac{8}{\pi} \biggl( K_{21} + K_{22} + K_{23} + K_{24} + \HA_0(\eta) K_{15} \biggr) \biggr] 
\biggr\} 
\nonumber \\ &&
+\frac{R_9}{9 \eta ^2 (1-y+\eta  y)} \HA_0(\eta )
-\frac{(1-\eta ) R_8}{9 \eta ^2 (1-y+\eta  y)} \bigl[ \HA_0(y) + \HA_1(y) \bigr]
\nonumber \\ &&
+ \frac{(1-\eta ) \big(5+104 \eta -13 \eta ^2\big)}{12 \eta ^2 \pi \sqrt{1-y}\sqrt{y}} \zeta_2
\biggl\{
        2 (1-\eta )
        + (1+\eta ) \HA_0(\eta )
\biggr\} 
~,
\\
F_5^{D_2}(y) &=&
-\frac{4 R_{10}}{9 \eta ^2}
+ \frac{4 \big(1+54 \eta -27 \eta ^2-4 \eta ^3\big)}{9 \eta ^2 \sqrt{1-y} \sqrt{y}} G_1(y)
\biggl\{
	2 (1-\eta ) 
	+ (1+\eta ) \HA_0(\eta )
\biggr\} 
\nonumber \\ &&
+\frac{10 (1+\eta ) \big(1+54 \eta -27 \eta ^2-4 \eta ^3\big)}{27 \eta ^2 \pi \sqrt{1-y} \sqrt{y}}
+\frac{4 \big(1+\eta +\eta ^2\big)\big(1+54 \eta -27 \eta ^2-4 \eta ^3\big)}{27 (1-\eta ) \eta ^2 \pi \sqrt{1-y} \sqrt{y} } \HA_0(\eta )
\nonumber \\ &&
+\frac{(1+\eta ) \big(1+54 \eta -27 \eta ^2-4 \eta ^3\big)}{9 (1-\eta )^2 \eta  \pi \sqrt{1-y} \sqrt{y} } \HA_0^2(\eta )
+ \frac{1+54 \eta -27 \eta ^2-4 \eta ^3}{9 \eta ^2 \sqrt{1-y} \sqrt{y}}
\biggl\{
        4 (1+\eta) 
\nonumber \\ &&
\times \biggl[ G_6(y) + G_7(y) 
	- \frac{8}{\pi} \biggl( K_{19} + K_{20} \biggr) \biggr]
	- (1-\eta)^2 \biggl[ G_{12}(y) + G_{13}(y) - K_{13} 
\nonumber \\ &&
- K_{14} 
	+ \HA_0(\eta) \biggl( G_4(y) - K_6 \biggr) 
	+ \frac{8}{\pi} \biggl( K_{21} + K_{22} + K_{23} + K_{24} + \HA_0(\eta) K_{15} \biggr) \biggr] 
\biggr\} 
\nonumber \\ &&
+\frac{2 R_{11}}{27 \eta ^2 (1-y+\eta  y)} \bigl[ \HA_0(\eta ) 
+ \HA_0(y) + \HA_1(y) \bigr]
- \frac{1+54 \eta -27 \eta ^2-4 \eta ^3}{6 \eta ^2 \pi \sqrt{1-y} \sqrt{y}} \zeta_2
\nonumber\\ && \times
\biggl\{
        2 (1-\eta ) 
        + (1+\eta ) \HA_0(\eta )
\biggr\} 
~,
\\
F_6^{D_2}(y) &=&
\frac{80 (1-\eta ) }{9-9 (1-\eta ) y} 
\big(
        \HA_0(\eta )
        +\HA_0(y)
        +\HA_1(y)
\big)
~,
\\
F_7^{D_2}(y) &=&
\frac{2 R_{12}}{27 \eta }
- \frac{2 \big(81-189 \eta -103 \eta ^2\big) }{27 \eta  \sqrt{1-y} \sqrt{y}} G_1(y)
\biggl\{
	2 (1-\eta )
	+ (1+\eta ) \HA_0(\eta )
\biggr\} 
\nonumber \\ &&
-\frac{5 (1+\eta ) \big(81-189 \eta -103 \eta ^2\big)}{81 \eta  \pi \sqrt{1-y} \sqrt{y}}
-\frac{2 \big(1+\eta +\eta ^2\big)\big(81-189 \eta -103 \eta ^2\big)}{81 (1-\eta ) \eta  \pi \sqrt{1-y} \sqrt{y}} \HA_ 0(\eta )
\nonumber \\ &&
-\frac{(1+\eta ) \big(81-189 \eta -103 \eta ^2\big)}{54 (1-\eta )^2 \pi \sqrt{1-y} \sqrt{y}} \HA_0^2(\eta )
- \frac{\big(81-189 \eta -103 \eta ^2\big)}{54 \eta \sqrt{1-y} \sqrt{y}}
\biggl\{
        4 (1+\eta) 
\nonumber \\ &&
\times \biggl[ G_6(y) + G_7(y) 
	- \frac{8}{\pi} \biggl( K_{19} + K_{20} \biggr) \biggr]
	- (1-\eta)^2 \biggl[ G_{12}(y) + G_{13}(y) - K_{13} - K_{14} 
\nonumber \\ &&
	+ \HA_0(\eta) \biggl( G_4(y) - K_6 \biggr) 
	+ \frac{8}{\pi} \biggl( K_{21} + K_{22} + K_{23} + K_{24} + \HA_0(\eta) K_{15} \biggr) \biggr] 
\biggr\} 
\nonumber \\ &&
+\frac{R_{13}}{27 \eta  (1-y+\eta  y)} \bigl[ \HA_0(\eta ) 
+ \HA_0(y) + \HA_1(y) \bigr]
+ \frac{81-189 \eta -103 \eta ^2}{36 \eta  \pi \sqrt{1-y} \sqrt{y}} \zeta_2
\nonumber\\ && \times
\biggl\{
        2 (1-\eta ) 
        + (1+\eta ) \HA_0(\eta )
\biggr\} 
~,
\\
F_{+}^{D_2}(z) &=&
\frac{8}{9 (1-z)} 
\biggl\{
        2 (1-\eta ) \bigl[ G_{10}(z) +G_{11}(z) \bigr]
        + \HA_0^2(\eta )
        +2 (1-\eta ) \HA_0(\eta ) G_{3}(z)
\biggr\}
~
\end{eqnarray}
The functions $G_i$ and $K_i$ are given in the appendix.
The additional polynomials read 
\begin{eqnarray}
R_1 &=& - (8 \eta -27) \bigl[ \eta (1 - z) - z \bigr]
~, 
\\
R_2 &=& (8 \eta -27) \bigl[ \eta ( 1 + z) + z \bigr]
~,
\\
R_3 &=& (8 \eta -27) \left[ 2 \eta + \left(\eta ^2-1\right) z \right]
~,
\\
R_4 &=& - (15 \eta +1) \bigl[ (1-\eta ) y + \eta \bigr]
~,
\\
R_5 &=& (15 \eta +1) \left[-\eta ^2+\left(-\eta ^2+2 \eta +1\right) y +\left(\eta ^2-1\right) y^2 \right]
~,
\\
R_6 &=& \eta ^2 (2 \eta -5)+\left(-3 \eta ^2+6 \eta +3\right) y +3 \left(\eta ^2-1\right) y^2
~,
\\
R_7 &=& - \left(13 \eta ^3-117 \eta ^2+99 \eta +5\right) \bigl[ (1 - \eta ) y + \eta \bigr]
~,
\\
R_8 &=& \left(13 \eta ^2-104 \eta -5\right) y \left( 1 -\eta ^2 + 2 \eta +\left(\eta ^2-1\right) y \right)
~,
\\
R_9 &=& \left(13 \eta ^3-117 \eta ^2+99 \eta +5\right) y \left( 1 -\eta ^2+2 \eta +\left(\eta ^2-1\right) y\right)
~,
\\
R_{10} &=& -\left(4 \eta ^3+27 \eta ^2-54 \eta -1\right) \bigl[ (1-\eta) y+\eta \bigr]
~,
\\
R_{11} &=& -2 \eta ^2 \left(4 \eta ^2+31 \eta -71\right)-3 \left(4 \eta ^5+19 \eta ^4-112 \eta ^3+80 \eta ^2+56 \eta +1\right) y 
\NN \\ &&
+3 \left(4 \eta ^5+27 \eta ^4-58 \eta ^3-28 \eta ^2+54 \eta +1\right) y^2
~,
\\
R_{12} &=& -\left(103 \eta ^2+189 \eta -81\right) \bigl[ (1-\eta ) y+\eta \bigr]
~,
\\
R_{13} &=& \eta  \left(112 \eta ^2+152 \eta -53\right)+\left(103 \eta ^4-17 \eta ^3-562 \eta ^2-27 \eta +81\right) y 
\NN \\ &&
+\left(-103 \eta ^4-189 \eta ^3+184 \eta ^2+189 \eta -81\right) y^2
~.
\end{eqnarray}
We note that although single sums have a different support other than $z \in \bigl[ 0,1 \bigr]$, for example
\begin{eqnarray}
        S_1 \left( \left\{ \frac{1}{1-\eta} \right\} ,N \right) = \int\limits_{0}^{1/(1-\eta)} \mathrm{d}z \frac{z^N -1}{z-1} ,
\end{eqnarray}  
the final result is defined on the usual support $x \in [0,1]$.
The contributions in other domains cancel analytically.
\section{Results}
\label{sec:6}

\vspace*{1mm}
\noindent
The renormalized 2- and 3-loop OMEs $\tilde{A}_{gg,Q}^{2(3)}$ (\ref{eq:renOME2}, \ref{eq:renOME3}) can be 
obtained from the different contributions to the renormalized masses, the expansion coefficients of the 
$\beta$-function and anomalous dimensions, together with the constant part of the unrenormalized 3-loop OME 
$\tilde{a}_{gg,Q}^{(3)}$ in the two-mass case. In the following we present this function both in $N$- and 
$z$-space and will give the corresponding results for $\tilde{A}_{gg,Q}^{2(3)}$ in 
Appendix~\ref{app:B}.
\subsection{\boldmath $N$-space}
\label{sec:61}

\vspace*{1mm}
\noindent
In Mellin $N$-space one obtains
\begin{eqnarray}
\lefteqn{\tilde{a}_{gg,Q}^{(3)} (N) =} \nonumber \\ && 
\tfrac{1}{2}\left(1+(-1)^N\right) \Biggl\{\textcolor{blue}{T_F^3} \biggl\{
 \frac{32}{3} \bigl( L_1^3 + L_2^3 \bigr)
+\frac{64}{3} L_1 L_2 \bigl( L_1 + L_2 \bigr)
+ 32 \zeta_2 \bigl( L_1 + L_2 \bigr) 
+\frac{128}{9} \zeta_3
\biggr\}
\nonumber \\ &&
+\textcolor{blue}{C_F T_F^2} 
\Biggl\{
         \frac{\big(2+N+N^2\big)^2}{(N-1) N^2 (N+1)^2 (N+2)} \Biggl[ 
24 L_1^3 + 24 L_2^3 +16 L_1 L_2 \bigl(L_1+L_2\bigr) 
\nonumber \\ &&
+48 H_ 0(\eta ) \bigl(L_1^2 - L_2^2\bigr) 
+16 \bigl(L_1^2+L_2^2\bigr) S_1
+32 S_1 \HA_0(\eta) \bigl(L_1-L_2\bigr)
+\biggl(48 \HA_0^2(\eta)+\frac{16}{3} S_1^2
\nonumber \\ &&
-16 S_2+40 \zeta_2\biggr) \bigl(L_1+L_2\bigr)
-\frac{32}{9} \HA_0^3(\eta) 
-\frac{64}{3} \HA_0^2(\eta) \HA_1(\eta)
+\frac{128}{3} \HA_0(\eta) \HA_{0,1}(\eta)
-\frac{352}{9} \zeta_3
\nonumber \\ &&
-\frac{128}{3} \HA_{0,0,1}(\eta)
+32 \biggl(\HA_0^2(\eta)-\frac{1}{3} S_2\biggr) S_1
+\frac{32}{27} S_1^3
-\frac{704}{27} S_3
+\frac{128}{3} S_{2,1}
+\frac{32}{3} \zeta_2 S_1
\nonumber \\ &&
-\frac{32}{3} \HA_0^2(\eta) \biggl( S_1\biggl(\frac{1}{1-\eta },N\biggr) + S_1\biggl(\frac{\eta }{\eta -1},N\biggr)  \biggr)
+\frac{64}{3} S_ {1,2}\biggl(\frac{1}{1-\eta },1-\eta ,N\biggr)
\nonumber \\ &&
-\frac{64}{3} \HA_0(\eta) \biggl( S_{1,1}\biggl(\frac{1}{1-\eta },1-\eta ,N\biggr) - S_{1,1}\biggl(\frac{\eta }{\eta -1},\frac{\eta -1}{\eta },N\biggr) \biggr)
\nonumber \\ &&
-\frac{64}{3} S_{1,1,1}\biggl(\frac{1}{1-\eta },1-\eta ,1,N\biggr)
-\frac{64}{3} S_{1,1,1}\biggl(\frac{\eta }{\eta -1},\frac{\eta -1}{\eta },1,N\biggr)
\nonumber \\ &&
+\frac{64}{3} S_{1,2}\biggl(\frac{\eta }{\eta -1},\frac{\eta -1}{\eta },N\biggr)
\Biggr]
+ \frac{P_{63} L_1+P_{64} L_2}{54 \eta  (N-1) N^4 (N+1)^4 (N+2)}
\nonumber \\ &&
-\frac{(1+\eta) \big(5-2 \eta +5 \eta ^2\big)}{4 \eta ^{3/2}} \biggl[
\frac{1}{4} \left( \HA_1\big(\sqrt{\eta }\big) + \HA_{-1}\big(\sqrt{\eta }\big) \right) \bigl(L_1-L_2\bigr)^2
\nonumber \\ &&
+\left( \HA_{0,1}\big(\sqrt{\eta }\big) + \HA_{0,-1}\big(\sqrt{\eta }\big) \right) \bigl(L_1-L_2\bigr)
+2 H_ {0,0,1}\big(\sqrt{\eta }\big)+2 H_ {0,0,-1}\big(\sqrt{\eta }\big)
\biggr]
\nonumber \\ &&
+\frac{1}{(N-1) N^3 (N+1)^3 (N+2)} \biggl[
\frac{P_{50}}{24 \eta} \bigl(L_1^2 + L_2^2\bigr)
+\frac{P_{49}}{12 \eta} L_1 L_2
+\frac{32}{9} P_{29} \bigl(L_1+L_2\bigr) S_1
\nonumber \\ &&
+\frac{32}{3} P_{29} \HA_0(\eta) \bigl(L_1-L_2\bigr)
+\frac{32}{27} P_{29} S_1^2
-\frac{32}{9} P_{29} S_2
+\frac{32}{3} P_{29} \HA_0^2(\eta)
-\frac{16}{9} \zeta_2 P_{45} 
\biggr]
\nonumber \\ &&
-\frac{16 P_{62}}{81 \eta  (N-1) N^4 (N+1)^4 (N+2) (2N-3) (2N-1)} S_1
\nonumber \\ &&
+\frac{16 P_{22}}{3 \eta  (N-1) N (N+1)^2 (N+2) (2N-3) (2N-1)} \HA_0(\eta )
\nonumber \\ &&
	+\frac{P_{67}}{243 \eta  (N-1) N^5 (N+1)^5 (N+2) (2N-3) (2N-1)}
\nonumber \\ &&
        - \frac{4 P_{42} (1-\eta )^{-N}}{3 \eta  (N-1) N^3 (N+1)^2 (N+2) (2N-3) (2N-1)}
	\biggl[
                 \HA_0^2(\eta )
\nonumber \\ &&
                +2 \HA_0(\eta ) S_1(1-\eta ,N)
                -2 S_2(1-\eta ,N)
                +2 S_{1,1}(1-\eta ,1,N)
        \biggr] 
\nonumber \\ &&
        - \frac{4 P_{41}}{3 (N-1) N^3 (N+1)^2 (N+2) (2N-3) (2N-1)} \biggl(\frac{\eta }{1-\eta }\biggr)^N
	\biggl[
                 \HA_0^2(\eta )
\nonumber \\ &&
                -2 \HA_0(\eta ) S_ 1\biggl(\frac{\eta -1}{\eta },N\biggr)
                -2 S_ 2\biggl(\frac{\eta -1}{\eta },N\biggr)
                +2 S_ {1,1}\biggl(\frac{\eta -1}{\eta },1,N\biggr)
        \biggr]
\nonumber \\ &&
        -\frac{2 (1+\eta ) P_{34} 2^{-2 N}}{3 \eta ^{3/2} (N-1) N (N+1)^2 (N+2) (2N-3) (2N-1)} \binom{2 N}{N}
	\biggl[
                 \HA_0^2(\eta ) \biggl( H_ {-1}\big(\sqrt{\eta }\big) 
\nonumber \\ &&
		+ H_ 1\big(\sqrt{\eta }\big) \biggr)
                -4 \HA_0(\eta ) \biggl( H_ {0,1}\big(\sqrt{\eta }\big) + H_ {0,-1}\big(\sqrt{\eta }\big) \biggr)
                +8 \biggl( H_ {0,0,1}\big(\sqrt{\eta }\big) + H_ {0,0,-1}\big(\sqrt{\eta }\big) \biggr)
        \biggr] 
\nonumber \\ &&
	-\frac{2^{2-2 N} P_{28}}{3 \eta  (N-1) N (N+1)^2 (N+2) (2N-3) (2N-1)} \binom{2 N}{N} 
        \sum_{i=1}^N  \frac{4^i \big(\frac{\eta }{\eta -1}\big)^i}{i \binom{2 i}{i}}  
	\biggl[
		\frac{1}{2} H_ 0(\eta )^2 
\nonumber \\ &&
		- H_ 0(\eta ) S_ 1\biggl(\frac{\eta -1}{\eta },i\biggr) - S_ 2\biggl(\frac{\eta -1}{\eta },i\biggr) + S_ {1,1}\biggl(\frac{\eta -1}{\eta },1,i\biggr) 
	\biggr] 
\nonumber \\ &&
        +\frac{2^{3-2 N} P_{36}}{3 \eta  (N-1) N (N+1)^2 (N+2) (2N-3) (2N-1)}  \binom{2 N}{N} 
        \sum_{i=1}^N { \frac{4^i}{i^2 \binom{2 i}{i}} \biggl( \frac{1}{i} - S_1(i) \biggr) } 
\nonumber \\ &&
	+\frac{ (1-\eta^2 ) 2^{4-2 N} P_{30}}{3 \eta  (N-1) N (N+1)^2 (N+2) (2N-3) (2N-1)} \binom{2 N}{N} H_ 0(\eta ) 
        \sum_{i=1}^N { \frac{4^i}{i^2 \binom{2 i}{i}} }
\nonumber \\ &&
        +\frac{2^{2-2 N} P_{38}}{3 \eta  (N-1) N (N+1)^2 (N+2) (2N-3) (2N-1)} \binom{2 N}{N} 
        \sum_{i=1}^N  \frac{4^i (1-\eta)^{-i}}{i \binom{2 i}{i}} 
	\biggl[
		\frac{1}{2} H_ 0(\eta )^2  
\nonumber \\ &&
		+ H_ 0(\eta ) S_ 1(1-\eta ,i) - S_ 2(1-\eta ,i) + S_{1,1}(1-\eta ,1,i)
	\biggr] 
\Biggr\}
\nonumber \\ &&
+\textcolor{blue}{C_A T_F^2} 
\Biggl\{
\frac{1}{(N-1) N (N+1) (N+2)} \biggl[\frac{16}{9} P_{17} L_1^3 + \frac{8}{9} P_{20} L_2^3 - \frac{8}{3} P_{15} L_1^2 L_2 - \frac{16}{3} P_{16} L_1 L_2^2 \biggr]
\nonumber \\ &&
-\frac{16}{3} \bigl(5 L_1^3 + 5 L_2^3 + 2 L_1^2 L_2 + 2 L_1 L_2^2 \bigr) S_1
+\frac{1}{(N-1) N^2 (N+1)^2 (N+2)} \biggl[\frac{P_{39}}{54 \eta} \bigl(L_1^2 + L_2^2\bigr) 
\nonumber \\ &&
- \frac{8}{27} P_{32} \bigl(L_1 + L_2\bigr) S_1 + \frac{P_{35}}{27 \eta} L_1 L_2 \biggr]
+ \bigr( L_1^2 + L_2^2 \bigl)\biggl[
                \frac{8}{3} H_ 1(\eta)
                +\frac{16 \big(6+85 N-85 N^2\big)}{27 (N-1) N} S_ 1
        \biggr]
\nonumber \\ &&
+\frac{(1+\eta) \big(4+11 \eta +4 \eta ^2\big)}{6 \eta ^{3/2}} \biggl[-\frac{1}{2} \bigl( \HA_1\big(\sqrt{\eta }\big)+ \HA_{-1}\big(\sqrt{\eta }\big) \bigr) \bigl(L_1-L_2\bigr)^2
\nonumber \\ &&
-2 \bigl( \HA_{0,1}\big(\sqrt{\eta }\big)+ \HA_{0,-1}\big(\sqrt{\eta }\big) \bigr) \bigl(L_1-L_2\bigr)
\biggr]
-L_1 L_2 \biggl[
\frac{64 \big(3-10 N+10 N^2\big)}{27 (N-1) N} S_1
\nonumber \\ &&
+\frac{16}{3} \HA_1(\eta)
\biggr]
+ \bigl( L_1^2 - L_2^2 \bigr) \biggl[
\frac{2 (1+2 N) P_{19}}{(N-1) N (N+1)^2 (N+2)} \HA_0(\eta )
-32 H_ 0(\eta ) S_1
\biggr]
\nonumber \\ &&
+ \bigl( L_1 + L_2 \bigr) \biggl[
\frac{2 P_{26}}{3 (N-1) N (N+1)^2 (N+2)} \HA_0^2(\eta )
                +\biggl(
                        \frac{224 \big(1+N+N^2\big)}{3 (N-1) N (N+1) (N+2)}
\nonumber \\ &&
                        -\frac{112}{3} S_ 1
                \biggr) \zeta_2
-\frac{64}{3} \HA_0^2(\eta) S_1
        \biggr]
	+ ( L_1 - L_2 )
	\biggl[
		\frac{4 P_{60}}{9 (N-1)^2 N^2 (N+1)^3 (N+2)^2} \HA_0(\eta )
\nonumber \\ &&
		+\frac{16}{3} \HA_{0,1}(\eta )
		-\frac{4 P_{21}}{9 (N-1) N (N+1)^2} \HA_0(\eta ) S_1
		-\frac{16}{3} \HA_0(\eta ) S_ 1^2
		-16 \HA_0(\eta ) S_ 2
	\biggr]
\nonumber \\ &&
	-\frac{2}{27 \eta  (N-1) N^3 (N+1)^3 (N+2)} \bigl(P_{51} L_1 - P_{52} L_2\bigr)
\nonumber \\ &&
	+ \frac{2 P_{66}}{3645 \eta  (N-1) N^4 (N+1)^4 (N+2) (2N-3) (2N-1)}
\nonumber \\ &&
+\frac{1}{45 \eta (N-1) N^2 (N+1)^2 (N+2) (2N-3) (2N-1)} \Biggl[
P_{53} (1-\eta)^{-N} S_ 2(1-\eta ,N)
\nonumber \\ &&
+P_{48} (1-\eta)^{-N} \biggl(
\frac{1}{2} \HA_0^2(\eta) 
+ \HA_0(\eta ) S_ 1(1-\eta,N)
+ S_{1,1}(1-\eta,1,N)
\biggr)
\nonumber \\ &&
+\biggl(\frac{\eta }{1-\eta }\biggr)^N \biggl\{P_{44} \biggl[
\frac{1}{2} \HA_0^2(\eta)
+ S_{1,1}\biggl(\frac{\eta -1}{\eta },1,N\biggr)
\biggr]
+P_{46} \biggl[
\HA_0(\eta ) S_ 1\biggl(\frac{\eta -1}{\eta },N\biggr)
\nonumber \\ &&
+ S_2\biggl(\frac{\eta -1}{\eta },N\biggr) 
\biggr]
\biggr\}
\Biggr]
        + \biggl( \frac{(1+\eta) 2^{-2 N} P_{54} }{90 \eta ^{3/2} (N-1) N (N+1)^2 (N+2) (2N-3) (2N-1)}  \binom{2 N}{N}
\nonumber \\ && 
-\frac{(1+\eta ) \big(5+22 \eta +5 \eta ^2\big)}{9 \eta ^{3/2}} S_1\biggr)
	\biggl[
                  \HA_0^2(\eta ) \bigl( \HA_{-1}\big(\sqrt{\eta }\big) + \HA_1\big(\sqrt{\eta }\big) \bigr)
\nonumber \\ && 
                - 4 \HA_0(\eta ) \bigl( \HA_{0,1}\big(\sqrt{\eta }\big) + \HA_{0,-1}\big(\sqrt{\eta }\big) \bigr)
                + 8 \bigl( \HA_{0,0,1}\big(\sqrt{\eta }\big) + \HA_{0,0,-1}\big(\sqrt{\eta }\big) \bigr)
        \biggr] 
\nonumber \\ &&
        +\frac{P_{40}}{45 \eta  (N-1) N (N+1) (N+2) (2N-3) (2N-1)} \HA_0(\eta )
\nonumber \\ &&
        +\biggl[
                \frac{P_{61}}{540 \eta  (N-1)^2 N^2 (N+1)^3 (N+2)^2}
                -\frac{(1+\eta ) P_{24} \HA_{-1}\big(\sqrt{\eta }\big)}{360 \eta ^{3/2} (N-1) N (N+1) (N+2)} 
        \biggr] \HA_0^2(\eta )
\nonumber \\ &&
	-\frac{4 P_{18}}{27 (N-1) N (N+1) (N+2)} \HA_0(\eta)
	\biggl[
		\HA_0(\eta )^2 + 6 \HA_0(\eta ) \HA_1(\eta ) - 12 \HA_{0,1}(\eta )
	\biggr]
\nonumber \\ &&
	-\frac{(1+\eta ) P_{24}}{360 \eta ^{3/2} (N-1) N (N+1) (N+2)} \HA_0(\eta)
	\biggl[
		\HA_0(\eta ) \HA_1\big(\sqrt{\eta }\big) - 4 \HA_{0,1}\big(\sqrt{\eta }\big) - 4 \HA_{0,-1}\big(\sqrt{\eta }\big)
	\biggr]
\nonumber \\ &&
        -\frac{256 \big(1+N+N^2\big)}{9 (N-1) N (N+1) (N+2)} \HA_{0,0,1}(\eta )
        -\frac{(1+\eta ) P_{25}}{45 \eta ^{3/2} (N-1) N (N+1) (N+2)} \biggl( \HA_{0,0,1}\big(\sqrt{\eta }\big) 
\nonumber \\ &&
	+ \HA_{0,0,-1}\big(\sqrt{\eta }\big) \biggr)
        +\biggl[
                \frac{8 P_{65}}{3645 \eta  (N-1) N^3 (N+1)^3 (N+2) (2N-3) (2N-1)}
\nonumber \\ &&
                +\frac{8 P_{37} \HA_0(\eta )}{45 \eta  (N-1) N^2 (N+1)^2 (N+2)}
                +\frac{2 P_{23} \HA_0^2(\eta )}{9 \eta  (N-1) N (N+1)^2}
                +\frac{32}{27} \HA_0^3(\eta )
		+\frac{128}{9} \HA_{0,0,1}(\eta )
\nonumber \\ &&
                +\frac{64}{9} \HA_0^2(\eta ) \HA_1(\eta )
		-\frac{128}{9} \HA_0(\eta ) \HA_{0,1}(\eta )
        \biggr] S_1
        -\frac{16 P_{14}}{15 \eta  (N-1) N (N+1)} \bigl( S_3 - S_{2,1} \bigr)
\nonumber \\ &&
        +\biggl[
                \frac{4 P_{27}}{135 \eta  (N-1) N^2 (N+1)^2 (N+2)}
                -\frac{16}{3} \HA_0^2(\eta )
        \biggr] S_1^2
\nonumber \\ &&
        -\biggl[
                \frac{4 P_{33}}{135 \eta  (N-1) N^2 (N+1)^2 (N+2)}
                -\frac{16 P_{13}}{15 \eta  (N-1) N (N+1)} \HA_0(\eta )
                +16 \HA_0^2(\eta )
        \biggr] S_2
\nonumber \\ &&
	-\frac{16  \big(1-7 N+4 N^2+4 N^3\big)}{15 \eta (N-1) N (N+1)}
	\biggl[
		\frac{1}{2} \HA_0^2(\eta) \biggl(\eta^2 S_ 1\biggl(\frac{1}{1-\eta },N\biggr) 
		+S_ 1\biggl(\frac{\eta }{\eta-1},N\biggr) 
               \biggr)
\nonumber \\ &&
		+\HA_0(\eta) \biggl(\eta^2 S_ {1,1}\biggl(\frac{1}{1-\eta },1-\eta ,N\biggr)
		- S_ {1,1}\biggl(\frac{\eta }{\eta-1},\frac{\eta-1}{\eta },N\biggr) 
                \biggr)
\nonumber \\ &&
		- S_ {1,2}\biggl(\frac{\eta }{\eta-1},\frac{\eta-1}{\eta },N\biggr)
		+ S_ {1,1,1}\biggl(\frac{\eta }{\eta-1},\frac{\eta-1}{\eta },1,N\biggr)
\nonumber \\ &&
		-\eta^2 S_ {1,2}\biggl(\frac{1}{1-\eta },1-\eta ,N\biggr)
		+\eta^2 S_ {1,1,1}\biggl(\frac{1}{1-\eta },1-\eta ,1,N\biggr)
	\biggr]
\nonumber \\ &&
+\frac{2^{-1-2 N}}{45 \eta ^2 (N-1) N (N+1)^2 (N+2) (2N-3) (2N-1)} \binom{2 N}{N} \Biggl[
P_{43} \sum_{i=1}^N \frac{4^i \big(\frac{\eta }{\eta -1}\big)^i}{i \binom{2 i}{i}} \biggl\{
\nonumber \\ &&
S_ 2\biggl(\frac{\eta-1 }{\eta },i\biggr) + \HA_0(\eta ) S_ 1\biggl(\frac{\eta-1 }{\eta },i\biggr) 
\biggr\}
+P_{47} \sum_{i=1}^N  \frac{4^i \big(\frac{\eta }{\eta -1}\big)^i}{i \binom{2 i}{i}} \biggl\{
S_ {1,1}\biggl(\frac{\eta -1}{\eta },1,i\biggr)
\nonumber \\ &&
+\frac{1}{2} \HA_0^2(\eta)
\biggr\}
+\eta P_{57} \sum_{i=1}^N  \frac{4^i (1-\eta)^{-i}}{i \binom{2 i}{i}} \biggl\{
\frac{1}{2} \HA_0^2(\eta ) + \HA_0(\eta ) S_ 1(1-\eta ,i) +  S_{1,1}(1-\eta ,1,i)
\biggr\} 
\nonumber \\ &&
+(1-\eta^2) P_{55} \HA_0(\eta) \sum_{i=1}^N { \frac{4^i}{i^2 \binom{2 i}{i}} }
+\eta P_{56} \sum_{i=1}^N { \frac{4^i (1-\eta)^{-i}}{i \binom{2 i}{i}} S_ 2(1-\eta ,i) }
+P_{59} \sum_{i=1}^N { \frac{4^i}{i^3 \binom{2 i}{i}} }
\nonumber \\ &&
+P_{58} \sum_{i=1}^N { \frac{4^i}{i^2 \binom{2 i}{i}} S_ 1(i) }
\Biggr]                   
        +\biggl[
                \frac{8 P_{31}}{27 (N-1) N^2 (N+1)^2 (N+2)}
                -\frac{1120}{27} S_1
        \biggr] \zeta_2
\nonumber \\ &&
        +\biggl[
                -\frac{128 \big(1+N+N^2\big)}{27 (N-1) N (N+1) (N+2)}
                +\frac{64}{27} S_1
        \biggr] \zeta_3
\Biggr\}\Biggr\}.
\label{EQ:aN}
\end{eqnarray}

\noindent
The polynomials $P_i$ read
\begin{eqnarray}
P_{13} &=&
\left(\eta ^2-1\right) \left(4 N^3+4 N^2-7 N+1\right)
~,
\\
P_{14} &=&
\left(\eta ^2+1\right) \left(4 N^3+4 N^2-7 N+1\right)
~,
\\
P_{15} &=&
N^4+2 N^3-11 N^2-16 N-12
~,
\\%
P_{16} &=&
N^4+2 N^3-6 N^2-9 N-6
~,
\\%
P_{17} &=&
2 N^4+4 N^3+25 N^2+17 N+24
~,
\\%
P_{18} &=&
3 N^4+6 N^3+13 N^2+10 N+16
~,
\\%
P_{19} &=&
3 N^4+9 N^3+15 N^2+7 N+10
~,
\\%
P_{20} &=&
5 N^4+10 N^3+49 N^2+32 N+48
~,
\\%
P_{21} &=&
92 N^4+65 N^3-152 N^2-179 N-90
~,
\\%
P_{22} &=&
\left(\eta ^2-1\right) \left(5 N^4+10 N^3+73 N^2+32 N+32\right)
~,
\\%
P_{23} &=& 
\left(5 \eta ^2-102 \eta +5\right) N^4
+\left(5 \eta ^2-48 \eta +5\right) N^3
-\left(5 \eta ^2-206 \eta + 5\right) N^2
\nonumber \\ &&
-\left(5 \eta ^2-244 \eta + 5\right) N
+164 \eta
~,
\\%
P_{24} &=&
3 \left(71 \eta ^2-46 \eta +71\right) N^4
+42 \left(17 \eta ^2-2 \eta +17\right) N^3
-\left(253 \eta ^2+1382 \eta +253\right) N^2
\nonumber \\ &&
-2 \left(593 \eta ^2+862 \eta +593\right) N
-128 \left(2 \eta ^2+13 \eta +2\right)
~,
\\%
P_{25} &=&
3 \left(111 \eta ^2+64 \eta +111\right) N^4
+18 \left(53 \eta ^2+32 \eta +53\right) N^3
-\left(373 \eta ^2+1712 \eta +373\right) N^2
\nonumber \\ &&
-2 \left(713 \eta ^2+1192 \eta +713\right) N
-128 \left(2 \eta ^2+13 \eta +2\right)
~,
\\%
P_{26} &=&
12 N^5+45 N^4+87 N^3+73 N^2+69 N+14
~,
\\%
P_{27} &=&
-140 \eta  N^5-190 \eta  N^4
+\left(63 \eta ^2+320 \eta +63\right) N^3
+2 \left(108 \eta ^2+535 \eta +108\right) N^2
\nonumber \\ &&
+\left(279 \eta ^2+700 \eta +279\right) N
-2 \left(9 \eta ^2-160 \eta +9\right)
~,
\\%
P_{28} &=&
-36 N^6
-36 \eta  N^5
+\left(5 \eta ^2-18 \eta +225\right) N^4
+2 \left(5 \eta ^2-108 \eta +9\right) N^3
\nonumber \\ &&
+\left(73 \eta ^2+246 \eta -495\right) N^2
+8 \left(4 \eta ^2+21 \eta +27\right) N
+32 \eta  (\eta +9)
~,
\\%
P_{29} &=&
4 N^6+3 N^5-50 N^4-129 N^3-100 N^2-56 N-24
~,
\\%
P_{30} &=&
9 N^6-55 N^4-2 N^3+142 N^2-46 N+8
~,
\\%
P_{31} &=&
99 N^6+297 N^5+631 N^4+767 N^3+1118 N^2+784 N+168
~,
\\%
P_{32} &=&
344 N^6+978 N^5+209 N^4-1032 N^3-817 N^2-210 N-96
~,
\\%
P_{33} &=&
-440 \eta  N^6
-1100 \eta  N^5
+270 \eta  N^4
+\left(63 \eta ^2+1640 \eta +63\right) N^3
\nonumber \\ &&
+\left(216 \eta ^2+2810 \eta +216\right) N^2
+3 \left(93 \eta ^2+700 \eta +93\right) N
-6 \left(3 \eta ^2-160 \eta +3\right)
~,
\\%
P_{34} &=&
-36 \eta  N^6
-36 \eta  N^5
+\left(5 \eta ^2+202 \eta +5\right) N^4
+2 \left(5 \eta ^2-104 \eta +5\right) N^3
\nonumber \\ &&
+\left(73 \eta ^2-322 \eta +73\right) N^2
+32 \left(\eta ^2+11 \eta +1\right) N
+32 \left(\eta ^2+8 \eta +1\right)
~,
\\%
P_{35} &=&
-9 \left(4 \eta ^2-93 \eta +4\right) N^6
-27 \left(4 \eta ^2-93 \eta +4\right) N^5
+\left(-36 \eta ^2+3589 \eta -36\right) N^4
\nonumber \\ &&
+3 \left(36 \eta ^2+1211 \eta +36\right) N^3
+\left(72 \eta ^2+5942 \eta +72\right) N^2
+4224 \eta  N
+768 \eta 
~,
\\%
P_{36} &=&
18 \left(\eta ^2+1\right) N^6
+36 \eta  N^5+\left(-115 \eta ^2+18 \eta -115\right) N^4
-2 \left(7 \eta ^2-108 \eta +7\right) N^3
\nonumber \\ &&
+\left(211 \eta ^2-246 \eta +211\right) N^2
-4 \left(31 \eta ^2+42 \eta +31\right) N
-16 \left(\eta ^2+18 \eta +1\right)
~,
\\%
P_{37} &=&
\left(\eta ^2-1\right) \left(25 N^6+75 N^5+25 N^4-96 N^3-122 N^2-93 N+6\right)
~,
\\%
P_{38} &=&
36 \eta ^2 N^6
+36 \eta  N^5
+\left(-225 \eta ^2+18 \eta -5\right) N^4
-2 \left(9 \eta ^2-108 \eta +5\right) N^3
\nonumber \\ &&
+\left(495 \eta ^2-246 \eta -73\right) N^2
-8 \left(27 \eta ^2+21 \eta +4\right) N
-32 (9 \eta +1)
~,
\\%
P_{39} &=&
9 \left(4 \eta ^2+171 \eta +4\right) N^6
+27 \left(4 \eta ^2+171 \eta +4\right) N^5
+\left(36 \eta ^2+11555 \eta +36\right) N^4
\nonumber \\ &&
-3 \left(36 \eta ^2-4925 \eta +36\right) N^3
+\left(-72 \eta ^2+20890 \eta -72\right) N^2
+14592 \eta  N
+3264 \eta 
~,
\\%
P_{40} &=&
\left(\eta ^2-1\right) \left(52 N^6+200 N^5-1925 N^4+2394 N^3-1447 N^2+622 N-3384\right)
~,
\\%
P_{41} &=&
18 N^7-(5 \eta +9) N^6-2 (5 \eta +48) N^5+(111-73 \eta ) N^4-8 (4 \eta -33) N^3
\nonumber \\ &&
-8 (4 \eta +21) N^2-96
~,
\\%
P_{42} &=&
18 \eta  N^7-(9 \eta +5) N^6-2 (48 \eta +5) N^5+(111 \eta -73) N^4
\nonumber \\ &&
+8 (33 \eta -4) N^3-8 (21 \eta +4) N^2 -96 \eta
~,
\\
P_{43} &=&
-800 N^8-8 (270 \eta +269) N^7
+4 \left(30 \eta ^2-1185 \eta +589\right) N^6
\nonumber \\ &&
-6 \left(2 \eta ^3+55 \eta ^2-1440 \eta -1409\right) N^5
+\left(147 \eta ^3-1005 \eta ^2+945 \eta -3703\right) N^4
\nonumber \\ &&
+\left(471 \eta ^3+6075 \eta ^2-915 \eta -7383\right) N^3
+\left(-1599 \eta ^3-1095 \eta ^2+10815 \eta +3839\right) N^2
\nonumber \\ &&
+\left(-3117 \eta ^3-6015 \eta ^2+2085 \eta +1351\right) N
-6 \left(91 \eta ^3+465 \eta ^2+645 \eta +127\right)
~,
\\
P_{44} &=&
-400 N^8-4 (128 \eta +219) N^7
-4 \left(3 \eta ^2+300 \eta -404\right) N^6
\nonumber \\ &&
+\left(-525 \eta ^2+2410 \eta +3419\right) N^5
-\left(489 \eta ^2+2750 \eta +3561\right) N^4
\nonumber \\ &&
-3 \left(157 \eta ^2-1958 \eta +637\right) N^3
+\left(1299 \eta ^2+4686 \eta +2875\right) N^2
\nonumber \\ &&
-2 \left(1581 \eta ^2+638 \eta +381\right) N
+48 \eta  (3 \eta -80)
~,
\\
P_{45} &=&
33 N^8+132 N^7+106 N^6-108 N^5-74 N^4+282 N^3+245 N^2+148 N+84
~,
\\
P_{46} &=&
 400 N^8+(512 \eta +876) N^7
+4 \left(3 \eta ^2+300 \eta -404\right) N^6
\nonumber \\ &&
+\left(525 \eta ^2-2410 \eta -3419\right) N^5
+\left(489 \eta ^2+2750 \eta +3561\right) N^4
\nonumber \\ &&
+3 \left(157 \eta ^2-1958 \eta +637\right) N^3
-\left(1299 \eta ^2+4686 \eta +2875\right) N^2
\nonumber \\ &&
+2 \left(1581 \eta ^2+638 \eta +381\right) N 
+48 (80-3 \eta ) \eta
~,
\\
P_{47} &=&
800 N^8+8 (270 \eta +269) N^7
-4 \left(30 \eta ^2-1185 \eta +589\right) N^6
\nonumber \\ &&
+6 \left(2 \eta ^3+55 \eta ^2-1440 \eta -1409\right) N^5
+\left(-147 \eta ^3+1005 \eta ^2-945 \eta +3703\right) N^4
\nonumber \\ &&
+\left(-471 \eta ^3-6075 \eta ^2+915 \eta +7383\right) N^3
+\left(1599 \eta ^3+1095 \eta ^2-10815 \eta -3839\right) N^2
\nonumber \\ &&
+\left(3117 \eta ^3+6015 \eta ^2-2085 \eta -1351\right) N
+6 \left(91 \eta ^3+465 \eta ^2+645 \eta +127\right)
~,
\\
P_{48} &=&
-400 \eta ^2 N^8
-4 \eta  (219 \eta +128) N^7
+4 \left(404 \eta ^2-300 \eta -3\right) N^6
\nonumber \\ &&
+\left(3419 \eta ^2+2410 \eta -525\right) N^5
-\left(3561 \eta ^2+2750 \eta +489\right) N^4
\nonumber \\ &&
-3 \left(637 \eta ^2-1958 \eta +157\right) N^3
+\left(2875 \eta ^2+4686 \eta +1299\right) N^2
\nonumber \\ &&
-2 \left(381 \eta ^2+638 \eta +1581\right) N
+48 (3-80 \eta )
~,
\\
P_{49} &=&
 -3 \left(5 \eta ^2+282 \eta +5\right) N^8
-12 \left(5 \eta ^2+282 \eta +5\right) N^7
-4 \left(15 \eta ^2+718 \eta +15\right) N^6
\nonumber \\ &&
+\left(30 \eta ^2+2716 \eta +30\right) N^5
+\left(75 \eta ^2+4486 \eta +75\right) N^4
\nonumber \\ &&
+\left(30 \eta ^2-868 \eta +30\right) N^3
-1280 \eta  N^2-1024 \eta  N -1024 \eta
~,
\\
P_{50} &=&
 3 \left(5 \eta ^2-422 \eta +5\right) N^8
+12 \left(5 \eta ^2-422 \eta +5\right) N^7
+12 \left(5 \eta ^2-326 \eta +5\right) N^6
\nonumber \\ &&
+\left(-30 \eta ^2+4196 \eta -30\right) N^5
-25 \left(3 \eta ^2-10 \eta +3\right) N^4
\nonumber \\ &&
-10 \left(3 \eta ^2+1718 \eta +3\right) N^3
-14400 \eta  N^2-8448 \eta  N -4352 \eta
~,
\\
P_{51} &=&
 \left(36 \eta ^2-93 \eta -36\right) N^8
+12 \left(12 \eta ^2-31 \eta -12\right) N^7
+16 \left(9 \eta ^2-376 \eta -9\right) N^6
\nonumber \\ &&
-6 \left(12 \eta ^2+2719 \eta -12\right) N^5
+\left(-180 \eta ^2-23011 \eta +180\right) N^4
\nonumber \\ &&
-6 \left(12 \eta ^2+3019 \eta -12\right) N^3
-6032 \eta  N^2+1376 \eta  N +1056 \eta
~,
\\
P_{52} &=&
 \left(36 \eta ^2+93 \eta -36\right) N^8
+12 \left(12 \eta ^2+31 \eta -12\right) N^7
+16 \left(9 \eta ^2+376 \eta -9\right) N^6
\nonumber \\ &&
+\left(-72 \eta ^2+16314 \eta +72\right) N^5
+\left(-180 \eta ^2+23011 \eta +180\right) N^4
\nonumber \\ &&
+\left(-72 \eta ^2+18114 \eta +72\right) N^3
+6032 \eta  N^2-1376 \eta  N -1056 \eta
~,
\\
P_{53} &=&
400 \eta ^2 N^8
+4 \eta  (219 \eta +128) N^7
-4 \left(404 \eta ^2-300 \eta -3\right) N^6
\nonumber \\ &&
+\left(-3419 \eta ^2-2410 \eta +525\right) N^5
+\left(3561 \eta ^2+2750 \eta +489\right) N^4
\nonumber \\ &&
+3 \left(637 \eta ^2-1958 \eta +157\right) N^3
-\left(2875 \eta ^2+4686 \eta +1299\right) N^2
\nonumber \\ &&
+2 \left(381 \eta ^2+638 \eta +1581\right) N+48 (80 \eta -3)
~,
\\
P_{54} &=&
 400 \left(\eta ^2-\eta +1\right) N^8
+4 \left(269 \eta ^2+\eta +269\right) N^7
-2 \left(589 \eta ^2-1744 \eta +589\right) N^6
\nonumber \\ &&
+\left(-4221 \eta ^2+66 \eta -4221\right) N^5
+2 \left(889 \eta ^2-874 \eta +889\right) N^4
\nonumber \\ &&
+12 \left(288 \eta ^2-503 \eta +288\right) N^3
-20 \left(56 \eta ^2+187 \eta +56\right) N^2
\nonumber \\ &&
+\left(883 \eta ^2+1082 \eta +883\right) N
+654 \eta ^2+2676 \eta+654
~,
\\
P_{55} &=&
 800 \left(\eta ^2+1\right) N^8
+8 \left(269 \eta ^2+270 \eta +269\right) N^7
-4 \left(589 \eta ^2-1185 \eta +589\right) N^6
\nonumber \\ &&
-6 \left(1409 \eta ^2+1442 \eta +1409\right) N^5
+7 \left(529 \eta ^2-114 \eta +529\right) N^4
\nonumber \\ &&
+3 \left(2461 \eta ^2+462 \eta +2461\right) N^3
-\left(3839 \eta ^2+12414 \eta +3839\right) N^2
\nonumber \\ &&
-\left(1351 \eta ^2+5202 \eta +1351\right) N 
+ 762 \eta ^2+3324 \eta +762
~,
\\
P_{56} &=&
-800 \eta ^3 N^8
-8 \eta ^2 (269 \eta +270) N^7
+4 \eta  \left(589 \eta ^2-1185 \eta +30\right) N^6
\nonumber \\ &&
+6 \left(1409 \eta ^3+1440 \eta ^2-55 \eta -2\right) N^5
+\left(-3703 \eta ^3+945 \eta ^2-1005 \eta +147\right) N^4
\nonumber \\ &&
+\left(-7383 \eta ^3-915 \eta ^2+6075 \eta +471\right) N^3
+\left(3839 \eta ^3+10815 \eta ^2-1095 \eta -1599\right) N^2
\nonumber \\ &&
+\left(1351 \eta ^3+2085 \eta ^2-6015 \eta -3117\right) N
-6 \left(127 \eta ^3+645 \eta ^2+465 \eta +91\right)
~,
\\
P_{57} &=&
800 \eta ^3 N^8+8 \eta ^2 (269 \eta +270) N^7
-4 \eta  \left(589 \eta ^2-1185 \eta +30\right) N^6
\nonumber \\ &&
-6 \left(1409 \eta ^3+1440 \eta ^2-55 \eta -2\right) N^5
+\left(3703 \eta ^3-945 \eta ^2+1005 \eta -147\right) N^4
\nonumber \\ &&
+\left(7383 \eta ^3+915 \eta ^2-6075 \eta -471\right) N^3
+\left(-3839 \eta ^3-10815 \eta ^2+1095 \eta +1599\right) N^2
\nonumber \\ &&
+\left(-1351 \eta ^3-2085 \eta ^2+6015 \eta +3117\right) N 
+ 6 \left(127 \eta ^3+645 \eta ^2+465 \eta +91\right)
~,
\\
P_{58} &=&
-800 \left(\eta ^4+1\right) N^8
-8 \left(269 \eta ^4+270 \eta ^3+270 \eta +269\right) N^7
+4 \big(589 \eta ^4-1185 
\nonumber \\ &&
\eta ^3+60 \eta ^2-1185 \eta +589\big) N^6
+\big(8454 \eta ^4+8628 \eta ^3-660 \eta ^2+8628 \eta +8454\big) N^5
\nonumber \\ &&
+\big(-3703 \eta ^4+1092 \eta ^3-2010 \eta ^2+1092 \eta -3703\big) N^4
-3 \big(2461 \eta ^4+148 \eta ^3
\nonumber \\ &&
-4050 \eta ^2+148 \eta +2461\big) N^3
+\left(3839 \eta ^4+9216 \eta ^3-2190 \eta ^2+9216 \eta +3839\right) N^2
\nonumber \\ &&
+\left(1351 \eta ^4-1032 \eta ^3-12030 \eta ^2-1032 \eta +1351\right) N 
\nonumber \\ &&
-6 \left(127 \eta ^4+736 \eta ^3+930 \eta ^2+736 \eta +127\right)
~,
\\
P_{59} &=&
800 \left(\eta ^4+1\right) N^8
+8 \left(269 \eta ^4+270 \eta ^3+270 \eta +269\right) N^7
-4 \big(589 \eta ^4-1185 \eta ^3
\nonumber \\ &&
+60 \eta ^2-1185 \eta +589\big) N^6
-6 \left(1409 \eta ^4+1438 \eta ^3-110 \eta ^2+1438 \eta +1409\right) N^5
\nonumber \\ &&
+\left(3703 \eta ^4-1092 \eta ^3+2010 \eta ^2-1092 \eta +3703\right) N^4
+3 \big(2461 \eta ^4+148 \eta ^3
\nonumber \\ &&
-4050 \eta ^2+148 \eta +2461\big) N^3
-\left(3839 \eta ^4+9216 \eta ^3-2190 \eta ^2+9216 \eta +3839\right) N^2
\nonumber \\ &&
+\left(-1351 \eta ^4+1032 \eta ^3+12030 \eta ^2+1032 \eta -1351\right) N 
\nonumber \\ &&
+ 6 \left(127 \eta ^4+736 \eta ^3+930 \eta ^2+736 \eta +127\right)
~,
\\
P_{60} &=&
9 N^9+84 N^8+723 N^7+2137 N^6+1907 N^5-716 N^4-2167 N^3-1229 N^2
\nonumber \\ &&
-400 N-132
~,
\\
P_{61} &=&
9 \left(71 \eta ^2+134 \eta +71\right) N^9
+3 \left(1353 \eta ^2+5642 \eta +1353\right) N^8
\nonumber \\ &&
+2 \left(3153 \eta ^2+74122 \eta +3153\right) N^7
-6 \left(767 \eta ^2-70930 \eta +767\right) N^6
\nonumber \\ &&
-3 \left(4811 \eta ^2-119250 \eta +4811\right) N^5
+3 \left(833 \eta ^2-59782 \eta +833\right) N^4
\nonumber \\ &&
+768 \left(19 \eta ^2-563 \eta +19\right) N^3
-12 \left(211 \eta ^2+16410 \eta +211\right) N^2
\nonumber \\ &&
-64 \left(111 \eta ^2+899 \eta +111\right) N 
+ 576 \left(\eta ^2-55 \eta +1\right)
~,
\\
P_{62} &=&
 92 \eta  N^{10}+\left(135 \eta ^2+274 \eta +135\right) N^9
+4 \left(135 \eta ^2-491 \eta +135\right) N^8
\nonumber \\ &&
+\left(2646 \eta ^2-3740 \eta +2646\right) N^7
+12 \left(423 \eta ^2-356 \eta +423\right) N^6
\nonumber \\ &&
+\left(4563 \eta ^2-302 \eta +4563\right) N^5
+32 \left(81 \eta ^2+112 \eta +81\right) N^4
\nonumber \\ &&
+16 \left(54 \eta ^2+533 \eta +54\right) N^3
+8328 \eta  N^2+4032 \eta  N + 864 \eta
~,
\\
P_{63} &=&
 -3 \left(45 \eta ^2+784 \eta -45\right) N^{10}
-15 \left(45 \eta ^2+784 \eta -45\right) N^9
\nonumber \\ &&
-8 \left(135 \eta ^2+1696 \eta -135\right) N^8
+\left(-270 \eta ^2+10528 \eta +270\right) N^7
\nonumber \\ &&
+5 \left(189 \eta ^2+2480 \eta -189\right) N^6
+\left(945 \eta ^2-52496 \eta -945\right) N^5
\nonumber \\ &&
+\left(270 \eta ^2-36832 \eta -270\right) N^4
+53664 \eta  N^3+71008 \eta  N^2+37632 \eta  N
\nonumber \\ &&
+12672 \eta
~,
\\
P_{64} &=&
 3 \left(45 \eta ^2-784 \eta -45\right) N^{10}
+15 \left(45 \eta ^2-784 \eta -45\right) N^9
\nonumber \\ &&
+8 \left(135 \eta ^2-1696 \eta -135\right) N^8
+2 \left(135 \eta ^2+5264 \eta -135\right) N^7
\nonumber \\ &&
+\left(-945 \eta ^2+12400 \eta +945\right) N^6
+\left(-945 \eta ^2-52496 \eta +945\right) N^5
\nonumber \\ &&
+\left(-270 \eta ^2-36832 \eta +270\right) N^4
+53664 \eta  N^3+71008 \eta  N^2+37632 \eta  N 
\nonumber \\ &&
+ 12672 \eta
~,
\\
P_{65} &=&
 20 \left(405 \eta ^2-10412 \eta +405\right) N^{10}
+\left(6561 \eta ^2-373928 \eta +6561\right) N^9
\nonumber \\ &&
+\left(-37422 \eta ^2+662146 \eta -37422\right) N^8
+\left(-14175 \eta ^2+1155334 \eta -14175\right) N^7
\nonumber \\ &&
+2 \left(8505 \eta ^2-213523 \eta +8505\right) N^6
+2 \left(8667 \eta ^2-495421 \eta +8667\right) N^5
\nonumber \\ &&
+10 \left(11907 \eta ^2+15026 \eta +11907\right) N^4
+12 \left(7722 \eta ^2+19067 \eta +7722\right) N^3
\nonumber \\ &&
-18 \left(243 \eta ^2-1316 \eta +243\right) N^2
+77760 \eta  N + 25920 \eta
~,
\\
P_{66} &=&
 2052 \left(21 \eta ^2+31 \eta +21\right) N^{12}
+324 \left(449 \eta ^2+589 \eta +449\right) N^{11}
\nonumber \\ &&
+\left(-143289 \eta ^2+4324133 \eta -143289\right) N^{10}
+\left(-619569 \eta ^2+7670353 \eta -619569\right) N^9
\nonumber \\ &&
-4 \left(45360 \eta ^2+86993 \eta +45360\right) N^8
+2 \left(227529 \eta ^2-7933945 \eta +227529\right) N^7
\nonumber \\ &&
-\left(18225 \eta ^2+21127667 \eta +18225\right) N^6
-\left(836973 \eta ^2+9493739 \eta +836973\right) N^5
\nonumber \\ &&
-50 \left(16605 \eta ^2-80419 \eta +16605\right) N^4
-24 \left(11421 \eta ^2-125029 \eta +11421\right) N^3
\nonumber \\ &&
-1225440 \eta  N^2-518400 \eta  N + 181440 \eta
~,
\\
P_{67} &=&
 12 \left(405 \eta ^2-3766 \eta +405\right) N^{14}
+48 \left(405 \eta ^2-3766 \eta +405\right) N^{13}
\nonumber \\ &&
+\left(8505 \eta ^2+20626 \eta +8505\right) N^{12}
-6 \left(7155 \eta ^2-116218 \eta +7155\right) N^{11}
\nonumber \\ &&
-\left(9315 \eta ^2+228902 \eta +9315\right) N^{10}
+\left(322866 \eta ^2-3020828 \eta +322866\right) N^9
\nonumber \\ &&
+\left(815427 \eta ^2-112666 \eta +815427\right) N^8
+\left(952074 \eta ^2+4787348 \eta +952074\right) N^7
\nonumber \\ &&
+45 \left(14967 \eta ^2+41806 \eta +14967\right) N^6
+2 \left(162243 \eta ^2-504122 \eta +162243\right) N^5
\nonumber \\ &&
+32 \left(2592 \eta ^2+35513 \eta +2592\right) N^4
+1629312 \eta  N^3+670752 \eta  N^2
\nonumber \\ &&
+86400 \eta  N -72576 \eta
~.
\end{eqnarray}
The expression for $\tilde{a}^{(3)}_{gg,Q}(N)$ exhibits potential poles at $N = 1/2$ and $N = 3/2$ due to rational
pre-factors, which have to be investigated. An expansion around the corresponding values in $N$ using {\tt 
HarmonicSums} shows, after some calculation, that these poles vanish for general values of $\eta$. 
In the case $\eta = 1$, the corresponding result had been obtained in Ref.~\cite{Ablinger:2014tla} before.
For the proof in the case $\eta \in ]0,1]$, 201 special replacement rules had to be derived and applied. A few of 
them are presented in 
Appendix~\ref{app:C}. 
\subsection{Momentum fraction space}
\label{sec:62}

\vspace*{1mm}
\noindent
In $z$-space,  $\tilde{a}_{gg,Q}^{(3)}$ receives three contributions,
the $\delta$-distribution, a $+$-distribution and a regular part, since it belongs to one of the
diagonal OMEs. Their Mellin transform reads
\begin{eqnarray}
\tilde{a}_{gg,Q}^{(3)} (N) &=& \int_{0}^{1} \text{d} z \ z^{N-1} \ \delta (1-z) \ \tilde{a}_{gg,Q}^{(3),\delta} (z) + \int_{0}^{1} \text{d} z \ \left( z^{N-1} - 1 \right) \ \tilde{a}_{gg,Q}^{(3),+} (z)
\NN \\ 
 &+& \int_{0}^{1} \text{d} z \ z^{N-1} \ \tilde{a}_{gg,Q}^{(3),\text{reg}} (z) ~.
\end{eqnarray}
In turn, the different terms can be obtained by a  Mellin inversion:


with the polynomials
\begin{eqnarray}
Q_{7} &=& 4853 \left(\eta ^2-\eta +1\right)+6 \left(3569 \eta ^2+15296 \eta +3569\right) z~,
\\
Q_{8} &=& 40 \left(9 \eta ^2+215 \eta +9\right)+\left(459 \eta ^2+4730 \eta +459\right) z-880 \eta  z^2~,
\\
Q_{9} &=& 95 \eta ^2+130 \eta +95+80 \left(\eta ^2+8 \eta +1\right) z+\left(68 \eta ^2-8 \eta +68\right) z^2~,
\\
Q_{10} &=& 3627 \eta ^2-22422 \eta +3627 +40 \left(209 \eta ^2+3130 \eta +209\right) z
\nonumber \\ &&
+840 \left(17 \eta ^2-2 \eta +17\right) z^2~,
\\
Q_{11} &=& 1400 \left(117 \eta ^2-1793 \eta +117\right)+7 \left(18873 \eta ^2-546250 \eta +18873\right) z
\nonumber \\ &&
+25 \left(1755 \eta ^2-826 \eta +1755\right) z^2~,
\\
Q_{12} &=& 1280 \eta +40 \left(9 \eta ^2-86 \eta +9\right) z+\left(459 \eta ^2+250 \eta +459\right) z^2 +440 \eta  z^3~,
\\
Q_{13} &=& 1230 \eta +10 \left(36 \eta ^2-551 \eta +36\right) z +\left(459 \eta ^2+1240 \eta +459\right) z^2 +1570 \eta  z^3~,
\\
Q_{14} &=& \left(1 - \eta ^2\right) \left(19950-1400 z-3969 z^2 - 4875 z^3\right)~,
\\
Q_{15} &=& -350 \left(513 \eta ^2+454 \eta +513\right)+12600 \left(\eta ^2+25 \eta +1\right) z+63 \bigl(567 \eta ^2
\nonumber \\ &&
-2150 \eta +567\bigr) z^2
+25 \left(1755 \eta ^2-826 \eta +1755\right) z^3~,
\\
Q_{16} &=& (z-1) \bigl[ 208 \left(5319 \eta ^2+24500 \eta +5319\right)+\left(30861 \eta ^2-36940750 \eta +30861\right)  z
\nonumber \\ &&
+25 \left(5265 \eta ^2-377902 \eta +5265\right) z^2 \bigr]
\\ 
Q_{17} &=&
-9 \eta ^2-760 \eta - 9 + 8 \left(18 \eta ^2-95 \eta +18\right) z
~,\\ 
Q_{18} &=&
-9 \eta ^2+100 \eta -9 +4 \left(36 \eta ^2+25 \eta +36\right) z
~,\\ 
Q_{19} &=&
-18 \left(\eta ^2+1\right)+\left(27 \eta ^2-320 \eta +27\right) z -320 \eta  z^2 
~,\\  
Q_{20} &=&
\eta ^2-\eta ^2 z+4 \left(\eta ^2-1\right) z^2
~,\\ 
Q_{21} &=&
\left(\eta ^2+1\right) \left(1-z+4 z^2\right)
~,\\  
Q_{22} &=&
\left(\eta ^2-1\right) \left(-141 +23 z + 12 z^2\right)
~,\\  
Q_{23} &=&
\left(\eta ^2-1\right) \left(2 -z + 16 z^2\right)
~,\\  
Q_{24} &=&
-24 \left(3 \eta ^2-8 \eta +3\right)-\left(109 \eta ^2+446 \eta +109\right) z +4 \left(11 \eta ^2-86 \eta +11\right) z^2
~,\\  
Q_{25} &=&
36 \left(\eta ^2+1\right)-5 \left(9 \eta ^2-32 \eta +9\right) z +8 \left(9 \eta ^2+20 \eta +9\right) z^2
~,\\  
Q_{26} &=&
18 \left(\eta ^2+1\right)-5 \left(9 \eta ^2-64 \eta +9\right) z +8 \left(9 \eta ^2+40 \eta +9\right) z^2
~,\\  
Q_{27} &=&
-7 \left(279 \eta ^2-538 \eta +279\right)+4 \left(153 \eta ^2+5296 \eta +153\right) z 
\nonumber \\ &&
+8 \left(27 \eta ^2-520 \eta +27\right) z^2
~,\\  
Q_{28} &=&
63 \left(23 \eta ^2+2874 \eta +23\right)-21 \left(1035 \eta ^2-21322 \eta +1035\right) z 
\nonumber \\ &&
+\left(8586 \eta ^2-66472 \eta +8586\right) z^2
~,\\  
Q_{29} &=&
20 \left(27 \eta ^2+208 \eta +27\right)+\left(117 \eta ^2-86 \eta +117\right) z+4 \left(117 \eta ^2-986 \eta +117\right) z^2 
\nonumber \\ && -2080 \eta  z^3
~,\\  
Q_{30} &=&
5 \eta +\left(\eta ^2-95 \eta +1\right) z +4 \left(2 \eta ^2-55 \eta +2\right) z^2 -80 \eta  z^3
~,\\  
Q_{31} &=&
-20 \eta -3 \left(3 \eta ^2-40 \eta +3\right) z +12 \left(3 \eta ^2+5 \eta +3\right) z^2 +20 \eta  z^3
~,\\  
Q_{32} &=&
\left(\eta ^2-1\right) \left(5 -17 z +10 z^2 + 2 z^3\right)
~,\\  
Q_{33} &=&
5 \left(27 \eta ^2+208 \eta +27\right)-3 \left(153 \eta ^2+220 \eta +153\right) z+30 \left(9 \eta ^2+26 \eta +9\right) z^2 
\nonumber \\ &&
+2 \left(27 \eta ^2-520 \eta +27\right) z^3
~,\\ 
Q_{34} &=&
5 \left(27 \eta ^2+104 \eta +27\right)-3 \left(153 \eta ^2+110 \eta +153\right) z+90 \left(3 \eta ^2-\eta +3\right) z^2 
\nonumber \\ &&
+\left(54 \eta ^2-520 \eta +54\right) z^3
~,\\  
Q_{35} &=&
42 \left(\eta ^2+159 \eta +1\right)-\left(183 \eta ^2+3088 \eta +183\right) z+3 \left(71 \eta ^2+1044 \eta +71\right) z^2 
\nonumber \\ &&
+\left(111 \eta ^2-6890 \eta +111\right) z^3
~,\\  
Q_{36} &=&
-19320 \eta +\left(-1953 \eta ^2+21806 \eta -1953\right) z +\left(612 \eta ^2-8896 \eta +612\right) z^2 
\nonumber \\ &&
+8 \left(27 \eta ^2+2155 \eta +27\right) z^3
~,\\  
Q_{37} &=&
7479 \eta ^2+1869560 \eta +7479 -\bigl(47655 \eta ^2+1947526 \eta+47655\bigr) z 
-\bigl(28593 \eta ^2
\nonumber \\ &&
-2351174 \eta +28593\bigr) z^2 +2 \left(55647 \eta ^2-1501024 \eta +55647\right) z^3
~,\\  
Q_{38} &=&
-1403 \left(\eta ^2-\eta +1\right)
-\left(9445 \eta ^2+10652 \eta +9445\right) z
+3 \big(4789 \eta ^2-10942 \eta
\nonumber \\ &&
 +4789\big) z^2
+\left(278 \eta ^2+47476 \eta +278\right) z^3
-4 \left(3023 \eta ^2-5606 \eta +3023\right) z^4
\nonumber \\ &&
+336 \left(11 \eta ^2-86 \eta +11\right) z^5
~.
\end{eqnarray}
In the above equations a series of functions, $F_k$, have been used. They further depend on the 
functions $G_k(y)$ and $K_k$, which are given in Appendix~\ref{app:D} and for which we suppress the 
$\eta$
dependence for brevity. The functions $F_k$ are given by 

The additional polynomials $R_k$ are given by
\begin{eqnarray}
R_{14} &=&
2 \eta  \left(5 \eta ^2+54 \eta +5\right)+\left(19 \eta ^2+82 \eta +19\right) (\eta -1)^2 y
~,\\
R_{15} &=&
-2 \eta ^2 \left(9 \eta ^2-82 \eta +9\right)-4 \eta  \left(7 \eta ^4-20 \eta ^3+90 \eta ^2-20 \eta +7\right) y
\nonumber \\ &&
-(\eta -1)^2 \left(19 \eta ^4+74 \eta ^3+198 \eta ^2+74 \eta +19\right) y^2
\nonumber \\ &&
+\left(\eta ^2-1\right)^2 \left(19 \eta ^2+26 \eta +19\right) y^3
~,\\
R_{16} &=&
\left(\eta ^2-1\right) \biggl[ 
18 \eta ^2
-10 \eta  \left(\eta ^2+10 \eta +1\right) y
-(\eta -1)^2 \left(19 \eta ^2+110 \eta +19\right) y^2
\nonumber \\ &&
+(\eta -1)^2 \left(19 \eta ^2+82 \eta +19\right) y^3
\biggr]
~,\\
R_{17} &=&
\eta  \left(\eta ^2+18 \eta +1\right)+\left(\eta ^2+10 \eta +1\right) (\eta -1)^2 y
~,\\
R_{18} &=&
\left(\eta ^2-1\right) y \biggl[
-\eta  \left(\eta ^2+10 \eta +1\right)
-\left(\eta ^2+11 \eta +1\right) (\eta -1)^2 y
\nonumber \\ &&
+\left(\eta ^2+10 \eta +1\right) (\eta -1)^2 y^2
\biggr]
~,\\
R_{19} &=&
20 \eta ^3
-\eta  \left(\eta ^4-8 \eta ^3+54 \eta ^2-8 \eta +1\right) y
\nonumber \\ &&
-(\eta -1)^2 \left(\eta ^4+11 \eta ^3+36 \eta ^2+11 \eta +1\right) y^2
\nonumber \\ &&
+\left(\eta ^2-1\right)^2 \left(\eta ^2+8 \eta +1\right) y^3
~,\\
R_{20} &=&
\eta  \left(83 \eta ^2-54 \eta +83\right)+2 \left(\eta ^2-53 \eta +1\right) (\eta -1)^2 y
~,\\
R_{21} &=&
\left(\eta ^2-1\right) y \biggl[
\eta  \left(-83 \eta ^2+268 \eta -83\right)
-\left(2 \eta ^2-185 \eta +2\right) (\eta -1)^2 y
\nonumber \\ &&
+2 \left(\eta ^2-53 \eta +1\right) (\eta -1)^2 y^2
\biggr]
~,\\
R_{22} &=&
112 \eta ^3
+\eta  \left(79 \eta ^4-110 \eta ^3-162 \eta ^2-110 \eta +79\right) y
\nonumber \\ &&
-(\eta -1)^2 \left(2 \eta ^4+139 \eta ^3+54 \eta ^2+139 \eta +2\right) y^2
\nonumber \\ &&
+2 \left(\eta ^2-1\right)^2 \left(\eta ^2+26 \eta +1\right) y^3
~,\\
R_{23} &=&
-2 \eta  \left(59 \eta ^2-150 \eta +59\right)
+(\eta -1)^2 \left(17 \eta ^2+302 \eta +17\right) y
~,\\
R_{24} &=&
-2 \eta ^2 \left(45 \eta ^2-122 \eta +45\right)
-8 \eta  \left(19 \eta ^4-56 \eta ^3+90 \eta ^2-56 \eta +19\right) y
\nonumber \\ &&
-(\eta -1)^2 \left(17 \eta ^4-86 \eta ^3+330 \eta ^2-86 \eta +17\right) y^2
\nonumber \\ &&
+\left(\eta ^2-1\right)^2 \left(17 \eta ^2-2 \eta +17\right) y^3
~,\\
R_{25} &=&
\left(\eta ^2-1\right) \biggl[
90 \eta ^2
+2 \eta  \left(59 \eta ^2-286 \eta +59\right) y
-(\eta -1)^2 \left(17 \eta ^2+454 \eta +17\right) y^2
\nonumber \\ &&
+(\eta -1)^2 \left(17 \eta ^2+302 \eta +17\right) y^3
\biggr]
~,\\
R_{26} &=&
-2 \eta  \left(177 \eta ^2-50 \eta +177\right)
+(\eta -1)^2 \left(21 \eta ^2+446 \eta +21\right) y
~,\\
R_{27} &=&
-2 \eta ^2 \left(75 \eta ^2+154 \eta +75\right)
+4 \eta  \left(-99 \eta ^4+176 \eta ^3+150 \eta ^2+176 \eta -99\right) y
\nonumber \\ &&
-(\eta -1)^2 \left(21 \eta ^4-658 \eta ^3-550 \eta ^2-658 \eta +21\right) y^2
\nonumber \\ &&
+\left(\eta ^2-1\right)^2 \left(21 \eta ^2-346 \eta +21\right) y^3
~,\\
R_{28} &=&
\left(\eta ^2-1\right) \biggl[
2 \eta  \left(177 \eta ^2-598 \eta +177\right) y
-(\eta -1)^2 \left(21 \eta ^2+842 \eta +21\right) y^2
\nonumber \\ &&
+150 \eta ^2+(\eta -1)^2 \left(21 \eta ^2+446 \eta +21\right) y^3
\biggr]
~,\\
R_{29} &=&
-2 \eta  \left(125 \eta ^2-882 \eta +125\right)
+(\eta -1)^2 \left(65 \eta ^2+1262 \eta +65\right) y
~,\\
R_{30} &=&
-4 \eta ^2 \left(63 \eta ^2-442 \eta +63\right)
-4 \eta  \left(95 \eta ^4-409 \eta ^3+1260 \eta ^2-409 \eta +95\right) y
\nonumber \\ &&
-(\eta -1)^2 \left(65 \eta ^4+382 \eta ^3+2898 \eta ^2+382 \eta +65\right) y^2
\nonumber \\ &&
+\left(\eta ^2-1\right)^2 \left(65 \eta ^2+502 \eta +65\right) y^3
~,\\
R_{31} &=&
\left(\eta ^2-1\right) \biggl[
2 \eta  \left(125 \eta ^2-946 \eta +125\right) y
-(\eta -1)^2 \left(65 \eta ^2+1642 \eta +65\right) y^2
\nonumber \\ &&
+252 \eta ^2+(\eta -1)^2 \left(65 \eta ^2+1262 \eta +65\right) y^3
\biggr]
~,\\
R_{32} &=&
\eta ^3+\eta +(\eta -1)^2 \left(\eta ^2+\eta +1\right) y
~,\\
R_{33} &=&
\left(\eta ^2-1\right) y \biggl[
-\eta  \left(\eta ^2+\eta +1\right)
-\left(\eta ^2-1\right)^2 y
+\left(\eta ^2+\eta +1\right) (\eta -1)^2 y^2
\biggr]
~,\\
R_{34} &=&
2 \eta ^3
-\left(\eta ^5+\eta ^4+\eta ^2+\eta \right) y
-(\eta -1)^2 \left(\eta ^4+2 \eta ^3+2 \eta +1\right) y^2
\nonumber \\ &&
+\left(\eta ^2-1\right)^2 \left(\eta ^2-\eta +1\right) y^3
~,\\
R_{35} &=&
\left(\eta ^4+\eta ^3+\eta +1\right) \sqrt{z}
~,\\
R_{36} &=&
-2 \left(\eta ^5+\eta \right)
-\left(\eta ^4+1\right) (\eta -1)^2 z
+\left(\eta ^2-1\right)^2 \left(\eta ^2-\eta +1\right) z^2
~,\\
R_{37} &=&
\left(\eta ^3+\eta ^2+\eta +1\right) \sqrt{z} \bigl[
        -\eta 
        -(\eta -1)^2 z
        +(\eta -1)^2 z^2
\bigr]
~,\\
R_{38} &=&
\left(\eta ^5+\eta ^4+\eta ^3+\eta ^2+\eta +1\right) \sqrt{z} \bigl[
        -\eta 
        -(\eta -1)^2 z
        +(\eta -1)^2 z^2
\bigr]
~,\\
R_{39} &=&
(z-1) \bigl[ 1 + \eta ^4 -5 \eta ^4 z +5 \eta ^3 z-z -4 z \eta ^3(1-z) +4 z \eta ^4(1-z) + 4 \eta ^4 z^2
\nonumber \\ &&
-4 \eta ^3 z^2\bigr]-z \eta (1-z)
~,\\
R_{40} &=&
(\eta +1) (z-1)^2 \bigl[
        -2 \left(\eta ^3+\eta \right)
        -\left(\eta ^2+1\right) (\eta -1)^2 z 
\nonumber \\ &&
        +\left(\eta ^2+\eta +1\right) (\eta -1)^2 z^2 
\bigr]
~,\\
R_{41} &=&
-73 \eta ^4-90 \eta ^3-90 \eta -73 +(\eta -1)^2 \left(73 \eta ^2+163 \eta +73\right) z
~,\\
R_{42} &=&
\left(\eta ^2-1\right) \biggl[
\eta  \left(269 \eta ^2+220 \eta +269\right)
+\bigl(219 \eta ^4-437 \eta ^3 
\nonumber \\ &&
  -491 \eta ^2-437 \eta +219 \bigr) z
-3 (\eta -1)^2 \left(146 \eta ^2+253 \eta +146\right) z^2
\nonumber \\ &&
+3 (\eta -1)^2 \left(73 \eta ^2+163 \eta +73\right) z^3
\biggr]
~,\\
R_{43} &=&
\eta  \left(269 \eta ^4+220 \eta ^3+220 \eta +269\right)
+\bigl(219 \eta ^6-437 \eta ^5-710 \eta ^4
\nonumber \\ &&
  -100 \eta ^3-710 \eta ^2-437 \eta +219\bigr) z
-3 (\eta -1)^2 \bigl(146 \eta ^4+253 \eta ^3
\nonumber \\ &&
  +180 \eta ^2+253 \eta +146\bigr) z^2
+3 \left(\eta ^2-1\right)^2 \left(73 \eta ^2+17 \eta +73\right) z^3
~,\\
R_{44} &=&
-25 \eta ^4-128 \eta ^3+210 \eta ^2-128 \eta -25 +2 (\eta -1)^2 \left(61 \eta ^2+226 \eta +61\right) y
~,\\
R_{45} &=&
-2 \eta  \left(75 \eta ^4+524 \eta ^3-1054 \eta ^2+524 \eta +75\right)
-\bigl(75 \eta ^6+486 \eta ^5
\nonumber \\ &&
  -2795 \eta ^4+3892 \eta ^3-2795 \eta ^2+486 \eta +75\bigr) y
-3 (\eta -1)^2 \bigl(47 \eta ^4
\nonumber \\ &&
  -176 \eta ^3-30 \eta ^2-176 \eta +47\bigr) y^2
+72 \left(\eta ^2-1\right)^2 \left(3 \eta ^2-8 \eta +3\right) y^3
~,\\
R_{46} &=&
\left(\eta ^2-1\right) \biggl[
2 \eta  \left(75 \eta ^2+704 \eta +75\right)
+3 \left(25 \eta ^4+88 \eta ^3-922 \eta ^2+88 \eta +25\right) y
\nonumber \\ &&
-9 (\eta -1)^2 \left(49 \eta ^2+258 \eta +49\right) y^2
+6 (\eta -1)^2 \left(61 \eta ^2+226 \eta +61\right) y^3
\biggr]
~,\\
R_{47} &=&
127 \eta ^4+736 \eta ^3+930 \eta ^2+736 \eta +127 -36 (\eta -1)^2 \left(\eta ^2+6 \eta +1\right) y
~,\\
R_{48} &=&
\left(\eta ^2-1\right) \biggr[
144 \eta ^2
+\left(127 \eta ^4+530 \eta ^3-1602 \eta ^2+530 \eta +127\right) y
\nonumber \\ &&
-(\eta -1)^2 \left(163 \eta ^2+770 \eta +163\right) y^2
+36 (\eta -1)^2 \left(\eta ^2+6 \eta +1\right) y^3
\biggl]
~,\\
R_{49} &=&
16 \eta ^2 \left(3 \eta ^2+160 \eta +3\right)
+\bigl(127 \eta ^6+252 \eta ^5-367 \eta ^4-5336 \eta ^3-367 \eta ^2
\nonumber \\ &&
  +252 \eta +127\bigr) y
-3 (\eta -1)^2 \left(115 \eta ^4+688 \eta ^3+1050 \eta ^2+688 \eta +115\right) y^2
\nonumber \\ &&
+2 \left(\eta ^2-1\right)^2 \left(109 \eta ^2+446 \eta +109\right) y^3
~,\\
R_{50} &=&
-993 \eta ^4-1324 \eta ^3+3690 \eta ^2-1324 \eta -993 
\nonumber \\ &&
+4 (\eta -1)^2 \left(581 \eta ^2+1706 \eta +581\right) y
~,\\
R_{51} &=&
-\left(\eta ^2-1\right) \biggl[
6 \eta  \left(127 \eta ^2+1262 \eta +127\right)
+\bigl(993 \eta ^4+1024 \eta ^3-15506 \eta ^2
\nonumber \\ &&
  +1024 \eta +993\bigr) y
-(\eta -1)^2 \left(3317 \eta ^2+10810 \eta +3317\right) y^2
\nonumber \\ &&
+4 (\eta -1)^2 \left(581 \eta ^2+1706 \eta +581\right) y^3
\biggr]
~,\\
R_{52} &=&
-2 \eta  \left(381 \eta ^4+1338 \eta ^3-2966 \eta ^2+1338 \eta +381\right)
+\bigl(-993 \eta ^6+1210 \eta ^5
\nonumber \\ &&
  +8521 \eta ^4-15588 \eta ^3+8521 \eta ^2+1210 \eta -993\bigr) y
+(\eta -1)^2 \bigl(655 \eta ^4
\nonumber \\ &&
  +1796 \eta ^3-2070 \eta ^2+1796 \eta +655\bigr) y^2
\nonumber \\ &&
+2 \left(\eta ^2-1\right)^2 \left(169 \eta ^2-574 \eta +169\right) y^3
~,\\
R_{53} &=&
\left(\eta ^2-1\right) \biggl[
-2 \eta  \left(11 \eta ^2-160 \eta +11\right)
-\left(33 \eta ^2-406 \eta +33\right) (\eta -1)^2 y
\nonumber \\ &&
+3 \left(11 \eta ^2-86 \eta +11\right) (\eta -1)^2 y^2
\biggr]
~,\\
R_{54} &=&
-2 \eta  \left(11 \eta ^4+32 \eta ^3-470 \eta ^2+32 \eta +11\right)
+\bigl(-33 \eta ^6+154 \eta ^5+161 \eta ^4
\nonumber \\ &&
  -2100 \eta ^3+161 \eta ^2+154 \eta -33\bigr) y
+9 \left(\eta ^2-1\right)^2 \left(11 \eta ^2-86 \eta +11\right) y^2
\nonumber \\ &&
-6 \left(\eta ^2-1\right)^2 \left(11 \eta ^2-86 \eta +11\right) y^3
~,\\
R_{55} &=&
-353 \eta ^4-1200 \eta ^3-750 \eta ^2-1200 \eta -353 +8 (\eta -1)^2 \bigl(31 \eta ^2
\nonumber \\ &&
  +121 \eta +31\bigr) y
~,\\
R_{56} &=&
-4 \eta  \left(57 \eta ^4+275 \eta ^3+300 \eta ^2+275 \eta +57\right)
+\bigl(-353 \eta ^6-296 \eta ^5+3025 \eta ^4
\nonumber \\ &&
  +2960 \eta ^3+3025 \eta ^2-296 \eta -353\bigr) y
+(\eta -1)^2 \bigl(811 \eta ^4+3128 \eta ^3+3690 \eta ^2
\nonumber \\ &&
  +3128 \eta +811\bigr) y^2
-2 \left(\eta ^2-1\right)^2 \left(229 \eta ^2+506 \eta +229\right) y^3
~,\\
R_{57} &=&
-\left(\eta ^2-1\right) \biggl[
4 \eta  \left(57 \eta ^2+155 \eta +57\right)
+\left(353 \eta ^4+26 \eta ^3-2222 \eta ^2+26 \eta +353\right) y
\nonumber \\ &&
-(\eta -1)^2 \left(601 \eta ^2+1958 \eta +601\right) y^2
+8 (\eta -1)^2 \left(31 \eta ^2+121 \eta +31\right) y^3
\biggr]
~,\\
R_{58} &=&
623 \eta ^4+4124 \eta ^3+1386 \eta ^2+4124 \eta +623 -54 (\eta -1)^2 \left(11 \eta ^2+74 \eta +11\right) y
~,\\
R_{59} &=&
\left(\eta ^2-1\right) \biggl[
28 \eta  \left(267 \eta ^2+1916 \eta +267\right)
+15 \bigl(623 \eta ^4+2514 \eta ^3-11458 \eta ^2
\nonumber \\ &&
  +2514 \eta +623\bigr) y
-15 (\eta -1)^2 \left(1217 \eta ^2+8062 \eta +1217\right) y^2
\nonumber \\ &&
+810 (\eta -1)^2 \left(11 \eta ^2+74 \eta +11\right) y^3
\biggr]
~,\\
R_{60} &=&
4 \eta  \left(1869 \eta ^4+12404 \eta ^3+12254 \eta ^2+12404 \eta +1869\right)
+\bigl(9345 \eta ^6+32808 \eta ^5
\nonumber \\ &&
  -150697 \eta ^4-109312 \eta ^3-150697 \eta ^2+32808 \eta +9345\bigr) y
\nonumber \\ &&
-45 (\eta -1)^2 \left(425 \eta ^4+3188 \eta ^3+3654 \eta ^2+3188 \eta +425\right) y^2
\nonumber \\ &&
+60 \left(\eta ^2-1\right)^2 \left(163 \eta ^2+1034 \eta +163\right) y^3
~,\\
R_{61} &=&
\eta ^3+\eta +(\eta -1)^2 \left(\eta ^2+\eta +1\right) y
~,\\
R_{62} &=&
\left(\eta ^2-1\right) y \biggl[
-\eta  \left(\eta ^2+\eta +1\right)
-\left(\eta ^2-1\right)^2 y
+\left(\eta ^2+\eta +1\right) (\eta -1)^2 y^2
\biggr]
~,\\
R_{63} &=&
2 \eta ^3
-\left(\eta ^5+\eta ^4+\eta ^2+\eta \right) y
-(\eta -1)^2 \left(\eta ^4+2 \eta ^3+2 \eta +1\right) y^2
\nonumber \\ &&
+\left(\eta ^2-1\right)^2 \left(\eta ^2-\eta +1\right) y^3
~.
\end{eqnarray}
We remark that in intermediary steps of the calculation also a lot of constants appear, which are no multiple 
zeta values, see also Appendix~\ref{app:D}. They all cancel in the result given above.

\subsection{Transformation to the \boldmath $\overline{\sf MS}$ scheme}
\label{sec:63}

\vspace*{1mm}
\noindent
Since there is a finite two-mass contribution $\tilde{A}_{gg,Q}^{(2)}$, which depends on both 
heavy quark masses, at 3-loop order the OME $\tilde{A}_{gg,Q}^{(3)}$ differs if calculated 
in the on-mass shell scheme ({\sf OMS}) or the $\overline{\sf MS}$ scheme. 
The change from the ${\sf OMS}$- to the $\overline{\sf MS}$-mass is best performed on the complete, 
renormalized OME, which we present in the following.

The relation between the ${\sf OMS}$-mass $m_i$ and the $\overline{\sf MS}$-mass 
$\overline{m}_i$ for, $i=1,2$ reads 
\begin{eqnarray}
	c_{m_i} = \frac{m_i}{\bar{m}_i} = 1 + \sum\limits_{k=1}^\infty a_s^k c_{m_i}^{(k)}~,
\end{eqnarray}
with the coefficients, cf. Section~\ref{sec:2},
\begin{eqnarray}
	c_{m_i}^{(1)} &=& 4 C_F \left( 1 - \frac{3}{4} L_i \right)~, \\
	c_{m_i}^{(2)} &=& 
C_F N_F T_F 
\Biggl[
        -\frac{71}{6}
        -8 \zeta_2
        +\frac{26}{3} L_i
        -2 L_i^2
\Biggr]
+C_A C_F 
\Biggl[
        \frac{1111}{24}
        -8 \zeta_2
        +24 \zeta_2 \ln(2)
        -6 \zeta_3
        -\frac{185}{6} L_i
\nonumber \\ &&
        +\frac{11}{2} L_i^2
\Biggr]
+C_F T_F 
\Biggl[
        -\frac{107}{3}
        +8 \zeta_2
        +\frac{52}{3} L_i
        -4 L_i^2
        +24 \zeta_2 r_i
        -12 r_i^2
        -8  r_i^2 \HA_0\big(r_i\big)
        +24 \zeta_2 r_i^3
\nonumber \\ &&
        -8 r_i^4 \HA_0^2\big(r_i\big) 
        -8 \zeta_2 r_i^4
        -8\HA_{1,0}\big(r_i\big)
\biggl(1- r_i-r_i^3+r_i^4\biggr)
        +8\HA_{-1,0}\big(r_i\big)
\biggl(1+r_i+r_i^3+r_i^4\biggr)
\biggr]
\nonumber \\ &&
+ C_F^2
\Biggl[
        -\frac{71}{8}
        +30 \zeta_2
        -48 \ln(2) \zeta_2
        +12 \zeta_3
        +\frac{9}{2} L_i
        +\frac{9}{2} L_i^2
\Biggr]
\end{eqnarray}
using the definition of $r_i$ Eq.~(\ref{eq:r12}). In the r.h.s. the use of the $\MS$-masses is 
implied.

One obtains
\begin{eqnarray}
A_{gg,Q}^{(2),\MS} \left( \frac{\bar{m}_1^2}{\mu^2},\frac{\bar{m}_2^2}{\mu^2} \right)
&=& A_{gg,Q}^{(2),{\sf OMS}} \left( \frac{\bar{m}_1^2}{\mu^2},\frac{\bar{m}_2^2}{\mu^2} \right)
- 2 \beta_{0,Q} \left( c_{m_1}^{(1)} + c_{m_2}^{(1)} \right) \\
A_{gg,Q}^{(3),\MS} \left( \frac{\bar{m}_1^2}{\mu^2},\frac{\bar{m}_2^2}{\mu^2} \right)
&=& A_{gg,Q}^{(3),{\sf OMS}} \left( \frac{\bar{m}_1^2}{\mu^2},\frac{\bar{m}_2^2}{\mu^2} \right)
	+ 2 \beta_{0,Q} \Bigl\{ \left. c_{m_1}^{(1)} \right.^2 + \left. c_{m_2}^{(1)} \right.^2 - 2 ( c_{m_1}^{(2)} + c_{m_2}^{(2)} ) \Bigl\}
\nonumber \\ && \hspace{-0.5cm}
+ 4 \beta_{0,Q}^2 \Bigl\{ c_{m_1}^{(1)} L_2 + c_{m_2}^{(1)} L_1 \Bigr\} + \frac{1}{2} \hat{\gamma}_{gg}^{(1)} \Bigl\{ c_{m_1}^{(1)} + c_{m_2}^{(1)} \Bigr\} 
\nonumber \\ && \hspace{-0.5cm}
+\frac{1}{2} \Bigl[ 2 \beta_{0,Q} ( \gamma_{gg}^{(0)} + 2 \beta_0 ) + \frac{1}{2} \gamma_{gq}^{(0)} \hat{\gamma}_{qg}^{(0)} + 8 \beta_{0,Q}^2 \Bigr] \Bigl\{ c_{m_1}^{(1)} L_1 + c_{m_2}^{(1)} L_2 \Bigr\}~.
\end{eqnarray}
\subsection{Numerical results}
\label{sec:64}

\begin{figure}[H]\centering
\includegraphics[width=0.7\textwidth]{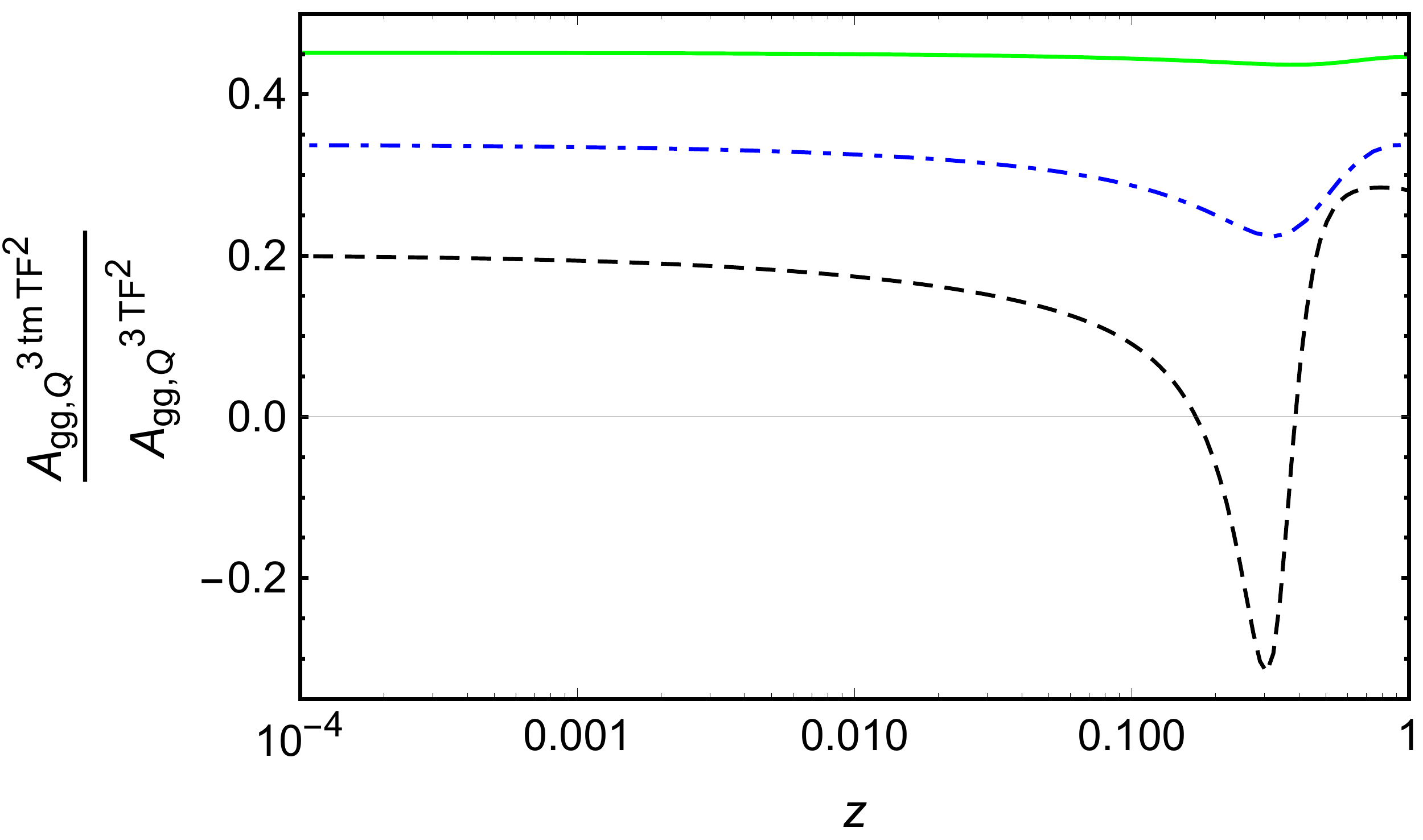}
\caption[]{\label{fig:num} \sf
The ratio of the two-mass (tm) contributions to the massive OME $A_{gg,Q}^{(3)\rm{tm}}$ to all contributions 
to $A_{gg,Q}^{(3)}$ of $O(T_F^2)$ as a function of $z$ and $\mu^2$.
Dashed line (black): $\mu^2 = 50~\GeV^2$.
Dash-dotted line (blue): $\mu^2 = 100~\GeV^2$.
Full line (green): $\mu^2 = 1000~\GeV^2$. Here the on-shell heavy quark masses $m_c = 1.59~\GeV$ and $m_b =
4.78~\GeV$ \cite{Alekhin:2012vu,Agashe:2014kda} have been used.}
\end{figure}

\vspace*{1mm}
\noindent
In Figure~\ref{fig:num} we compare the 3-loop two-mass effects contributing to $A_{gg,Q}$ to the complete 
effect of the ${\cal O}(T_F^2)$ term due to heavy quarks for a series of $\mu^2$ values as a function of $z$
in the open interval $[0,1[$. 

\noindent
The contribution of the two-mass term to the whole $T_F^2$-contribution is significant. At lower values of 
$\mu^2$ the ratio in Figure~\ref{fig:num} shows a profile varying with the momentum fraction $z$. It flattens at 
large $\mu^2$ due to the
dominating logarithms and reaches values of $O(0.4)$ at $\mu^2 \simeq 1000$~GeV$^2$.

\section{Conclusions}
\label{sec:7}

\vspace*{1mm}
\noindent
We have calculated the two-mass 3-loop contributions to the massive OME $A_{gg,Q}$ in analytic
form both in Mellin $N$- and $z$-space for a general mass ratio $\eta$. The OME contributes to 
the two-mass variable flavor number scheme. The close values of the charm and bottom quark masses
make it necessary to use this extended scheme. The relative contribution of the two-mass contributions 
to  the whole massive $T_F^2$ contributions of $A_{gg,Q}^{(3)}$ are significant and they amount to values 
of $O(0.4)$ for $\mu^2 \simeq 1000$~GeV$^2$. 

The OME has been first calculated in $N$-space by direct integration of the contributing Feynman 
integrals, which made one Mellin-Barnes representation necessary. The problem was thus turned into
a nested summation problem, in which the mass ratio $\eta$ appeared as fixed parameter in the ground
field. The corresponding sums could be calculated using the packages {\tt Sigma, EvaluateMultiSums} and {\tt 
SumProduction}, being the largest and most demanding computation we have ever performed as a summation project. 
For the 
infinite sums the limit $N \rightarrow \infty$ was performed using procedures of the
package {\tt HarmonicSums}. The overall computational time in the summation part amounted to four to five 
months, including runs needed for code optimization.
The $N$-space result contains harmonic sums, generalized harmonic sums due to the $\eta$ dependence,
and (inverse) binomial extensions thereof. The Mellin variable $N$ also occurs as exponent in $\eta$-ratios.
We proved analytically that the evanescent poles at $N =1/2$ and $N=3/2$ vanish.
The package {\tt HarmonicSums} provides algorithms to calculate the inverse Mellin transform of the $N$-space
expressions, which are needed for a series of phenomenological and experimental applications. This is the
case because not all parameterizations of parton densities have a simple Mellin space representation,
even not at the starting scale $Q_0^2$, cf.~\cite{Blumlein:2011zu}. The $z$-space representation can finally 
be given in terms of general iterated integrals over root-valued letters, also containing the parameter
$\eta$. These can be reduced to (poly)logarithms of involved arguments, up to one integral in some cases.
We were choosing this representation to obtain a fast numerical implementation. The corresponding integrals
can in principle be performed within the $G$-iterated integrals. However, corresponding fast numerical 
implementations would have still to be worked out for part of these functions.
We have checked that our general $N$-results and those in $z$-space are in accordance with the
moments we have calculated before for $N = 2,4,6$ using different techniques. 

With this contribution only one further two-mass OME, $A_{Qg}^{(3)}$, has to be calculated to complete all 
two-mass quantities
of the VFNS to 3-loop order. At 2-loop order the study of the two-mass VFNS has already been performed
in Ref.~\cite{VFNS2}.

During the calculation we obtained a series of analytic integrals, which are listed in the appendix. They 
are of use in further 3-loop two-mass calculations. One more result of the present analytic calculation is 
that special numbers, appearing in intermediary steps, and which are not multiple zeta values, cancel in the final 
result. This is as well the case for one singular Mellin transform due to the behaviour $\propto N^2$, which
cancels between different Feynman diagrams. 
\appendix
\section{Momentum Integrals}
\label{app:A}

\vspace*{1mm}
\noindent
The appearing tensor integrals are mapped to scalar integrals using the following relations
\begin{eqnarray}
 \int \frac{\dd^d k}{(2\pi)^d}  k_{\mu_1} k_{\mu_2} f(k^2) 
&=&  \frac{g_{\mu_1\mu_2}}{D} \int \frac{\dd^d k}{(2\pi)^d}  k^2 f(k^2), \\
\int \frac{\dd^d k}{(2\pi)^d}  k_{\mu_1} k_{\mu_2} k_{\mu_3} k_{\mu_4} f(k^2)  
&=& \frac{S_{\mu_1\mu_2\mu_3\mu_4}}{D(D+2)}  \int \frac{\dd^d k}{(2\pi)^d} (k^2)^2 f(k^2),
\nonumber\\
\\
\int \frac{\dd^d k}{(2\pi)^d}  k_{\mu_1} k_{\mu_2} k_{\mu_3} k_{\mu_4} k_{\mu_5} k_{\mu_6} f(k^2)  
&=& \frac{S_{\mu_1\mu_2\mu_3\mu_4\mu_5\mu_6}}{D(D+2)(D+4)} 
\int \frac{\dd^d k}{(2\pi)^d}  (k^2)^3 f(k^2),
\end{eqnarray}
with the symmetric tensors
\begin{eqnarray}
S_{\mu_1\mu_2\mu_3\mu_4} &=& g_{\mu_1\mu_2}g_{\mu_3\mu_4} + g_{\mu_1\mu_3}g_{\mu_2\mu_4} 
+ g_{\mu_1\mu_4} g_{\mu_2\mu_3} 
\\
S_{\mu_1\mu_2\mu_3\mu_4\mu_5\mu_6} &=& 
g_{\mu_1\mu_2} 
\left[
  g_{\mu_3\mu_4} g_{\mu_5\mu_6} 
+ g_{\mu_3\mu_5} g_{\mu_4\mu_6}
+ g_{\mu_3\mu_6} g_{\mu_4\mu_5}\right]
\nonumber\\
&&
+g_{\mu_1\mu_3} 
\left[
  g_{\mu_2\mu_4} g_{\mu_5\mu_6} 
+ g_{\mu_2\mu_5} g_{\mu_4\mu_6}
+ g_{\mu_2\mu_6} g_{\mu_4\mu_5}\right]
\nonumber\\
&&
+g_{\mu_1\mu_4} 
\left[
  g_{\mu_2\mu_3} g_{\mu_5\mu_6} 
+ g_{\mu_2\mu_5} g_{\mu_3\mu_6}
+ g_{\mu_2\mu_6} g_{\mu_3\mu_5}\right]
\nonumber\\
&&
+g_{\mu_1\mu_5} 
\left[
  g_{\mu_2\mu_3} g_{\mu_4\mu_6} 
+ g_{\mu_2\mu_4} g_{\mu_3\mu_6}
+ g_{\mu_2\mu_6} g_{\mu_3\mu_4}\right]
\nonumber\\
&&
+g_{\mu_1\mu_6} 
\left[
  g_{\mu_2\mu_3} g_{\mu_4\mu_5} 
+ g_{\mu_2\mu_4} g_{\mu_3\mu_5}
+ g_{\mu_2\mu_5} g_{\mu_3\mu_4}\right].
\end{eqnarray}
Furthermore, integrals in which the local operator insertion contributes are calculated using
\begin{eqnarray}
 \int \frac{\dd^d k}{(2\pi)^d}  \left( k.\Delta + R_0 p.\Delta  \right)^N f(k^2)  
&=&  (\Delta.p)^N \ R_0^N \ \int \frac{\dd^d k}{(2\pi)^d} f(k^2) , \\
 \int \frac{\dd^d k}{(2\pi)^d} \ p.k \ \left( k.\Delta + R_0 p.\Delta  \right)^N f(k^2)  
&=&  \frac{N}{D} (\Delta.p)^N \ R_0^{N-1} \ \int \frac{\dd^d k}{(2\pi)^d}  k^2 f(k^2) , \\
  \int \frac{\dd^d k}{(2\pi)^d} \ (p.k)^2 \ \left( k.\Delta + R_0 p.\Delta  \right)^N f(k^2)  
&=&  \frac{N(N-1)}{D(D+2)} (\Delta.p)^N \ R_0^{N-2} \ \int \frac{\dd^d k}{(2\pi)^d}  (k^2)^2 f(k^2) , \\
 \int \frac{\dd^d k}{(2\pi)^d} \ (p.k)^3 \ \left( k.\Delta + R_0 p.\Delta  \right)^N f(k^2)  
&=&  
\frac{N(N-1)(N-2)}{D(D+2)(D+4)} (\Delta.p)^N \ R_0^{N-3} \ \int \frac{\dd^d k}{(2\pi)^d} (k^2)^3 f(k^2). 
\nonumber\\
\end{eqnarray}
Other terms vanish, since they turn out to be $\propto \Delta.\Delta = 0$.
\section{The OMEs \boldmath $A_{gg,Q}^{(2),(3)}$ in $N$- and $z$-space}
\label{app:B}

\vspace*{1mm}
\noindent
The OMEs $A_{gg,Q}^{(2),(3)}$ are given in $N$- and $z$-space by~:
\begin{eqnarray}
\tilde{A}_{gg,Q}^{(2)}(N) &=& 
2 \beta_{0,Q}^2 L_1 L_2,\\
\tilde{A}_{gg,Q}^{(3)}(z) &=& 
2 \beta_{0,Q}^2 L_1 L_2 \delta(1-z)
\end{eqnarray}
and
\begin{eqnarray}
\tilde{A}_{gg,Q}^{(3)}(N) &=&
-\textcolor{blue}{T_F^3} \biggl\{
\frac{32}{3} ( L_1^3 + L_2^3 ) 
+\frac{128}{9} L_1 L_2 ( L_1 + L_2 )
+32 \zeta_2 ( L_1 + L_2 )
+\frac{128}{9} \zeta_3
\biggr\}
\nonumber \\ &&
+ \textcolor{blue}{C_F T_F^2} \biggl\{
-\frac{184 \big(2+N+N^2\big)^2}{9 (N-1) N^2 (N+1)^2 (N+2)} ( L_1^3 + L_2^3 )
\nonumber \\ &&
-\frac{32 \big(2+N+N^2\big)^2}{3 (N-1) N^2 (N+1)^2 (N+2)} L_1 L_2 ( L_1 + L_2 )
\nonumber \\ &&
+( L_1^2 + L_2^2 )
\biggl[
        \frac{8 P_{68}}{9 (N-1) N^3 (N+1)^3 (N+2)}
        -\frac{16 \big(2+N+N^2\big)^2}{3 (N-1) N^2 (N+1)^2 (N+2)} S_1
\biggr]
\nonumber \\ &&
+ \frac{256}{3} L_1 L_2
+ ( L_1 + L_2 )
\biggl[
        -\frac{32 P_{70}}{3 (N-1) N^4 (N+1)^4 (N+2)}
\nonumber \\ &&
        -\frac{40 \big(2+N+N^2\big)^2}{(N-1) N^2 (N+1)^2 (N+2)} \zeta_2
\biggr]
+\frac{4 P_{71}}{9 \eta  (N-1) N^5 (N+1)^5 (N+2)}
\nonumber \\ &&
-\frac{32 (1-\eta^2)}{3 \eta } \HA_0(\eta )
-\frac{16(1-\eta^2 )}{3} \HA_0^2(\eta )
\nonumber \\ &&
-\frac{32 \big(1-\sqrt{\eta }\big)^2 \big(1-\eta^2\big) \big(1+\sqrt{\eta }+\eta \big)}{3 \eta ^2} \biggl[ H_0(\eta ) H_1\big(\sqrt{\eta }\big) - 2 H_{0,1}\big(\sqrt{\eta }\big) \biggr]
\nonumber \\ &&
+\frac{32 \big(1+\sqrt{\eta }\big)^2 \big(1-\eta^2\big) \big(1-\sqrt{\eta }+\eta \big)}{3 \eta ^2} \biggl[ H_0(\eta ) H_{-1}\big(\sqrt{\eta }\big) - 2 H_{0,-1}\big(\sqrt{\eta }\big) \biggr]
\nonumber \\ &&
+\biggl[
        \frac{16 P_{69}}{9 (N-1) N^3 (N+1)^3 (N+2)}
        -\frac{32 \big(2+N+N^2\big)^2}{3 (N-1) N^2 (N+1)^2 (N+2)} S_1
\biggr] \zeta_2
\nonumber \\ &&
-\frac{160 \big(2+N+N^2\big)^2}{9 (N-1) N^2 (N+1)^2 (N+2)} \zeta_3
\biggr\}
\nonumber \\ &&
+ \textcolor{blue}{C_A T_F^2} \biggl\{
- ( L_1^3 + L_2^3 ) 
\biggl[
         \frac{800 \big(1+N+N^2\big)}{27 (N-1) N (N+1) (N+2)}
        -\frac{400}{27} S_1
\biggr]
\nonumber \\ &&
-L_1 L_2 ( L_1 + L_2 )
\biggl[
	 \frac{256 \big(1+N+N^2\big)}{9 (N-1) N (N+1) (N+2)}
	-\frac{128}{9} S_1
\biggr]
\nonumber \\ &&
- ( L_1^2 + L_2^2 ) 
\biggl[
         \frac{4 P_{74}}{27 (N-1) N^2 (N+1)^2 (N+2)}
        -\frac{1040}{27} S_1
\biggr]
\nonumber \\ &&
- ( L_1 + L_2 ) 
\biggl[
         \frac{16 P_{75}}{27 (N-1) N^3 (N+1)^3 (N+2)}
        -\frac{32 (47+56 N)}{27 (N+1)} S_1
\nonumber \\ &&
        +\biggl(
                 \frac{224 \big(1+N+N^2\big)}{3 (N-1) N (N+1) (N+2)}
                -\frac{112}{3} S_1
        \biggr) \zeta_2
\biggr]
-\frac{16 P_{76}}{81 (N-1) N^4 (N+1)^4 (N+2)}
\nonumber \\ &&
+\frac{32 P_{72}}{81 (N-1) N (N+1)^2} S_1
+\frac{16}{3 (N+1)} S_1^2
-\frac{16 (1+2 N)}{3 (N+1)} S_2
\nonumber \\ &&
-
\biggl[
         \frac{8 P_{73}}{27 (N-1) N^2 (N+1)^2 (N+2)}
        -\frac{1120}{27} S_1
\biggr] \zeta_2
\nonumber \\ &&
-\biggl[
         \frac{896 \big(1+N+N^2\big)}{27 (N-1) N (N+1) (N+2)}
        -\frac{448}{27} S_1
\biggr] \zeta_3
\biggr\}
~,
\end{eqnarray}
with the polynomials
\begin{eqnarray}
P_{68} &=& 66 N^8+264 N^7+202 N^6-246 N^5-257 N^4+396 N^3+335 N^2+220 N+156,
 \\ 
P_{69} &=& 
3 \left(-12 \eta ^{3/2}+4 \eta ^2-12 \sqrt{\eta }+15\right) N^8
+12 \left(-12 \eta ^{3/2}+4 \eta ^2-12 \sqrt{\eta }+15\right) N^7
\nonumber \\ &&
+2 \left(-72 \eta ^{3/2}+24 \eta ^2-72 \sqrt{\eta }+77\right) N^6
-12 \left(-6 \eta ^{3/2}+2 \eta ^2-6 \sqrt{\eta }+11\right) N^5
\nonumber \\ &&
-2 \left(-90 \eta ^{3/2}+30 \eta ^2-90 \sqrt{\eta }+67\right) N^4
+\left(72 \eta ^{3/2}-24 \eta ^2+72 \sqrt{\eta }+258\right) N^3
\nonumber \\ &&
+245 N^2+148 N+84~,
\\
P_{70} &=& 2 N^{10}+10 N^9+22 N^8+16 N^7-47 N^6-173 N^5-154 N^4
\nonumber \\ &&
-27 N^3+27 N^2+24 N+12~,
\\
P_{71} &=& 288 \eta +\left(72 \eta ^2+247 \eta +72\right) N^{12}
+6 \left(72 \eta ^2+247 \eta +72\right) N^{11}
\nonumber \\ &&
+\left(936 \eta ^2+2923 \eta +936\right) N^{10}
+2 \left(360 \eta ^2+659 \eta +360\right) N^9
\nonumber \\ &&
-\left(360 \eta ^2+1523 \eta +360\right) N^8
-2 \left(504 \eta ^2-1583 \eta +504\right) N^7
\nonumber \\ &&
-9 \left(72 \eta ^2-1001 \eta +72\right) N^6
-2 \left(72 \eta ^2-2417 \eta +72\right) N^5
\nonumber \\ &&
-792 \eta  N^4+1224 \eta  N^2+864 \eta  N
~,
\\
P_{72} &=& 328 N^4+256 N^3-247 N^2-175 N+54
~,
\\
P_{73} &=& 99 N^6+297 N^5+631 N^4+767 N^3+1118 N^2+784 N+168
~,
\\
P_{74} &=& 171 N^6+513 N^5+1159 N^4+1463 N^3+2102 N^2+1456 N+312~,
\\
P_{75} &=& 15 N^8+60 N^7+572 N^6+1470 N^5+2135 N^4+1794 N^3+722 N^2-24 N-72,
\nonumber\\
\\ 
P_{76} &=& 3 N^{10}+15 N^9+3316 N^8+12778 N^7+22951 N^6+23815 N^5+14212 N^4
\nonumber \\ &&
+3556 N^3-30 N^2+288 N+216
~.
\end{eqnarray}
The corresponding contributions in $z$-space, again separated into $\delta$-, $+$ and regular components, read
\begin{eqnarray}
\tilde{A}_{gg,Q}^{(3),\delta}(z) &=&
\textcolor{blue}{T_F^3} \biggl\{
-\frac{32}{3} ( L_1^3 + L_2^3 )
-\frac{128}{9} L_1 L_2 (L_1 + L_2)
-32 (L_1 + L_2) \zeta_2
-\frac{128}{9} \zeta_3
\biggr\}
\nonumber \\ &&
+ \textcolor{blue}{C_F T_F^2} \biggl\{
\frac{176}{3} \big( L_1^2 + L_2^2 \big)
+\frac{256}{3} L_1 L_2
-\frac{64}{3} ( L_1 + L_2 )
+\frac{4 \big(72+247 \eta +72 \eta ^2\big)}{9 \eta }
\nonumber \\ &&
-\frac{32 (1-\eta^2 )}{3 \eta } \HA_0(\eta )
-\frac{16(1-\eta^2)}{3} \HA_0^2(\eta )
\nonumber \\ &&
-\frac{32 \big(1-\sqrt{\eta }\big)^2 (1-\eta^2 ) \big(1+\sqrt{\eta }+\eta \big)}{3 \eta ^2} \biggl[ \HA_0(\eta ) \HA_1\big(\sqrt{\eta }\big) - 2 \HA_{0,1}\big(\sqrt{\eta }\big) \biggr]
\nonumber \\ &&
+\frac{32 \big(1+\sqrt{\eta }\big)^2 (1-\eta^2 ) \big(1-\sqrt{\eta }+\eta \big)}{3 \eta ^2} \biggl[ \HA_0(\eta ) \HA_{-1}\big(\sqrt{\eta }\big) - 2 \HA_{0,-1}\big(\sqrt{\eta }\big) \biggr]
\nonumber \\ &&
+\frac{16}{3} \big(15+4 \eta ^2-12 \sqrt{\eta }-12 \eta ^{3/2}\big) \zeta_2
\biggr\}
\nonumber \\ &&
+ \textcolor{blue}{C_A T_F^2} \biggl\{
-\frac{76}{3} \big( L_1^2 + L_2^2 \big)
-\frac{80}{9} \big( L_1 + L_2 \big)
-\frac{16}{27}
-\frac{88}{3} \zeta_2
\biggr\}
+ \tilde{a}_{gg,Q}^{(3),\delta}(z)
~,
\\
\tilde{A}_{gg,Q}^{(3),+}(z) &=&
 \frac{ \textcolor{blue}{C_A T_F^2} }{1-z} \biggl\{
-\frac{400}{27} \big( L_1^3 + L_2^3 \big)
-\frac{128}{9} L_1 L_2 ( L_1 + L_2 )
-\frac{1040}{27} \big( L_1^2 + L_2^2 \big)
\nonumber \\ &&
-(L_1+L_2) 
\biggl[
         \frac{1792}{27}
        +\frac{112}{3} \zeta_2
\biggr]
-\frac{10496}{81}
-\frac{32}{3} \HA_0
-\frac{1120}{27} \zeta_2
-\frac{448}{27} \zeta_3
\biggr\}
\nonumber \\ &&
+ \tilde{a}_{gg,Q}^{(3),+}(z)
~,
\\
\tilde{A}_{gg,Q}^{(3),\text{reg}}(z) &=&
 \textcolor{blue}{C_F T_F^2} \biggl\{
-\big(L_1^3+L_2^3\big)
\biggl[
	 \frac{184 (1-z) \big(4+7 z+4 z^2\big)}{27 z}
        +\frac{368}{9} (1+z) \HA_0
\biggr]
\nonumber \\ &&
-L_1 L_2 (L_1+L_2) 
\biggl[
         \frac{32 (1-z) \big(4+7 z+4 z^2\big)}{9 z}
        +\frac{64}{3} (1+z) \HA_0
\biggr]
\nonumber \\ &&
+\big(L_1^2+L_2^2\big)
\biggl[
	 \frac{16 (1-z) \big(59-382 z-157 z^2\big)}{27 z}
        -\frac{8}{9} \big(85+151 z-8 z^2\big) \HA_0
\nonumber \\ &&
        -\frac{104}{3} (1+z) \HA_0^2
        -\frac{16 (1-z) \big(4+7 z+4 z^2\big)}{9 z} \HA_1
        -\frac{32}{3} (1+z) \HA_{0,1}
        +\frac{32}{3} (1+z) \zeta_2
\biggr]
\nonumber \\ &&
+(L_1+L_2) 
\biggl[
        \frac{64 (1-z) \big(1-9 z-3 z^2\big)}{z}
        -128 (2+3 z) \HA_0
        -16 (3+5 z) \HA_0^2
\nonumber \\ &&
        -\frac{32}{3} (1+z) \HA_0^3
        -\biggl(
                 \frac{40 (1-z) \big(4+7 z+4 z^2\big)}{3 z}
                +80 (1+z) \HA_0
        \biggr) \zeta_2
\biggr]
\nonumber \\ &&
+\frac{64 (1-z) \big(5-64 z-19 z^2\big)}{3 z}
-128 (5+7 z) \HA_0
-64 (2+3 z) \HA_0^2
\nonumber \\ &&
-\frac{16}{3} (3+5 z) \HA_0^3
-\frac{8}{3} (1+z) \HA_0^4
+\biggl[
        \frac{32 (1-z) \big(41-184 z-67 z^2\big)}{27 z}
\nonumber \\ &&
        -\frac{16}{9} \big(31+61 z-8 z^2\big) \HA_0
        -\frac{112}{3} (1+z) \HA_0^2
        -\frac{32 (1-z) \big(4+7 z+4 z^2\big)}{9 z} \HA_1
\nonumber \\ &&
        -\frac{64}{3} (1+z) \HA_{0,1}
\biggr] \zeta_2
+\frac{64}{3} (1+z) \zeta_2^2
-\biggl[
         \frac{160 (1-z) \big(4+7 z+4 z^2\big)}{27 z}
\nonumber \\ &&
        +\frac{320}{9} (1+z) \HA_0
\biggr] \zeta_3
\biggr\}
+ \textcolor{blue}{C_A T_F^2} \biggl\{
\frac{400 \big(z^3-z^2+2z-1\big)}{27 z} \big(L_1^3+L_2^3\big)
\nonumber \\ &&
+\frac{128 \big(z^3-z^2+2z-1\big)}{9 z} L_1 L_2 (L_1+L_2)
\nonumber \\ &&
+\big(L_1^2+L_2^2\big)
\biggl[
	\frac{104 \big(23z^3-19z^2+29z-23\big)}{27 z}
        -\frac{208}{9} (1+z) \HA_0
\biggr]
\nonumber \\ &&
+(L_1+L_2) 
\biggl[
        \frac{32 \big(-139+157 z-137 z^2+175 z^3\big)}{27 z}
        -\frac{32}{9} (13+22 z) \HA_0
\nonumber \\ &&
        -\frac{32}{3} (1+z) \HA_0^2
        -\frac{32}{3} z \HA_1
        +\frac{112 \big(z^3-z^2+2z-1\big)}{3 z} \zeta_2
\biggr]
\nonumber \\ &&
+\frac{32 \big(1187z^3-949z^2+881z-791\big)}{81 z}
-\frac{32}{27} (62+161 z) \HA_0
-\frac{16}{9} (13+22 z) \HA_0^2
\nonumber \\ &&
-\frac{32}{9} (1+z) \HA_0^3
-\frac{64 \big(-3+3 z+4 z^2\big) \HA_1}{9 z}
+\frac{16}{3} z \HA_1^2
\nonumber \\ &&
+\biggl[
        \frac{112 \big(23z^3-19z^2+29z-23\big)}{27 z}
        -\frac{224}{9} (1+z) \HA_0
\biggr] \zeta_2
\nonumber \\ &&
+\frac{448 \big(z^3-z^2+2z-1\big)}{27 z} \zeta_3
\biggr\}
+ \tilde{a}_{gg,Q}^{(3),\text{reg}}(z)
~.
\end{eqnarray}
\section{Some Identities between \boldmath $G$-Functions}
\label{app:C}

\vspace*{1mm}
\noindent
In the following we list a few special identities for $\eta$-dependent $G$-functions and related quantities
which appear
in expanding $\tilde{a}^{(3)}_{gg,Q}(N)$ around $N = 1/2$ and $3/2$. One, e.g., obtains
\begin{eqnarray}
&& G\left(
        \left\{\frac{1}{\eta + z - z\eta}, \frac{1}{z}, \frac{1}{1-z}, \frac{1}{z}, \frac{1}{z}\right\},1\right) 
\nonumber\\ &&=
\frac{1}{1-\eta} \Biggl[
         \HA_{0,1,0,0,0}(\eta )
        +\HA_{0,1,0,1,0}(\eta )
        +\HA_{0,1,1,0,0}(\eta )
        +\HA_{0,1,1,1,0}(\eta )
        +\HA_{1,1,0,0,0}(\eta )
        +\HA_{1,1,0,1,0}(\eta )
\nonumber\\ &&
        +\HA_{1,1,1,0,0}(\eta )
        +\HA_{1,1,1,1,0}(\eta )
        +\Biggl(\HA_{0,1,0}(\eta ) 
        +\HA_{0,1,1}(\eta ) 
        +\HA_{1,1,0}(\eta )
        +\HA_{1,1,1}(\eta )\Biggr) \zeta_2
        +6 \zeta_5
\Biggr]
\nonumber\\ &&
+  \frac{6}{5(1-\eta)} \HA_1(\eta ) \zeta_2^2,
\\
&& G\left(\left\{ 
\frac{\sqrt{x}}{4-3x},\frac{1}{4-3x},\frac{1}{1-x}\right\},1\right)
\nonumber\\ 
&& =
\frac{\pi^2}{27} 
- G\left(\left\{ 
\frac{1}{4-3x},
\frac{1}{1-x},
\frac{\sqrt{x}}{4-3x}
\right\},1\right)
- G\left(\left\{ 
\frac{1}{4-3x},
\frac{\sqrt{x}}{4-3x},
\frac{1}{1-x}
\right\},1\right)
-\frac{4}{9} i \pi \ln(2) 
\nonumber\\ &&
-\frac{4}{9} \ln(2)\ln(3)
-\frac{2 \pi^2}{27 \sqrt{3}} \ln(2 + \sqrt{3})
+\frac{8}{9 \sqrt{3}} i \pi \ln(2) \ln(2+\sqrt{3})
+\frac{8}{9 \sqrt{3}} \ln(2) \ln(3) \ln(2+\sqrt{3})
\nonumber\\ &&
-\frac{2}{9} \Li_2(4) + \frac{4}{9 \sqrt{3}} \ln(2+\sqrt{3}) \Li_2(4), 
\\
&& G\left(\left\{-\frac{1}{2 + \sqrt{3+x}},\frac{1}{1-x},\frac{1-x}{x}\right\},1\right)
= 8 \sqrt{3}
- 4 \pi 
- \frac{2 \pi^2}{3} 
+ \frac{4 \pi^2}{\sqrt{3}} 
\nonumber\\ &&
- 9 \sqrt{3} 
G\left(\left\{\frac{1}{4 - 3x}, \frac{1}{1-x}, \frac{\sqrt{x}}{4-3x}\right\},1\right)
- 9 \sqrt{3} 
G\left(\left\{\frac{1}{4 - 3x}, \frac{\sqrt{x}}{4-3x}, \frac{1}{1-x}\right\},1\right)
\nonumber\\ &&
+ 9 \sqrt{3} 
G\left(\left\{\frac{\sqrt{x}}{4 - 3x}, \frac{1}{4-3x}, \frac{3}{1+\sqrt{4-3x}}\right\},1\right)
+ \frac{1}{2} 
 {_4F_3}\left[\begin{array}{c} -\frac{1}{2}, 1, 1, 1 \\ \;\;\; 2, 2, 
2\end{array};\frac{1}{4}\right]
+ 32 \ln(2) 
\nonumber\\ &&
+ 16 \sqrt{3} \ln(2) 
- 8 i \sqrt{3} \pi \ln(2)
+ 4 \pi^2 \ln(2)
- \frac{1}{2} \ln(2)
 {_4F_3}\left[\begin{array}{c} -\frac{1}{2}, 1, 1, 1 \\ \;\;\; 2, 2, 
2\end{array};\frac{1}{4}\right]
- 32 \ln^2(2) 
\nonumber\\ &&
- 4 \sqrt{3} \ln(3) 
- \frac{1}{3} \pi^2 \ln(3)
- 8 \sqrt{3} \ln(2) \ln(3)
+ 8 \ln^2(2) \ln(3) 
- 16 \ln(2 + \sqrt{3}) 
\nonumber\\ && 
- \frac{8}{3} \pi^2 \ln(2 + \sqrt{3})
+ 16 i \pi \ln(2) \ln(2+\sqrt{3}) 
+ 8 \ln(3) \ln(2 + \sqrt{3}) 
+ 8 \ln(2) \ln(3) \ln(2 + \sqrt{3})
\nonumber\\ &&
+ 8 \ln^2(2+\sqrt{3})
- 2 \ln(3) \ln^2(2+\sqrt{3})\ln(3) 
- 4 \sqrt{3} \Li_2(4) 
+ 8 \ln(2+\sqrt{3}) \Li_2(4) 
\nonumber\\ &&
- 16 \Li_2\left(2,\frac{1}{4}(2+\sqrt{3})\right)
+ 4 \ln(2) \Li_2\left(2,\frac{1}{4}(2+\sqrt{3})\right)
- 14 \zeta_3.
\end{eqnarray}
Also special constants contribute, e.g.:
\begin{eqnarray}
&& \Li_2(7 - 4\sqrt{3}) = - \frac{1}{2} \ln^2(2) - \frac{1}{2} \ln(2) \ln(3)
- \frac{1}{8} \ln^2(3) + \ln(2) \ln(2 + \sqrt{3}) 
+ \frac{1}{2} \ln(3) \ln(2 + \sqrt{3})
\nonumber\\ &&
- \frac{1}{2} \ln^2(2 + \sqrt{3}) - \Li_2\left(\frac{1}{2}-\frac{1}{\sqrt{3}}\right),
\\
&& \Li_2\left(\frac{1}{4+2\sqrt{2}}\right) = \frac{17}{24} \pi^2 - \frac{15}{8} \ln^2(2)
+ \frac{3}{2} \ln(2) \ln(\sqrt{2}-1)
+ \frac{1}{2} \ln^2(\sqrt{2}-1)
-  6 \Li_2\left(\frac{1}{\sqrt{2}}\right),
\nonumber\\ 
\\
&& \Li_2(4) = \zeta_2 - 2 i \pi \ln(2) - \ln^2(2) - 2 \Li_2\left(-\frac{1}{2}\right),
\\
&& \ln(3 - 2 \sqrt{2}) = 2 \ln(\sqrt{2}-1).
\end{eqnarray}
The hypergeometric $_4F_3$-constant can be expressed by
\begin{eqnarray}
{_4F_3}\left[\begin{array}{c} -\frac{1}{2}, 1, 1, 1 \\ \;\;\; 2, 2,
2\end{array};\frac{1}{4}\right]
&=& \frac{416}{27} - \frac{68}{3 \sqrt{3}} - \frac{256}{9} \ln(2) + \frac{32}{3} \ln^2(2)
+ \frac{128}{9} \ln(2 + \sqrt{2})
\nonumber\\ &&
- \frac{32}{3} \ln(2) \ln(2 +\sqrt{3})
+ \frac{8}{3} \ln^2(2 + \sqrt{3}) - \frac{16}{3} \Li_2\left(\frac{1}{2} \left(1 - 
\frac{\sqrt{3}}{2}\right)\right).
\nonumber\\
\end{eqnarray}
Similar expressions of this kind often appear in iterated integrals over root-valued alphabets, cf. also 
\cite{Ablinger:2014yaa}.
There are many  other, partly lengthy,  relations more, which we all had to reduce to a suitable 
level to prove that the evanescent poles at $N = 1/2$ and $N=3/2$ vanish. 
\section{Representation of the functions \boldmath $G_l$ and $K_l$}
\label{app:D}

\vspace*{1mm}
\noindent
Before the absorption of a few rational pre-factors in $N$, all emerging integrals first written
in $G$-functions can be expressed in terms of polylogarithms at algebraic arguments in $z$ and $\eta$.
In cases it leads to simplifications, we also use arcus- and area-functions instead of logarithms,
which belong to the harmonic (poly)logarithms of complex-valued argument.

The different functions $G_l \equiv G_l(z,\eta)$ and constants $K_l = K_l(\eta)$ are given by
\begin{eqnarray}
G_1 &=& G\left[\left\{\sqrt{(1-x)x}\right\}, z\right] 
=
\frac{1}{2} \sqrt{1-z}
   z^{3/2}-\frac{1}{4}
   \sqrt{1-z}
   \sqrt{z}-\frac{1}{4} \arctan\left(\frac{\sqrt{1-z}}{
   \sqrt{z}}\right)+\frac{\pi
   }{8}
\nonumber\\ &&
\\
G_2 &=& G\left[\left\{\frac{1}{z+\eta(1-x)}\right\},z\right]
=
\frac{\ln(z+\eta(1-z))-\ln(\eta)}{1-\eta}
\\
G_3 &=& G\left[\left\{\frac{1}{1 - z(1-\eta)}\right\}, z\right]
=
-\frac{\ln(1 - z (1-\eta))}{1-\eta}
\\
G_4 &=& G\left[\left\{\frac{\sqrt{x(1-x)}}{1-x(1-\eta)}\right\},z\right]
=
\frac{1}{(1-\eta)^2}\Biggl[
\frac{1}{2} \pi  (\eta +1)-(1-\eta )\sqrt{(1-z) z}
\nonumber\\ &&
-(\eta +1) \arctan\left(\frac{\sqrt{1-z}}{\sqrt{z}}\right)
   -2 \sqrt{\eta } 
\arctan\left(\frac{\sqrt{\eta }
   \sqrt{z}}{\sqrt{1-z}}\right)
\Biggr]
\\
G_5 &=& G\left[\left\{-\frac{\sqrt{x(1-x)}}{x(1-\eta)+\eta}\right\},z\right]
= \frac{1}{(1-\eta)^2}\Biggl[
-\frac{\pi}{2}(1+\eta) - \sqrt{z(1-z)}(1-\eta)
\nonumber\\ &&
+2 \sqrt{\eta} \arctan\left(\frac{\sqrt{z}}{\sqrt{(1-z)\eta}}\right)
+(1+\eta) \arctan\left(\frac{\sqrt{1-z}}{\sqrt{z}}\right)
\Biggr]
\\
G_6 &=& G\left[\left\{\sqrt{(1-x)x},\frac{1}{1-x}\right\}, z\right]
=
\frac{1}{4} \ln (1-z) 
        \sqrt{(1-z) z}(1-2z)
+\Biggl[
        \arcsin \big(
                \sqrt{1-z}\big)
\nonumber\\ &&
        -\frac{1}{4} i \ln (1-z)
\Biggr] \ln \big(
        i \sqrt{1-z}
        +\sqrt{z}\big)
-\frac{1}{2} \arcsin \big(
        \sqrt{1-z}
\big)
\ln \left(
                -1+\big(
                        \sqrt{z}
                        +i \sqrt{1
                        -z
                        }
                \big)^2\right)
\nonumber\\ &&
+\frac{1}{48} \Biggl[
        -3 \pi 
        +6 \arcsin \big(
                \sqrt{1-z}\big)
        -12 i \arcsin^2 \big(
                \sqrt{1-z}\big)
        +6 \sqrt{(1-z) z}(1+2z)
\nonumber\\ &&
        +12 i \zeta_2
        -12 i \text{Li}_2\left(
                \frac{1}{\big(
                        i \sqrt{1-z}
                        +\sqrt{z}
                \big)^2}\right)
\Biggr]
+\frac{1}{4} \pi  \ln (2)
\\
G_7 &=& G\left[\left\{\sqrt{(1-x)x},\frac{1}{x}\right\}, z\right]
=
\frac{i}{2} \text{Li}_2\left(-\sqrt{1-z}-i
   \sqrt{z}\right)
-\frac{i}{2}
   \text{Li}_2\left(1-\sqrt{1-z}-i\sqrt{z}\right)
\nonumber\\ &&
+\ln(z) \left(\frac{1}{2}
   \sqrt{1-z} z^{3/2}-\frac{1}{4} \sqrt{(1-z)
   z}+\frac{1}{4} \arcsin\left(\sqrt{z}\right)\right)
\nonumber\\ &&
+\frac{3}{8}
   \sqrt{1-z} \sqrt{z}+\frac{i}{4} \arctan^2\left(\frac{\sqrt{z}}{\sqrt{1-z}}\right)
+ \arctan\left(\frac{\sqrt{z}}{\sqrt{1-z}}\right)
\nonumber\\ &&
\times \left(\frac{1}{8}-\frac{1}{2}
   \ln\left(\sqrt{1-z}+i
   \sqrt{z}+1\right)\right) 
   +\frac{i \zeta_2}{4}
-\frac{1}{4}
   \sqrt{1-z} z^{3/2}
\\
G_8 &=& G\left[\left\{\frac{1}{x + \eta(1-x)}, \frac{1}{1-x}\right\}, z\right]
= -\frac{1}{1-\eta} \Biggl[
\ln(1 - z) \ln(z+\eta(1-z))
\nonumber\\ &&
+ \Li_2((1-\eta)(1-z)) 
- \Li_2((1-\eta))
\Biggr]
\\
G_9 &=& G\left[\left\{\frac{1}{x+\eta(1-x)},\frac{1}{x}\right\}, z\right]
=
\frac{1}{1-\eta}
\Biggl[
\text{Li}_2\left(-\frac{z (1-\eta )}{\eta}\right)
+\ln(z) (\ln((1-\eta ) z+\eta)
\nonumber\\ &&
-\ln(\eta ))
\Biggr]
\nonumber\\
\\
G_{10} &=& G\left[\left\{\frac{1}{1 - x(1-\eta)},\frac{1}{1-x}\right\}, z\right]
=
\frac{1}{1-\eta}
\Biggl[
\text{Li}_2\left(
-\frac{(1-z) (1-\eta)}{\eta }
\right)
\nonumber\\ &&
+\ln(1-z) (\ln(1-(1-\eta) z)
-\ln(\eta))
-\text{Li}_2\left(-\frac{1-\eta }{\eta}\right)
\Biggr]
\\
G_{11} &=& G\left[\left\{\frac{1}{1 - x(1-\eta)}, \frac{1}{x}\right\}, z\right]
=
-\frac{1}{1-\eta} \Biggl[
\ln(z) \ln(1 - z (1 - \eta)) + \Li_2(z(1-\eta))
\Biggr]
\\
G_{12} &=& G\left[\left\{\frac{\sqrt{(1-x)x}}{1-x(1-\eta)}, \frac{1}{1-x}\right\}, z\right]
=
\frac{1}{(1-\eta)^2} \Biggl[
-i \Biggl[
\eta 
   \text{Li}_2\left(-\left(\sqrt{1-z}+i \sqrt{z}\right)^2\right)
\nonumber\\ &&
+\sqrt{\eta }
   \text{Li}_2\left(\frac{\left(1-\frac{i
   \sqrt{z}}{\sqrt{1-z}}\right) \sqrt{\eta
   }}{\sqrt{\eta }-1}\right)
-\sqrt{\eta }
   \text{Li}_2\left(\frac{\left(\frac{i
   \sqrt{z}}{\sqrt{1-z}}+1\right) \sqrt{\eta
   }}{\sqrt{\eta }-1}\right)
\nonumber\\ &&
-\sqrt{\eta }
   \text{Li}_2\left(\frac{\left(1-\frac{i
   \sqrt{z}}{\sqrt{1-z}}\right) \sqrt{\eta
   }}{\sqrt{\eta }+1}\right)
+\sqrt{\eta }
   \text{Li}_2\left(\frac{\left(\frac{i
   \sqrt{z}}{\sqrt{1-z}}+1\right) \sqrt{\eta
   }}{\sqrt{\eta
   }+1}\right)
\nonumber\\ &&
+\text{Li}_2\left(-\left(\sqrt{1-z} + i \sqrt{z}\right)^2\right)\Biggr]
+(\eta -1) \sqrt{(1-z) z}
+i (\eta +1) \arcsin^2\left(\sqrt{z}\right)
\nonumber\\ &&
+(1-\eta ) \arcsin\left(\sqrt{z}\right)
+2 (\eta +1) \ln(2) \arcsin\left(\sqrt{z}\right)
\nonumber\\ &&
+\ln(1-z) \Biggl[
(1-\eta ) \sqrt{(1-z) z}
+2\sqrt{\eta } \arctan\left(\frac{\sqrt{\eta z}}{\sqrt{1-z}}\right)
\Biggr]
+\ln
   \left(\frac{1-\sqrt{\eta }}{\sqrt{\eta
   }+1}\right) \Bigl(\pi  \sqrt{\eta }
\nonumber\\ &&
-2
   \sqrt{\eta } \arctan\left(\frac{\sqrt{1-z}}{\sqrt{z}}\right)
   \Bigr)-\frac{1}{2} i (\eta +1) \zeta_2
\Biggr]
\\
G_{13} &=&  G\left[\left\{\frac{\sqrt{(1-x)x}}{1-x(1-\eta)},\frac{1}{x}\right\},z\right]
=
-  \frac{i (1+\eta ) \pi  \arcsin \big(\sqrt{z}\big)}{(1-\eta )^2}
-i \frac{(1+\eta ) \arcsin \big(\sqrt{z}\big)^2}{(1-\eta )^2}
\nonumber \\ &&
-\frac{2 (1+\eta ) \arcsin \big(\sqrt{z}\big) \ln (2)}{(1-\eta )^2}
+\frac{1}{ (1-\eta )^2} 
\biggl[
        - i (1+\eta ) \text{Li}_2\left(\frac{1}{\big(\sqrt{1-z}+i \sqrt{z}\big)^2}\right)
\nonumber \\ &&
        + ( 1 - \eta )\sqrt{(1-z) z}
        + ( 1 + \eta ) i \zeta_2
        +2 i \sqrt{\eta } \text{Li}_2\left(-\frac{i \sqrt{\eta } \sqrt{z}}{\sqrt{1-z}}\right)
        -2 i \sqrt{\eta } \text{Li}_2\left(\frac{i \sqrt{\eta } \sqrt{z}}{\sqrt{1-z}}\right)
\nonumber \\ &&
        + i \sqrt{\eta } \text{Li}_2\left(\frac{\sqrt{\eta } \big(1-\frac{i \sqrt{z}}{\sqrt{1-z}}\big)}{-1+\sqrt{\eta }}\right)
        - i \sqrt{\eta } \text{Li}_2\left(\frac{\sqrt{\eta } \big(1-\frac{i \sqrt{z}}{\sqrt{1-z}}\big)}{1+\sqrt{\eta }}\right)
        - i \sqrt{\eta } \text{Li}_2\left(\frac{\sqrt{\eta } \big(1+\frac{i \sqrt{z}}{\sqrt{1-z}}\big)}{-1+\sqrt{\eta }}\right)
\nonumber \\ &&
        + i \sqrt{\eta } \text{Li}_2\left(\frac{\sqrt{\eta } \big(1+\frac{i \sqrt{z}}{\sqrt{1-z}}\big)}{1+\sqrt{\eta }}\right)
\biggr]
+\arctan \left(\frac{\sqrt{1-z}}{\sqrt{z}}\right)
\biggl[
	 \frac{4 \ln \big(1-\sqrt{\eta }\big)}{(1-\eta )^2} \sqrt{\eta }
        -\frac{2 \sqrt{\eta } \ln (1-\eta )}{(1-\eta )^2}
\biggr]
\nonumber \\ &&
+\arctan \left(\frac{\sqrt{z}}{\sqrt{1-z}}\right)
\biggl[
	\frac{1}{1-\eta }
        +\frac{2 i (1+\eta ) \arcsin \big(\sqrt{z}\big)}{(1-\eta )^2}
        +\frac{(1+\eta ) \ln (z)}{(1-\eta )^2}
\biggr]
\nonumber \\ &&
-\frac{2 \pi  \ln \big(1-\sqrt{\eta }\big)}{(1-\eta )^2} \sqrt{\eta }
-\frac{2 \ln (z)}{(1-\eta )^2} \big( \arctan \left( \frac{\sqrt{\eta } \sqrt{z}}{\sqrt{1-z}} \right) \big) \sqrt{\eta }
+\frac{\pi  \sqrt{\eta } \ln (1-\eta )}{(1-\eta )^2}
\nonumber \\ &&
-\biggl[
         \frac{(1+\eta ) \arcsin \big(\sqrt{z}\big)}{(1-\eta )^2}
        +\frac{\sqrt{(1-z) z}}{1-\eta }
\biggr] \ln (z),
\\ 
G_{14} &=& G\left[\left\{-\frac{\sqrt{(1-x)x}}{x(1-\eta) + \eta}, \frac{1}{1-x}\right\}, z\right] 
=
-
\frac{(1+\eta ) \ln(2) \pi }{(1-\eta )^2}
+i \frac{(1+\eta ) \arcsin \big(\sqrt{1-z}\big)^2}{(1-\eta )^2}
\nonumber \\ &&
+\frac{1}{ (-1+\eta )^2} 
\biggl[
        \frac{1}{2} ( 1 - \eta ) \pi 
        +  i \eta \text{Li}_2\left(-\frac{1}{1-2 z-2 i \sqrt{(1-z) z}}\right)
        - ( 1-\eta ) \frac{\sqrt{z}}{\sqrt{1-z}}
\nonumber \\ &&
        + ( 1-\eta) \frac{z^{3/2}}{\sqrt{1-z}}
        - i ( 1+\eta) \zeta_2
        + i \sqrt{\eta } \text{Li}_2\left(\frac{1-\frac{i \sqrt{z}}{\sqrt{1-z}}}{1-\sqrt{\eta }}\right)
        - i \sqrt{\eta } \text{Li}_2\left(\frac{1-\frac{i \sqrt{z}}{\sqrt{1-z}}}{1+\sqrt{\eta }}\right)
\nonumber \\ &&
        - i \sqrt{\eta } \text{Li}_2\left(\frac{1+\frac{i \sqrt{z}}{\sqrt{1-z}}}{1-\sqrt{\eta }}\right)
        + i \sqrt{\eta } \text{Li}_2\left(\frac{1+\frac{i \sqrt{z}}{\sqrt{1-z}}}{1+\sqrt{\eta }}\right)
        + i \text{Li}_2\left(-\frac{1}{1-2 z-2 i \sqrt{(1-z) z}}\right)
\biggr]
\nonumber \\ &&
+\frac{1+\eta}{(1-\eta)^2} \arcsin \big(\sqrt{1-z}\big)
\biggl[
	2 \ln(2)
	+i \pi
\biggr]
+ \frac{2 \sqrt{\eta}}{(1-\eta)^2} \arctan \big(\frac{\sqrt{z}}{\sqrt{1-z}}\big)
\biggl[
	-  i \pi  
\nonumber \\ &&
        - 2 \ln \big(1-\sqrt{\eta }\big)
        +  \ln (1-\eta )
\biggr]
+\arctan \big(\frac{\sqrt{1-z}}{\sqrt{z}}\big)
\biggl[
	-\frac{1}{1-\eta }
        -i \frac{2 (1+\eta ) \arcsin \big(\sqrt{1-z}\big)}{(1-\eta )^2}
\nonumber \\ &&
        -\frac{(1+\eta ) \ln (1-z)}{(1-\eta )^2}
\biggr]
+\frac{2 \pi  \ln \big(1-\sqrt{\eta }\big)}{(1-\eta )^2} \sqrt{\eta }
-\frac{2 \sqrt{\eta } \ln (1-z)}{(1-\eta )^2}  \arctan \left(\frac{\sqrt{z}}{\sqrt{\eta } \sqrt{1-z}}\right)
\nonumber \\ &&
-\frac{2 \sqrt{\eta } \pi }{(1-\eta )^2} \ln \left(1-\frac{i \sqrt{z}}{\sqrt{1-z}}\right)
+\biggl[
        \frac{(1+\eta ) \arcsin \big(\sqrt{1-z}\big)}{(1-\eta )^2}
        +\frac{\sqrt{(1-z) z}}{1-\eta }
\biggr] \ln (1-z),
\\
G_{15} &=& G\left[\left\{-\frac{\sqrt{(1-x)x}}{\eta+x(1-\eta)},\frac{1}{x}\right\}, z\right]
=
-i \frac{(1+\eta ) \arcsin^2 \big(\sqrt{z}\big)}{(1-\eta )^2}
+\frac{1}{6 (1-\eta )^2} 
\biggl[
        3 (1-\eta) \pi 
\nonumber \\ &&
        - 6 i \eta \text{Li}_2\left(1-2 z+2 i \sqrt{(1-z) z}\right)
        +6 (1-\eta) \sqrt{(1-z) z}
        +6 (1 -6\sqrt{\eta} +\eta) i \zeta_2
\nonumber \\ &&
        -12 i \sqrt{\eta } \text{Li}_2\left( -\frac{i \sqrt{z}}{\sqrt{\eta } \sqrt{1-z}} \right)
        +12 i \sqrt{\eta } \text{Li}_2\left( \frac{i \sqrt{z}}{\sqrt{\eta } \sqrt{1-z}} \right)
        +6 i \sqrt{\eta } \text{Li}_2\left( \frac{1-\frac{i \sqrt{z}}{\sqrt{1-z}}}{1+\sqrt{\eta }} \right)
\nonumber \\ &&
        +6 i \sqrt{\eta } \text{Li}_2\left( \frac{1+\frac{i \sqrt{z}}{\sqrt{1-z}}}{1-\sqrt{\eta }} \right)
        -6 i \sqrt{\eta } \text{Li}_2\left( \frac{1+\frac{i \sqrt{z}}{\sqrt{1-z}}}{1+\sqrt{\eta }} \right)
        -6 i \sqrt{\eta } \text{Li}_2\left( \frac{i \big(i+\frac{\sqrt{z}}{\sqrt{1-z}}\big)}{-1+\sqrt{\eta }} \right)
\nonumber \\ &&
        -6 i \text{Li}_2 \left(1-2 z+2 i \sqrt{(1-z) z} \right)
\biggr]
+ \frac{1+\eta}{(1-\eta )^2} \arcsin \big(\sqrt{z}\big) 
\biggl[
	 2  \ln(2)
        -i \pi
\biggr]
\nonumber \\ &&
+ \frac{2}{(1-\eta )^2} \arctan \left(\frac{\sqrt{z}}{\sqrt{1-z}}\right)
\biggl[
	  i (1+\eta ) \arcsin \big(\sqrt{z}\big)
        +  i \pi  \sqrt{\eta }
        + 2 \sqrt{\eta } \ln \big(1-\sqrt{\eta }\big) 
\nonumber \\ &&
        -  \sqrt{\eta } \ln (1-\eta )
\biggr]
+\frac{1}{(1-\eta )^2} \arctan \left(\frac{\sqrt{1-z}}{\sqrt{z}}\right)
\biggl[
	-1+\eta +2 i \sqrt{\eta } \pi 
        +(1+\eta ) \ln (z)
\biggr]
\nonumber \\ &&
-\frac{2 \pi }{(1-\eta )^2} \sqrt{\eta } \ln \big(1-\sqrt{\eta }\big)
+\frac{2 \sqrt{\eta } }{(1-\eta )^2} \ln (z) \arctan \left(\frac{\sqrt{z}}{\sqrt{\eta } \sqrt{1-z}}\right) 
-\frac{\pi  \sqrt{\eta } \ln (1-z)}{(1-\eta )^2}
\nonumber \\ &&
+ \frac{1}{(1-\eta)^2}
\biggl[
         (1+\eta ) \arcsin \big(\sqrt{z}\big)
        +\frac{1}{2} \big(-(1+\eta ) \pi 
	+2 (-1+\eta ) \sqrt{(1-z) z}\big)
\biggr] \ln (z).
\end{eqnarray}
Furthermore, the functions $K_l(\eta) \equiv K_l$ contribute. For the more complicated among them 
we first obtained a longer representation, which finally could be reduced. In these cases we present both 
representations, since they contain relations between polylogarithms. Structures like this are particularly 
obtained by integrating using {\tt Mathematica}. The comparison of both these cases my be helpful in other
calculations to obtain more compact results.  

Here the constants $c_1$ to $c_8$ are given by
\begin{eqnarray}
c_1 &=& G\left[\left\{\frac{\sqrt{(1-x)x}}{1+x},\frac{1}{x},\sqrt{(1-x)x}\right\},1\right] 
\nonumber\\ 
&=&
-\frac{17}{24}
+i \frac{1}{2\sqrt{2}} \pi  \ln^2 \big(
        \sqrt{2}-1\big)
+\frac{1}{2\sqrt{2}} \ln (2) \text{Li}_2\left(
        \frac{1}{2} \left(
                1+\frac{1}{\sqrt{2}}\right)\right) 
+\Biggl[
        \frac{21}{16}-\frac{7}{8 \sqrt{2}}\Biggr] \zeta_3
\nonumber\\ &&
+\frac{5}{4} \ln(2)
+\frac{1}{16\sqrt{2}}(1+2\pi i)\ln^2 (2)
+\frac{1}{\sqrt{2}} \ln^3 (2)
+\Biggl[
        -\frac{15}{32}-\frac{1}{16} \frac{1}{\sqrt{2}}-\frac{1}{8} \frac{1}{\sqrt{2}} \ln (2)\Biggr] \zeta_2
\nonumber\\ &&
-\frac{1}{\sqrt{2}}
\Biggl[
        \zeta_2
        +\frac{1}{4}(1-2 \pi i)\ln (2)
        +\frac{3}{8} \ln^2 (2)
\Biggr] \ln \big(\sqrt{2}-1\big)
-\frac{1}{6} \frac{1}{\sqrt{2}} \ln^3 \big(\sqrt{2}-1\big)
\nonumber\\ && +\frac{1}{2\sqrt{2}}\Biggl[
        1-2 \ln (2)\Biggr] 
\text{Li}_2\left(
        \frac{1}{\sqrt{2}}\right)
-\sqrt{2} \text{Li}_3\left(
        \frac{1}{\sqrt{2}}\right)
-\frac{1}{\sqrt{2}} \text{Li}_3\left(
        \frac{1}{2} \left(
                1-\frac{1}{\sqrt{2}}\right)\right)
\nonumber\\ &&
+\frac{1}{\sqrt{2}} \text{Li}_3\left(
        1-\frac{1}{\sqrt{2}}\right)
+\frac{1}{\sqrt{2}} \text{Li}_3\left(
        1+\frac{1}{\sqrt{2}}\right)
+\frac{1}{\sqrt{2}} \text{Li}_3\left(
        -\frac{2}{\sqrt{2}-1}\right)
\\
c_2 &=& G\left[\left\{\frac{\sqrt{(1-x)x}}{1+x},\frac{1}{1-x},\sqrt{(1-x)x}\right\},1\right] 
\nonumber\\ &=&
-
\frac{1}{24}
-\frac{1}{2 \sqrt{2}} \ln (2) \text{Li}_2\left(\frac{1}{2} \left(1+\frac{1}{\sqrt{2}}\right)\right) 
+\frac{1}{4 \sqrt{2}} \ln (2) \text{Li}_2\left(\big(\sqrt{2}-1\big)^4\right)
+\biggl[
        \frac{1}{32} \big(3+\sqrt{2}\big)
\nonumber \\ &&
        +\frac{7}{8 \sqrt{2}} \ln(2)
        -\sqrt{2} \ln \big(\sqrt{2}-1\big)
        -\frac{3}{\sqrt{2}} \ln \big(1+\sqrt{2}\big)
\biggr] \zeta_2
+\frac{21 \zeta_3}{32} \big(2+\sqrt{2}\big) 
+\biggl[
        -\frac{1}{4}
\nonumber \\ &&
        +\frac{1}{4 \sqrt{2}} (1-2 i \pi ) \ln \big(\sqrt{2}-1\big) 
\biggr] \ln (2)
+\biggl[
        -\frac{1}{16 \sqrt{2}} (1+2 i \pi ) 
        +\frac{7}{8} \frac{1}{\sqrt{2}} \ln \big(\sqrt{2}-1\big)
\biggr] \ln (2)^2
\nonumber \\ &&
-\frac{7}{6 \sqrt{2}} \ln (2)^3
-\frac{1}{2 \sqrt{2}} i \pi \ln \big(\sqrt{2}-1\big)^2
+\frac{1}{6 \sqrt{2}} \ln \big(\sqrt{2}-1\big)^3
-\biggl[
        \frac{1}{2 \sqrt{2}} + \frac{\ln (2)}{\sqrt{2}}
\biggr] \text{Li}_2\left(\frac{1}{\sqrt{2}}\right)
\nonumber \\ &&
-\frac{1}{\sqrt{2}} \ln (2) \text{Li}_2\left(\big(\sqrt{2}-1\big)^2\right)
-\sqrt{2} \text{Li}_3\left(\frac{1}{\sqrt{2}}\right)
+\frac{1}{\sqrt{2}} \text{Li}_3\left(\frac{1}{2} \left(1-\frac{1}{\sqrt{2}}\right)\right)
\nonumber \\ &&
-\frac{1}{\sqrt{2}} \text{Li}_3\left(1-\frac{1}{\sqrt{2}}\right)
-\frac{1}{\sqrt{2}} \text{Li}_3\left(1+\frac{1}{\sqrt{2}}\right)
-\frac{1}{\sqrt{2}} \text{Li}_3\left(-\frac{2}{-1+\sqrt{2}}\right)
\\
c_3 &=& G\left[\left\{\frac{\sqrt{(1-x)x}}{1+x},\sqrt{(1-x)x},\frac{1}{x}\right\}, 1\right] 
\nonumber\\ &=&
\frac{317}{144}
-\frac{1}{2\sqrt{2}} \ln (2) \text{Li}_2\left(\frac{1}{2} \left(1+\frac{1}{\sqrt{2}}\right)\right) 
-\Biggl[
        \frac{21}{32}+\frac{5}{16 \sqrt{2}}\Biggr] \zeta_3
-\frac{5}{4} \ln (2)
-\frac{1}{16\sqrt{2}} 
\nonumber\\ &&
\times (1 - 2 \pi i)\ln^2 (2)
+\frac{1}{12\sqrt{2}}  \ln^3(2)
+\Biggl[
        -\frac{15}{32}+\frac{1}{16} \frac{1}{\sqrt{2}}-\frac{9}{8} \ln(2)-\frac{1}{8} 
\frac{1}{\sqrt{2}} 
\ln (2)\Biggr] \zeta_2
\nonumber\\ && 
+ \frac{1}{4\sqrt{2}}
\Biggl[
       (1+ 2 \pi i) \ln(2)+ \ln^2 (2)\Biggr] 
\ln \big(
        \sqrt{2}-1\big)
+\frac{1}{2\sqrt{2}}\Biggl[
        i \pi +\ln (2)\Biggr] \ln^2 \big(\sqrt{2}-1\big)
\nonumber\\ &&
-\frac{1}{\sqrt{2}}\Biggl[
        \frac{1}{2}-\ln (2)\Biggr] \text{Li}_2\left(
        \frac{1}{\sqrt{2}}\right)
+\frac{1}{\sqrt{2}} \text{Li}_3\left(
        \frac{1}{\sqrt{2}}\right)
-\frac{1}{\sqrt{2}} \text{Li}_3\left(
        \frac{1}{2} \left(
                1+\frac{1}{\sqrt{2}}\right)\right)
\nonumber\\ &&
+\frac{1}{\sqrt{2}} \text{Li}_3\left(
        1+\frac{1}{\sqrt{2}}\right)
-\frac{1}{\sqrt{2}} \text{Li}_3\left(
        \frac{1}{1+\sqrt{2}}\right)
+\frac{1}{\sqrt{2}} \text{Li}_3\left(
        \frac{2}{1+\sqrt{2}}\right)
\\
c_4 &=& G\left[\left\{\frac{\sqrt{(1-x)x}}{1+x},\sqrt{(1-x)x},\frac{1}{1-x}\right\}, 1\right] 
\nonumber\\ &=&
-
\frac{155}{144}
+\frac{1}{4} \ln(2)
-\frac{3}{4}\ln^2(2)
+\frac{1}{16\sqrt{2}}(1-2\pi i) \ln^2(2)
+\frac{1}{2\sqrt{2}} \ln(2)\text{Li}_2\left(
        \frac{1}{2} \left(
                1+\frac{1}{\sqrt{2}}\right)\right) 
\nonumber\\ &&
+\Biggl[
        \frac{15}{32}+\frac{9}{8} \ln(2)-\frac{1}{16} \frac{1}{\sqrt{2}}+\frac{1}{8\sqrt{2}} \ln(2) 
\Biggr] \zeta_2
+\Biggl[
        -\frac{21}{32}+\frac{3}{32 \sqrt{2}}\Biggr] \zeta_3
-\frac{1}{4\sqrt{2}} \Biggl[
(1+ 2 \pi i)\ln(2) 
\nonumber\\ &&
+ 2 \ln^2(2) 
\Biggr] \ln\big(\sqrt{2}-1\big)
-\frac{1}{2\sqrt{2}}\Bigl[\ln(2) + i \pi 
      \Bigr] \ln^2\big(\sqrt{2}-1\big)
+\frac{1}{2\sqrt{2}} \text{Li}_2\left(
        \frac{1}{\sqrt{2}}\right)
\nonumber\\ &&
+\frac{1}{\sqrt{2}} \text{Li}_3\left(
        \frac{1}{\sqrt{2}}\right)
+\frac{1}{\sqrt{2}} \text{Li}_3\left(
        \frac{1}{2} \left(
                1+\frac{1}{\sqrt{2}}\right)\right)
-\frac{1}{\sqrt{2}} \text{Li}_3\left(
        1+\frac{1}
        {\sqrt{2}}\right)
\nonumber\\ &&
+\frac{1}{\sqrt{2}} \text{Li}_3\left(
        \frac{1}{1+\sqrt{2}}\right)
-\frac{1}{\sqrt{2}} \text{Li}_3\left(
        \frac{2}{1+\sqrt{2}}\right),
\\
c_5 &=& G \left[\left\{\frac{\sqrt{(1-x)x}}{x-2},\frac{1}{1-x},\sqrt{(1-x)x}\right\}, 1\right] 
\nonumber \\  
&=& -c_1 -\frac{3 \zeta_2}{16 \sqrt{2}} \biggl[ -4 + 5\sqrt{2} + 16 \ln(2) + 16 \ln(\sqrt{2}-1) \biggr]
,
\\
c_6 &=& G \left[\left\{\frac{\sqrt{(1-x)x}}{x-2},\frac{1}{x},\sqrt{(1-x)x}\right\}, 1\right] 
\nonumber \\ 
&=& -c_2 + \frac{3 \zeta_2}{16\sqrt{2}} \biggl[ -4 + \sqrt{2} + 24 \ln(2) 
+ 16 \ln(\sqrt{2}-1) \biggr]
,
\\
c_7 &=& G \left[\left\{\frac{\sqrt{(1-x)x}}{x-2},\sqrt{(1-x)x},\frac{1}{1-x}\right\}, 1\right] 
= -c_3 + \frac{3 \zeta_2}{16} \Bigl( 2\sqrt{2} - 3 \Bigr) \Bigl( 4\ln(2) - 1 \Bigr)
,
\\
c_8 &=& G \left[\left\{\frac{\sqrt{(1-x)x}}{x-2},\sqrt{(1-x)x},\frac{1}{x}\right\}, 1\right] 
= -c_4 - \frac{3 \zeta_2}{16} \Bigl( 2\sqrt{2} - 3 \Bigr) \Bigl( 4\ln(2) - 1 \Bigr)
.
\end{eqnarray}
The following set of constants contributes in the first expressions for $K_l$ given above.
\begin{eqnarray}
\label{eq:LISET}
&& \ln(2), 
\pi, 
\ln(\sqrt{2}-1), 
\zeta_3, 
\Li_2\left((\sqrt{2}-1)^2\right),
\Li_2\left((\sqrt{2}-1)^4\right),
\Li_3\left(\sqrt{2}-1\right),
\nonumber\\
&&
\Li_3\left(2(\sqrt{2}-1)\right),
\Li_3\left(\frac{1}{2}\left(1+\frac{1}{\sqrt{2}}\right)\right),
\Li_3\left(1+ \frac{1}{\sqrt{2}}\right),
\end{eqnarray}
with $\zeta_2 = \pi^2/6$. The new constants, most of which are not multiple zeta values \cite{Blumlein:2009cf},
however, finally cancel. 
The first expressions were obtained by integrating using {\tt Mathematica} and applying functional identities
between (poly)logarithms \cite{LEWIN}.
For the second expression,
we used relations built in {\tt HarmonicSums}. The cancellation is due to special value relations
of polylogarithms. The corresponding relations may also be numerically verified, e.g. by using
{\tt PSLQ} \cite{PSLQ}.

\noindent
We note the relation
\begin{eqnarray}
\label{eq:REL1}
\frac{5\zeta_2}{4} 
-\frac{\ln^2(2)}{4}
- \ln(2) \ln\left(1+\frac{1}{\sqrt{2}}\right)
+ 2 \Li_2\left(-\frac{1}{\sqrt{2}}\right) 
- \Li_2((\sqrt{2}-1)^2)
+ \Li_2(-(\sqrt{2}-1)^2) = 0.
\nonumber\\
\end{eqnarray}
Abel's relation for $x = 1 -1/\sqrt{2}$ and $y = -1/\sqrt{2}$, Euler's relation and the mirror relation, 
cf.~\cite{LEWIN}, 
\begin{eqnarray}
 \Li_2\left(\frac{xy}{(1-x)(1-y)}\right) &=& 
 \Li_2\left(\frac{x}{1-y}\right)
+\Li_2\left(\frac{y}{1-x}\right) - \Li_2(x) - \Li_2(y) -\ln(1-x) \ln(1-y),
\nonumber\\
\\
\Li_2\left(1-z\right) &=& - \Li_2(z) - \ln(z) \ln(1-z) + \zeta_2,
\\
\Li_2(-z) &=& \frac{1}{2} \Li_2(z^2) - \Li_2(z), 
\end{eqnarray}
allow to rewrite
\begin{eqnarray}
\Li_2(-(\sqrt{2}-1)^2) &=&
-\frac{5}{4} \zeta_2
-\frac{1}{4} \ln^2(2)
+\ln(2) \ln\big(
        1+\sqrt{2}\big)
-2 \text{Li}_2\left(
        -\frac{1}{\sqrt{2}}\right)
+\text{Li}_2\left(
        \big(
                \sqrt{2}-1\big)^2\right),
\nonumber\\
\end{eqnarray}
which proofs (\ref{eq:REL1}). The relation
\begin{eqnarray}
\label{eq:REL2}
  \HA_{0, -1, -1}(z)
- \HA_{0, -1,  1}(z) 
- \HA_{0,  1, -1}(z) 
+ \HA_{0,  1,  1}(z) 
- \frac{1}{2} 
  \HA_{0,  1,  1}\left(z^2\right) = 0
\end{eqnarray}
holds. It is obtained by first considering
\begin{eqnarray}
  \HA_{-1, -1}(x)
- \HA_{-1,  1}(x) 
- \HA_{ 1, -1}(x) 
+ \HA_{ 1,  1}(x) = \frac{1}{2} \ln^2(1-x^2).
\end{eqnarray}
The integration of the left-letter $1/x$ then proofs (\ref{eq:REL2}). Both relations play a role in deriving
the constants $c_1$ to $c_4$.

Furthermore, one may use the relations 
\begin{eqnarray}
\Li_2\left(\frac{1}{2}\left(1+\frac{1}{\sqrt{2}}\right)\right) &=&
-\frac{9}{8} \ln^2(2)
+\frac{3}{2} \ln(2) \ln\big(
        \sqrt{2}-1\big)
+\frac{3}{2} \ln^2\big(
        \sqrt{2}-1\big)
\nonumber\\ &&
-\text{Li}_2\left(
        \big(
                \sqrt{2}-1\big)^2\right) + \frac{1}{2}\text{Li}_2\left(
        \big(
                \sqrt{2}-1\big)^4\right)
+\zeta_2,
\\
\Li_2\left(\frac{1}{\sqrt{2}}\right) &=& \frac{7}{8} \zeta_2 - \frac{1}{8} \ln^2(2) + \frac{1}{2} \ln(2) 
\ln(\sqrt{2}-1) - 
\Li_2\left((\sqrt{2}-1)^2\right)
\nonumber\\ &&
+ \frac{1}{4} \Li_2\left((\sqrt{2}-1)^4\right),
\\
\Li_2(\sqrt{2}-1) &=& \Li_2\left(\frac{1}{\sqrt{2}}\right) - \frac{1}{4} \zeta_2 + \frac{1}{8} 
\ln^2(2) - \frac{1}{2} 
\ln(2) \ln(\sqrt{2}-1) 
\nonumber\\ &&
- \frac{1}{2} \ln^2(\sqrt{2}-1),
\\
\Li_2(\sqrt{2}(\sqrt{2}-1)) &=& \frac{5}{4}\zeta_2 - \frac{1}{8} \ln^2(2) - \frac{1}{2}\ln^2(\sqrt{2}-1)
- \Li_2\left(\frac{1}{\sqrt{2}}\right),
\\
\Li_3\left(\frac{1}{\sqrt{2}}\right) &=& 
-\frac{5}{8} \zeta_2 \ln(2)
+ \frac{1}{48} \ln^3(2) - \zeta_2 \ln(\sqrt{2}-1) + 
 \frac{1}{4} \ln(2) \ln^2(\sqrt{2}-1) 
\nonumber\\ &&
+ \frac{1}{3} \ln^3(\sqrt{2}-1)
+ \Li_3\left(\sqrt{2}-1\right) + \Li_3\left(\sqrt{2}(\sqrt{2}-1)\right) - \frac{25}{32} \zeta_3,
\nonumber\\
\\
\Li_3\left(\sqrt{2}(\sqrt{2}-1)\right) &=&
\zeta_2 \ln(2) + \frac{1}{8} i \pi \ln^2(2) - \frac{1}{48} \ln^3(2)
+ 2 \zeta_2 \ln(\sqrt{2}-1) 
\nonumber\\ &&
+ \frac{1}{2} i\pi \ln(2) \ln(\sqrt{2}-1)
-\frac{1}{8} \ln^2(2) \ln(\sqrt{2}-1) 
+ \frac{1}{2} i\pi \ln^2(\sqrt{2}-1)
\nonumber\\ &&
- \frac{1}{4} \ln(2) \ln^2(\sqrt{2}-1)
-\frac{1}{6} \ln^3(\sqrt{2}-1)
+\Li_3\left(1+\frac{1}{\sqrt{2}}\right)
\end{eqnarray}
to rewrite some of the polylogarithms above. One may finally use the relation
\begin{eqnarray}
\Li_3\left(\frac{1}{z}\right) &=& \Li_3(z) + \frac{1}{6} \ln^3(z) - \frac{1}{2} i \pi \ln^2(z) - 2 \zeta_2 
\ln(z),~~~z \in [0,1]
\end{eqnarray}
to rewrite the last two $\Li_3$-functions in (\ref{eq:REL1}) in a more uniform way in terms of
\begin{eqnarray}
\Li_3(2 \sqrt{2}(\sqrt{2} - 1))~~~~~\text{and}~~~~~\Li_3(\sqrt{2}(\sqrt{2}-1)).
\end{eqnarray}
Thus the arguments of the four trilogs contributing differ by a relative factor of $\sqrt{2}$.
One may as well rewrite $\Li_2((\sqrt{2}-1)^2)$ and $\Li_2((\sqrt{2}-1)^4)$ into $\Li_2(2(\sqrt{2}-1))$ and 
$\Li_2(2 \sqrt{2}(\sqrt{2}-1))$ and then obtain the set
\begin{eqnarray}
\label{eq:LISET1}
&& \ln(2), 
\pi, 
\ln(\sqrt{2}-1), 
\zeta_3, 
\Li_2\left(2(\sqrt{2}-1)\right),
\Li_2\left(2\sqrt{2}(\sqrt{2}-1)\right),
\nonumber\\
&&
\Li_3\left(\sqrt{2}-1\right),
\Li_3\left(\sqrt{2}(\sqrt{2}-1)\right),
\Li_3\left(2(\sqrt{2}-1)\right),
\Li_3\left(2\sqrt{2}(\sqrt{2}-1)\right).
\end{eqnarray}
	
In intermediary steps of the calculation also the following $\eta$-dependent constants and functions occur, 
which we list for completeness and the use in other calculations. Here we also simplified some relations 
given in \cite{Ablinger:2017err}.
\begin{eqnarray}
G_{16} &=& G\left(\left\{\frac{\sqrt{x}}{1-x},\frac{1}{x}\right\},z\right)
=
\HA_{-1}\big(\sqrt{z}\big) \HA_0(z)
+\HA_0(z) \HA_1\big(\sqrt{z}\big)
-2 \HA_{0,1}\big(\sqrt{z}\big)
-2 \HA_{0,-1}\big(\sqrt{z}\big)
\NN \\ &&
+2 \sqrt{z} \big(2-\HA_0(z)\big)~,
\\
G_{17} &=&  G\left(\left\{\frac{\sqrt{x}}{1-x},\frac{1}{x}\right\},\frac{1}{z}\right)
=
 \frac{4}{\sqrt{z}}
+\HA_0(z) \biggl[
         \frac{2}{\sqrt{z}}
        -\HA_{-1}\big(
                \sqrt{z}\big)
\biggr]
-6 \zeta_2
-\HA_0(z) \HA_1\big(
        \sqrt{z}\big)
\NN \\ &&
+2 \HA_{0,1}\big(
        \sqrt{z}\big)
+2 \HA_{0,-1}\big(
        \sqrt{z}\big)~,
\\
G_{18} &=& G\left(\left\{\frac{\sqrt{x}}{1-x},\frac{1}{x},\frac{1}{x}\right\},z\right)
=
-8 \sqrt{z}
+2 \HA_0(z) \biggl[
        2 \sqrt{z}
        - \HA_{0,1}\big(\sqrt{z}\big)
        - \HA_{0,-1}\big(\sqrt{z}\big)
\biggr] 
\NN \\ &&
+\frac{1}{2} \HA_0^2(z) \biggl[
        -2 \sqrt{z}
        + \HA_1\big(\sqrt{z}\big)
        + \HA_{-1}\big(\sqrt{z}\big)
\biggr] 
\NN \\ &&
+4 \HA_{0,0,1}\big(\sqrt{z}\big)
+4 \HA_{0,0,-1}\big(\sqrt{z}\big)~,
\\
G_{19} &=& G\left(\left\{\frac{1}{x},\frac{\sqrt{x}}{1-x},\frac{1}{x}\right\},z\right)
=
16 \sqrt{z}
+2 \HA_0(z) \biggl[
        -2 \sqrt{z}
        +\HA_{0,1}\big(\sqrt{z}\big)
        +\HA_{0,-1}\big(\sqrt{z}\big)
\biggr] 
\NN \\ &&
-8 \HA_{0,0,1}\big(\sqrt{z}\big)
-8 \HA_{0,0,-1}\big(\sqrt{z}\big)~,
\\
G_{20} &=& G\left(\left\{\frac{1}{1-x},\frac{\sqrt{x}}{1-x},\frac{1}{x}\right\},z\right)
=
-16 \sqrt{z}
+ 4 \HA_{-1}(\sqrt{z})
+\HA_0(z) \biggl[
        4 \sqrt{z}
        -2 \HA_{-1}\big(\sqrt{z}\big)
\NN \\ &&
        -\frac{1}{2} \HA_{-1}^2\big(\sqrt{z}\big)
\biggr] 
+ \HA_1\big(\sqrt{z}\big) \biggl[
        4
        +\HA_0(z) \bigl\{-2+\HA_{-1}\big(\sqrt{z}\big)\bigr\}
\biggr] 
+\frac{1}{2} \HA_0(z) \HA_1^2\big(\sqrt{z}\big)
\NN \\ && +2 \Biggl(
        2
	- \HA_{1}(z)
\Biggr) 
\biggl[ \HA_{0,1}\big(\sqrt{z}\big) + \HA_{0,-1}\big(\sqrt{z}\big) \biggr]
-2 \HA_0(z) \HA_{-1,1}\big(\sqrt{z}\big)
+2 \HA_{0,1,1}\big(\sqrt{z}\big)
\NN \\ &&
-2 \HA_{0,1,-1}\big(\sqrt{z}\big)
+2 \HA_{0,-1,1}\big(\sqrt{z}\big)
-2 \HA_{0,-1,-1}\big(\sqrt{z}\big)~,
\\
G_{21} &=& G\left(\left\{\frac{\sqrt{x}}{1-x},\frac{1}{1-x},\frac{1}{x}\right\}, z\right)
=
8 \sqrt{z}
+\bigl[
        -4 \sqrt{z}
        +2 \big(1+\sqrt{z}\big) \HA_{-1}\big(\sqrt{z}\big)
\nonumber \\ &&
        -\frac{1}{2} \HA_{-1}^2\big(\sqrt{z}\big)
\biggr] \HA_0(z)
+\bigl[
         2 \big(1-\sqrt{z}\big)
        -\HA_{-1}\big(\sqrt{z}\big)
\bigr] \HA_0(z) \HA_1\big(\sqrt{z}\big)
+\frac{1}{2} \HA_0(z) \HA_1^2\big(\sqrt{z}\big)
\nonumber \\ &&
-\bigl[
         4 \big(1-\sqrt{z}\big)
        +2 \HA_1\big(\sqrt{z}\big)
        +2 \HA_{-1}\big(\sqrt{z}\big)
\bigr] \HA_{0,1}\big(\sqrt{z}\big)
-\bigl[
         4 \big(1+\sqrt{z}\big)
        -2 \HA_1\big(\sqrt{z}\big)
\nonumber \\ &&
        -2 \HA_{-1}\big(\sqrt{z}\big)
\bigr] \HA_{0,-1}\big(\sqrt{z}\big)
+2 \HA_0(z) \HA_{-1,1}\big(\sqrt{z}\big)
+2 \HA_{0,1,1}\big(\sqrt{z}\big)
+2 \HA_{0,1,-1}\big(\sqrt{z}\big)
\nonumber\\ &&
-2 \HA_{0,-1,1}\big(\sqrt{z}\big)
-2 \HA_{0,-1,-1}\big(\sqrt{z}\big)~,
\\
G_{22} &=& G\left(\left\{\frac{1}{1-x},\frac{1}{x},\frac{1}{x}\right\},\frac{1}{z}\right)
=
\frac{1}{6} \HA_0^3(z)
+\frac{1}{2} \HA_0^2(z) \HA_1(z)
-\HA_0(z) \HA_{0,1}(z)
+\HA_{0,0,1}(z)~,
\nonumber\\
\\
G_{23} &=& G\left(\left\{\frac{\sqrt{x}}{1-x},\frac{1}{x},\frac{1}{x}\right\},\frac{1}{z}\right)
=
- \frac{8}{\sqrt{z}}
-2 \HA_0(z) \bigl[
         \frac{2}{\sqrt{z}}
        + \HA_{0,1}\big(
                \sqrt{z}\big)
        + \HA_{0,-1}\big(
                \sqrt{z}\big)
\bigr] 
\NN \\ &&
- \frac{1}{2} \HA_0^2(z) \biggl[
        \frac{2}{\sqrt{z}}
        -  \HA_1\big(\sqrt{z}\big)
        - \HA_{-1}\big(\sqrt{z}\big)
\biggr]
+4 \HA_{0,0,1}\big(\sqrt{z}\big)
+4 \HA_{0,0,-1}\big(\sqrt{z}\big)~.
\end{eqnarray}

\noindent
Furthermore, one has
\begin{eqnarray}
K_{29} &=& G\left(\left\{\sqrt{x (1-x)},\frac{\sqrt{x (1-x)}}{x \eta -\eta-x},\frac{1}{1-x}\right\},1\right)
=
-\frac{35-16 \eta +53 \eta ^2}{144 (1-\eta )^3}
+\frac{\sqrt{\eta }}{4(1-\eta)^2} \HA_ {-1}\big(\sqrt{\eta }\big) 
\NN\\ &&
\times \HA_ 0(\eta )
\biggl[
        1 
        +\frac{1}{4} \HA_ {-1}\big(\sqrt{\eta }\big)
\biggr] 
+ \frac{1}{4(1-\eta)^2} \HA_ 0(\eta ) 
\biggl[
        \frac{(3-\eta ) \eta ^2 \HA_ 1(\eta )}{ (1-\eta )^2}
        + \sqrt{\eta } \HA_ 1\big(\sqrt{\eta }\big) \bigl\{
                \sqrt{\eta } 
            - 1 
\NN \\ && 
                +\frac{1}{2} \HA_ {-1}\big(\sqrt{\eta }\big)
        \bigr\}
        -\frac{\sqrt{\eta }}{4} \HA_ 1^2\big(\sqrt{\eta }\big)
        - \sqrt{\eta } (1+\sqrt{\eta }) \HA_ {-1}\big(\sqrt{\eta }\big)
        + \frac{\sqrt{\eta }}{4} \HA_ {-1}^2\big(\sqrt{\eta }\big)
        - \sqrt{\eta } \HA_ {-1,1}\big(\sqrt{\eta }\big)
\biggr] 
\NN \\ &&
+\frac{\sqrt{\eta }}{4(1-\eta)^2} \HA_ 0(\eta ) \HA_ 1\big(\sqrt{\eta }\big) 
\biggl[
        1
        -\frac{1}{2} \HA_ {-1}\big(\sqrt{\eta }\big)
\biggr]
-\frac{\sqrt{\eta }}{16 (1-\eta )^2} \HA_ 0(\eta ) \HA_ 1^2\big(\sqrt{\eta }\big)
\NN \\ &&
-\frac{(3-\eta ) \eta ^2 }{4 (1-\eta )^4} \HA_{0,1}(\eta )
- \frac{1}{2 (1-\eta )^2} \HA_ {0,1}\big(\sqrt{\eta }\big) 
\biggl[
         \eta 
        -\sqrt{\eta } \HA_ 1\big(\sqrt{\eta }\big)
\biggr] 
+ \frac{1}{2(1-\eta)^2} \HA_ {0,-1}\big(\sqrt{\eta }\big)
\NN \\ && \times
\biggl[
        \eta
        - \sqrt{\eta } \HA_ {-1}\big(\sqrt{\eta }\big)
\biggr]
+\frac{\sqrt{\eta }}{4 (1-\eta )^2} 
\biggl[
\HA_ 0(\eta ) \HA_ {-1,1}\big(\sqrt{\eta }\big)
-2 \HA_ {0,1,1}\big(\sqrt{\eta }\big)
+2 \HA_ {0,-1,-1}\big(\sqrt{\eta }\big)
\biggr]
\NN \\ &&
+\frac{\zeta_2}{4(1-\eta)^2} \biggl[
        \sqrt{\eta} \bigl\{
        -3 \HA_ 1(\eta )
        + \HA_ 1\big(\sqrt{\eta }\big)
        + \HA_ {-1}\big(\sqrt{\eta }\big)
        \bigr\}
        +\frac{3 - 5 \eta + 13 \eta^2 - 3 \eta^3}{4 (1-\eta )^2}
\NN \\ &&
        - 6 \ln(2) \frac{ 1 - \eta - \eta^2 + \eta^3 }{4 (1-\eta )^2}
\biggr] 
+\frac{7 (1+\eta ) \zeta_3}{32 (1-\eta )^2}~,
\\
K_{30} &=& G\left(\left\{\sqrt{x (1-x)},\frac{\sqrt{x (1-x)}}{x \eta -\eta-x},\frac{1}{x}\right\},1\right)
=
\frac{53-16 \eta +35 \eta ^2}{144 (1-\eta )^3}
\nonumber \\ &&
+\frac{1}{(1-\eta )^4}
\biggl[
-\frac{1}{8} \eta  (1+\eta ) \HA_0^2(\eta )
-\frac{1}{4} (3-\eta ) \eta ^2 \bigl[ \HA_0(\eta ) \HA_1(\eta ) - \HA_{0,1}(\eta ) \bigr]
\nonumber \\ &&
+\frac{1}{16} \big(3-13 \eta +5 \eta ^2-3 \eta ^3\big) \zeta_2
\biggr]
+\frac{1}{(1-\eta )^2}
\biggl[
\biggl(
        \frac{1}{4} \eta  \HA_{-1}\big(\sqrt{\eta }\big)
        -\frac{1}{8} \sqrt{\eta } \HA_{-1}^2\big(\sqrt{\eta }\big)
\biggr) \HA_0(\eta )
\nonumber \\ &&
+\frac{1}{16} \sqrt{\eta } H_{-1}\big(\sqrt{\eta }\big) \HA_0^2(\eta )
+\biggl(
        -\frac{1}{4} \eta  \HA_0(\eta )
        +\frac{1}{16} \sqrt{\eta } \HA_0^2(\eta )
\biggr) \HA_1\big(\sqrt{\eta }\big)
+\frac{1}{8} \sqrt{\eta } \HA_0(\eta ) \HA_1^2\big(\sqrt{\eta }\big)
\nonumber \\ &&
+\biggl(
        \frac{\eta }{2}
        -\frac{1}{2} \sqrt{\eta } \HA_1\big(\sqrt{\eta }\big)
\biggr) \HA_{0,1}\big(\sqrt{\eta }\big)
+\biggl(
        -\frac{\eta }{2}
        +\frac{1}{2} \sqrt{\eta } \HA_{-1}\big(\sqrt{\eta }\big)
\biggr) \HA_{0,-1}\big(\sqrt{\eta }\big)
\nonumber \\ &&
-\frac{1}{2} \sqrt{\eta } \bigl[ \HA_{0,0,1}\big(\sqrt{\eta }\big) + \HA_{0,0,-1}\big(\sqrt{\eta }\big) - \HA_{0,1,1}\big(\sqrt{\eta }\big) + \HA_{0,-1,-1}\big(\sqrt{\eta }\big) \bigr] 
\nonumber \\ &&
+\biggl(
        \frac{3}{8} (1+\eta ) \ln(2)
        +\frac{1}{4} \sqrt{\eta } \bigl[ 3 \HA_0(\eta ) + 3 \HA_1(\eta ) - \HA_1\big(\sqrt{\eta }\big) - \HA_{-1}\big(\sqrt{\eta }\big) \bigr]
\biggr) \zeta_2
\nonumber \\ &&
+\frac{7}{32} (1+\eta ) \zeta_3
\biggr]~,
\\
K_{31} &=& G\left(\left\{\frac{1}{x},\frac{1}{x \eta-\eta-x},\frac{1}{1-x}\right\},1\right)
=
\frac{1}{1-\eta}
\biggl[
 \frac{1}{2} \HA_0^2(\eta ) \HA_1(\eta )
+\HA_0(\eta ) \HA_1^2(\eta )
\nonumber \\ &&
-2 \HA_1(\eta ) \HA_{0,1}(\eta )
+2 \HA_{0,1,1}(\eta )
+2 \HA_1(\eta ) \zeta_2
-\HA_{0,0,1}(\eta )
-\zeta_3
\biggr]~,
\\
K_{32} &=& G\left(\left\{\frac{1}{x},\frac{1}{x \eta -\eta-x},\frac{1}{x}\right\},1\right)
= \frac{1}{1-\eta } \biggl[
-\frac{1}{3} \HA_0^3(\eta )
-\HA_0^2(\eta ) \HA_1(\eta )
-\HA_0(\eta ) \HA_1^2(\eta )
\NN \\ &&
+2 \HA_1(\eta ) \HA_{0,1}(\eta )
+2 \HA_{0,0,1}(\eta )
-2 \HA_{0,1,1}(\eta )
-2 \biggl\{ \HA_0(\eta ) 
+ \HA_1(\eta ) \biggr\} \zeta_2
\biggr]~,
\\
K_{33} &=& G\left(\left\{\sqrt{x (1-x)},\frac{\sqrt{x (1-x)}}{\eta x-x+1},\frac{1}{1-x}\right\},1\right)
=
-\frac{53-16 \eta +35 \eta ^2}{144 (1-\eta )^3}
\nonumber \\ &&
+\frac{1}{(1-\eta )^4}
\biggl[
 \frac{1}{8} \eta  (1+\eta ) \HA_0^2(\eta )
-\frac{1}{4} (-3+\eta ) \eta ^2 \bigl[ \HA_0(\eta ) \HA_1(\eta ) - \HA_{0,1}(\eta ) \bigr]
\nonumber \\ &&
+\frac{1}{16} \big(3-5 \eta +13 \eta ^2-3 \eta ^3\big) \zeta_2
\biggr]
+\frac{1}{(1-\eta )^2}
\biggl[
\biggl(
        -\frac{1}{4} \eta  \HA_{-1}\big(\sqrt{\eta }\big)
        +\frac{1}{8} \sqrt{\eta } \HA_{-1}^2\big(\sqrt{\eta }\big)
\biggr) \HA_0(\eta )
\nonumber \\ &&
-\frac{1}{16} \sqrt{\eta } \HA_{-1}\big(\sqrt{\eta }\big) \HA_0^2(\eta )
+\biggl(
        \frac{1}{4} \eta  \HA_ 0(\eta )
        -\frac{1}{16} \sqrt{\eta } \HA_0^2(\eta )
\biggr) \HA_1\big(\sqrt{\eta }\big)
-\frac{1}{8} \sqrt{\eta } \HA_0(\eta ) \HA_ 1^2\big(\sqrt{\eta }\big)
\nonumber \\ &&
+\biggl(
        -\frac{\eta }{2}
        +\frac{1}{2} \sqrt{\eta } \HA_ 1\big(\sqrt{\eta }\big)
\biggr) \HA_{0,1}\big(\sqrt{\eta }\big)
+\biggl(
        \frac{\eta }{2}
        -\frac{1}{2} \sqrt{\eta } \HA_{-1}\big(\sqrt{\eta }\big)
\biggr) \HA_{0,-1}\big(\sqrt{\eta }\big)
\nonumber \\ &&
+\frac{1}{2} \sqrt{\eta } \bigl[ \HA_{0,0,1}\big(\sqrt{\eta }\big) + \HA_{0,0,-1}\big(\sqrt{\eta }\big) - \HA_ {0,1,1}\big(\sqrt{\eta }\big) + \HA_ {0,-1,-1}\big(\sqrt{\eta }\big) \bigr]
\nonumber \\ &&
+\biggl(
        \frac{3}{8} (1+\eta ) \ln(2)
        -\frac{1}{2} \sqrt{\eta } \bigl[ H_ 1\big(\sqrt{\eta }\big) + H_ {-1}\big(\sqrt{\eta }\big) \bigr]
\biggr) \zeta_2
-\frac{7}{32} (1+\eta ) \zeta_3
\biggr]~,
\\
K_{34} &=& G\left(\left\{\sqrt{x (1-x)},\frac{\sqrt{x (1-x)}}{\eta x-x+1},\frac{1}{x}\right\},1\right)
=
 \frac{35-16 \eta +53 \eta ^2}{144 (1-\eta )^3}
\nonumber \\ &&
+\frac{1}{(1-\eta )^4}
\biggl[
-\frac{1}{4} (3-\eta ) \eta ^2 \big[ \HA_0(\eta ) \HA_1(\eta ) - H_{0,1}(\eta ) \big]
+\frac{1}{16} \big(3-13 \eta +5 \eta ^2-3 \eta ^3\big) \zeta_2
\biggr]
\nonumber \\ &&
+\frac{1}{(1-\eta )^2}
\biggl[
\biggl(
        \frac{1}{4} \eta  \HA_{-1}\big(\sqrt{\eta }\big)
        -\frac{1}{8} \sqrt{\eta } \HA_{-1}^2\big(\sqrt{\eta }\big)
\biggr) \HA_0(\eta )
-\frac{1}{4} \eta  \HA_0(\eta ) \HA_1\big(\sqrt{\eta }\big)
\nonumber \\ &&
+\frac{1}{8} \sqrt{\eta } \HA_0(\eta ) \HA_1^2\big(\sqrt{\eta }\big)
+\biggl(
        \frac{\eta }{2}
        -\frac{1}{2} \sqrt{\eta } \HA_1\big(\sqrt{\eta }\big)
\biggr) \HA_{0,1}\big(\sqrt{\eta }\big)
\nonumber \\ &&
+\biggl(
        -\frac{\eta }{2}
        +\frac{1}{2} \sqrt{\eta } \HA_{-1}\big(\sqrt{\eta }\big)
\biggr) \HA_{0,-1}\big(\sqrt{\eta }\big)
+\frac{ \sqrt{\eta } }{2} \bigl[ \HA_{0,1,1}\big(\sqrt{\eta }\big) - \HA_{0,-1,-1}\big(\sqrt{\eta }\big) \bigr]
\nonumber \\ &&
+\biggl(
        -\frac{3}{8} (1+\eta ) \ln(2)
        +\frac{1}{2} \sqrt{\eta } \bigl[ \HA_1\big(\sqrt{\eta }\big) + \HA_{-1}\big(\sqrt{\eta }\big) \bigr]
\biggl) \zeta_2
-\frac{7}{32} (1+\eta ) \zeta_3
\biggr]
~,
\\
K_{35} &=& G\left(\left\{\frac{1}{1-x(1-\eta)},\frac{1}{1-x}\right\},1\right) 
=
\frac{1}{1-\eta} \biggl[
\frac{1}{2} \HA_0(\eta )^2
+\HA_0(\eta ) \HA_1(\eta )
-\HA_{0,1}(\eta )
+\zeta_2
\biggr]~,
\nonumber\\
\\
K_{36} &=& G\left(\left\{\frac{\sqrt{x (1-x)}}{x\eta -\eta-x},\frac{1}{x}\right\},1\right)
=
\frac{\pi }{2 (1-\eta )^2} \biggl[
        1
        +2 \ln (2) ( 1 + \eta )
        -\eta 
        + 2 \sqrt{\eta } \biggl\{
        \HA_0(\eta )
        + \HA_1(\eta )
\NN \\ &&
        - \HA_1\big(\sqrt{\eta }\big)
        - \HA_{-1}\big(\sqrt{\eta }\big)
        \biggr\}
\biggr]~,
\\
K_{37} &=& G\left(\left\{\frac{\sqrt{x (1-x)}}{x \eta -\eta-x},\frac{1}{1-x}\right\},1\right)
=
\frac{\pi }{(1-\eta )^2} \biggl[
        \frac{1+3 \eta}{2}
        +\sqrt{\eta } \biggl\{
                -2 \sqrt{\eta }
                -\HA_1(\eta )
                +\HA_1\big(\sqrt{\eta }\big)
\NN \\ &&
                +\HA_{-1}\big(\sqrt{\eta }\big)
        \biggr\} 
        -(1+\eta ) \ln (2)
\biggr]~,
\\
K_{38} &=& G\left(\left\{\frac{1}{\eta x-x+1},\frac{1}{x}\right\},1\right)
=
-\frac{1}{1-\eta} \biggl[
\HA_0(\eta ) \HA_1(\eta )
-\HA_{0,1}(\eta )
+\zeta_2
\biggr]~,
\\
K_{39} &=& G\left(\left\{\sqrt{x (1-x)},\frac{\sqrt{x (1-x)}}{x \eta -\eta-x}\right\},1\right)
=
-\frac{1+\eta +\eta ^2}{6 (1-\eta )^3}
+\frac{1}{(1-\eta )^2}
\biggl[
 \frac{1}{8} \sqrt{\eta } \bigl[ \HA_0(\eta ) \HA_{-1}\big(\sqrt{\eta }\big) 
\nonumber \\ &&
+ \HA_0(\eta ) \HA_1\big(\sqrt{\eta }\big) - 2 \HA_{0,1}\big(\sqrt{\eta }\big) - 2 \HA_{0,-1}\big(\sqrt{\eta }\big) \bigr]
-\frac{3}{16} \big(1-4 \sqrt{\eta }+\eta \big) \zeta_2
\biggr]
\nonumber \\ &&
-\frac{\eta  (1+\eta )}{4 (1-\eta )^4} \HA_0(\eta ) ~,
\\
K_{40} &=& G\left(\left\{\frac{1}{x},\frac{1}{\eta x-x+1}\right\},1\right)
=
\frac{1}{1-\eta } \biggl[
        \HA_0(\eta ) \HA_1(\eta )
        -\HA_{0,1}(\eta )
        +\zeta_2
\biggr]~,
\\
K_{41} &=& G\left(\left\{\frac{1}{x \eta -\eta-x},\frac{1}{x}\right\},1\right)
=
\frac{1}{1-\eta } \biggr[
        \frac{1}{2} \HA_0^2(\eta )
        +\HA_0(\eta ) \HA_1(\eta )
        -\HA_{0,1}(\eta )
        +\zeta_2
\biggr]~,
\\
K_{42} &=& G\left(\left\{\frac{1}{x \eta-\eta-x},\frac{1}{1-x}\right\},1\right)
=
-\frac{1}{1-\eta} \biggl[
\HA_0(\eta ) \HA_1(\eta )
-\HA_{0,1}(\eta )
+\zeta_2
\biggr]~,
\\
K_{43} &=& G\left(\left\{\frac{1}{x},\frac{1}{x \eta-\eta -x}\right\},1\right)
=
-\frac{1}{1-\eta} \biggl[
\frac{1}{2} \HA_0^2(\eta )
+\HA_0(\eta ) \HA_1(\eta )
-\HA_{0,1}(\eta )
+\zeta_2
\biggr]~,
\\
K_{44} &=& G\left(\left\{\frac{\sqrt{x (1-x)}}{\eta x-x+1},\frac{1}{x}\right\},1\right) 
=
\frac{\pi }{2 (1-\eta )^2} \biggl[
        1 -\eta  + 4 \sqrt{\eta } \HA_{-1}\big(\sqrt{\eta }\big)
        -2 \ln(2) ( 1 + \eta )
\biggr],
\nonumber\\
\\
K_{45} &=& G\left(\left\{\frac{\sqrt{x (1-x)}}{\eta x-x+1},\frac{1}{1-x}\right\},1\right) 
=
\frac{\pi }{2 (1-\eta )^2} \biggl[
        1 -\eta - 2 \sqrt{\eta } \bigl\{ 2 \HA_{-1}\big(\sqrt{\eta }\big) 
\nonumber \\ &&
	- \HA_0(\eta) \bigr\}
        +2 \ln(2) ( 1 + \eta )
\biggr]~,
\\
K_{46} &=& G\left(\left\{\frac{1}{\frac{1}{1-\eta}-x},\frac{1}{x+1}\right\},1\right)
=
- \ln(2) \bigl\{ \HA_0(\eta ) - \HA_0(2-\eta ) \bigr\}
-\HA_{0,1}\left(\frac{2 (1-\eta )}{2-\eta }\right)
\NN \\ &&
+\HA_{0,1}\left(\frac{1-\eta }{2-\eta }\right)~,
\\
K_{47} &=& G\left(\left\{\frac{1}{\eta x-x+1},\frac{1}{x-1}\right\},1\right)
=-\frac{1}{1-\eta} \biggl[
        \HA_0(\eta ) \HA_1(\eta )
        -\HA_{0,1}(\eta )
        +\zeta_2
        +\frac{1}{2} \HA_0^2(\eta )
\biggr],
\\
K_{48} &=& G\left(\left\{\sqrt{x (1-x)},\frac{\sqrt{x (1-x)}}{\eta x-x+1}\right\},1\right)
=
-\frac{1+\eta +\eta ^2}{6 (1-\eta )^3}
-\frac{(3-\eta ) \eta ^2}{4 (1-\eta )^4} \HA_0(\eta )
\nonumber \\ &&
+\frac{1}{(1-\eta )^2}
\biggl[
\biggl(
        -\frac{\eta }{4}
        +\frac{1}{8} \sqrt{\eta } \HA_{-1}\big(\sqrt{\eta }\big)
\biggr) \HA_0(\eta )
+\frac{1}{8} \sqrt{\eta } \bigl[ \HA_0(\eta ) \HA_1\big(\sqrt{\eta }\big) - 2 \HA_{0,1}\big(\sqrt{\eta }\big) 
\nonumber \\ &&
- 2 \HA_{0,-1}\big(\sqrt{\eta }\big) \bigr]
+\frac{3}{16} (1+\eta ) \zeta_2
\biggr]
~.
\end{eqnarray}
There are further functions emerging. We define the following variables in order to obtain a more compact 
representation.
\begin{eqnarray}
y &=& \frac{\sqrt{z}-i \sqrt{1-z}}{\sqrt{z}+i \sqrt{1-z}}, \\
u &=& i \sqrt{\frac{z}{1-z}}, \\
\lambda &=& \frac{\sqrt{1-\eta}-i \sqrt{\eta}}{\sqrt{1-\eta}+i \sqrt{\eta}}, \\
\rho &=& \frac{1-\sqrt{\eta}}{1+\sqrt{\eta}}, \\
\chi &=& \frac{1}{\eta} \left(\sqrt{1-\eta} \sqrt{1-z}-\sqrt{1-z+\eta z}\right)^2, \\
\omega &=& \left(\sqrt{z+\eta-\eta z}-i \sqrt{1-\eta} \sqrt{1-z}\right)^2, \\
\xi &=& \left(\sqrt{\eta-\eta z}-i \sqrt{1-\eta+\eta z}\right)^2.
\end{eqnarray}
In some of the following functions a $_3F_2$-function contributes. It can be rewritten, 
cf.~\cite{MARICHEV}, by
\begin{eqnarray}
_3F_2\left[\begin{array}{c} -\frac{1}{2}, \frac{3}{2}, \frac{3}{2} \\ 
\;\;\; \frac{5}{2}, \frac{5}{2}\end{array};z\right] &=&
-\frac{9}{8 \sqrt{2} z^{3/2}}(1+i) \biggl[
-\text{Li}_2\left( -\sqrt{1-z}-i \sqrt{z}\right)
+\text{Li}_2\left(1-\sqrt{1-z}-i \sqrt{z}\right)
\nonumber \\ &&
+\frac{1}{2} \ln^2\left(\sqrt{1-z}+i \sqrt{z}\right)
+\frac{1}{4} \ln  \left(\sqrt{1-z}+i \sqrt{z}\right)
- \frac{\zeta_2}{2}
\nonumber \\ &&
-\ln\left(1+\sqrt{1-z}+i \sqrt{z}\right) \ln \left(\sqrt{1-z}+i\sqrt{z}\right)
\biggr]
-\frac{9 \sqrt{1-z} (3-2 z)}{32 z}, 
\nonumber \\
\end{eqnarray}
which is valid for all values of $z$.

Now we present a few $G$-functions of weight {\sf w = 2} that depend on $\eta$, with $0<\eta<1$. 
A few of them are valid for $0<z<1-\eta$ only.

\begin{align}
G_{24} =& G\left(\left\{\frac{1}{1-x},\sqrt{1-x} \sqrt{1-x-\eta}\right\},z\right) \;=\; 
\frac{1}{8} \sqrt{1-\eta } (3 \eta -2)
+\frac{\eta^2}{4} \Biggl\{
\frac{1}{2} \ln \left(1-\sqrt{1-\eta }\right)
\nonumber \\ & 
+\arcsin^2\left(\frac{1}{\sqrt{\eta}}\right)
+2 i \left[
\ln \left(1+\sqrt{1-\eta}\right)
-\ln (\eta )
+\ln (2)\right] \arcsin\left(\frac{1}{\sqrt{\eta }}\right)
\nonumber \\ &
+\text{Li}_2\left(\frac{\eta -2 \sqrt{1-\eta }-2}{\eta}\right)
-\text{Li}_2\left(-\frac{\left(\sqrt{1-z}+\sqrt{1-z-\eta}\right)^2}{\eta }\right)
\nonumber \\ & 
-\frac{1}{2} \ln \left(\sqrt{1-z}-\sqrt{1-z-\eta}\right)
+\left[\ln \left(1+\sqrt{1-\eta}\right)
-\frac{\ln (\eta )}{2}
+\frac{i \pi }{2}\right] \ln (1-z)
\nonumber \\ &
-\arcsin^2\left(\sqrt{\frac{1-z}{\eta}}\right)
-2 i \ln \left(\sqrt{1-z-\eta}+\sqrt{1-z}\right) \arcsin\left(\sqrt{\frac{1-z}{\eta}}\right)
\nonumber \\ & 
+i \ln\left(\frac{\eta^2}{4 (1-z)}\right) \arcsin\left(\sqrt{\frac{1-z}{\eta}}\right)
\Biggr\}
+\frac{1}{4} \sqrt{1-\eta } (\eta -2) \ln (1-z)
\nonumber \\ & 
-\frac{1}{8} \sqrt{1-z} \sqrt{1-z-\eta} (3 \eta +2 z-2) \, .
\\
G_{25} =& G\left(\left\{\sqrt{1-x} \sqrt{1-x-\eta},\frac{1}{1-x}\right\},z\right) \;=\;
\frac{4}{9} i \sqrt{\eta } (1-z)^{3/2} \, _3F_2\left[\begin{array}{c} -\frac{1}{2}, \frac{3}{2}, \frac{3}{2} \\ \;\;\; \frac{5}{2}, \frac{5}{2}\end{array};\frac{1-z}{\eta }\right]
\nonumber \\ &
-\frac{4}{9} i \sqrt{\eta } \, _3F_2\left[\begin{array}{c} -\frac{1}{2}, \frac{3}{2}, \frac{3}{2} \\ \;\;\; \frac{5}{2}, \frac{5}{2}\end{array};\frac{1}{\eta }\right]
-\frac{i}{4} \eta ^2 \ln(1-z) \arcsin\left(\sqrt{\frac{1-z}{\eta}}\right)
\nonumber \\ &
-\frac{1}{4} \sqrt{1-z} \sqrt{1-z-\eta} (\eta +2 z-2) \ln (1-z) \, .
\\
G_{26} =& G\left(\left\{\frac{1}{x},\sqrt{1-x} \sqrt{1-x-\eta}\right\},z\right) \;=\; 
\frac{\eta^2}{4} \Biggl\{
\frac{3}{2} \ln ^2\left(\sqrt{1-\eta }+1\right)
+\text{Li}_2\left(\frac{1}{2}+\frac{1}{2} \sqrt{1-\eta }\right)
\nonumber \\ & \hspace{20mm}
+\text{Li}_2\left(\frac{\sqrt{1-z}+\sqrt{1-z-\eta}}{\sqrt{1-z} \left(1-\sqrt{1-\eta }\right)}\right)
+\text{Li}_2\left(\frac{\sqrt{1-z}+\sqrt{1-z-\eta}}{\sqrt{1-z} \left(1+\sqrt{1-\eta }\right)}\right)
-\frac{\pi^2}{6}
\nonumber \\ & \hspace{20mm}
-\text{Li}_2\left(\frac{1+\sqrt{1-\eta }}{1-\sqrt{1-\eta }}\right)
-\text{Li}_2\left(\frac{1}{2}+\frac{1}{2} \sqrt{1-\frac{\eta }{1-z}}\right)
+\biggl[
2\ln \left(1-\sqrt{1-\eta }\right)
\nonumber \\ & \hspace{20mm}
-\ln (2 \eta z)
-\ln \left(1+\sqrt{1-\frac{\eta }{1-z}}\right)
-i \pi
\biggr] \ln \left(1+\sqrt{1-\eta }\right)
-\frac{1}{2} \text{Li}_2(z)
\nonumber \\ & \hspace{20mm}
-\frac{1}{2} \ln ^2\left(1+\sqrt{1-\frac{\eta }{1-z}}\right)
+\biggl[
-\ln \left(1-\sqrt{1-\eta }\right)
+\ln \left(\frac{2 \eta  z}{1-z}\right)
+i \pi
\nonumber \\ & \hspace{20mm}
-\ln \left(1-\sqrt{1-\frac{\eta }{1-z}}\right)
\biggr] \ln \left(1+\sqrt{1-\frac{\eta }{1-z}}\right)
\Biggr\}
+\sqrt{1-\eta } \left(\frac{1}{2}-\frac{\eta }{4}\right) \biggl[
\nonumber \\ & \hspace{20mm}
-\ln \left(\frac{\sqrt{1-z}+\eta -\sqrt{1-\eta } \sqrt{1-z-\eta}}{\sqrt{1-z}-\eta +\sqrt{1-\eta } \sqrt{1-z-\eta}}\right)
-\ln \left(\frac{1-\sqrt{1-z}}{1+\sqrt{1-z}}\right)
\nonumber \\ & \hspace{20mm}
+\ln \left(\frac{\eta  z}{4 (1-\eta )}\right)
+\frac{3}{2}
\biggr]
+\left(\eta -1-\frac{\eta^2}{8}\right) \ln \left(\frac{\sqrt{1-z}+\sqrt{1-z-\eta}}{1+\sqrt{1-\eta }}\right)
\nonumber \\ & \hspace{20mm}
+\sqrt{1-z} \sqrt{1-z-\eta} \left(\frac{3 \eta }{8}+\frac{z-3}{4}\right).
\\
G_{27} = &G\left(\left\{\sqrt{1-x} \sqrt{1-x-\eta},\frac{1}{1-x-\eta}\right\},z\right) \;=\;
\frac{4}{9} (1-\eta )^{3/2} \sqrt{\eta } \, _3F_2\left[\begin{array}{c} -\frac{1}{2}, \frac{3}{2}, \frac{3}{2} \\ \;\;\; \frac{5}{2}, \frac{5}{2}\end{array};\frac{\eta -1}{\eta }\right]
\nonumber \\ & 
-\frac{4}{9} \sqrt{\eta } (1-z-\eta)^{3/2} \, _3F_2\left[\begin{array}{c} -\frac{1}{2}, \frac{3}{2}, \frac{3}{2} \\ \;\;\; \frac{5}{2}, \frac{5}{2}\end{array};-\frac{1-z-\eta}{\eta}\right]
\nonumber \\ & 
-\frac{\eta^2}{4} \ln (1-z-\eta) \arcsinh\left(\frac{\sqrt{1-z-\eta}}{\sqrt{\eta}}\right)
-\frac{\eta^2}{4} \ln (1-\eta) \arctanh\left(\sqrt{1-\eta}\right)
\nonumber \\ &
+\frac{\eta^2}{4} \ln (1-\eta) \left[\arcsinh\left(\sqrt{\frac{1-\eta}{\eta}}\right)
+\arctanh\left(\sqrt{\frac{1-z-\eta}{1-z}}\right)\right]
\nonumber \\ &
-\frac{1}{4} \sqrt{1-z} \sqrt{1-z-\eta} (\eta +2 z-2) \ln \left(1-\frac{z}{1-\eta }\right) \, .
\end{align}
A larger set of functions is valid for $0<z<1$~:
\begin{align}
G_{28} =& G\left(\left\{\frac{1}{1-x},-\frac{\sqrt{x} \sqrt{1-x}}{x+\eta-x \eta}\right\},z\right) 
\;=\;
\frac{\pi }{2 (\eta -1)}
+\frac{\sqrt{1-z} \sqrt{z}}{1-\eta }
+\frac{1}{1-\eta}\arctan\left(\sqrt{\frac{1-z}{z}}\right)
\nonumber \\ & 
+\frac{\sqrt{\eta }}{(1-\eta)^2} \Biggl\{
-\pi  \ln (1-\eta)
-\pi  \ln (1-z)
-2 \ln \left(\rho\right) \arctan\left(\sqrt{\frac{1-z}{z}}\right)
\nonumber \\ & 
-i \text{Li}_2\left(\frac{1-u}{1-\sqrt{\eta}}\right)
+i \text{Li}_2\left(\frac{1+u}{1-\sqrt{\eta}}\right)
+i \text{Li}_2\left(\frac{1-u}{1+\sqrt{\eta}}\right)
-i \text{Li}_2\left(\frac{1+u}{1+\sqrt{\eta}}\right)
\Biggr\}
\nonumber \\ & 
+\frac{(\eta +1)}{(1-\eta )^2} \Biggl\{
-i \text{Li}_2(y)
-2 i \arcsin\left(\sqrt{1-z}\right) \arctan\left(\sqrt{\frac{z}{1-z}}\right)
+ i \zeta_2
\nonumber \\ & 
-i \arcsin^2\left(\sqrt{1-z}\right)
+\left[\frac{1}{2} \ln (1-z)+\ln (2)\right] \left(\pi -2 \arcsin\left(\sqrt{1-z}\right)\right)
\Biggr\} \, . 
\\
G_{29} =& G\left(\left\{\frac{1}{1-x},\sqrt{1-x} \sqrt{1-\eta+\eta x}\right\},z\right) \;=\;
\frac{1}{4 \eta ^{3/2}} \Biggl\{
i \text{Li}_2\left(\frac{1}{\lambda}\right)
-i \text{Li}_2\left(-\xi\right)
\nonumber \\ & 
+\frac{1}{4} \arctan\left(\frac{2 \eta -1}{2 \sqrt{1-\eta } \sqrt{\eta }}\right)
+2 i \arcsin\left(\sqrt{\eta }\right) \arctan\left(\sqrt{\frac{1-\eta}{\eta}}\right)
\nonumber \\ & 
-i \arcsin^2\left(\sqrt{\eta } \sqrt{1-z}\right)
+\frac{1}{4} \arctan\left(\frac{1-2 \eta  (1-z)}{2 \sqrt{\eta (1-z) (1-\eta +\eta z)}}\right)
\nonumber \\ & 
+\ln (4 \eta  (1-z)) \left(
\arcsin\left(\sqrt{\eta } \sqrt{1-z}\right)
-\arcsin\left(\sqrt{\eta }\right)
\right)
+i \arcsin^2\left(\sqrt{\eta }\right)
\nonumber \\ & 
-2 i \arcsin\left(\sqrt{\eta } \sqrt{1-z}\right)
\arctan\left(\sqrt{\frac{1-\eta+\eta z}{\eta (1-z)}}\right)
\Biggr\}
-\frac{\sqrt{1-\eta } (2 \eta -3)}{8 \eta }
\nonumber \\ & 
-\frac{1}{8 \eta }\sqrt{1-z} \sqrt{1-\eta+\eta z} (3-2 \eta +2 \eta  z)
-\frac{1}{4 \eta }\sqrt{1-\eta } (2 \eta -1) \ln (1-z) \, .
\\
G_{30} =& G\left(\left\{\sqrt{1-x} \sqrt{1-x+\eta x},\frac{1}{1-x}\right\},z\right) \;=\;
\frac{\eta^2}{8 (1-\eta)^{3/2}} \Biggl\{
-2 \arcsinh^2\left(\sqrt{\frac{1-\eta}{\eta}}\right)
\nonumber \\ & 
+\left[1-4 \ln(2)-4 \ln\left(\frac{\eta+\sqrt{1-\eta }-1}{\eta}\right)\right] \arcsinh\left(\sqrt{\frac{1-\eta}{\eta}}\right)
\nonumber \\ & 
+2 \text{Li}_2\left(-\frac{\eta +2 \sqrt{1-\eta }-2}{\eta }\right)
+2 \arcsinh^2\left(\frac{\sqrt{1-\eta } \sqrt{1-z}}{\sqrt{\eta }}\right)
\nonumber \\ & 
-\ln (1-z) \left[
2 \ln \left(\sqrt{1-\eta} \sqrt{1-z}+\sqrt{1-z+\eta z}\right)
-\ln (\eta )
\right]
\nonumber \\ &
-2 \text{Li}_2\left(\chi\right)
+\arcsinh\left(\frac{\sqrt{1-\eta } \sqrt{1-z}}{\sqrt{\eta }}\right) \big(4 \ln\left(1-\chi\right)-1\big)
\Biggr\}
\nonumber \\ & 
-\frac{\eta+2}{8 (\eta-1)}
+\sqrt{1-z} \sqrt{1-z+\eta z} \frac{(\eta +2 \eta  z-2 z+2)}{8 (\eta -1)}
\nonumber \\ & 
-\sqrt{1-z} \sqrt{1-z+\eta z} \frac{(2-\eta +2 \eta  z-2 z)}{4 (\eta -1)} \ln (1-z) \, .
\\
G_{31} =& G\left(\left\{\sqrt{1-x} \sqrt{x+\eta-x \eta},\frac{1}{1-x}\right\},z\right) \;=\; 
\frac{1}{4 (1-\eta )^{3/2}} \Biggl\{
i \arcsin^2\left(\sqrt{1-\eta }\right)
\nonumber \\ & 
+\arcsin\left(\sqrt{1-\eta }\right) \left[\ln(1-\eta )+2 i \arctan\left(\frac{\sqrt{\eta}}{\sqrt{1-\eta}}\right)-\frac{1}{2}+2 \ln(2)\right]
\nonumber \\ & 
+i \text{Li}_2\left(-\frac{1}{\lambda}\right)
-i \text{Li}_2\left(\omega\right)
+\arcsin\left(\sqrt{1-\eta } \sqrt{1-z}\right) \left[\frac{1}{2}-2 \ln \left(1-\omega\right)\right]
\nonumber \\ & 
+\ln(1-z) \arctan\left(\sqrt{\frac{(1-\eta) (1-z)}{z+\eta-\eta z}}\right)
-i \arcsin^2\left(\sqrt{1-\eta } \sqrt{1-z}\right)
\Biggr\}
\nonumber \\ & 
-\frac{\sqrt{\eta } (2 \eta +1)}{8 (1-\eta)}
-\frac{(2 \eta-2 \eta z+2 z-1)}{4 (1-\eta )} \sqrt{1-z} \sqrt{z+\eta-\eta z}  \ln (1-z)
\nonumber \\ & 
+\frac{(2 \eta-2 \eta z+2 z+1)}{8 (1-\eta )} \sqrt{1-z} \sqrt{z+\eta-\eta  z} \, .
\\
G_{32} =& G\left(\left\{\sqrt{1-x} \sqrt{1-\eta+\eta x},\frac{1}{1-x}\right\},z\right) \;=\;
-\frac{4}{9} (1-z)^{3/2} \, _3F_2\left[\begin{array}{c} -\frac{1}{2}, \frac{3}{2}, \frac{3}{2} \\ \;\;\; \frac{5}{2}, \frac{5}{2}\end{array};\eta -z \eta \right]
\nonumber \\ & 
+\frac{4}{9} \, _3F_2\left[\begin{array}{c} -\frac{1}{2}, \frac{3}{2}, \frac{3}{2} \\ \;\;\; \frac{5}{2}, \frac{5}{2}\end{array};\eta \right]
-\frac{1}{4 \eta}\sqrt{1-z} \sqrt{1-\eta+\eta z} (1-2 \eta+2 \eta z) \ln (1-z)
\nonumber \\ & 
+\frac{1}{4 \eta^{3/2}} \ln(1-z) \arcsin\left(\sqrt{\eta } \sqrt{1-z}\right) \, .
\\
G_{33} =& G\left(\left\{\sqrt{x} \sqrt{1-x},\frac{1}{x+\eta-x \eta}\right\},z\right) \;=\;
\frac{1}{8 (1-\eta )} \Biggl\{
\pi  \ln (1-\eta )
+4 i \pi  \arcsin\left(\frac{1}{\sqrt{1-\eta }}\right)
\nonumber \\ & 
-2 \ln \left(\rho\right) \arcsin\left(\frac{1}{\sqrt{1-\eta }}\right)
-i \text{Li}_2\left(\rho\right)
-i \text{Li}_2\left(\frac{1}{\rho}\right)
+2 i \arcsin^2\left(\sqrt{1-z}\right)
\nonumber \\ & 
+2 \left[\arcsin\left(\sqrt{1-z}\right)
-\arcsin\left(\frac{1}{\sqrt{1-\eta }}\right)
\right] 
\ln \left(1+\frac{y}{\rho} \right)
+i \text{Li}_2\left(-\frac{y}{\rho} \right)
\nonumber \\ & 
+2 \left[\arcsin\left(\frac{1}{\sqrt{1-\eta }}\right)
+\arcsin\left(\sqrt{1-z}\right)
\right] 
 \ln \left(1+\rho y \right)
+i \text{Li}_2\left(- \rho y\right)
\nonumber \\ & 
-2 \left(\sqrt{1-z} \sqrt{z} (1-2 z)+\arcsin\left(\sqrt{1-z}\right)\right) \ln (z+\eta-\eta  z)
\nonumber \\ & 
-\ln (\eta ) \left[-2 \sqrt{1-z} \sqrt{z} (1-2 z)-2 \arctan\left(\sqrt{\frac{1-z}{z}}\right)+\pi \right]
\nonumber \\ & 
- 9i \zeta_2
-2 \pi  \ln (2)
\Biggl\}
+\frac{\pi  \left(1-\sqrt{\eta }\right)}{16 \left(1+\sqrt{\eta}\right)^3}
-\frac{\eta ^2+6 \eta +1}{8 (1-\eta )^3} \arcsin\left(\sqrt{1-z}\right)
\nonumber \\ & 
+\left[\frac{\sqrt{\eta } (1+\eta)}{2 (1-\eta )^3}
-\frac{i}{2 (1-\eta )} \arcsin\left(\frac{1}{\sqrt{1-\eta }}\right)\right] \arctan\left(\frac{\sqrt{\eta} \sqrt{1-z}}{\sqrt{z}}\right) 
\nonumber \\ & 
+\frac{\sqrt{1-z} \sqrt{z} (\eta +2 \eta  z-2 z+3)}{8 (1-\eta )^2} \, .
\\
G_{34} =& G\left(\left\{\sqrt{x} \sqrt{1-x},\frac{1}{1-x+\eta x}\right\},z\right) \;=\;
-\frac{\pi \left(1-\sqrt{\eta}\right)}{16 \left(1+\sqrt{\eta}\right)^3}
+\frac{1}{1-\eta} \Biggl\{
-\frac{\pi}{8} \ln(1-\eta)
\nonumber \\ & 
+\ln \left(\rho\right) \left[\frac{i}{4} \ln \left(\frac{2}{1+\sqrt{\eta}}\right)
+\frac{i}{8} \ln(1-\eta)
+\frac{\pi}{8}\right]
+\ln (2) \left[\frac{\pi }{4}-\frac{i}{4} \ln(\rho)\right]
\nonumber \\ & 
+\frac{i}{8} \text{Li}_2\left(-\rho\right)
+\frac{i}{8} \text{Li}_2\left(-\frac{1}{\rho}\right)
+\sqrt{z} \sqrt{1-z} \left(\frac{1}{4}-\frac{z}{2}\right) \ln (1-z+\eta z)
\nonumber \\ & 
-\frac{i}{8} \text{Li}_2\left(\rho y\right)
-\frac{i}{8} \text{Li}_2\left(\frac{y}{\rho}\right)
-\frac{i}{8} \ln \left(\rho\right) \biggl[
\ln \left(1-\rho y\right)
-\ln \left(1-\frac{y}{\rho}\right)
\biggr]
\nonumber \\ & 
-\frac{1}{4} \ln \left(\rho\right) \arctan\left(\sqrt{\frac{1-z}{\eta z}}\right)
+\arcsin\left(\sqrt{1-z}\right) \biggl[
\frac{1}{4} \ln (1-z+\eta z)
\nonumber \\ & 
-\frac{1}{4} \ln \left(1-\rho y\right)
-\frac{1}{4} \ln \left(1-\frac{y}{\rho}\right)
+\frac{1}{2} \ln(y)
+i \arctan\left(\sqrt{\frac{1-z}{z}}\right)\biggr]
\nonumber \\ & 
-\frac{1}{4} i \arcsin^2\left(\sqrt{1-z}\right)
+\frac{3 i \zeta_2}{8}
\Biggr\}
+\frac{\sqrt{z} \sqrt{1-z}}{4 (1-\eta )^2} \left(\frac{3 \eta }{2}+\left(1-\eta\right) z+\frac{1}{2}\right)
\nonumber \\ & 
+\frac{\eta^2+6 \eta+1}{8 (1-\eta)^3} \arcsin\left(\sqrt{1-z}\right)
-\frac{\sqrt{\eta} (1+\eta)}{2 (1-\eta)^3} \arctan\left(\sqrt{\frac{1-z}{\eta z}}\right) \, .
\end{align}
\begin{align}
G_{35} =& G\left(\left\{\sqrt{x} \sqrt{1-x},-\frac{\sqrt{x} \sqrt{1-x}}{x+\eta-\eta 
x}\right\},z\right) 
\;=\;
\frac{\sqrt{\eta }}{(1-\eta )^2} \Biggl\{
\frac{1}{4} \text{Li}_2(\rho )
+\frac{1}{8} \text{Li}_2\left(-\frac{y}{\rho }\right)
-\frac{1}{8} \text{Li}_2(-y \rho )
\nonumber \\ & 
+\frac{1}{16} \ln^2(\rho)
+\frac{i \pi}{8} \ln (\rho )
+\frac{1}{2} \sqrt{1-z} \sqrt{z} (2 z-1) \arctan\left(\frac{\sqrt{z}}{\sqrt{\eta } \sqrt{1-z}}\right)
-\frac{\zeta_2}{4}
\nonumber \\ & 
+\left[\frac{\pi }{4}
-\frac{i}{4} \ln (\rho )\right] \arcsin\left(\sqrt{z}\right)
\Biggr\}
+\frac{z^3}{3 (1-\eta )}
+\frac{5 \eta  z^2-3 z^2}{8 (1-\eta )^2}
-\frac{\eta  z}{2 (1-\eta )^3}
\nonumber \\ & 
+\frac{\eta (1+\eta)}{4 (1-\eta )^4} \left[\ln \left(1+\frac{1-(1-\eta ) (1-2 z)}{\eta }\right)-\ln (2)\right]
+\frac{1+\eta}{(1-\eta )^2} \Biggl\{
-\frac{z}{8}
\nonumber \\ & 
-\frac{\pi}{8} \arcsin\left(\sqrt{z}\right)
+\sqrt{1-z} \sqrt{z} (2 z-1) \left[\frac{1}{4} \arctan\left(\sqrt{\frac{1-z}{z}}\right)-\frac{\pi }{8}\right]
\nonumber \\ & 
-\frac{1}{8} \arctan^2\left(\sqrt{\frac{1-z}{z}}\right)
+\frac{3 \zeta_2}{16}
\Biggr\} \, .
\\
G_{36} =& G\left(\left\{\sqrt{x} \sqrt{1-x},\frac{\sqrt{x} \sqrt{1-x}}{1-x+\eta x}\right\},z\right) 
\;=\;
\frac{\sqrt{\eta }}{(1-\eta )^2} \Biggl\{
-\frac{1}{8} \text{Li}_2\left(-\rho\right)
+\frac{1}{8} \text{Li}_2\left(-\frac{1}{\rho}\right)
\nonumber \\ & 
+\frac{1}{8} \text{Li}_2\left(\rho y\right)
-\frac{1}{8} \text{Li}_2\left(\frac{y}{\rho}\right)
+\frac{1}{2} (1-2 z) \sqrt{1-z} \sqrt{z} \arctan\left(\sqrt{\frac{\eta z}{1-z}}\right)
\nonumber \\ & 
+\frac{i}{4} \ln \left(\rho\right) \arcsin\left(\sqrt{z}\right)
\Biggr\}
+\frac{z (3-z)}{4 (1-\eta )^2}+\frac{8 z^3-9 z^2-3 z}{24 (1-\eta )}-\frac{z}{2 (1-\eta )^3}
\nonumber \\ & 
+\frac{1+\eta}{(1-\eta )^2} \Biggl\{
(1-2 z) \sqrt{1-z} \sqrt{z} \left[\frac{1}{4} \arctan\left(\sqrt{\frac{1-z}{z}}\right)-\frac{\pi}{8}\right]
\nonumber \\ & 
+\frac{1}{8} \arctan^2\left(\sqrt{\frac{1-z}{z}}\right)
+\frac{\pi}{8} \arctan\left(\sqrt{\frac{z}{1-z}}\right)
-\frac{3\zeta_2}{16}
\Biggr\}
\nonumber \\ & 
-\frac{\eta (1+\eta)}{4 (1-\eta)^4} \ln(1-z+\eta z) \, .
\end{align}
\begin{align}
G_{37} =& G\left(\left\{\frac{1}{1-x+\eta x},\sqrt{1-x} \sqrt{1-x+\eta x}\right\},z\right) \;=\;
-\frac{\eta^2}{(1-\eta )^{5/2}} \Biggl\{
\frac{1}{16} \ln\left(2-\eta+2 \sqrt{1-\eta}\right)
\nonumber \\ & 
+\frac{1}{4} \arcsin^2\left(\frac{1}{\sqrt{\eta }}\right)
+\frac{i}{2} \left[-\ln \left(1-\sqrt{1-\eta }\right)+\ln (\eta )-\ln (2)\right] \arcsin\left(\frac{1}{\sqrt{\eta }}\right)
\nonumber \\ &
+\frac{1}{4} \text{Li}_2\left(\frac{\eta +2 \sqrt{1-\eta }-2}{\eta }\right)
+\frac{i}{2} \arcsin\left(\frac{\sqrt{1-z+\eta z}}{\sqrt{\eta }}\right) \ln \left(1+\chi\right)
\nonumber \\ & 
-\frac{1}{8} \ln \left(\sqrt{1-\eta } \sqrt{1-z}+\sqrt{1-z+\eta z}\right)
-\frac{1}{4} \arcsin^2\left(\frac{\sqrt{1-z+\eta z}}{\sqrt{\eta }}\right)
\nonumber \\ & 
+\frac{1}{4} \ln\big(1-z+\eta z\big) \biggl[
-\ln\left(1-\eta +\sqrt{1-\eta}\right)
+\frac{\ln(\eta)}{2}
+\frac{1}{2} \ln(1-\eta)
-\frac{i \pi }{2}
\biggr]
\nonumber \\ & 
-\frac{1}{4} \text{Li}_2\left(-\chi\right)
\Biggr\}
+\frac{3 \eta-2}{8 (1-\eta )^2}
-\frac{(2-\eta )}{4 (1-\eta)^2} \ln\big(1-z+\eta z\big)
\nonumber \\ & 
+\sqrt{1-z} \sqrt{1-z+\eta z} \frac{(2-3 \eta +2 \eta z-2 z)}{8 (1-\eta)^2} \, . 
\\
G_{38} =& G\left(\left\{\frac{1}{x+\eta-x \eta},\sqrt{1-x} \sqrt{x+\eta-x \eta}\right\},z\right) 
\;=\;
\frac{1}{4 (1-\eta )^{5/2}} \Biggl\{
-i \arcsin^2\left(\sqrt{1-\eta }\right)
\nonumber \\ & 
-\frac{1}{2} \arctan\left(\sqrt{\frac{1-\eta}{\eta}}\right)
+2 i \arcsin\left(\sqrt{1-\eta }\right) \arctan\left(\sqrt{\frac{1-\eta}{\eta}}\right)
\nonumber \\ & 
+i \arcsin^2\left(\sqrt{1-\eta} \sqrt{1-z}\right)
+\frac{1}{2} \arctan\left(\sqrt{\frac{(1-\eta) (1-z)}{z+\eta-\eta z}}\right)
\nonumber \\ & 
-i \text{Li}_2\left(\lambda\right)
+i \text{Li}_2\left(-\frac{1}{\omega}\right)
-2 \arcsin\left(\sqrt{1-\eta } \sqrt{1-z}\right) \ln \left(1+\frac{1}{\omega}\right)
\nonumber \\ & 
+2 \ln(2) \arcsin\left(\sqrt{1-\eta }\right)
+\arcsin\left(\sqrt{1-\eta }\right) \ln(z+\eta-\eta z)
\Biggr\}
\nonumber \\ &
+\frac{\sqrt{\eta } (3-2 \eta )}{8 (1-\eta )^2}
+\sqrt{1-z} \sqrt{z+\eta-\eta  z} \left(\frac{2 \eta -3}{8 (1-\eta )^2}+\frac{z}{4 (1-\eta )}\right)
\nonumber \\ & 
+\frac{(1-2 \eta) \sqrt{\eta}}{4 (1-\eta )^2} \big(\ln(z+\eta-\eta z)-\ln (\eta)\big) \, .
\end{align}
\begin{align}
G_{39} =& G\left(\left\{\sqrt{1-x} \sqrt{1-\eta+\eta x},\frac{1}{\eta  (x-1)+1}\right\},z\right) 
\;=\;
\frac{1}{4 \eta ^{5/2}} \Biggl\{
-i \arcsin^2\left(\sqrt{1-\eta+\eta z}\right)
\nonumber \\ & 
-\ln (1-\eta) \arctan\left(\sqrt{\frac{1-\eta}{\eta}}\right)
-\frac{1}{2} \ln (1-\eta) \arctan\left(\frac{2 \eta-1}{2 \sqrt{1-\eta} \sqrt{\eta}}\right)
\nonumber \\ & 
+2 \arcsin\left(\sqrt{1-\eta }\right) \left[\frac{1}{2} \ln (1-\eta)
+i \arctan\left(\sqrt{\frac{\eta}{1-\eta}}\right)-\frac{1}{4}+\ln(2)\right]
\nonumber \\ & 
+i \text{Li}_2\left(-\frac{1}{\lambda}\right)
-i \text{Li}_2\left(\xi\right)
+2 \arcsin\left(\sqrt{1-\eta+\eta z}\right) \left[\frac{1}{4}-\ln \left(1-\xi\right)\right]
\nonumber \\ & 
-\frac{1}{2} \ln (1-\eta ) \arctan\left(\frac{1-2 \eta  (1-z)}{2 \sqrt{\eta (1-z) (1-\eta+\eta z)}}\right)
+i \arcsin^2\left(\sqrt{1-\eta}\right)
\nonumber \\ & 
+\ln (1-\eta+\eta z) \arctan\left(\sqrt{\frac{1-\eta+\eta z}{\eta (1-z)}}\right)
\Biggr\}
-\frac{\sqrt{1-\eta } (2 \eta +1)}{8 \eta^2}
\nonumber \\ & 
+\frac{1}{4 \eta ^2} (1-2 \eta +2 \eta z) \sqrt{1-z} \sqrt{1-\eta+\eta z} \ln \left(1+\frac{\eta z}{1-\eta }\right)
\nonumber \\ & 
+\frac{1}{8 \eta^2} (1+2 \eta -2 \eta z) \sqrt{1-z} \sqrt{1-\eta+\eta z}  \, .
\end{align}
\begin{align}
G_{40} =& G\left(\left\{\frac{1}{x+\eta -\eta x},-\frac{\sqrt{x} \sqrt{1-x}}{x+\eta-\eta 
x}\right\},z\right) 
\;=\;
\frac{\sqrt{\eta }}{(1-\eta )^3} \Biggl\{
-\frac{i}{2} \ln ^2\left(1-\sqrt{\eta }\right)
+\pi  \ln \left(1+\sqrt{\eta }\right)
\nonumber \\ & 
-i \text{Li}_2\left(\frac{1}{1-\sqrt{\eta }}\right)
-i \text{Li}_2\left(1-\sqrt{\eta }\right)
+i \text{Li}_2\left(\frac{1}{2}-\frac{u}{2 \sqrt{\eta }}\right)
-i \text{Li}_2\left(\frac{1}{2}+\frac{u}{2 \sqrt{\eta }}\right)
\nonumber \\ & 
+i \text{Li}_2\left(\frac{1-u}{1-\sqrt{\eta }}\right)
-i \text{Li}_2\left(\frac{1+u}{1-\sqrt{\eta }}\right)
-i \text{Li}_2\left(\frac{1-u}{1+\sqrt{\eta }}\right)
+i \text{Li}_2\left(\frac{1+u}{1+\sqrt{\eta }}\right)
\nonumber \\ & 
+\ln (4 \eta ) \arctan\left(\frac{\sqrt{\eta } \sqrt{1-z}}{\sqrt{z}}\right)
+\left[\ln \left(\eta+\frac{z}{1-z}\right)+2\right] \arctan\left(\frac{\sqrt{z}}{\sqrt{\eta } \sqrt{1-z}}\right)
\nonumber \\ & 
+2 \ln (\rho ) \arctan\left(\sqrt{\frac{1-z}{z}}\right)
+\pi  \ln (1-z)
-\frac{\pi}{2} \ln (\eta )
+ 2 i \zeta_2
-\pi  \ln (2)
\Biggr\}
\nonumber \\ & 
+\frac{1+\eta}{(1-\eta )^3} \Biggl\{
\frac{\pi}{2} \ln (1-\eta )
+\frac{i}{2} \text{Li}_2\left(\frac{1}{\rho }\right)
+\frac{i}{2} \text{Li}_2(\rho )
-\frac{i}{2} \text{Li}_2\left(-\frac{1}{y \rho }\right)
-\frac{i}{2} \text{Li}_2\left(-\frac{\rho }{y}\right)
\nonumber \\ & 
+\arcsin\left(\frac{1}{\sqrt{1-\eta}}\right) \left[
\ln (\rho )
+\ln \left(1+\frac{1}{\rho  y}\right)
-\ln \left(1+\frac{\rho }{y}\right)
\right]
-\frac{\pi}{2} \ln (z+\eta -\eta  z)
\nonumber \\ & 
+\arccos\left(\sqrt{z}\right) \left[
\ln \left(1+\frac{1}{\rho y}\right)
+\ln \left(1+\frac{\rho}{y}\right)
\right]
-i \arccos^2\left(\sqrt{z}\right)
-\frac{\pi }{2}
-\pi \ln (2)
\nonumber \\ & 
-2 i \arcsin\left(\frac{1}{\sqrt{1-\eta}}\right) \arctan\left(\frac{\sqrt{\eta } \sqrt{1-z}}{\sqrt{z}}\right)
+\arctan\left(\sqrt{\frac{1-z}{z}}\right)
-\frac{3 i \zeta_2}{2}
\Biggr\}
\nonumber \\ & 
-\frac{\sqrt{1-z} \sqrt{z}}{(1-\eta )^2} \, .
\end{align}
\begin{align}
G_{41} =& G\left(\left\{\frac{1}{1-x+\eta x},\frac{\sqrt{x} \sqrt{1-x}}{1-x+\eta 
x}\right\},z\right) \;=\;
\frac{i \sqrt{\eta }}{(1-\eta )^3} \Biggl\{
-i \ln \left(\frac{1}{\eta }+\frac{z}{1-z}\right) \arctan\left(\sqrt{\frac{\eta z}{1-z}}\right)
\nonumber \\ & 
+\text{Li}_2\left(\frac{1}{2}-\frac{u \sqrt{\eta }}{2}\right)
-\text{Li}_2\left(\frac{1}{2}+\frac{u \sqrt{\eta }}{2}\right)
-\text{Li}_2\left(\frac{\sqrt{\eta} (1-u)}{1+\sqrt{\eta }}\right)
+\text{Li}_2\left(\frac{\sqrt{\eta} (1+u)}{1+\sqrt{\eta }}\right)
\nonumber \\ & 
+\text{Li}_2\left(-\frac{\sqrt{\eta} (1-u)}{1-\sqrt{\eta }}\right)
-\text{Li}_2\left(-\frac{\sqrt{\eta} (1+u)}{1-\sqrt{\eta }}\right)
-2 i \arctan\left(\sqrt{\frac{\eta z}{1-z}}\right)
\nonumber \\ & 
-i \ln \left(\frac{\eta }{4}\right) \left[
\frac{\pi}{2}
-\arctan\left(\sqrt{\frac{1-z}{\eta z}}\right)
\right]
- i\ln (\rho ) \left[
2 \arctan\left(\sqrt{\frac{1-z}{z}}\right)
-\pi 
\right]
\Biggr\}
\nonumber \\ & 
+\frac{1+\eta}{(1-\eta)^3} \Biggl\{
\left[-\frac{\pi }{2}-i \arcsinh\left(\sqrt{\frac{\eta }{1-\eta }}\right)\right] \ln (\rho )
+\pi  \ln \left(1-\sqrt{\eta }\right)
-\pi  \ln (2)
\nonumber \\ & 
+\frac{i}{2} \text{Li}_2\left(-\frac{1}{\rho}\right)
+\frac{i}{2} \text{Li}_2(-\rho)
-\frac{i}{2} \text{Li}_2\left(\frac{1}{y \rho}\right)
-\frac{i}{2} \text{Li}_2\left(\frac{\rho }{y}\right)
+\frac{1}{2} \arcsin(1-2 z)
\nonumber \\ & 
+\ln \left(1-\frac{1}{\rho  y}\right) \left[\arccos\left(\sqrt{z}\right)
-i \arcsinh\left(\sqrt{\frac{\eta }{1-\eta }}\right)\right]
-i \arccos^2\left(\sqrt{z}\right)
-\frac{\pi }{4}
\nonumber \\ & 
+\ln \left(1-\frac{\rho }{y}\right) \left[\arccos\left(\sqrt{z}\right)+i \arcsinh\left(\sqrt{\frac{\eta }{1-\eta }}\right)\right]
-\frac{\pi}{2} \ln (\eta  z-z+1)
\nonumber \\ & 
+\arcsinh\left(\sqrt{\frac{\eta }{1-\eta }}\right) \left[\pi -2 \arctan\left(\sqrt{\frac{1-z}{\eta z}}\right)\right]
+\frac{3 i \zeta_2}{2}
\Biggr\}
+\frac{\sqrt{1-z} \sqrt{z}}{(1-\eta )^2} .
\end{align}
We have also $G$-functions of weight {\sf w = 3}.
\begin{align}
G_{42} =& G\left(\left\{\sqrt{x} \sqrt{1-x},\sqrt{x} \sqrt{1-x},\frac{1}{x}\right\},z\right) \;=\;
\frac{1}{\big(1-2 \sqrt{1-z} \sqrt{z}\big)^4} \biggl(
-\frac{1}{8}-3 z+z^2+4 z^3-2 z^4
\nonumber \\ & 
+\sqrt{z(1-z)} \big(1+4 z-4 z^2\big)
\biggr)
\Biggl\{\frac{35}{128} \zeta_3
        -\frac{3 \zeta_2}{64}
        +\frac{3 \zeta_2}{16} \ln(2) 
        +\frac{z}{8}
        -2 z^2
        +\frac{7}{3} z^3
        -\frac{3}{4} z^4
\nonumber \\ & 
        +\left(-\frac{\pi}{4} \ln(2)+\frac{\pi}{16}-\frac{\sf C}{2}\right)
\left[(1-2 z) \sqrt{z(1-z)}
                +\arctan\left(\frac{1-2 \sqrt{z(1-z)}}{1-2 z}\right)
\right]
\nonumber \\ & 
        +\frac{z}{2} (1-z) (1-2 z)^2 \left[
                \frac{\ln(2)}{2}
                +H_1\left(\frac{1-2 \sqrt{z(1-z)}}{1-2 z}\right)
                +H_{\{4,1\}}\left(\frac{1-2 \sqrt{z(1-z)}}{1-2 z}\right)
        \right]
\nonumber \\ & 
        +\sqrt{z(1-z)} (1-2 z) \left[
                H_{\{4,0\},1}\left(\frac{1-2 \sqrt{z(1-z)}}{1-2 z}\right)
                +H_{\{4,0\},\{4,1\}}\left(\frac{1-2 \sqrt{z(1-z)}}{1-2 z}\right)
        \right]
\nonumber \\ & 
        +H_{\{4,0\},\{4,0\},\{4,1\}}\left(\frac{1-2 \sqrt{z(1-z)}}{1-2 z}\right)
        +H_{\{4,0\},\{4,0\},1}\left(\frac{1-2 \sqrt{z(1-z)}}{1-2 z}\right)
\nonumber \\ & 
        +\left(\frac{\ln(2)}{2}-\frac{1}{4}\right)
\biggl[\sqrt{z(1-z)} (1-2 z) \arctan\left(\frac{1-2 \sqrt{z(1-z)}}{1-2 z}\right)
\nonumber \\ & 
                +\frac{1}{2} \arctan^2\left(\frac{1-2 \sqrt{z(1-z)}}{1-2 z}\right)
        \biggr]
\Biggr\}.
\\
G_{43} =& G\left(\left\{\sqrt{x} \sqrt{1-x},\sqrt{x} \sqrt{1-x},\frac{1}{1-x}\right\},z\right) 
\;=\;
\frac{1}{\big(1-2 \sqrt{z(1-z)}\big)^4} \biggl(
        \frac{1}{8}
        +3 z
        -z^2
        -4 z^3
        +2 z^4
\nonumber \\ & 
        -\sqrt{z(1-z)} \big(1+4 z-4 z^2\big)
\biggr)
\Biggl\{-\frac{21}{128} \zeta_3
        -\frac{\pi ^2}{128}
        +\frac{\pi ^2}{32} \ln(2)
        -\frac{z}{8}
        +\frac{z^2}{2}
        +\frac{2}{3} z^3
        -\frac{3}{4} z^4
\nonumber \\ & 
        +\left(
                -\frac{\pi}{4} \ln(2)+\frac{\pi}{16}+\frac{\sf C}{2}
        \right)
\left[(1-2 z) \sqrt{z(1-z)}
                +\arctan\left(\frac{1-2 \sqrt{z(1-z)}}{1-2 z}\right)
        \right]
\nonumber \\ & 
        +\frac{z}{2} (1-z) (1-2 z)^2 \left[
                \frac{\ln(2)}{2}
                -H_{-1}\left(\frac{1-2 \sqrt{z(1-z)}}{1-2 z}\right)
                +H_{\{4,1\}}\left(\frac{1-2 \sqrt{z(1-z)}}{1-2 z}\right)
        \right]
\nonumber \\ & 
        -\sqrt{z(1-z)} (1-2 z) \left[
                H_{\{4,0\},-1}\left(\frac{1-2 \sqrt{z(1-z)}}{1-2 z}\right)
                -H_{\{4,0\},\{4,1\}}\left(\frac{1-2 \sqrt{z(1-z)}}{1-2 z}\right)
        \right]
\nonumber \\ & 
        -H_{\{4,0\},\{4,0\},-1}\left(\frac{1-2 \sqrt{z(1-z)}}{1-2 z}\right)
        +H_{\{4,0\},\{4,0\},\{4,1\}}\left(\frac{1-2 \sqrt{z(1-z)}}{1-2 z}\right)
\nonumber \\ & 
        +\left(
                \frac{\ln(2)}{2}-\frac{1}{4}
        \right)
\biggl[\sqrt{z(1-z)}(1-2 z) \arctan\left(\frac{1-2 \sqrt{z(1-z)}}{1-2 z}\right) 
\nonumber \\ & 
                +\frac{1}{2} \arctan^2\left(\frac{1-2 \sqrt{z(1-z)}}{1-2 z}\right)
        \biggr]
\Biggr\}.
\end{align}
Here also cyclotomic harmonic polylogarithms \cite{Ablinger:2011te} contribute, which are
characterized by letters of the type
\begin{eqnarray}
\frac{x^l}{\Phi_k(x)}~~~\text{or}~~~\frac{1}{x},~~0 \leq l < \varphi(k),
\end{eqnarray}
where $\Phi_k$ denotes the $k$th cyclotomic polynomial and $\varphi(k)$ Euler's totient function. In the above
iterated integrals the letters 
\begin{eqnarray}
\frac{1}{1+x^2},~~~~\frac{x}{1+x^2}
\end{eqnarray}
contribute according to the index sets $\{4,0\}$ and $\{4,1\}$ and {\sf C} denotes Catalan's constant.
 
Some of the cyclotomic polylogarithms 
appearing above can be expressed in terms of standard polylogarithms as follows
\begin{eqnarray}
H_{\{4,0\}}(z) &=& \arctan(z), \\ 
H_{\{4,0\},\{4,0\}}(z) &=& \frac{1}{2} \arctan^2(z), \\ 
H_{\{4,0\},\{4,1\}}(z) &=& 
\left[\frac{1}{2} \ln\left(1+z^2\right)+\ln(2)\right] \left[\frac{\pi }{4}-\frac{1}{2} \arctan\left(\frac{1}{z}\right)\right]
\nonumber \\ &&
-\frac{i}{4} \text{Li}_2\left(\frac{1}{2}-i \frac{z}{2}\right)
+\frac{i}{4} \text{Li}_2\left(\frac{1}{2}+i \frac{z}{2}\right), \\
H_{\{4,0\},1}(z) &=&
\ln(1-z) \left[\arctan\left(\frac{1}{z}\right)-\frac{\pi}{4}\right]
-\frac{i}{2} \text{Li}_2\left(\frac{1}{2}+\frac{i}{2}\right)
+\frac{i}{2} \text{Li}_2\left(\frac{1}{2}-\frac{i}{2}\right)
\nonumber \\ &&
+\frac{i}{2} \text{Li}_2\left(\left(\frac{1}{2}+\frac{i}{2}\right) (1-z)\right)
-\frac{i}{2} \text{Li}_2\left(\left(\frac{1}{2}-\frac{i}{2}\right) (1-z)\right), \\
H_{\{4,0\},-1}(z) &=&
\ln(1+z) \left[\frac{3 \pi}{4}-\arctan\left(\frac{1}{z}\right)\right]
-\frac{i}{2} \text{Li}_2\left(\frac{1}{2}+\frac{i}{2}\right)
+\frac{i}{2} \text{Li}_2\left(\frac{1}{2}-\frac{i}{2}\right)
\nonumber \\ &&
+\frac{i}{2} \text{Li}_2\left(\left(\frac{1}{2}+\frac{i}{2}\right) (1+z)\right)
-\frac{i}{2} \text{Li}_2\left(\left(\frac{1}{2}-\frac{i}{2}\right) (1+z)\right).
\end{eqnarray}
Expressions in terms of polylogarithms for the other cyclotomic harmonic polylogarithms 
$H_{\{4,0\},\{4,0\},1}(z)$, $H_{\{4,0\},\{4,0\},-1}(z)$ and 
$H_{\{4,0\},\{4,0\},\{4,1\}}(z)$ can also be found, but they are larger and will not be shown here.

\vspace*{5mm}
\noindent
{\bf Acknowledgment.}\\
This work was supported in part by the Austrian Science Fund (FWF) grant SFB F50 (F5009-N15).
The Feynman diagrams have been drawn using {\tt Axodraw} \cite{Vermaseren:1994je}.

\end{document}